\documentclass[a4paper,12pt]{article}

\usepackage{amsmath,amsbsy,amssymb,amsfonts,bm}
\usepackage{graphics,mathrsfs,cite, color, xcolor}
\usepackage{tcolorbox}

\makeatletter
\catcode`\@=11
\@addtoreset{equation}{section}

\makeatother

\allowdisplaybreaks[3]

\renewcommand{\thefootnote}{\arabic{footnote}}
\newcommand{\gs}{g_s}
\newcommand{\ls}{l_s}
\newcommand{\Exp}[1]{\operatorname{e}^{#1}}
\newcommand{\Ad}{\operatorname{Ad}}
\newcommand{\rank}{\operatorname{rank}}
\newcommand{\ii}{\operatorname{i}}

\newcommand{\tr}{\operatorname{tr}}
\newcommand{\abs}[1]{{\lvert #1 \rvert}}
\newcommand{\rmd}{\mathrm{d}}
\newcommand{\nn}{\nonumber}

\newcommand{\DD}{\mathrm{D}}
\newcommand{\FF}{\mathrm{F}}
\newcommand{\MM}{\mathrm{M}}
\newcommand{\KKM}{\mathrm{KKM}}

\newcommand{\NS}{\mathrm{NS}}
\newcommand{\PP}{\mathrm{P}}

\newcommand{\sfM}{\mathsf{M}}
\newcommand{\rmT}{{\tiny\mathrm{T}}}
\newcommand{\ten}{\mathsf{A}}

\newcommand{\lZ}{{\mathbb Z}}
\newcommand{\vV}{\bm{V}}

\newcommand{\cA}{\mathcal A}\newcommand{\cB}{\mathcal B}
\newcommand{\cD}{\mathcal D}
\newcommand{\cE}{\mathcal E}\newcommand{\cF}{\mathcal F}
\newcommand{\cG}{\mathcal G}\newcommand{\cH}{\mathcal H}
\newcommand{\cJ}{\mathcal J}
\newcommand{\cK}{\mathcal K}\newcommand{\cL}{\mathcal L}
\newcommand{\cM}{\mathcal M}
\newcommand{\cP}{\mathcal P}
\newcommand{\cQ}{\mathcal Q}
\newcommand{\cS}{\mathcal S}
\newcommand{\cV}{\mathcal V}

\newcommand{\cZ}{\mathcal Z}

\newcommand{\sfa}{\mathsf{a}}\newcommand{\sfb}{\mathsf{b}}
\newcommand{\sfc}{\mathsf{c}}\newcommand{\sfd}{\mathsf{d}}

\newcommand{\sfg}{\mathsf{g}}
\newcommand{\sfi}{\mathsf{i}}\newcommand{\sfj}{\mathsf{j}}
\newcommand{\sfk}{\mathsf{k}}\newcommand{\sfl}{\mathsf{l}}
\newcommand{\sfm}{\mathsf{m}}\newcommand{\sfn}{\mathsf{n}}
\newcommand{\sfp}{\mathsf{p}}
\newcommand{\sfq}{\mathsf{q}}\newcommand{\sfr}{\mathsf{r}}

\newcommand{\sfE}{\mathsf{E}}
\newcommand{\sfG}{\mathsf{G}}

\newcommand{\GL}{\mathrm{GL}}
\newcommand{\Gl}{\mathrm{gl}}
\newcommand{\Oo}{\mathrm{o}}
\newcommand{\SO}{\mathrm{SO}}

\newcommand{\Spin}{\mathrm{Spin}}
\newcommand{\SL}{\mathrm{SL}}
\newcommand{\Sl}{\mathrm{sl}}
\newcommand{\Sp}{\mathrm{Sp}}
\newcommand{\SU}{\mathrm{SU}}

\allowdisplaybreaks[4]

\newcommand{\half}{{{\textstyle\frac{1}{2}}}}

\newcommand{\bpm}{\begin{pmatrix}}
\newcommand{\epm}{\end{pmatrix}}

\newcommand{\comm}[2]{\left[#1,#2\right]}

\newcommand\bcP{{\bar{\cP}}}

\newcommand\hcL{{\hat{\cal L}}}

\newcommand\hA{\widehat{A}}
\newcommand\hB{\widehat{B}}
\newcommand\hL{\widehat{L}}
\newcommand\hM{\widehat{M}}
\newcommand\hN{\widehat{N}}
\newcommand\hP{\widehat{P}}
\newcommand\hQ{\widehat{Q}}
\newcommand\hR{\widehat{R}}
\newcommand\hS{\widehat{S}}

\newcommand\hK{\widehat{K}}

\newcommand{\hcLo}{\hat{\cal L}^{\scriptscriptstyle 0}}

\def\breta{\bar{\eta}}

\def\brPhi{{{\bar{\Phi}}}}

\def\bra{\bar{a}}
\def\brb{\bar{b}}
\def\brc{\bar{c}}
\def\bre{\bar{e}}

\def\brm{{\bar{m}}}
\def\brn{{\bar{n}}}
\def\brp{{\bar{p}}}
\def\brq{{\bar{q}}}

\def\brP{{\bar{P}}}

\def\brV{{\bar{V}}}

\def\hbrm{\hat{\bar{m}}}
\def\hbrn{\hat{\bar{n}}}
\def\hbrp{\hat{\bar{p}}}
\def\hbrq{\hat{\bar{q}}}

\newcommand{\OO}{\mathrm{O}}
\newcommand{\ODD}{\OO(D,D)}
\newcommand{\ODDG}{\OO(D,D+\dim\,G)}

\newcommand{\ODG}{{\OO(1,D{-1}+ \text{dim}\, G)}}
\newcommand{\OG}{{\OO(\text{dim}\, G)}}
\newcommand{\OoD}{{\OO(D{-1},1)}}

\begin{document}

\begin{titlepage}
\renewcommand{\thefootnote}{\fnsymbol{footnote}}

\vspace*{-0.5cm}

\begin{center}
{\Large \bf Effective Action for Non-Geometric Fluxes\\
from \\
\vskip 0.3cm
Duality Covariant Actions}%
\end{center}
\vspace{1cm}

\centerline{
\textsc{Kanghoon Lee $^{a}$}%
\footnote{E-mail address: kanghoon.lee1@gmail.com} 
\ \ \ 
\textsc{Soo-Jong Rey $^{a,b}$}%
\footnote{E-mail address: rey.soojong@gmail.com}
\ \ \ 
\textsc{Yuho Sakatani $^{a,c}$}%
\footnote{E-mail address: yuho.sake@gmail.com}%
}

\vspace{0.2cm}

\begin{center}
${}^a${\it Fields, Gravity \& Strings, CTPU}\\
{\it Institute for Basic Sciences, Daejeon  34047 \rm KOREA}

${}^b${\it School of Physics \& Astronomy and Center for Theoretical Physics,}\\
{\it Seoul National University, Seoul 08826 \rm KOREA}\\

${}^c${\it Department of Physics, Kyoto Prefectural University of Medicine,}\\
{\it Kyoto 606-0823, \rm JAPAN}
\end{center}

\vspace*{1cm}

\begin{abstract}
The (heterotic) double field theories and the exceptional field theories are manifestly duality covariant formulations, describing low-energy limit of various superstring and M-theory compactifications. These field theories are known to be reduced to the standard descriptions by introducing appropriately parameterized generalized metric and by applying suitably chosen section conditions. In this paper, we apply these formulations to non-geometric backgrounds. We introduce different parameterizations for the generalized metric in terms of the dual fields which are pertinent to non-geometric fluxes. Under certain simplifying assumptions, we construct new effective action for non-geometric backgrounds. We then study the non-geometric backgrounds sourced by exotic branes and find their $U$-duality monodromy matrices. The charge of exotic branes obtained from these monodromy matrices agrees with the charge obtained from the non-geometric flux integral. 
\end{abstract}
\thispagestyle{empty}
\setcounter{footnote}{0}
\end{titlepage}

\newpage
\tableofcontents

\vskip1cm
\rightline{\sl I do not wish, at this stage, to examine the logical justification of this form of}
\rightline{\sl argumentation; for the present, I am considering it as a \textsc{practice},}
\rightline{\sl which we can observe in the habits of men and animals.}
\rightline{\textsc{bertrand russell}, 'Philosophy'. }

\newpage

\section{Introduction}
\label{sec:introduction}

Recently, a significant progress has been achieved for novel formulations of supergravity in which duality symmetries in string and M-theory compactification are manifest. 
They include the double field theory (DFT) \cite{Siegel:1993xq,Siegel:1993th,Siegel:1993bj,Hull:2009mi,Hull:2009zb,Hohm:2010jy,Hohm:2010pp}, the exceptional field theory (EFT) \cite{Berman:2010is,Berman:2011pe,Berman:2011cg,Berman:2011jh,Berman:2012vc,Hohm:2013pua,Hohm:2013vpa,Hohm:2013uia,Aldazabal:2013via,Godazgar:2014nqa,Hohm:2014fxa,Rosabal:2014rga,Musaev:2014lna,Hohm:2015xna,Abzalov:2015ega,Musaev:2015ces,Berman:2015rcc,Ciceri:2016dmd,Baguet:2016jph} (see also \cite{West:2001as,West:2003fc,West:2004st,Hull:2007zu,Pacheco:2008ps,Hillmann:2009ci,Park:2013gaj,Aldazabal:2013mya} for closely related attempts) as well as the generalized geometry \cite{Hitchin:2004ut,Gualtieri:2003dx,Grana:2008yw,Coimbra:2011nw,Coimbra:2011ky,Coimbra:2012af}.  
One important advantage of these formulations is that they can treat wide variety of spacetimes, such as non-geometric backgrounds \cite{Dabholkar:2002sy,Hull:2004in,Shelton:2005cf,Dabholkar:2005ve}, that are not globally describable in the conventional formulation of supergravity. 
As pointed out in \cite{deBoer:2010ud,deBoer:2012ma}, non-geometric backgrounds arise quite naturally in superstring theories. 
Backgrounds sourced by exotic branes \cite{Elitzur:1997zn,Blau:1997du,Hull:1997kb,Obers:1997kk,Obers:1998fb,Eyras:1999at,LozanoTellechea:2000mc} are concrete examples. 
As an application of DFT and related formulations such as the $\beta$-supergravity \cite{Andriot:2011uh,Andriot:2012wx,Andriot:2012an,Blumenhagen:2012nk,Blumenhagen:2012nt,Blumenhagen:2013aia,Andriot:2013xca,Andriot:2014qla}, a background of a particular exotic brane, so-called a $5^2_2$-brane, was studied in \cite{deBoer:2010ud,Bergshoeff:2011se,Kikuchi:2012za,deBoer:2012ma,Hassler:2013wsa,Geissbuhler:2013uka,Kimura:2013fda,Kimura:2013zva,Chatzistavrakidis:2013jqa,Kimura:2013khz,Andriot:2014uda,Kimura:2014upa,Okada:2014wma} and the exotic $5^2_2$-brane was identified with a magnetic source of the non-geometric $Q$-flux \cite{Hassler:2013wsa,Andriot:2014uda,Okada:2014wma}. 

One reason why the exotic $5^2_2$-brane received special attention is that the non-geometric $Q$-flux, which is intrinsic to the $5^2_2$-brane background, is related to a $T$-duality monodromy, and the much developed DFTs efficiently describe such background. 
It is known that backgrounds of other exotic branes possess other non-geometric fluxes that are related to the $Q$-flux via $U$-duality transformations \cite{Hull:1994ys,Obers:1998fb}. 
In order to describe such non-geometric backgrounds, variants of the $\beta$-supergravity, which can describe the background of an exotic $p$-brane (called a $p^{7-p}_3$-brane) or a $1^6_4$-brane, was proposed in \cite{Sakatani:2014hba}. 
There, each of these exotic branes was identified as the magnetic sources of a non-geometric $P$-flux \cite{Aldazabal:2006up,Aldazabal:2008zza,Aldazabal:2010ef} or a non-geometric $Q$-flux associated with a $6$-vector, $\beta^{m_1\cdots m_6}$ \cite{Sakatani:2014hba}. 
However, the reformulation of \cite{Sakatani:2014hba} is applicable only to a limited situation; coexistence of different non-geometric fluxes are not allowed and existence of isometries are assumed. 
In fact, EFT, a manifestly $E_{d(d)}$ $U$-duality covariant formulation of the supergravity, is a more suitable formulation, and indeed, backgrounds of the exotic $5^3$-brane, $5^2_2$-brane, and the $5^2_3$-brane were studied in $\SL(5)$ EFT \cite{Malek:2012pw,Blair:2014zba}. 
One of the main purposes of this paper is to systematically identify the non-geometric fluxes in $E_{d(d)}$ EFT for the cases of $4\leq d\leq 7$. 

The goal of this paper is to develop effective actions for a certain class of non-geometric flux backgrounds in Type II string and M-theories. 
Our starting point is the duality covariant action in an extended field theory, such as the manifestly $U$-duality covariant EFT. 
Since the $U$-duality orbit is of infinite order, there are in practice infinitely many possible parameterization of the $U$-duality group. 
The key idea is to identify the most effective parameterization for a given set of non-geometric flux background. 
Note that our non-geometric parameterization is efficient for backgrounds with only non-geometric fluxes. 
For backgrounds with both geometric and non-geometric fluxes, such as the truly non-geometric backgrounds of \cite{Dibitetto:2012rk}, a more general treatment will be required.%
\footnote{Note that the section condition or the strong constraint in DFT/EFT can be relaxed through the generalized Scherk-Schwarz reduction \cite{Grana:2012rr}, which provides all the fluxes in the maximal and half-maximal gauged supergravity \cite{Dibitetto:2012rk}. In this paper we will restrict ourselves to the usual section condition, and the non-geometric fluxes considered in this paper are included in the same duality orbit with geometric fluxes. However, extension of the non-geometric fluxes to the gauged DFT/EFT would be straightforward via generalized Scherk-Schwarz reduction.}

Our construction can be extended to non-geometric flux backgrounds in heterotic string theories.  
Heterotic string exhibits $\OO(D,D+16)$ or $\ODDG$ duality group, where $G$ is the heterotic Yang-Mills group, $E_8\times E_8$ or $SO(32)$, and the corresponding heterotic DFT \cite{Siegel:1993th,Siegel:1993xq,Hohm:2011ex} provides a duality manifest description of the effective field theory. 
Again, the key idea is to identify the most effective parameterization. 
Through the non-geometric parameterization of heterotic generalized vielbein, we construct heterotic $Q$-flux which includes Chern-Simons like term and an additional non-geometric bi-vector flux associated with the heterotic Yang-Mills field strength. The corresponding non-geometric effective action can be constructed from $\ODDG$ gauged DFT \cite{Grana:2012rr,Berman:2013cli,Cho:2015lha}. If we take the maximal Abelian reduction of heterotic Yang-Mills gauge symmetry, $G = U(1)^{16}$, the non-geometric gauged DFT reduces to the non-geometric parameterization by Blumenhagen and Sun \cite{Blumenhagen:2014iua}. 

This paper is organized as follows. 
In section \ref{sec:general}, after reviewing some elements of Lie algebra, we explain the general construction of the generalized metric or vielbein. 
In section \ref{sec:EFT-M}, we consider the EFT in terms of the M-theory. 
We show two different parameterizations of the generalized vielbein; the conventional geometric parameterization and the dual non-geometric parameterization. 
Using the two different parameterizations, we write down two different eleven-dimensional effective actions. 
We also consider the dimensional reduction to the type IIA theory, and obtain the non-geometric fluxes in the type IIA theory. 
EFT in terms of the type IIB theory is discussed in section \ref{sec:EFT-IIB} and ten-dimensional action for the non-geometric fluxes in the type IIB theory is obtained. 
In section \ref{sec:HDFT}, we find a parameterization of heterotic DFT relevant for non-geometric fluxes. 
In section \ref{sec:applications}, the relation between the non-geometric fluxes and exotic branes are discussed. 
Discussions and future directions are given in section \ref{sec:discussion}. We relegated much of technical details to the Appendices. 
In appendix \ref{app:notations}, we fix our notations. 
In appendix \ref{app:action}, we show the explicit calculations of the EFT action. 
In appendix \ref{app:gauged-DFT}, we review double-vielbein formalism for $\ODDG$ gauged DFT.   
In appendix \ref{app:exotic}, exotic branes in type II/M-theory are reviewed briefly.

\section{General framework}
\label{sec:general}

In extended field theories, such as DFT or EFT, it is known that, in a Borel gauge, fundamental fields are packaged into the generalized metric or the generalized vielbein. 
In this section, we review the formal definition of generalized metric used in \cite{Berman:2011jh,Tumanov:2014pfa} (but in a slightly different manner), as a coset representative of $G/K$ where $G$ is the duality group and $K$ is the maximal compact subgroup of $G$. 
We then show how to parameterize the generalized metric for the well-known examples of DFT and the Einstein gravity. 

\subsection{Parameterization of Lie algebra}
We first recall three decomposition methods for a real semi-simple Lie algebra $\mathfrak{g}$.\footnote{Here, we suppose $\mathfrak{g}$ is a split real Lie algebra, considering the applications to DFT, $\mathfrak{g}=\Oo(d,d)$, and EFT, $\mathfrak{g}=e_{d(d)}$. Application to a non-split case is considered in section \ref{sec:HDFT}.} 

The first is known as triangular decomposition. 
Associated with $\mathfrak{g}$ is the Cartan matrix $A_{ij}$ ($i,j=1,\dotsc,\rank\mathfrak{g}$) that has the structure
\begin{align}
 A_{ii}=2\,,\quad A_{ij}\in \lZ_{\leq 0}\quad (i\neq j)\,, \quad A_{ij}=0 \quad \Leftrightarrow \quad A_{ji}=0\,,\quad \det A_{ij} >0 \,,
\end{align}
where $\lZ_{\leq 0}$ denotes non-positive integers. 
In $\mathfrak{g}$, consider the Chevalley basis generators $\{H_i,\,E_i,\,F_i\}$, which obey the properties
\begin{align}
\begin{split}
 &[H_i,\,H_i]=0\,,\qquad 
 [H_i,\,E_j]=A_{ji}\,E_j\,,\qquad 
 [H_i,\,F_j]=-A_{ji}\,F_j\,,\qquad 
 [E_i,\,F_j]=\delta_{ij}\,H_i\,, 
\\
 &\hskip1.5cm [\underbrace{E_i,\,[\cdots,\,[E_i}_{1-A_{ji}},\,E_j]\cdots]]=0\,,\qquad \quad 
 [\underbrace{F_i,\,[\cdots,\,[F_i}_{1-A_{ji}},\,F_j]\cdots]]=0\,. 
\end{split}
\end{align}
It is known that the generators $\{H_i,\,E_i,\,F_i\}$, together with the commutators of $E_i$ or $F_i$\,,
\begin{align}
 [E_{i_1},\,[\cdots,\,[E_{i_{k-1}} ,\,E_{i_k}]\cdots]]\,\qquad \text{and} \qquad
 [F_{i_1},\,[\cdots,\,[F_{i_{k-1}} ,\,F_{i_k}]\cdots]]\,, 
\end{align}
form a complete set of basis of $\mathfrak{g}$\,. In the Chevalley basis, the generators $H_i$ for $i = 1, \dotsc, \rank \mathfrak{g}$ form the Cartan subalgebra $\mathfrak{h}$, the generator $E_i$ is associated with the positive simple root $\alpha_i\in \mathfrak{h}^*$ with $\alpha_i(H_j)=A_{ij}$, and the generator $F_i$ is associated with the negative simple root $-\alpha_i$\,. 
We denote the space of positive root by $\Delta_+$ and the space of negative root by $\Delta_-$, respectively. 
For an arbitrary positive root $\alpha\in \Delta_+$, 
\begin{align}
\alpha=\sum_{n=1}^k\alpha_{i_n}\, , 
\end{align}
we can construct the associated generator as $k$-tuple left-commutator
\begin{align}
 E_\alpha \equiv [E_{i_1},\,[\cdots,\,[E_{i_{k-1}} ,\,E_{i_k}]\cdots]]\,.
\end{align}
For the corresponding negative root $- \alpha \in \Delta_-$, we also construct the associated generator as $k$-tuple right-commutator 
\begin{align}
 F_\alpha \equiv [ [\cdots [ F_{i_k},F_{i_{k-1}}],\,\cdots],\,F_{i_1}]\,.
\end{align}
Denote the space spanned by $E_\alpha$ and $F_\alpha$ ($\alpha\in \Delta_+$) as $\mathfrak{n}_+$ and $\mathfrak{n}_-$, respectively. Then, we obtain the triangular decomposition by decomposing the Lie algebra $\mathfrak{g}$ as
\begin{align}
 \mathfrak{g} = \mathfrak{n}_- \oplus \mathfrak{h} \oplus \mathfrak{n}_+ \,.
\end{align}

The second method is known as the Cartan decomposition. 
Define the Cartan involution $\theta$ by
\begin{align}
 \theta(H_i) = - H_i \,,\qquad
 \theta(E_i) = - F_i \,,\qquad
 \theta(F_i) = - E_i \,. 
\label{eq:Chevalley-involution}
\end{align}
From the distributive property that $\theta([s,\,t])=[\theta(s),\,\theta(t)]$ for $s,t\in \mathfrak{g}$, it follows that 
\begin{align}
 \theta(E_\alpha) = - F_\alpha \,,\qquad \theta(F_\alpha) = - E_\alpha \qquad \text{for every} \qquad
 \alpha\in \Delta_+\,. 
\end{align}
Redefining the generators as
\begin{align}
 S_\alpha \equiv E_\alpha + F_\alpha \, \qquad \text{and} \qquad 
 J_\alpha \equiv E_\alpha - F_\alpha \qquad \text{for every} \qquad
 \alpha\in \Delta_+\,, 
\end{align}
we can diagonalize the Cartan involution as
\begin{align}
 \theta(H_i) = - H_i \,,\qquad
 \theta(S_\alpha) = - S_\alpha \,,\qquad
 \theta(J_\alpha) = + J_\alpha \,, 
\end{align}
and classify the generators according to the parity under the involution $\theta$:
\begin{align}
 \mathfrak{k} = \{ s \in \mathfrak{g} \mid \theta(s)= +s \}=\text{span}(J_\alpha) \,\qquad \text{and} \qquad 
 \mathfrak{p} = \{ s \in \mathfrak{g} \mid \theta(s)=-s \}=\text{span}(H_i,\,S_\alpha)\,.
\nonumber
\end{align}
We are thus decomposing the Lie algebra $\mathfrak{g}$ as 
\begin{align}
 \mathfrak{g} = \mathfrak{k} \oplus \mathfrak{p} \,,
\end{align}
obtaining the Cartan decomposition. Since the number of the positive roots is $(\dim\mathfrak{g} - \rank\mathfrak{g})/2$, we have
\begin{align}
 \dim \mathfrak{k} = \frac{\dim\mathfrak{g} - \rank\mathfrak{g}}{2} \,,\qquad 
 \dim \mathfrak{p} = \frac{\dim\mathfrak{g} + \rank\mathfrak{g}}{2} \,. 
\end{align}
Although the commutator in $\mathfrak{p}$ is not closed (since it has the odd parity under $\theta$), the Lie commutators in $\mathfrak{k}$ yields a subalgebra, sometimes called the Cartan-involution-invariant subalgebra, which coincides with the maximal compact subalgebra of $\mathfrak{g}$\,. 

The third method is known as the Iwasawa decomposition, the decomposition we shall be using in the present paper. There are two possible types of Iwasawa decomposition. The positive decomposition is defined by
\begin{align}
 \mathfrak{g} = \mathfrak{k}\oplus \mathfrak{h} \oplus \mathfrak{n}_+ \,,
\label{eq:iwasawa+}
\end{align}
where $\mathfrak{b}_+ \equiv \mathfrak{h} \oplus \mathfrak{n}_+$ is referred to as the positive Borel subalgebra. 
The negative decomposition is defined by
\begin{align}
 \mathfrak{g} = \mathfrak{n}_- \oplus \mathfrak{h} \oplus \mathfrak{k} \, ,
\label{eq:iwasawa-}
\end{align}
where $\mathfrak{b}_- \equiv \mathfrak{n}_-\oplus \mathfrak{h}$\, is referred to as the negative Borel subalgebra. 

Associated to the Lie algebra $\mathfrak{g}$, we construct the corresponding Lie group $\mathfrak{G}$ as the exponential map. We can realize group element $g \in \mathfrak{G}$ in any of the above decomposition of $\mathfrak{g}$. 
In particular, we can straightforwardly extend the definition of the Cartan involution $\theta$ to an arbitrary group element $g \in \mathfrak{G}$, and then define an anti-involution $\sharp$ by
\begin{align}
 g^\sharp \equiv \theta(g^{-1})\,,\quad (ab)^{\sharp}=b^\sharp a^\sharp \qquad 
 \mbox{where} \qquad g,\,a,\,b\in \mathfrak{G} \,.
 \label{dualinvolution}
\end{align}

In section \ref{sec:EFT-M} and \ref{sec:EFT-IIB}, we take the Lie algebra $\mathfrak{g} = e_{d(d)}$ and its Lie group $\mathfrak{G} = E_{d(d)}$ as the duality symmetry (summarized in Table \ref{table:constants0}). 
Suppose that the (generalized) momenta $Z_M$ ($M=1,\dotsc,D\equiv \dim l_1$), which generate abelian translations ($[Z_M,\,Z_N]=0$) in the extended space $X$ of the $U$-duality action, are in the fundamental representation $l_1$ of the Lie group $\mathfrak{G}$ \cite{West:2003fc},
\begin{align}
 [h,\,Z_M] = - \sum_N (\rho_h)_M{}^N\, Z_N \qquad (h\in \mathfrak{g}) \,. 
\label{eq:l1rep-matrix}
\end{align}
\begin{table}[t]
\centering
 \begin{tabular}{|c||c|c|c|c|}\hline
 $d$ & 4 & 5 & 6 & 7 \cr\hline\hline
 $\mathfrak{G} = E_{d(d)}$ & $\SL(5)$ & $\SO(5,5)$ & $E_{6(6)}$ & $E_{7(7)}$ \cr\hline
 $D = \dim l_1$& 10 & 16 & 27 & 56 \cr\hline
\end{tabular}
\caption{Duality group $\mathfrak{G}$ for various toroidal compactification of M-theory to $\mathbb{R}^{11-d}$.}
\label{table:constants0}
\end{table}
Here, $\rho_h$ is the matrix realization for the element $h\in \mathfrak{g}$ in the $l_1$-representation. 
Defining $\overline{Z}^M\equiv -\theta\bigl(Z_M\bigr)$, we obtain from \eqref{eq:l1rep-matrix} the following commutator:
\begin{align}
 \bigl[\theta(h),\,\overline{Z}^M\bigr] = -\sum_N (\rho_h)_M{}^N\, \overline{Z}^N \qquad (h\in \mathfrak{g}) \,.
\end{align}
To render the position of indices consistent, we introduce the fundamental forms, $\delta_{MN}$ and $\delta^{MN}$, whose components are equal to $\delta_M^N$ (and are not generalized tensors), and define the dual matrix realization $(\bar{\rho}_h)^M{}_N\equiv \delta^{MK}\,(\rho_h)_K{}^L\,\delta_{LN}$. 
We then obtain
\begin{align}
 \bigl[h,\,\overline{Z}^M\bigr] = -(\bar{\rho}_{\theta(h)})^M{}_N\, \overline{Z}^N \qquad (h\in \mathfrak{g}) \,.
\end{align}
We also introduce a natural $\mathfrak{G}$-invariant scalar product $(v,\,\bar{w})\equiv v^M\,\bar{w}_M$ for an element $v$ of the representation $l_1$ spanned by $A_M$ and $\bar{w}$ of the dual representation $\bar{l}_1$ spanned by $\overline{Z}^M$ (see Appendix A in \cite{Berman:2011jh}). 
This scalar product is defined so as to satisfy invariance under adjoint action
\begin{align}
 ([h,\,v],\,\bar{w}) + (v,\,[h,\,\bar{w}]) =0 \qquad (h\in \mathfrak{g}) \,,
\end{align}
equivalently,
\begin{align}
 (\Ad_g\cdot\,v,\,\Ad_g\cdot\,\bar{w}) = (v,\,\bar{w})\,, 
\end{align}
where $\Ad_g\cdot\,v\equiv g\,v\,g^{-1}$ $(g\in \mathfrak{G})$\,.
We will normalize the abelian generators $Z_M$ such that the scalar product becomes the identity matrix,
\begin{align}
 \bigl(Z_M,\,\overline{Z}^N\bigr) = \delta_M^N \,,
\end{align}
and then from
\begin{align}
 (\Ad_g\cdot\,v,\,\Ad_g\cdot\,\bar{w}) 
 = (\Exp{-[h,\,\cdot\,]}\,v,\,\Exp{-[h,\,\cdot\,]}\,\bar{w})
 = v^M\,\bar{w}_N\,(\Exp{\rho_h})_M{}^K\,(\Exp{\bar{\rho}_{\theta(h)}})^N{}_L\,(Z_K,\,\overline{Z}^L)\,,
\end{align}
for $g=\Exp{-h}\in \mathfrak{G}$, $v=v^M\,Z_M$, and $\bar{w}=\bar{w}_M\,\overline{Z}^M$, we have
\begin{align}
 (\Exp{\rho_h})_M{}^K\,(\Exp{\bar{\rho}_{\theta(h)}})^N{}_K = \delta_M^N\,.
\end{align}
Hence, we obtain
\begin{align}
 (\Exp{\rho_h})_M{}^N = (\Exp{-\bar{\rho}_{\theta(h)}})^N{}_M = \bigl(\Exp{\bar{\rho}_{h^\sharp}}\bigr)^N{}_M = \delta^{NL}\,\bigl(\Exp{\rho_{h^\sharp}}\bigr)_L{}^K\,\delta_{KM}\,,
\end{align}
where we defined $h^\sharp\equiv -\theta(h)$ for $h\in \mathfrak{g}$\,, and used the dual representation $(\bar{\rho}_h)^N{}_M=\delta^{NL}\,(\rho_h)_L{}^K\,\delta_{KM}$ in the last equality. 
This relation shows that the anti-involution $\sharp$ defined in (\ref{dualinvolution}), sometimes called the generalized transpose, acts as the matrix transpose in the matrix realization of Lie algebra $\mathfrak{g}$.

\subsection{The generalized metric}
We next study the geometry of extended space $X$ associated with the duality transformation group $\mathfrak{G}$. 
We shall define the generalized metric $\cM_{MN}$ of $X$ and explain how to parameterize $\cM_{MN}$ in terms of appropriate physical fields (see \cite{Berman:2011jh,Tumanov:2014pfa}). 
We first define a bilinear form
\begin{align}
 \langle v,\, w \rangle \equiv -\bigl(v,\,\theta(w)\bigr) = -v^M\,w^N\,\bigl(Z_M,\,\theta(Z_N)\bigr)\,,
\label{eq:metric}
\end{align}
for generalized vectors, $v=v^M\,Z_M$ and $w=w^M\,Z_M$, in the $l_1$-representation. 
From the identities, $\bigl(Z_M,\,\theta(Z_N)\bigr)=-\bigl(Z_M,\,\overline{Z}^N\bigr)=-\delta_M^N=-\delta_{MN}$, we see that the metric \eqref{eq:metric} is symmetric and positive-definite,
\begin{align}
 \langle v,\, w \rangle=\langle w,\, v\rangle=v^M\,w^N\,\delta_{MN} \,.
\end{align}
However, as $\delta_{MN}$ is not a generalized tensor, this metric is not $\mathfrak{G}$-invariant. 
Indeed, for general element $h\in\mathfrak{g}$, we find that the adjoint action
\begin{align}
 \bigl\langle [h,\,v],\, w \bigr\rangle + \bigl\langle v,\, [h,\,w] \bigr\rangle
=-\bigl([h,\,v],\, \theta(w)\bigr) - \bigl(v,\, [\theta(h),\,\theta(w)]\bigr) 
=-\bigl([h-\theta(h),\,v],\, \theta(w)\bigr) 
\nonumber
\end{align}
is nonzero. 
However, it is invariant under the maximal compact subgroup, $\mathfrak{K}$, of $\mathfrak{G}$, since $h=\theta(h)$ for $h\in \mathfrak{k}$. 

Starting from this (constant) positive-definite metric and a group element $g \in \mathfrak{G}$, we now define the generalized metric from the generalized bilinear form
\begin{align}
 \cM(v,\,w) \equiv \cM_{MN}\,v^M\,w^N 
 \equiv \langle \Ad_{g^{-1}}\cdot\, v ,\, \Ad_{g^{-1}}\cdot\, w \rangle \,. 
\label{eq:gen-met-def}
\end{align}
The generalized bilinear form is positive-definite by construction, and it is defined to be $\mathfrak{G}$-invariant. 
We assume that the generalized metric $\cM_{MN}$ varies over the spacetime, so the group element $g \in \mathfrak{G}$ should be spacetime-dependent as well. 
Denoting $g =\Exp{-h}$ ($h\in\mathfrak{g}$) and $\bar{w}_M\equiv \delta_{MN}\,w^N$, we have
\begin{align}
 \theta\bigl(\Ad_{g^{-1}}\cdot\, w\bigr) 
 = \Exp{[\theta(h),\,\cdot\,]} \overline{Z}^M\,\bar{w}_M
 = (\Exp{\bar{\rho}_h})^M{}_N\,\overline{Z}^N\,\bar{w}_M \,,
\end{align}
and the inner product in \eqref{eq:gen-met-def} becomes
\begin{align}
 \cM(v,\,w) &= v^M\,(\Exp{\rho_h})_M{}^K\,(\Exp{\bar{\rho}_{\theta(h)}})^N{}_L\,\bar{w}_N\,(Z_K,\,\overline{Z}^L) 
\nn\\
 &= v^M\,(\Exp{\rho_h})_M{}^K\,(\Exp{\bar{\rho}_h})^N{}_K\,\bar{w}_N
  = v^M\,(\Exp{\rho_h})_M{}^K\,(\Exp{\rho_{h^\sharp}})_K{}^L\,\delta_{LN}\,w^N \,. 
\end{align}
Introducing the generalized vielbein as $\cE_M{}^A \equiv (\Exp{\rho_h})_M{}^A$, we can express the generalized metric in the conventional form,
\begin{align}
 \cM_{MN} = \cE_M{}^A \,\cE_N{}^B \,\delta_{AB} \,. 
\label{eq:gen-met}
\end{align}
Here, the indices $A,B$ run over $1,\dotsc,D = \dim l_1$, which play the same role as the original indices $M,N$ but are interpreted as ``flat indices.'' 
As $\langle\,\cdot\,,\,\cdot\,\rangle$ is $\mathfrak{K}$-invariant, two generalized metrics constructed from $g \in \mathfrak{G}$ and $g \cdot k$ ($k \in \mathfrak{K}$), respectively, have the same structure. 
Thus, the generalized metric can be parameterized by a coset representative of $\mathfrak{G}/ \mathfrak{K}$, and so the number of the independent parameters is given by $\dim(\mathfrak{G}/ \mathfrak{K})=\dim \mathfrak{G}-\dim \mathfrak{K}$. 

For an explicit construction of the generalized metric, we find it convenient to use the Iwasawa decomposition \eqref{eq:iwasawa+} and parameterize the representative $g\in \mathfrak{G}/\mathfrak{H}$, where $\mathfrak{H}$ is the Cartan subgroup, in terms of functions, $h^i(x)$ and $A^\alpha(x)$, associated with generators of the positive Borel subalgebra $\mathfrak{b}_+$, and the $\mathfrak{K}$ equivalence class:
\begin{align}
 g(x) = \Exp{\sum_i h^i(x)\,H_i}\,\Exp{\sum_{\alpha\in\Delta_+} A^\alpha(x)\,E_\alpha}\,k(x)
 \quad \sim \quad \Exp{\sum_i h^i(x)\,H_i}\,\Exp{\sum_{\alpha\in\Delta_+} A^\alpha(x)\,E_\alpha} \,, 
\end{align}
Here, $k(x)$ denotes an element in the compact subgroup $\mathfrak{K}$ and $x$ refers to the coordinate system adopted. 
We can then obtain the generalized metric from the following generalized vielbein:
\begin{align}
 \cE_M{}^A(x) = \bigl(\Exp{h^i(x)\,\rho_{H_i}}\,\Exp{\sum_{\alpha\in\Delta_+} A^\alpha(x)\,\rho_{E_\alpha}}\bigr)_M{}^A \,.
\label{eq:vielbein1}
\end{align}

Note that the generalized metric $\cM_{MN}$ is invariant under the anti-involution $g\to g^\sharp$ (i.e.~symmetric), while the generalized vielbein is not. 
Using the above decomposition, we have
\begin{align}
 g^\sharp(x) &= k^\sharp(x)\,\Exp{\sum_{\alpha\in\Delta_+} A^\alpha(x)\,F_\alpha}\,\Exp{h^i(x)\,H_i} 
\nn\\
 &= \Exp{\widetilde{h}^i(x)\,H_i}\,\Exp{\sum_{\alpha\in\Delta_+} \widetilde{A}^\alpha(x)\,F_\alpha}\,\widetilde{k}(x) 
 \quad \sim \quad \Exp{\widetilde{h}^i(x)\,H_i}\,\Exp{\sum_{\alpha\in\Delta_+} \widetilde{A}^\alpha(x)\,F_\alpha}
\end{align}
with certain functions $\widetilde{h}^i(x)$, $\widetilde{A}^\alpha(x)$ and $\widetilde{k}(x)\in \mathfrak{K}$, whose relation to $h^i(x)$, $A^\alpha(x)$, and $k(x)$ is in general complicated. 
This expression for $g^\sharp$ corresponds to the alternative Iwasawa decomposition \eqref{eq:iwasawa-}, so we can obtain the generalized vielbein in terms of the functions associated with the generators of negative Borel subalgebra $\mathfrak{b}_-$\,:
\begin{align}
 \cE_M{}^A(x) = \bigl(\Exp{\sum_i \widetilde{h}^i(x)\,\rho_{H_i}}\,\Exp{\sum_{\alpha\in\Delta_+} \widetilde{A}^\alpha(x)\,\rho_{F_\alpha}}\bigr)_M{}^A \,.
\label{eq:vielbein2}
\end{align}

The key idea of this paper is that the above replacement $g\to g^\sharp$, which does not change the generalized metric, generally corresponds to the replacement from the conventional geometric parameterization of the generalized metric $\cM_{MN}$ to the dual ``non-geometric'' parameterization of it. 
A transformation between the conventional and the dual parameterization is sometimes referred to as the exotic duality transformation \cite{Bergshoeff:2011se,Sakatani:2014hba,Kimura:2016xzd}. In this paper, we will show that the exotic duality transformation is identifiable with the generalized transpose. 

It remains to confirm the tensorial property of the generalized metric. 
As we mentioned above, the flat bilinear form $\langle v,\, w \rangle= v^M\,w^N\,\delta_{MN}$ was not $\mathfrak{G}$-invariant. 
However, the generalized bilinear form $\cM(v,\,w) = \langle \Ad_{g^{-1}}\cdot\, v ,\, \Ad_{g^{-1}}\cdot\, w \rangle$ is invariant under $\mathfrak{G}$. 
This constrains the transformation rule for the group element $g$ (i.e.~the generalized vielbein). It then follows that, as the the $\mathfrak{K}$-invariance of $\delta_{MN}$, the transformation rule of the generalized vielbein generally has the following form:
\begin{align}
 \cE_M{}^A \to g_M{}^N\, \cE_N{}^B\,k_B{}^A\qquad \text{for} \qquad g\in \mathfrak{G},\, \ \ k\in \mathfrak{K} \,. 
\end{align}

\subsection{Example: Double Field Theory}
\label{subsubsection:DFT}

Before presenting our new results, we first illustrate the above general consideration for the DFT. 
In this case, the T-duality group is $\mathfrak{G}=\OO(d,d)$. 
We can decompose the generators of $\mathfrak{g} = \Oo(d,d)$ into representations of the $\GL(d)$, the $\Gl(d)$-generators $K^a{}_b$, $R^{ab}=R^{[ab]}$, $R_{ab}=R_{[ab]}$ ($a,b=1,\dotsc,d$), which obey the following commutation relations:
\begin{align}
\begin{aligned}{}
 [K^a{}_b,\,K^c{}_d]&= \delta_b^c\,K^a{}_d - \delta_d^a\,K^c{}_b\,,& \quad 
 [R^{ab},\,R_{cd}]&= 4\,\delta^{[a}_{[c}\,K^{b]}{}_{d]}\,,
\\
 [K^a{}_b,\,R^{cd}] &= \delta^c_b\, R^{ad} + \delta^d_b\, R^{ca}\,,&\quad 
 [K^a{}_b,\,R_{cd}] &= -\delta_c^a\, R_{bd} - \delta_d^a\, R_{cb}\,.
\end{aligned}
\end{align}
The Cartan subalgebra $\mathfrak{h}$ is generated by the diagonal components of $K^a{}_b$: $H_a \equiv K^a{}_a$ (no summation). 
The Cartan involution is given by
\begin{align}
 \theta(K^a{}_b)=-K^b{}_a\,\qquad \text{and} \qquad \theta(R^{ab}) = -R_{ab}\,,
\end{align}
and the Cartan-involution-invariant subgroup is generated by
\begin{align}
 J_{ab} \equiv K^a{}_b - K^b{}_a \, \qquad \text{and} \qquad T_{ab}\equiv R^{ab} - R_{ab}\,. 
\end{align}
In particular, the (anti)chiral combinations, $M^\pm_{ab}\equiv (J_{ab}\pm T_{ab})/2$\,, satisfy the algebra for $o(d)\times o(d)$:
\begin{align}
 [M^\pm_{ab},\,M^\pm_{cd}] = 2\,\delta^d_{[a}\, M^\pm_{a]c} - 2\,\delta^c_{[a}\, M^\pm_{b]d} \,,\qquad 
 [M^+_{ab},\,M^-_{cd}] = 0\,.
\end{align}
The positive and negative Borel subalgebras, $\mathfrak{b}_+$ and $\mathfrak{b}_-$, are spanned by $\{H_a,\,K^a{}_b\ (a<b),\, R^{ab}\}$ and $\{H_a,\,K^a{}_b\ (a>b),\,R_{ab}\}$, respectively.

In DFT, we take the fundamental (i.e.~vector) representation, whose matrix realization is given by the matrices,
\begin{align}
 (\rho_{K^c{}_d})_A{}^B = \begin{pmatrix} \delta_a^c\,\delta^b_d & 0 \cr 0 & -\delta^a_d\,\delta_b^c
 \end{pmatrix}\,, \quad
 (\rho_{R^{cd}})_A{}^B = \begin{pmatrix} 0 & 2\,\delta^{cd}_{ab} \cr 0 & 0 
 \end{pmatrix}\,,\quad 
 (\rho_{R_{cd}})_A{}^B = \begin{pmatrix} 0 & 0 \cr -2\,\delta^{ab}_{cd} & 0 
 \end{pmatrix}\,,
\end{align}
where $\delta^{cd}_{ab}\equiv \delta_{[a}^{[c}\,\delta_{b]}^{d]}$ (see Appendix \ref{app:notations} for our conventions). 
The commutators with the generalized momenta $Z_M=(P_m,\,\widetilde{P}^m)$ are given by
\begin{align}
\begin{aligned}{}
 [K^a{}_b,\,P_c] &= - \delta^a_c\, P_b\,,&\qquad 
 [K^a{}_b,\,\widetilde{P}^c] &= \delta^c_b\, \widetilde{P}^a\,,
\\
 [R^{ab},\,P_c] &= -2\,\delta_c^{[a}\,\widetilde{P}^{b]}&\,,\qquad 
 [R^{ab},\,\widetilde{P}^c] &= 0 \,,
\\
 [R_{ab},\,P_c] &= 0\,,&\qquad
 [R_{ab},\,\widetilde{P}^c] &= 2\,\delta^c_{[a}\,P_{b]}\,. 
\end{aligned}
\end{align}
Note that the variable $Z^M$ defined by
\begin{align}
 Z^M\equiv \eta^{MN}\,Z_N\,\qquad \text{and} \qquad \bigl(\eta^{MN}\bigr)\equiv \begin{pmatrix}
 0 & \delta^m_n \\ \delta_m^n & 0 \end{pmatrix} \,,
\end{align}
is in the same representation as $\overline{Z}^M$. We thus see that the $\OO(d,d)$-invariance of $(Z_M,\,\overline{Z}^N)=\delta_M^N$ is equivalent to the $\OO(d,d)$-invariance of another metric, $(\!(Z_M,\,Z_N)\!)\equiv \eta_{MN}$, which is commonly used in DFT. 
We also have the $K=\OO(d)\times \OO(d)$-invariant metric $\delta_{AB}$. 

We define the generalized vielbein in the gauge of positive Borel subalgebra by
\begin{align}
 \cE_M{}^A(x) = \bigl(\Exp{\sum_a h^a(x)\,\rho_{K^a{}_a}}\,\Exp{\sum_{a<b}h_a{}^b(x)\,\rho_{K^a{}_b}}\,\Exp{\frac{1}{2}\,\sum_{a, b} B_{ab}(x)\,\rho_{R^{ab}}}\bigr)_M{}^A\,. 
\end{align}
Here, $B_{ab}(x)$ is an anti-symmetric tensor field, which is identified with the Kalb-Ramond field. 
If we define $(E_M{}^A)(x) \equiv \Exp{\sum_a h^a(x)\,\rho_{K^a{}_a}}\,\Exp{\sum_{a<b}h_a{}^b(x)\,\rho_{K^a{}_b}}$ and $\cB^{(2)}\equiv {1 \over 2} \sum_{a,b} \,B_{ab}(x)\,\rho_{R^{ab}}$\,, we have
\begin{align}
 \cE_M{}^A(x) = \bigl(E (x) \Exp{\cB^{(2)}(x)} \bigr)_M{}^A\,
 \end{align}
 where
 \begin{align}
 (E_M{}^A)(x)=\begin{pmatrix} e_m{}^a(x) & 0 \cr 0 & (e^{-\rmT})^m{}_a(x) \end{pmatrix}\, \qquad \text{and} \qquad 
 \cB^{(2)}(x) =\begin{pmatrix} 0 & B_{ab}(x) \cr 0 & 0 
 \end{pmatrix} \,. 
\end{align}
Here, $e_m{}^a(x)$ is an upper-triangular matrix, to be identified with the (gauge-fixed) vielbein in $d$-dimensions, and $(e^{-\rmT})$ is the inverse of the transpose of the vielbein. 
This generalized vielbein yields the conventional generalized metric in DFT:
\vskip0.5cm
\begin{tcolorbox}
\vskip-0.5cm
\begin{align}
 \bigl(\cM_{MN}\bigr) &= \begin{pmatrix} 
 G_{mn}-B_{mk}\,G^{kl}\,B_{ln} & B_{mk}\,G^{kn} \cr -G^{mk}\,B_{kn} & G^{mn}
 \end{pmatrix} 
\nn\\
 &= \begin{pmatrix} 
 \delta_m^k & B_{mk} \cr 0 & \delta^m_k
 \end{pmatrix}\,\begin{pmatrix} 
 G_{kl} & 0 \cr 0 & G^{kl}
 \end{pmatrix}\,\begin{pmatrix} 
 \delta^l_n & 0 \cr -B_{ln} & \delta^l_n 
 \end{pmatrix}\,,
\label{eq:DFT-conventional}
\end{align}
\end{tcolorbox}
\noindent
where $G_{mn}\equiv e_m{}^a\,e_n{}^b\,\delta_{ab}$\,. 

Upon the anti-involution, $g\to g^\sharp$, the generalized vielbein takes the lower-triangular form, parameterized by 
\begin{align}
 \cE_M{}^A(x) = \bigl(E(x) \Exp{{\bm\beta}^{(2)}(x)}\bigr)_M{}^A \,,
 \end{align}
 where
 \begin{align} 
 (E_M{}^A)(x) =\begin{pmatrix} \widetilde{e}_m{}^a(x) & 0 \cr 0 & (\widetilde{e}^{-\rmT})^m{}_a(x) \end{pmatrix}\, \qquad \text{and} \qquad 
 {\bm\beta}^{(2)}(x) =\begin{pmatrix} 0 & 0 \cr -\beta^{ab}(x) & 0 \end{pmatrix}\,,
\end{align}
where $\widetilde{e}_m{}^a(x)$ is a lower-triangular matrix. 
In this case, the generalized metric becomes
\vskip0.5cm
\begin{tcolorbox}
\vskip-0.5cm
\begin{align}
 \bigl(\cM_{MN}\bigr) &= \begin{pmatrix} 
 \widetilde{G}_{mn} & \widetilde{G}_{mk}\,\beta^{kn} \cr -\beta^{mk}\,\widetilde{G}_{kn} & \widetilde{G}^{mn}-\beta^{mk}\,\widetilde{G}_{kl}\,\beta^{ln}
 \end{pmatrix} 
\nn\\
 &= \begin{pmatrix} 
 \delta_m^k & 0 \cr -\beta^{mk} & \delta^m_k
 \end{pmatrix}\,\begin{pmatrix} 
 \widetilde{G}_{kl} & 0 \cr 0 & \widetilde{G}^{kl}
 \end{pmatrix}\,\begin{pmatrix} 
 \delta^l_n & \beta^{ln} \cr 0 & \delta_l^n 
 \end{pmatrix} \,,
\label{eq:DFT-non-geometric}
\end{align}
\end{tcolorbox}
\noindent
where $\widetilde{G}_{mn}\equiv \widetilde{e}_m{}^a\,\widetilde{e}_n{}^b\,\delta_{ab}$\,. 
These dual variables were first introduced and extensively studied in \cite{Shapere:1988zv,Giveon:1988tt,Duff:1989tf,Tseytlin:1990nb,Giveon:1994fu}.

\subsection{Example: Einstein Gravity}
\label{subsubsection:EG}

It is illuminating to compare the above results for DFT with the case of pure Einstein gravity. In Einstein gravity, the generators $R_{ab}$ and $R^{ab}$ are absent, the Cartan-involution-invariant subgroup is simply generated by the local Lorentz $\OO(d)$ rotations, and there is no important difference between the gauges of positive and negative Borel subalgebras.
 
Indeed, as it is well-known, when we consider decomposing the spacetime into space and time, there are two natural parameterizations into upper or lower triangular decomposition:
\vskip0.5cm 
\begin{tcolorbox}
\vskip-0.5cm
\begin{align}
\begin{split}
 \underline{\text{Arnowitt-Deser-Misner \cite{Arnowitt:1962hi}:}}&\quad (g_{mn}) = \begin{pmatrix} 1 & N^k \cr 0 & \delta_i^k \end{pmatrix} 
 \begin{pmatrix} -N^2 & 0 \cr 0 & h_{kl} \end{pmatrix}
 \begin{pmatrix} 1 & 0 \cr N^l & \delta^l_j \end{pmatrix}\,,
\\
 \underline{\text{Landau-Lifschitz \cite{Landau:1982dva}:}} \quad &\quad (g_{mn}) = \begin{pmatrix} 1 & 0 \cr -g_i & \delta_i^k \end{pmatrix} 
 \begin{pmatrix} g_{00} & 0 \cr 0 & \gamma_{kl} \end{pmatrix}
 \begin{pmatrix} 1 & -g_j \cr 0 & \delta^l_j \end{pmatrix}\,.
\end{split}
\end{align}
\end{tcolorbox}
\noindent
These two parameterizations are related simply by a usual local Lorentz transformation. 
In comparison, the situation is different in the DFT case. 
In order to relate two parameterizations \eqref{eq:DFT-conventional} and \eqref{eq:DFT-non-geometric}, we need to use a non-trivial $\OO(d)\times \OO(d)$ subgroup of the $T$-duality group. 
In general, the parameterization \eqref{eq:DFT-conventional} is suited for the conventional geometric backgrounds, while \eqref{eq:DFT-non-geometric} is suited for non-geometric backgrounds, such as $T$-folds. As such, we will refer to the latter, negative Borel subalgebra parameterization as the non-geometric parameterization.

\subsection{Effective action for non-geometric fluxes}

By definition, the actions of the extended field theories are independent of the explicit parameterization of the generalized metric. However, once we parameterize the generalized metric in terms of appropriate physical fields, we can straightforwardly construct the effective actions appropriate for describing dynamics of these field excitations. 

As is well known in DFT \cite{Hohm:2010pp} or EFT \cite{Berman:2011jh}, parameterizing the generalized metric in terms of the conventional supergravity fields, we can derive the conventional supergravity action from DFT or EFT action. 
For example, if we choose the conventional, geometric parameterization and impose the section constraint $\widetilde{\partial}^m=0$, we find that the DFT action is reduced to
\begin{align}
 \cL = \Exp{-2\phi}\,\Bigl(R(G) + 4\,\abs{\rmd\phi}^2 -\frac{1}{2}\,\abs{H^{(3)}}^2 \Bigr) \,,
\end{align}
where $\phi$ is the conventional string dilaton field defined by the T-duality invariant dilaton of DFT, $\Exp{-2d}\equiv \abs{G}^{1/2} \Exp{-2\phi}$, and the three-form $H^{(3)}\equiv \rmd B^{(2)}$, called the $H$-flux, is the field strength for the Kalb-Ramond two-form potential $B^{(2)}$. 

On the other hand, if we choose the dual, non-geometric parameterization \eqref{eq:DFT-non-geometric}, we reduce the DFT action to the so-called $\beta$-supergravity \cite{Andriot:2011uh,Andriot:2012wx,Andriot:2012an,Andriot:2013xca,Andriot:2014qla}. 
Although the full expression is complicated, with the simplifying assumption that indices of $\beta^{mn}$ contracted with $\partial_m$ always vanishes and the constraint $\widetilde{\partial}^m=0$, the DFT action is reduced to the form
\begin{align}
 \widetilde{\cL} = \Exp{-2\widetilde{\phi}}\,\Bigl(R(\widetilde{G}) + 4\,\abs{\rmd\widetilde{\phi}}^2 -\frac{1}{2}\,\abs{Q^{(1,2)}}^2 \Bigr) \,. 
\end{align}
Here, the tilde signifies the non-geometric parameterization, and $\widetilde{\phi}$ is the dual dilaton field defined by $\Exp{-2d}\equiv \abs{\widetilde{G}}^{1/2} \Exp{-2\widetilde{\phi}}$\,. 
Further, we defined
\begin{align}
 \abs{Q^{(1,2)}}^2\equiv \frac{1}{2}\,\widetilde{G}^{m_1n_1}\,\widetilde{G}_{m_2n_2}\,\widetilde{G}_{m_3n_3}\,Q_{m_1}{}^{m_2m_3}\,Q_{n_1}{}^{n_2n_3}\,,\quad 
 Q_k{}^{mn} \equiv \partial_k \beta^{mn}\,.
\end{align}
The mixed-symmetry tensor,\footnote{This behaves as a tensor only under the simplifying assumption \cite{Andriot:2012wx}.} $Q_k{}^{mn}$, is called the non-geometric $Q$-flux. 
In this paper, we further generalize the $\beta$-supergravity starting from the (heterotic) DFT or EFT.

\section{Non-geometric Fluxes in EFT: M-Theory}
\label{sec:EFT-M}

In this section, we consider the eleven-dimensional supergravity of M-theory compactified on a $d$-torus, $\mathbb{T}^d$, equivalently, the ten-dimensional type IIA supergravity compactified on a $(d-1)$-torus, $\mathbb{T}^{d-1}$. 
This theory possesses the $U$-duality transformation symmetry, and the EFT provides the manifestly $U$-duality covariant formulation.
To construct the EFT, we consider an exceptional spacetime with the following generalized coordinates:
\begin{align}
 (X^I)= (x^\mu,\, Y^M) \quad (\mu,\nu=0,\dotsc,n-1,\,M=1,\dotsc, D) \,,
\end{align}
where $n\equiv (11-d)$ is the dimension of the uncompactified, external spacetime and $D$ is the dimension of a fundamental representation of the exceptional group $E_{d(d)}$ whose value for each $n$ is shown in Table \ref{table:constants}.
\begin{table}[t]
\centering
 \begin{tabular}{|c||c|c|c|c|c|}\hline
 $n$ & 7 & 6 & 5 & 4 \cr\hline\hline
 $E_{d(d)}$ & $\SL(5)$ & $\SO(5,5)$ & $E_{6(6)}$ & $E_{7(7)}$ \cr\hline
 $K_d$ & $\SO(5)$ & $\SO(5)\times \SO(5)$ & $\Sp(4)$ & $\SU(8)$
\cr\hline
 $D$& 10 & 16 & 27 & 56 \cr\hline
 $\alpha_n$ &  3 &  4 &  6 & 12 \cr\hline
\end{tabular}
\caption{The $U$-duality groups, their maximal compact subgroups and the integers, $D$ and $\alpha_n$, for various noncompact dimensions, $4\leq n\leq 7$.}
\label{table:constants}
\end{table}
In this paper, we consider the cases of noncompact dimensions $n=4,5,6,7$, equivalently, cases of compact dimensions $d=7,6,5,4$. 

The EFT actions for $n=4,5,6,7$ are presented in \cite{Hohm:2013vpa,Hohm:2013uia,Abzalov:2015ega,Musaev:2015ces} (see also \cite{Berman:2015rcc} for $n=9$, \cite{Hohm:2015xna} for $n=8$, and \cite{Hohm:2014fxa} for $n=3$). 
For simplicity, we focus on the following parts of the action, which are the relevant parts for our purposes:
\begin{align}
\begin{split}
 S_{\text{EFT}} &= \int \rmd^nx\,\rmd^DY\,\cL_{\text{EFT}} \,
\qquad \text{where} \qquad
 \cL_{\text{EFT}} = \cL_{\text{EH}} + \cL_{\text{scalar}} + \cL_{\text{pot}} \,,
\\
 \cL_{\text{EH}}&= e R \,, 
 \\
 \cL_{\text{scalar}} &= \frac{e}{4\alpha_n}\,g^{\mu\nu}\,\partial_\mu \cM_{MN}\,\partial_\nu \cM^{MN} \,,
\\
 \cL_{\text{pot}}&= \frac{e}{4\alpha_n}\,\cM^{MN}\,\partial_M\cM^{KL}\,\partial_N \cM_{KL} 
   -\frac{e}{2}\,\cM^{MN}\,\partial_N \cM^{KL}\,\partial_L \cM_{MK} 
   + e\,\partial_M \ln e \,\partial_N\cM^{MN}
\\
 &\quad 
   + e\,\cM^{MN}\, \partial_M \ln e\, \partial_N \ln e 
   +\frac{e}{4}\,\cM^{MN}\,\partial_M g^{\mu\nu}\,\partial_N g_{\mu\nu}\,. 
\end{split}
\end{align}
Here, $e$ abbreviates $\abs{\det g_{\mu\nu}}^{1/2}$, $R$ is the Ricci scalar of the external metric $g_{\mu\nu}$, and $\alpha_n$ is the integer shown in Table \ref{table:constants}. 
Note that the potential part in the EFT action is fully taken into account by $\cL_{\text{pot}}$. 

In the EFT, to render the gauge algebra closed, we will impose the section condition of the form, $Y^{MN}{}_{PQ}\,\partial_M(\cdots)\,\partial_N(\cdots)=0$, where $Y^{MN}{}_{PQ}$ for each EFT is given in \cite{Berman:2012vc,Hohm:2013pua}.
\footnote{The section condition of DFT can be relaxed in the flux formulation \cite{Aldazabal:2011nj,Geissbuhler:2011mx,Grana:2012rr} or in the approach of \cite{Lee:2015qza}, and the section condition of EFT may be also relaxed in these approaches.} 
As is well-known, there are two natural routes to solve for the section conditions: the M-theory section or the type IIB section \cite{Blair:2013gqa,Hohm:2013vpa}, where all background fields and gauge parameters depend only on $d$ coordinates $x^i$ or $d-1$ coordinates $x^\sfm$, respectively. 
In this section, we study the M-theory section and parameterize the generalized metric in terms of the conventional/dual fields in eleven dimensions. We relegate the parameterization in the type IIB section to section \ref{sec:EFT-IIB}. 

In the M-theory section, we decompose the internal $D$-dimensional coordinates $Y^M$ into some representations of $\SL(d)$. 
Explicitly, for each $n$, we introduce the following coordinates \cite{West:2003fc,Berman:2011jh}:
\begin{align}
\begin{split}
 n=7:&\quad (Y^M)= (x^i,\,y_{ij})\quad (i,j=7,8,9,\sfM) \,,
\\
 n=6:&\quad (Y^M)= (x^i,\,y_{ij},\,y_{i_1\cdots i_5})\quad (i,j=6,\dotsc,9,\sfM) \,,
\\
 n=5:&\quad (Y^M)= (x^i,\,y_{ij},\,y_{i_1\cdots i_5})\quad (i,j=5,\dotsc,9,\sfM) \,,
\\
 n=4:&\quad (Y^M)= (x^i,\,y_{ij},\,y_{i_1\cdots i_5},\,z_i)\quad (i,j=4,\dotsc,9,\sfM)\,,
\end{split}
\end{align}
where the conventional M-theory circle direction, denoted by $x^\sfM$, is one of the internal coordinates $x^i$. 
The section condition is satisfied when all fields are functions only of $x^i$, the physical coordinates on the $d$-torus. 
So, $\partial/\partial y_{ij}=\partial/\partial y_{i_1\cdots i_5}=\partial/\partial z_i=0$. 

\subsection{Parameterization of the generalized vielbein}
\label{sec:vielbein-M}

We now examine parameterization of the generalized metric (or vielbein) in the M-theory section of the EFT. 
The generalized metric in the $\SL(5)$ EFT was first obtained in \cite{Berman:2010is} (which in turn is based on the earlier work \cite{Duff:1990hn}) as
\begin{align}
 (M_{MN}) = \begin{pmatrix}
 G_{ij} +\frac{1}{2}\,A_{ikl}\,A^{kl}{}_j & -\frac{1}{\sqrt{2}}\, A_i{}^{j_1j_2} \\
 -\frac{1}{\sqrt{2}}\,A^{i_1i_2}{}_j & G^{i_1i_2,\,j_1j_2}\end{pmatrix}\,. 
\end{align}
Subsequently, the same generalized metric (up to an overall factor) was presented in \cite{Berman:2011jh} in the context of $E_{11}$ program \cite{West:2001as,West:2003fc}, and its extensions to $E_{d(d)}$ EFT with $5\leq d\leq 7$ were also presented (see also \cite{Hatsuda:2012vm,Hatsuda:2013dya} for $d=4,5$). 
The parameterization given in \cite{Berman:2011jh} was obtained by choosing the positive (or upper-triangular) Borel gauge. If we instead choose the negative (or lower triangular) Borel gauge, we can parameterize the generalized metric using the so-called dual $\Omega$-fields (the explicit form of $\Omega$-fields for $\SL(5)$ EFT is given in \cite{Malek:2012pw,Blair:2014zba}, which we repeat below). As the $\Omega$-fields are related to the non-geometric fluxes, we refer to the latter as non-geometric parameterization. 

In the rest of this subsection, we present two parameterizations of the generalized vielbein, i.e., the conventional parameterization and the non-geometric parameterization, for $4\leq d\leq 7$ (or $4\leq n\leq 7$). 
Using these parameterizations, we define the non-geometric fluxes in M-theory and construct the eleven-dimensional effective actions that are useful for describing these non-geometric fluxes. 

\subsubsection{$n=7$: $\mathfrak{G} = \SL(5)$}

For the $\mathfrak{g} = \Sl(5)$ Lie algebra, we decompose the 24 generators to\footnote{We relegate their commutators in Appendix \ref{app:Edd-algebra}.} \cite{Berman:2011jh}
\begin{align}
 K^a{}_b\,,\qquad R^{a_1a_2a_3}\,,\qquad R_{a_1a_2a_3}\qquad (a,b =7,8,9,\sfM)\,,
\end{align}
where $K^a{}_b$ are the $\Gl(4)$ generators and $R^{a_1a_2a_3}$ and $R_{a_1a_2a_3}$ are the generators that transform as totally antisymmetric under $\Gl(4)$. So, we are decomposing 24 generators into $16+4+4$ generators. 
Using this decomposition, a group element $g$ of $\mathfrak{G} = \SL(5)$ can be parameterized as
\begin{align}
 g = \Exp{h_a{}^b K^a{}_b} \Exp{\frac{1}{3!} a_{a_1a_2a_3} R^{a_1a_2a_3}} \Exp{\frac{1}{3!}\omega^{b_1b_2b_3} R_{b_1b_2b_3}} \quad \in \quad \mathfrak{G} = \SL(5) \,.
\end{align}
This element can always be rewritten in the form of positive Borel gauge:
\begin{align}
 g = \Exp{\sum_{a\leq b}h_a{}^b K^a{}_b} \Exp{\frac{1}{3!}A_{a_1a_2a_3} R^{a_1a_2a_3}}\,k \qquad \text{where} \qquad k\in \mathfrak{H} = \SO(5) \,. 
\end{align}
It turns out that the SO(5) element $k$ does not contribute to the generalized metric. Disregarding it, the number of independent parameters are $10+4$, which is equal to the dimension of the coset space $\mathfrak{G} / \mathfrak{H} = \SL(5)/\SO(5)$. 
We can identify the parameters, $e_i{}^b\equiv (\Exp{h})_i{}^b\in \GL(4)/\SO(4)$ and $A_{a_1a_2a_3}$, as the vielbein and the 3-form potential on the $4$-torus, respectively. 
Note that the left index of the matrix $(\Exp{h})$ is changed from $a$ to $i$ in order to interpret it as the curved index. 

From the formulas \eqref{eq:gen-met} and \eqref{eq:vielbein1} and the matrix representations \eqref{eq:rho-K-M}--\eqref{eq:M-representation}, the generalized vielbein and the metric become \cite{Berman:2010is}
\begin{align}
\begin{split}
 &\cM_{MN}\equiv \abs{G}^{\frac{1}{5}}\,M_{MN}\,, \qquad 
 \cE_M{}^A\equiv \abs{G}^{\frac{1}{10}} \, E_M{}^A \,,\qquad 
 \bigl(\abs{G}\equiv \det G_{ij}\,,\quad G_{ij}\equiv e_i{}^a\,e_j{}^b\,\delta_{ab} \bigr) \,,
\\
 &(E_M{}^A)\equiv \widehat{E}\, \Exp{\boldsymbol{A}^{(3)}} 
 = \begin{pmatrix}
 e_i{}^a & -\frac{1}{\sqrt{2}}\, A_{i a_1a_2} \\
 0 & e^{i_1i_2}_{\,\,a_1a_2} 
 \end{pmatrix} \,,
\\
 & (M_{MN}) = \bigl(E_M{}^A\,E_N{}^B\,\delta_{AB}\bigr)
 = \begin{pmatrix}
 G_{ij} +\frac{1}{2}\,A_{ikl}\,A^{kl}{}_j & -\frac{1}{\sqrt{2}}\, A_i{}^{j_1j_2} \\
 -\frac{1}{\sqrt{2}}\, A^{i_1i_2}{}_j & G^{i_1i_2,\,j_1j_2} \end{pmatrix} \,,
\end{split}
\end{align}
where
\begin{align}
\begin{split}
 &(\widehat{E}_M{}^A) 
  \equiv \begin{pmatrix}
  e_i{}^a & 0 \\
  0 & e^{i_1i_2}_{\,\,a_1a_2} 
 \end{pmatrix}\,,\quad 
 e^{i_1i_2}_{\,\,a_1a_2}\equiv (e^{-\rmT})^{i_1}{}_{[a_1}\,(e^{-\rmT})^{i_2}{}_{a_2]} \,,
\\
 &\boldsymbol{A}^{(3)} 
  \equiv \frac{1}{3!}\,A_{abc}\,\rho_{R^{abc}} =\begin{pmatrix}
   0 & -\frac{1}{\sqrt{2}}\, A_{a b_1b_2} \\
   0 & 0
  \end{pmatrix}\,,\quad 
 \delta_{AB} \equiv \begin{pmatrix}
  \delta_{ab} & 0 \\
  0 & \delta^{a_1a_2,\, b_1b_2} 
 \end{pmatrix}\,,
\\
 &G^{i_1\cdots i_n,\,j_1\cdots j_n} \equiv \delta^{i_1\cdots i_n}_{k_1\cdots k_n} \,G^{k_1j_1}\cdots G^{k_nj_n}\,, \qquad 
 \delta^{a_1a_2,\, b_1b_2} \equiv \delta^{a_1a_2}_{c_1c_2}\,\delta^{c_1 b_1}\,\delta^{c_2 b_2} \,, 
\end{split}
\end{align}
and the indices are changed using the vielbein (e.g.~$A_{i a_1a_2}\equiv e_i{}^c\, A_{c a_1a_2}$) and raised or lowered using the metric $G_{ij}$ and its inverse. See Appendix \ref{app:notations} for further details of our conventions. 

If we do not choose the Borel gauge, we can generally parameterize the $\SL(5)$ generalized metric as \cite{Malek:2012pw,Blair:2014zba}
\begin{align}
 &(E_M{}^A)\equiv \widehat{E}\, \Exp{\boldsymbol{A}^{(3)}} \Exp{\boldsymbol{\Omega}^{(3)}} 
 = \begin{pmatrix}
 e_i{}^a + \frac{1}{2}\,A_{i c_1c_2}\,\Omega^{c_1c_2 a} & -\frac{1}{\sqrt{2}}\, A_{i a_1a_2} \\
 -\frac{1}{\sqrt{2}}\, \Omega^{i_1i_2 a} & e^{i_1i_2}_{\,\,a_1a_2} 
 \end{pmatrix}\,,
\\
 &{\scriptsize (M_{MN}) = \begin{pmatrix}
 G_{ij} +\frac{1}{2} A_{ikl} A^{kl}{}_j + A_{(i\vert kl} \Omega^{kl}{}_{\vert j)} +\frac{1}{4} A_{ikl} \Omega^{klm} \Omega_m{}^{pq} A_{pqj} 
 & -\frac{1}{\sqrt{2}} \bigl(A_i{}^{j_1j_2}+\Omega_i{}^{j_1j_2}+\frac{1}{2} A_{ikl} \Omega^{klm} \Omega_m{}^{j_1j_2}\bigr) \\
 & \\
 -\frac{1}{\sqrt{2}} \bigl(A^{i_1i_2}{}_j+\Omega^{i_1i_2}{}_{j}+\frac{1}{2} \Omega^{i_1i_2k} \Omega_{klm} A^{lm}{}_j\bigr) 
 & G^{i_1i_2,\,j_1j_2}+\frac{1}{2} \Omega^{i_1i_2}{}_k \Omega^{kj_1j_2}\end{pmatrix}}\,,
\end{align}
where we defined the $\Omega$-matrix:
\begin{align}
 \boldsymbol{\Omega}^{(3)} &\equiv \frac{1}{3!}\,\Omega^{c_1c_2c_3}\,\rho_{R_{c_1c_2c_3}}
 = \begin{pmatrix}
 0 & 0 \\
 -\frac{1}{\sqrt{2}}\, \Omega^{a_1a_2 b} & 0 
 \end{pmatrix} \,.
\end{align}

Choosing $\Omega^{ijk}=0$ or $A_{ijk}=0$, we obtain two alternative parameterizations for the generalized metric, 
\begin{align}
 (\cM_{MN}) &= \abs{G}^{\frac{1}{5}}\,\begin{pmatrix}
 \ \ G_{ij} +\frac{1}{2}\,A_{ikl}\,A^{kl}{}_j \ \ & \ \ \ -\frac{1}{\sqrt{2}}\, A_i{}^{j_1j_2} \ \  \\
 \ \ -\frac{1}{\sqrt{2}}\, A^{i_1i_2}{}_j \ \ & \ \ \ G^{i_1i_2,\, j_1j_2} \ \ \end{pmatrix}
\\
 &= \abs{\widetilde{G}}^{\frac{1}{5}}\,\begin{pmatrix}
 \widetilde{G}_{ij} & -\frac{1}{\sqrt{2}}\, \Omega_i{}^{j_1j_2} \\
 -\frac{1}{\sqrt{2}}\, \Omega^{i_1i_2}{}_{j} & \widetilde{G}^{i_1i_2,\, j_1j_2}+\frac{1}{2}\,\Omega^{i_1i_2}{}_k\,\Omega^{kj_1j_2}\end{pmatrix} \,.
\end{align}
The first expression is the conventional, geometric parameterization, while the second expression is the non-geometric parameterization. 
From these two parameterizations, we obtain the following relation between the standard fields and the dual fields:
\begin{align}
 \widetilde{G}_{ij}
 = \frac{\abs{G}^{1/9}}{\abs{\sfE}^{1/9}}\,\sfE_{ij} \,, \qquad 
 \Omega^{ij_1j_2}= (\sfE^{-1})^{ik}\,G^{j_1k_1}\,G^{j_2k_2}\,A_{kk_1k_2}\,,
 \label{eq:Omega-def1}
 \end{align}
 where
 \begin{align}
 \sfE_{ij} \equiv G_{ij}+\frac{1}{2}\,A_{ikl}\,A^{kl}{}_j \, . 
\end{align}
Further, associated to the two parameterizations, the external metric is also expressed in two alternative ways:
\begin{align}
 g_{\mu\nu} = \abs{G}^{\frac{1}{n-2}}\,\sfg_{\mu\nu}
            = \abs{\widetilde{G}}^{\frac{1}{n-2}}\,\widetilde{\sfg}_{\mu\nu} \,.
\label{eq:Omega-def2}
\end{align}
We confirm that $\sfg_{\mu\nu}$ and $G_{ij}$ are components of the conventional metric in the eleven-dimensional supergravity, denoted by $\sfG_{\hat{\mu}\hat{\nu}}$ ($\hat{\mu},\hat{\nu}=0,\dotsc,9,\sfM$). 

\subsubsection{$n=6$: $\mathfrak{G} = \SO(5,5)$}

The generalized metric or vielbein generally has the overall factor,
\begin{align}
 \cM_{MN}\equiv \abs{G}^{\frac{1}{n-2}}\,M_{MN}\,, \qquad \text{equivalently}, \qquad 
 \cE_M{}^A\equiv \abs{G}^{\frac{1}{2(n-2)}} \, E_M{}^A \,,
\end{align}
that comes from the second term in the right-hand-side of \eqref{eq:rho-K-M}. 
In the following, we focus on the parameterizations of $M_{MN}$ and $E_M{}^A$. 

In the present case of $\mathfrak{G} = \SO(5,5)$, we can similarly parameterize the generalized vielbein as \cite{Berman:2010is}
\begin{align}
 &(E_M{}^A) \equiv \widehat{E}\, \Exp{\boldsymbol{A}^{(3)}} 
 = \begin{pmatrix}
  \qquad e_i{}^a & \qquad - \frac{1}{\sqrt{2}}\, A_{i a_1a_2} & \qquad \frac{5}{\sqrt{5!}}\, A_{i [a_1a_2}\, A_{a_3a_4a_5]} \\
  & & \\
\qquad   0 & e^{i_1i_2}_{\,\,a_1a_2} & -\frac{20}{\sqrt{5!}\sqrt{2}}\, \delta^{i_1i_2}_{[a_1a_2} A_{a_3a_4a_5]} \\
  & & \\
\qquad  0 & 0 & e^{i_1\cdots i_5}_{\,\,a_1\cdots a_5}
 \end{pmatrix}\,,
\end{align}
or as
\begin{align}
 &(E_M{}^A) \equiv \widehat{E}\, \Exp{\boldsymbol{\Omega}^{(3)}} 
 = \begin{pmatrix}
  \widetilde{e}_i{}^a & 0 & 0 \\
  & & \\
 -\frac{1}{\sqrt{2}}\, \Omega^{i_1i_2 a} & \widetilde{e}^{i_1i_2}_{\,\,a_1a_2} & 0 
 \\
 & & \\
  \frac{5}{\sqrt{5!}}\, \Omega^{[i_1i_2i_3}\, \Omega^{i_4i_5]a} & \quad -\frac{20}{\sqrt{5!}\sqrt{2}}\, \delta_{a_1a_2}^{[i_1i_2} \Omega^{i_3i_4i_5]} & \quad \widetilde{e}^{i_1\cdots i_5}_{\,\,a_1\cdots a_5}
 \end{pmatrix}\,,
\end{align}
where we defined
\begin{align}
 \widehat{E} &\equiv 
   \begin{pmatrix}
    \qquad e_i{}^a & 0 & 0 \quad \\
    \qquad 0 & \qquad e^{i_1i_2}_{\,\,a_1a_2} & 0 \quad \\
    \qquad 0 & 0 & \qquad e^{i_1\cdots i_5}_{\,\,a_1\cdots a_5} \quad
   \end{pmatrix}\,, 
\\
 \boldsymbol{A}^{(3)}
 &\equiv \frac{1}{3!}\,A_{c_1c_2c_3}\,\rho_{R^{c_1c_2c_3}} = \begin{pmatrix}
 0 & - \frac{1}{\sqrt{2}}\, A_{a b_1b_2} & 0 \\
 0 & 0 & -\frac{20}{\sqrt{5!}\sqrt{2}}\, \delta^{a_1a_2}_{[b_1b_2}\, A_{b_3b_4b_5]} \\
 0 & 0 & 0 
 \end{pmatrix} \,,
\\
 \boldsymbol{\Omega}^{(3)} &\equiv \frac{1}{3!}\,\Omega^{c_1c_2c_3}\,\rho_{R_{c_1c_2c_3}} 
 = \begin{pmatrix}
 0 & 0 & 0 \\
 -\frac{1}{\sqrt{2}}\, \Omega^{a_1a_2 b} & 0 & 0 \\
 0 & -\frac{20}{\sqrt{5!}\sqrt{2}}\, \delta_{b_1b_2}^{[a_1a_2}\, \Omega^{a_3a_4a_5]} & 0 
 \end{pmatrix} \,.
\end{align}
We can again redundantly parameterize the generalized vielbein as
\begin{align}
 E_M{}^A\equiv \widehat{E} \Exp{\boldsymbol{A}^{(3)}}\Exp{\boldsymbol{\Omega}^{(3)}}\,. 
\end{align}

\subsubsection{$n=5$: $\mathfrak{G} = E_{6(6)}$}

In the case $\mathfrak{G} = E_{6(6)}$, we can parameterize the generalized vielbein as \cite{Berman:2010is}
\begin{align}
 E_M{}^A &\equiv \widehat{E}\, \Exp{\boldsymbol{A}^{(6)}}\Exp{\boldsymbol{A}^{(3)}} 
 = \begin{pmatrix}
  e_i{}^a & \qquad - \frac{1}{\sqrt{2}}\, A_{i a_1a_2} & \qquad \frac{1}{\sqrt{5!}}\,\bigl(A_{ia_1\cdots a_5}+5\,A_{i [a_1a_2}\, A_{a_3a_4a_5]}\bigr) \\
  & & \\
  0 & \qquad e^{i_1i_2}_{\,\,a_1a_2} & \qquad -\frac{20}{\sqrt{5!}\sqrt{2}}\, \delta^{i_1i_2}_{[a_1a_2}\,A_{a_3a_4a_5]} \\
  & & \\
  0 & \qquad 0 & \qquad e^{i_1\cdots i_5}_{\,\,a_1\cdots a_5}
 \end{pmatrix}\,,
\end{align}
or as
\begin{align}
 E_M{}^A &\equiv \widehat{E}\, \Exp{\boldsymbol{\Omega}^{(6)}}\Exp{\boldsymbol{\Omega}^{(3)}} 
  = \begin{pmatrix}
  e_i{}^a & \qquad 0 & \qquad 0 \\
 & & \\
 -\frac{1}{\sqrt{2}}\, \Omega^{i_1i_2 a} & \quad e^{i_1i_2}_{\,\,a_1a_2} & \quad 0 \\
 & & \\
 \frac{1}{\sqrt{5!}}\,\bigl(-\Omega^{i_1\cdots i_5 a}+5\,\Omega^{[i_1i_2i_3}\, \Omega^{i_4i_5]a}\bigr) & \quad -\frac{20}{\sqrt{5!}\sqrt{2}}\, \delta_{a_1a_2}^{[i_1i_2}\,\Omega^{i_3i_4i_5]} & \quad e^{i_1\cdots i_5}_{\,\,a_1\cdots a_5}
  \end{pmatrix},
\end{align}
where we defined
\begin{align}
 \widehat{E} &\equiv 
 \begin{pmatrix}
  e_i{}^a  & \quad 0 & \quad 0 \\
  0 & \quad e^{i_1i_2}_{\,\,a_1a_2} & \quad 0 \\
  0 & \quad 0 & \quad e^{i_1\cdots i_5}_{\,\,a_1\cdots a_5}
 \end{pmatrix}\,, 
\\
 \boldsymbol{A}^{(3)}
 &\equiv \frac{1}{3!}\,A_{c_1c_2c_3}\,\rho_{R^{c_1c_2c_3}} = \begin{pmatrix}
 0 & - \frac{1}{\sqrt{2}}\, A_{a b_1b_2} & 0 \\
 0 & 0 & -\frac{20}{\sqrt{5!}\sqrt{2}}\, \delta^{a_1a_2}_{[b_1b_2}\, A_{b_3b_4b_5]} \\
 0 & 0 & 0 
 \end{pmatrix} \,,
\\
  \boldsymbol{A}^{(6)}
 &\equiv \frac{1}{6!}\,A_{c_1\cdots c_6}\,\rho_{R^{c_1\cdots c_6}} = \begin{pmatrix}
 0 & \qquad 0 & \qquad \frac{1}{\sqrt{5!}}\, A_{a b_1\cdots b_5} \\
 0 & \qquad 0 & \qquad 0 \\
 0 & \qquad 0 & \qquad 0 
 \end{pmatrix} \,,
\\
 \boldsymbol{\Omega}^{(3)} &\equiv \frac{1}{3!}\,\Omega^{c_1c_2c_3}\,\rho_{R_{c_1c_2c_3}} = \begin{pmatrix}
 0 & 0 & 0 \\
 -\frac{1}{\sqrt{2}}\, \Omega^{a_1a_2 b} & 0 & 0 \\
 0 & -\frac{20}{\sqrt{5!}\sqrt{2}}\, \delta_{b_1b_2}^{[a_1a_2}\, \Omega^{a_3a_4a_5]} & 0 
 \end{pmatrix} \,,
\\
 \boldsymbol{\Omega}^{(6)} &\equiv -\frac{1}{6!}\,\Omega^{c_1\cdots c_6}\,\rho_{R_{c_1\cdots c_6}} = \begin{pmatrix}
 0 & \qquad 0 & \qquad 0 \\
 0 & \qquad 0 & \qquad 0 \\
 -\frac{1}{\sqrt{5!}}\, \Omega^{a_1\cdots a_5 b} & \qquad 0 & \qquad 0 
 \end{pmatrix} \,.
\end{align}
We remark that the normalization of the 6-form is different from that used in \cite{Berman:2010is} by a factor 2. 
Note also that, in the middle expression of the last line, the minus sign is introduced in order to make the exotic duality, $A_{c_1\cdots c_6}\leftrightarrow\Omega^{c_1\cdots c_6}$, coincides with the matrix transpose. Stated differently, the negative sign comes from the fact that the Cartan involution \eqref{eq:Cartan-involution-M} for $R^{c_1\cdots c_6}$ appears with the positive sign, $\theta(R^{c_1\cdots c_6})= + R_{c_1\cdots c_6}$. 

\subsubsection{$n=4$: $\mathfrak{G} = E_{7(7)}$}

In the $E_{7(7)}$ case, we can parameterize the generalized vielbein as \cite{Berman:2011jh}
\begin{align}
 (E_M{}^A) \equiv \widehat{E}\,\Exp{\boldsymbol{A}^{(6)}}\Exp{\boldsymbol{A}^{(3)}} 
\qquad \text{or}\qquad
 (E_M{}^A) \equiv \widehat{E}\,\Exp{\boldsymbol{\Omega}^{(6)}}\Exp{\boldsymbol{\Omega}^{(3)}} \,,
\end{align}
where we defined
\begin{align}
 \widehat{E} &\equiv
 \begin{pmatrix}
  e_i{}^a & \quad 0 & \quad 0 & \quad 0 \\
  0 & \quad e^{i_1i_2}_{\,\,a_1a_2} & \quad 0 & \quad 0 \\
  0 & \quad 0 & \quad e_{i_1\cdots i_5}^{\,\,a_1\cdots a_5} & \quad 0 \\
  0 & \quad 0 & \quad 0 & \quad \abs{e}^{-1}\,e^i{}_a
 \end{pmatrix}\,, 
\\
 \boldsymbol{A}^{(3)}
 &\equiv \frac{1}{3!}\,A_{c_1c_2c_3}\,\rho_{R^{c_1c_2c_3}}
 = {\footnotesize
\begin{pmatrix}
 0 & - \frac{1}{\sqrt{2}}\, A_{a b_1b_2} & 0 & 0 \\
 & & & \\
 0 & 0 & -\frac{20}{\sqrt{5!}\sqrt{2}}\, \delta^{a_1a_2}_{[b_1b_2}\,A_{b_3b_4b_5]} & 0 \\
 & & & \\
 0 & 0 & 0 & -\frac{1}{2\sqrt{5!}}\,\epsilon^{a_1\cdots a_5c_1c_2}\,A_{c_1c_2 b}\\
 & & & \\
 0 & 0 & 0 & 0
 \end{pmatrix}}\,,
\\
  \boldsymbol{A}^{(6)}
 &\equiv \frac{1}{6!}\,A_{c_1\cdots c_6}\,\rho_{R^{c_1\cdots c_6}}
 = {\footnotesize
\begin{pmatrix}
\quad 0 & \qquad 0 & \qquad \frac{1}{\sqrt{5!}}\, A_{a b_1\cdots b_5} & \qquad 0 \\
 & & & \\
\quad 0 & \qquad 0 & \qquad 0 & \qquad \frac{2}{6!\sqrt{2}}\,\delta_b^{[a_1}\,\epsilon^{a_2]c_1\cdots c_6}\,A_{c_1\cdots c_6} \\
 & & & \\
\quad 0 & \qquad 0 & \qquad 0 & \qquad 0 \\
 & & & \\
\quad 0 & \qquad 0 & \qquad 0 & \qquad 0
 \end{pmatrix}}\,,
\\
 \boldsymbol{\Omega}^{(3)} &\equiv \frac{1}{3!}\,\Omega^{c_1c_2c_3}\,\rho_{R_{c_1c_2c_3}}= {\footnotesize
 \begin{pmatrix}
 0 & 0 & 0 & 0 \\
 & & & \\
 -\frac{1}{\sqrt{2}}\, \Omega^{a_1a_2 b} & 0 & 0 & 0 \\
 & & & \\
 0 & -\frac{20}{\sqrt{5!}\sqrt{2}}\, \delta_{b_1b_2}^{[a_1a_2}\, \Omega^{a_3a_4a_5]} & 0 & 0 \\
 & & & \\
 0 & 0 & -\frac{1}{2\sqrt{5!}}\,\Omega^{a c_1c_2}\,\epsilon_{c_1c_2 b_1\cdots b_5} & 0
 \end{pmatrix}}\,,
\\
 \boldsymbol{\Omega}^{(6)} &\equiv -\frac{1}{6!}\,\Omega^{c_1\cdots c_6}\,\rho_{R_{c_1\cdots c_6}}
 = {\footnotesize
 \begin{pmatrix}
 0 & 0 & 0 & \qquad 0 \\
 & & & \\
 0 & 0 & 0 & \qquad 0 \\
 & & & \\
 -\frac{1}{\sqrt{5!}}\, \Omega^{a_1\cdots a_5 b} & 0 & 0 & \qquad 0\\
 & & & \\
 0 & \frac{2}{6!\sqrt{2}}\,\delta^a_{[b_1}\, \epsilon_{b_2]c_1\cdots c_6}\,\Omega^{c_1\cdots c_6} & 0 & \qquad 0
 \end{pmatrix}}\,.
\end{align}
We remark that the parameterizations for $E_{d(d)}$ with $4\leq d\leq 6$ are obtainable by a truncation of those for $E_{7(7)}$.

\subsection{Eleven-dimensional effective action}
\label{sec:11d-action}

The eleven-dimensional effective action is obtained by solving the section condition such that the eleven-dimensional coordinates are given by $(x^{\hat{\mu}})\equiv (x^\mu,\,x^i)$; see Appendix \ref{app:action} for the detailed derivation. 
For instance, consider the $E_{7(7)}$ EFT in the geometric parameterization. The action becomes
\begin{align}
 \cL&= \abs{\sfG}^{\frac{1}{2}}\,\biggl( R(\sfG) \nn \\
&- \sfg^{\mu\nu} \, \left [ \frac{G^{i_1i_2i_3,\,j_1j_2j_3}}{2\cdot 3!}\,\partial_\mu A_{i_1i_2i_3}\, \partial_\nu A_{j_1j_2j_3} 
   + \frac{ G^{i_1\cdots i_6,\,j_1\cdots j_6}}{2\cdot 6!}\,\cF_{\mu,\, i_1\cdots i_6}\,\cF_{\nu,\, j_1\cdots j_6} \right]
\nn\\
 &- \frac{1 }{2\cdot 4!}\, G^{i_1\cdots i_4,\, j_1\cdots j_4} \, F_{i_1\cdots i_4}\,F_{j_1\cdots j_4}
    - \frac{1}{2\cdot 7!}\, G^{i_1\cdots i_7,\, j_1\cdots j_7}\,F_{i_1\cdots i_7}\,F_{j_1\cdots j_7}  \biggr) \,,
\end{align}
where
\begin{align}
\begin{split}
 R(\sfG)&\equiv R(\sfg) 
       + \sfg^{\mu\nu}\, \left[ \frac{1}{4} \partial_\mu G^{ij}\, \partial_\nu G_{ij}
       +\frac{1}{4}\,
       \partial_\mu\ln\abs{G}\,\partial_\nu\ln\abs{G} \right]
\\
 &+ R(G) +G^{ij}\,\left[ \, \frac{1}{4}\, \partial_i \sfg^{\mu\nu}\,\partial_j \sfg_{\mu\nu} \, + \, \frac{1}{4}\, \partial_i \ln \abs{\sfg}\, \partial_j \ln \abs{\sfg} \, \right] \,,
\\
 F_{i_1\cdots i_4}&\equiv 4\,\partial_{[i_1} A_{i_2i_3i_4]}\,, \nn \\ 
 F_{i_1\cdots i_7} &\equiv 7\,\partial_{[i_1} A_{i_2\cdots i_7]} + \tfrac{35}{2} \, A_{[i_1i_2i_3}\,F_{i_4i_5i_6i_7]} \,, 
\\
 \cF_{\mu,\, k_1\cdots k_6} &\equiv \partial_\mu A_{k_1\cdots k_6} - 10 \, A_{[k_1k_2k_3\vert}\,\partial_\mu A_{\vert k_4k_5k_6]}\,. 
\end{split}
\end{align}
Note that $R(\sfG)$ is equal to the Ricci scalar associated with the eleven-dimensional metric,
\begin{align}
 \bigl(\sfG_{\hat{\mu}\hat{\nu}}\bigr) \equiv \begin{pmatrix}
 \sfg_{\mu\nu} & 0 \cr 0 & G_{ij}
 \end{pmatrix}\,,
\end{align}
and the off-diagonal components are absent since we neglected some external parts of the EFT action. 
Note also that the above result generalizes the result of \cite{Berman:2011jh}, where only the potential part was calculated. 
Further, the 7-form, $F_{i_1\cdots i_7}$, is the Hodge dual of the 4-form, $F_{\mu_1\cdots \mu_4}=4\,\partial_{[\mu_1}A_{\mu_2\mu_3\mu_4]}$, and is independent of the internal components, $A_{i_1i_2i_3}$\,. 

On the other hand, in the non-geometric parameterization, the effective action becomes
\vskip0.5cm
\begin{tcolorbox}
\vskip-0.5cm
\begin{align}
 \cL &= \abs{\widetilde{\sfG}}^{\frac{1}{2}} \,\Bigl( R(\widetilde{\sfG}) 
 - \frac{1}{2}\,\abs{S^{(1,\,3)}}^2 - \frac{1}{2}\,\abs{S^{(1,\,6)}}^2\Bigr) \,,
\label{eq:non-geometric-action-11d}
\end{align}
where
\begin{align}
\begin{split}
 \abs{S^{(1,\,p)}}^2&\equiv \, \widetilde{\sfG}^{\hat{\mu}\hat{\nu}}\, \left[ \frac{1}{p!}\,\widetilde{G}_{i_1\cdots i_p,\, j_1\cdots j_p}\,S_{\hat{\mu}}{}^{i_1\cdots i_p}\,S_{\hat{\nu}}{}^{j_1\cdots j_p} \right]  \nn \\
 S_{\hat{\mu}}{}^{i_1\cdots i_3} &\equiv \, \partial_{\hat{\mu}} \Omega^{i_1\cdots i_3}\,, \nn \\
 S_{\hat{\mu}}{}^{i_1\cdots i_6} &\equiv \, \partial_{\hat{\mu}} \Omega^{i_1\cdots i_6}+10\,\Omega^{[i_1i_2i_3}\,\partial_{\hat{\mu}} \Omega^{i_4i_5i_6]}
 \,.
\end{split}
\end{align}
\end{tcolorbox}
\noindent
Here, we defined the non-geometric fluxes, to be referred to as the $S$-fluxes. 
This is one of the main results of this paper. 
In deriving \eqref{eq:non-geometric-action-11d}, 
we used the simplifying assumption,
\begin{align}
 \Omega^{ijk}\, \partial_i(\cdots) =0\,,\qquad 
 \partial_i \Omega^{ijk} =0\,, \qquad 
 \Omega^{ij_1\cdots j_5}\, \partial_i (\cdots) =0\,,\qquad 
 \partial_i \Omega^{ij_1\cdots j_5} =0\, .
 \label{assume} 
\end{align}

\subsection{Reduction to the type IIA theory}
\label{sec:type-IIA-action}

It is well-known that the type IIA supergravity can be derived from the eleven-dimensional supergravity by the following Kaluza-Klein decomposition of fields:
\begin{align}
\begin{split}
 &\rmd s^2 
 = \Exp{-\frac{2}{3}\,\phi}\, \textsc{G}_{{\boldsymbol{\hat\mu}}{\boldsymbol{\hat\nu}}}\,\rmd x^{\boldsymbol{\hat\mu}}\,\rmd x^{\boldsymbol{\hat\nu}} 
 + \Exp{\frac{4}{3}\,\phi}\,\bigl(\rmd x^\sfM +C_{\boldsymbol{\hat\mu}}\,\rmd x^{\boldsymbol{\hat\mu}}\bigr)\,\bigl(\rmd x^\sfM +C_{\boldsymbol{\hat\nu}}\,\rmd x^{\boldsymbol{\hat\nu}}\bigr)
\,,
\\
 &A_{{\boldsymbol{\hat\mu}}_1{\boldsymbol{\hat\mu}}_2{\boldsymbol{\hat\mu}}_3} =C_{{\boldsymbol{\hat\mu}}_1{\boldsymbol{\hat\mu}}_2{\boldsymbol{\hat\mu}}_3}\,,\qquad A_{{\boldsymbol{\hat\mu}}_1{\boldsymbol{\hat\mu}}_2\sfM} = -B_{{\boldsymbol{\hat\mu}}_1{\boldsymbol{\hat\mu}}_2} \,,\qquad 
 A_{{\boldsymbol{\hat\mu}}_1\cdots {\boldsymbol{\hat\mu}}_6}= - B_{{\boldsymbol{\hat\mu}}_1\cdots {\boldsymbol{\hat\mu}}_6}\,,
\\
 &A_{{\boldsymbol{\hat\mu}}_1\cdots {\boldsymbol{\hat\mu}}_5\sfM}= C_{{\boldsymbol{\hat\mu}}_1\cdots {\boldsymbol{\hat\mu}}_5} +5\,C_{[{\boldsymbol{\hat\mu}}_1{\boldsymbol{\hat\mu}}_2{\boldsymbol{\hat\mu}}_3}\,B_{{\boldsymbol{\hat\mu}}_4{\boldsymbol{\hat\mu}}_5]} \,,
\label{eq:11-to-10}
\end{split}
\end{align}
where ${\boldsymbol{\hat\mu}},{\boldsymbol{\hat\nu}} =0,\dotsc,9$. 

In the non-geometric parameterization, we consider an analogous Kaluza-Klein decomposition of fields:
\begin{align}
\begin{split}
 &\rmd \widetilde{s}^2 
 = \Exp{-\frac{2}{3}\,\widetilde{\phi}}\, \textsc{G}_{{\boldsymbol{\hat\mu}}{\boldsymbol{\hat\nu}}}\,\bigl(\rmd x^{\boldsymbol{\hat\mu}} +\gamma^{\boldsymbol{\hat\mu}}\,\rmd x^{\sfM}\bigr)\,
  \bigl(\rmd x^{\boldsymbol{\hat\nu}} +\gamma^{\boldsymbol{\hat\nu}}\,\rmd x^{\sfM}\bigr)
   + \Exp{\frac{4}{3}\,\widetilde{\phi}}\,\bigl(\rmd x^{\sfM}\bigr)^2\,,
\\
 &\Omega^{{\boldsymbol{\hat\mu}}_1{\boldsymbol{\hat\mu}}_2{\boldsymbol{\hat\mu}}_3}= \gamma^{{\boldsymbol{\hat\mu}}_1{\boldsymbol{\hat\mu}}_2{\boldsymbol{\hat\mu}}_3}\,,\qquad \Omega^{{\boldsymbol{\hat\mu}}_1{\boldsymbol{\hat\mu}}_2\sfM} = - \beta^{{\boldsymbol{\hat\mu}}_1{\boldsymbol{\hat\mu}}_2} \,,\qquad 
 \Omega^{{\boldsymbol{\hat\mu}}_1\cdots {\boldsymbol{\hat\mu}}_6}= - \beta^{{\boldsymbol{\hat\mu}}_1\cdots {\boldsymbol{\hat\mu}}_6}\,,
\\
 &\Omega^{{\boldsymbol{\hat\mu}}_1\cdots {\boldsymbol{\hat\mu}}_5\sfM}= \gamma^{{\boldsymbol{\hat\mu}}_1\cdots {\boldsymbol{\hat\mu}}_5}-5\,\gamma^{[{\boldsymbol{\hat\mu}}_1{\boldsymbol{\hat\mu}}_2{\boldsymbol{\hat\mu}}_3}\,\beta^{{\boldsymbol{\hat\mu}}_4{\boldsymbol{\hat\mu}}_5]} \,.
\label{eq:dual-11-to-10}
\end{split}
\end{align}
In the matrix notation, the two decompositions of the metric can be compared as
\begin{align}
\begin{split}
 (\sfG_{\hat{\mu}\hat{\nu}})&=
 \begin{pmatrix}
 \delta_{\boldsymbol{\hat\mu}}^{\boldsymbol{\hat\rho}} & C_{\boldsymbol{\hat\mu}} \\
 0 & 1
 \end{pmatrix}
 \begin{pmatrix}
 \Exp{-\frac{2}{3}\,\phi}\,\textsc{G}_{{\boldsymbol{\hat\rho}}{\boldsymbol{\hat\sigma}}} & 0 \\
 0 & \Exp{\frac{4}{3}\,\phi}
 \end{pmatrix}
 \begin{pmatrix}
 \delta_{\boldsymbol{\hat\nu}}^{\boldsymbol{\hat\sigma}} & \ 0 \\
 C_{\boldsymbol{\hat\nu}} & \ 1
 \end{pmatrix} \,,
\\
\\
 (\widetilde{\sfG}_{\hat{\mu}\hat{\nu}})&=
 \begin{pmatrix}
 \delta_{\boldsymbol{\hat\mu}}^{\boldsymbol{\hat\rho}} & \ \ 0 \\
 \gamma^{\boldsymbol{\hat\rho}} & \ \ 1
 \end{pmatrix}
 \begin{pmatrix}
 \Exp{-\frac{2}{3}\,\widetilde{\phi}}\,\widetilde{\textsc{G}}_{{\boldsymbol{\hat\rho}}{\boldsymbol{\hat\sigma}}} & 0 \\
 0 & \Exp{\frac{4}{3}\,\widetilde{\phi}}
 \end{pmatrix}
 \begin{pmatrix}
 \delta_{\boldsymbol{\hat\nu}}^{\boldsymbol{\hat\rho}} & 
\ \gamma^{\boldsymbol{\hat\rho}} \\
 0 & \ 1
 \end{pmatrix} \,. 
\end{split}
\end{align}

The non-geometric effective action \eqref{eq:non-geometric-action-11d}, up to a total derivative term, becomes
\vskip0.5cm
\begin{tcolorbox}
\vskip-0.5cm
\begin{align}
 \cL &= \abs{\widetilde{\textsc{G}}}^{\frac{1}{2}}\,\biggl[ \Exp{-2\widetilde{\phi}}\Bigl( R(\widetilde{\textsc{G}}) +4\,\abs{\rmd\widetilde{\phi}}^2 - \frac{1}{2}\,\abs{Q^{(1,\,2)}}^2 \Bigr)
\nn\\
 &\qquad\quad - \frac{1}{2} \Exp{-4\widetilde{\phi}} 
 \bigl(\abs{P^{(1,\,1)}}^2 +\abs{P^{(1,\,3)}}^2 + \abs{P^{(1,\,5)}}^2\bigr)
 \nn \\
 & \qquad \quad - \frac{1}{2} \Exp{-6\widetilde{\phi}}\abs{Q^{(1,\,6)}}^2 
 \biggr] \,. 
\label{eq:IIA-non-geometric}
\end{align}
\end{tcolorbox}
\noindent
Here, we defined the non-geometric $Q$- and $P$-fluxes as
\begin{align}
\begin{split}
 &Q_{\boldsymbol{\hat\mu}}{}^{mn}\equiv \partial_{\boldsymbol{\hat\mu}}\beta^{mn}\,,\nn \\
 & 
 P_{\boldsymbol{\hat\mu}}{}^m\equiv \partial_{\boldsymbol{\hat\mu}}\gamma^m \,, \nn \\
& P_{\boldsymbol{\hat\mu}}{}^{m_1m_2m_3}\equiv \partial_{\boldsymbol{\hat\mu}}\gamma^{m_1m_2m_3}-3\,\gamma^{[m_1}\,Q_{\boldsymbol{\hat\mu}}{}^{m_2m_3]}\,,
\\
 &P_{\boldsymbol{\hat\mu}}{}^{m_1\cdots m_5}\equiv \partial_{\boldsymbol{\hat\mu}}\gamma^{m_1\cdots m_5}-10\,\gamma^{[m_1m_2m_3}\,Q_{\boldsymbol{\hat\mu}}{}^{m_4m_5]}\,,
\\
 &Q_{\boldsymbol{\hat\mu}}{}^{m_1\cdots m_6}\equiv \partial_{\boldsymbol{\hat\mu}}\beta^{m_1\cdots m_6}-10\,\gamma^{[m_1m_2m_3}\,\partial_{\boldsymbol{\hat\mu}}\gamma^{m_4m_5m_6]}+6\,\gamma^{[m_1}\,P_{\boldsymbol{\hat\mu}}{}^{m_2\cdots m_6]}\,,
\end{split}
\end{align}
and
\begin{align}
\begin{split}
 \abs{\rmd\widetilde{\phi}}^2 \ \ &\equiv \widetilde{\textsc{G}}^{{\boldsymbol{\hat\mu}}{\boldsymbol{\hat\nu}}}\,\partial_{\boldsymbol{\hat\mu}}\widetilde{\phi}\,\partial_{\boldsymbol{\hat\nu}}\widetilde{\phi}\,, \nn \\
 \abs{Q^{(1,\,p)}}^2 &\equiv \frac{1}{p!}\,\widetilde{\textsc{G}}^{{\boldsymbol{\hat\mu}}{\boldsymbol{\hat\nu}}}\,\widetilde{\textsc{G}}_{m_1\cdots m_p,\,n_1\cdots n_p}\,Q_{\boldsymbol{\hat\mu}}{}^{m_1\cdots m_p}\,Q_{\boldsymbol{\hat\nu}}{}^{n_1\cdots n_p}\,,
\\
 \abs{P^{(1,\,p)}}^2 &\equiv \frac{1}{p!}\,\widetilde{\textsc{G}}^{{\boldsymbol{\hat\mu}}{\boldsymbol{\hat\nu}}}\,\widetilde{\textsc{G}}_{m_1\cdots m_p,\,n_1\cdots n_p}\,P_{\boldsymbol{\hat\mu}}{}^{m_1\cdots m_p}\,P_{\boldsymbol{\hat\nu}}{}^{n_1\cdots n_p}\,. 
\end{split}
\end{align}
We also used the identity,
\begin{align}
 \abs{\widetilde{\sfG}}^{\frac{1}{2}}\,R(\widetilde{\sfG}) 
 &= \abs{\widetilde{\textsc{G}}}^{\frac{1}{2}}\,\Bigl(\Exp{-2\widetilde{\phi}}\bigl[R(\widetilde{\textsc{G}}) +4\,\widetilde{\textsc{G}}^{{\boldsymbol{\hat\mu}}{\boldsymbol{\hat\nu}}}\,\partial_{\boldsymbol{\hat\mu}}\widetilde{\phi}\,\partial_{\boldsymbol{\hat\nu}}\widetilde{\phi}\bigr] - \frac{\Exp{-4\widetilde{\phi}}}{2}\,\abs{P^{(1,\,1)}}^2 \Bigr)
\nn\\
 & + \partial_{\boldsymbol{\hat\mu}} \biggl(\frac{14}{3}\,\abs{\widetilde{\textsc{G}}}^{\frac{1}{2}}\,\Exp{-2\widetilde{\phi}}\,\widetilde{\textsc{G}}^{{\boldsymbol{\hat\mu}}{\boldsymbol{\hat\nu}}}\,\partial_{\boldsymbol{\hat\nu}}\phi\biggr) \,.
\end{align}

Note that the terms proportional to $\Exp{-2\widetilde{\phi}}$ in the action \eqref{eq:IIA-non-geometric} match with the action of the $\beta$-supergravity \cite{Andriot:2011uh,Andriot:2012wx,Andriot:2012an,Andriot:2013xca,Andriot:2014qla} once the simplifying assumptions (\ref{assume}) are made. 
Moreover, \eqref{eq:IIA-non-geometric} generalizes the actions for the $P$-fluxes and the $Q^{(1,\,6)}$-flux obtained in \cite{Sakatani:2014hba} with non-trivial dilaton dependence taking into account.

\section{Non-geometric fluxes in EFT: type IIB section}
\label{sec:EFT-IIB}
We now turn to type IIB EFTs.  As previously emphasized in \cite{Schnakenburg:2001he,West:2005gu,Hohm:2013pua} the type IIB supergravity is also derivable from the $U$-duality covariant formulation. 
In particular, within the $\SL(5)$ EFT, a solution of the section condition that corresponds to the type IIB theory was found in \cite{Blair:2013gqa}. 
In the following, we present both the conventional parameterization and the non-geometric parameterization of the generalized vielbein in terms of fields in the ten-dimensional type IIB supergravity. 

In the type IIB case, we introduce the following generalized coordinates in which $\SL(d-1)$ and $\SL(2)$ covariance are manifest \cite{Schnakenburg:2001he,West:2005gu,Tumanov:2014pfa}:
\begin{align}
\begin{split}
 n=7:&\quad (Y^M)= (x^\sfm,\, y_\sfm^\alpha,\, y_{\sfm_1\sfm_2\sfm_3}) \qquad \qquad \qquad (\alpha=1,2,\ \sfm=7,\ 8,\ 9)\,,
\\
 n=6:&\quad (Y^M)= (x^\sfm,\, y_\sfm^\alpha,\, y_{\sfm_1\sfm_2\sfm_3})
 \qquad \qquad \qquad (\alpha=1,2,\ \sfm=6,\dotsc,9)\,,
\\
 n=5:&\quad (Y^M)= (x^\sfm,\, y_\sfm^\alpha,\, y_{\sfm_1\sfm_2\sfm_3},\,y^\alpha_{\sfm_1\cdots \sfm_5})\qquad \ \ (\alpha=1,2,\ \sfm=5,\dotsc,9)\,,
\\
 n=4:&\quad (Y^M)= (x^\sfm,\, y_\sfm^\alpha,\, y_{\sfm_1\sfm_2\sfm_3},\,y^\alpha_{\sfm_1\cdots \sfm_5},\,z_\sfm)\quad (\alpha=1,2,\ \sfm=4,\dotsc,9)\,.
\end{split}
\end{align}
Here, $y_{\sfm_1\sfm_2\sfm_3}$ and $y^\alpha_{\sfm_1\cdots \sfm_5}$ are totally antisymmetric coordinates and $z_\sfm$ is related to $x_{\sfm_1\cdots \sfm_6,\,\sfm}$ adopted in \cite{Tumanov:2014pfa} by $z_\sfm\equiv \frac{1}{\sqrt{6!}}\,\epsilon^{\sfm_1\cdots \sfm_6}\,x_{\sfm_1\cdots \sfm_6,\,\sfm}$\,. 

The conventional parameterization of the generalized metric in the type IIB section is shown in \cite{Tumanov:2014pfa} (in the context of the $E_{11}$ program). By taking a suitable truncation of it, we can obtain the generalized vielbein in $E_{d(d)}$ EFT for various $d$. 

In the type IIB section, corresponding to the curved indices $\sfm$, we introduce the flat indices as $\sfa$ and the curved and flat indices are changed with the vielbein, $e_\sfm{}^\sfa$. 

\subsection{Parameterizations of the generalized vielbein}
\label{sec:vielbein-IIB}

Here, in a way analogous to those given for M theory in \ref{sec:vielbein-M}, we construct the type IIB parameterizations of the generalized vielbein. 
We consider the cases $4\leq n\leq 7$ (or $4\leq d\leq 7$), and in all cases, we use the following matrices:
\begin{align}
 &(\epsilon_{\alpha\beta}) = \begin{pmatrix} 0 & 1\cr -1 & 0 \end{pmatrix} = (\epsilon^{\alpha\beta}) \,,
\\
\nn \\
 &\bigl(\cV_\alpha{}^{\beta}\bigr) \equiv \begin{cases}
 \Exp{\phi/2}\,\begin{pmatrix}
 \Exp{-\phi} & C^{(0)} \\
 0 & 1
 \end{pmatrix} & \quad \text{(geometric)} \\
 \Exp{\widetilde{\phi}/2}\,\begin{pmatrix}
 \Exp{-\widetilde{\phi}} & \quad 0 \\
 \gamma^{(0)} & \quad 1
 \end{pmatrix} & \text{(non-geometric)}
\end{cases}
 ,
\label{eq:cV-def}
\end{align}
where $\phi$ is the dilaton and $C^{(0)}$ is the Ramond-Ramond 0-form potential and $\widetilde{\phi}$ and $\gamma^{(0)}$ are their non-geometric duals. 
We also define the metric,
\begin{align}
 (m_{\alpha\beta})\equiv \cV\,\cV^{\rmT} 
 = \begin{cases}
 \Exp{\phi}\,\begin{pmatrix}
 \Exp{-2\phi} + (C^{(0)})^2 & C^{(0)} \\
 C^{(0)} & 1
 \end{pmatrix} & \quad \text{(geometric)} \\
 \Exp{\widetilde{\phi}}\,\begin{pmatrix}
\ \Exp{-2\widetilde{\phi}} & \ \ \gamma^{(0)} \\
 \ \gamma^{(0)} & \ \ 1 + (\gamma^{(0)})^2
 \end{pmatrix} & \text{(non-geometric)}
\end{cases} , 
\end{align}
and may denote it by $\widetilde{m}_{\alpha\beta}$ for the non-geometric parameterization. 
We also introduce the pair of the Kalb-Ramond $B$-field and the Ramond-Ramond 2-form as well as their dual bi-vectors,
\begin{align}
\bigl(B^{\alpha}_{\sfa\sfb}\bigr)\equiv \begin{pmatrix} B_{\sfa\sfb}\cr C_{\sfa\sfb}
\end{pmatrix} \,\qquad \text{and} \qquad
 \bigl(\beta_{\alpha}^{\sfa\sfb}\bigr) \equiv \begin{pmatrix} \beta^{\sfa\sfb}\cr \gamma^{\sfa\sfb} \end{pmatrix} \,. 
\end{align}
The Ramond-Ramond four-form always appears in the $S$-duality-invariant combination, 
\begin{align}
D^{(4)}= C^{(4)}+ \frac{1}{2}\,B^{(2)}\wedge C^{(2)}\,, 
\end{align}
and its dual four-vector field in the combination, 
\begin{align}
\eta^{\sfa_1\cdots \sfa_4}\equiv \gamma^{\sfa_1\cdots \sfa_4}-3\,\beta^{[\sfa_1\sfa_2}\,\gamma^{\sfa_3\sfa_4]}.
\end{align}
Finally, the 6-forms and the 6-vectors appear with the combination,
\begin{align}
 \bigl(B^{(6)\,\alpha}\bigr) \equiv \begin{pmatrix} C^{(6)} \cr B^{(6)} \end{pmatrix} \,\qquad \text{and} \qquad 
 \bigl(\beta^{\sfa_1\cdots \sfa_6}_{\alpha} \bigr) \equiv \begin{pmatrix} \gamma^{\sfa_1\cdots \sfa_6} \cr \beta^{\sfa_1\cdots \sfa_6} \end{pmatrix} \,. 
\end{align}

As in the case of the M-theory section, the generalized metric and vielbein appear with the factor,
\begin{align}
 \cM_{MN} \equiv \abs{G}^{\frac{1}{n-2}}\,M_{MN}\,\qquad \text{and} \qquad
 M_{MN} \equiv E_M{}^A\,E_N{}^B\,\delta_{AB} \,. 
\end{align}
In the following, we present the explicit parameterization for $E_M{}^A$ only. 

\subsubsection{$n=7$: $\mathfrak{G} = \SL(5)$}

In the case $\mathfrak{G} = \SL(5)$, the generalized vielbein parameterizations are given by
\begin{align}
 (E_M{}^A) \equiv \widehat{E}\, \Exp{\cB^{(2)}} \Exp{v} 
\qquad \text{or}\qquad
 (E_M{}^A) \equiv \widehat{E}\, \Exp{\beta^{(2)}} \Exp{v} \,,
\end{align}
where
\begin{align}
\begin{split}
 &\widehat{E} = \begin{pmatrix}
 e_\sfm{}^\sfb & 0 & 0 \\
 0 & \delta^\alpha_\beta \, e^\sfm{}_\sfb & 0 \\
 0 & 0 & e^{\sfm_1\sfm_2\sfm_3}_{\,\,\sfb_1\sfb_2\sfb_3}
 \end{pmatrix} ,
\qquad 
 \Exp{v}=
 \begin{pmatrix}
 \delta_\sfa^\sfb & 0 & 0 \\
 0 & \cV_{\alpha}{}^{\beta}\,\delta^\sfa_\sfb & 0 \\
 0 & 0 & \delta^{\sfa_1\sfa_2\sfa_3}_{\sfb_1\sfb_2\sfb_3}
 \end{pmatrix} ,
\\
 &\cB^{(2)}= \frac{1}{2}\,B_{\sfc_1\sfc_2}^{\gamma}\,\rho_{R^{\sfc_1\sfc_2}_{\gamma}}=
 \begin{pmatrix}
 0 & B^{\beta}_{\sfa\sfb} & 0 \\
 0 & 0 & \frac{3}{\sqrt{3!}}\,\epsilon_{\alpha\gamma}\,\delta^\sfa_{[\sfb_1}\,B_{\sfb_2\sfb_3]}^{\gamma} \\
 0 & 0 & 0
 \end{pmatrix} , 
\\
 &\beta^{(2)}
 = \frac{1}{2}\,\beta^{\sfc_1\sfc_2}_{\gamma}\, \rho_{R^{\gamma}_{\sfc_1\sfc_2}} = \begin{pmatrix}
 0 & 0 & 0 \\
 -\beta^{\sfa\sfb}_{\alpha} & 0 & 0 \\
 0 & \frac{3}{\sqrt{3!}}\,\epsilon^{\beta\gamma}\,\delta^{[\sfa_1}_{\sfb}\,\beta^{\sfa_2\sfa_3]}_{\gamma} & 0 
 \end{pmatrix} .
\end{split}
\end{align}

\subsubsection{$n=6$: $\mathfrak{G} = \SO(5,5)$}

In this case, the generalized vielbein parameterizations are given by
\begin{align}
 (E_M{}^A) \equiv \widehat{E}\, \Exp{\cD^{(4)}} \Exp{\cB^{(2)}} \Exp{v} 
\qquad \text{or}\qquad
 (E_M{}^A) \equiv \widehat{E}\, \Exp{\eta^{(4)}} \Exp{\beta^{(2)}} \Exp{v} \,,
\end{align}
with
\begin{align}
\begin{split}
 &\widehat{E} = \begin{pmatrix}
 e_\sfm{}^\sfb & 0 & 0 \\
 0 & \delta_{\alpha}^{\beta}\,e^\sfm{}_\sfb & 0 \\
 0 & 0 & e^{\sfm_1\sfm_2\sfm_3}_{\,\,\sfb_1\sfb_2\sfb_3} 
 \end{pmatrix} ,\qquad
 \Exp{v} = \begin{pmatrix}
 \delta_\sfa^\sfb & 0 & 0 \\
 0 & \cV_\alpha{}^{\beta}\,\delta^\sfa_\sfb & 0 \\
 0 & 0 & \delta^{\sfa_1\sfa_2\sfa_3}_{\sfb_1\sfb_2\sfb_3} 
 \end{pmatrix} ,
\\
 &\cB^{(2)} = \frac{1}{2}\,B_{\sfc_1\sfc_2}^{\gamma}\,\rho_{R^{\sfc_1\sfc_2}_{\gamma}}
 = \begin{pmatrix}
 0 & B^{\beta}_{\sfa \sfb} & 0 \\
 0 & 0 & \frac{3}{\sqrt{3!}}\,\epsilon_{\alpha\gamma}\, \delta^\sfa_{[\sfb_1}\,B_{\sfb_2\sfb_3]}^{\gamma} \\
 0 & 0 & 0 
 \end{pmatrix} ,
\\
 &\cD^{(4)}= \frac{1}{4!}\,D_{\sfc_1\cdots\sfc_4}\,\rho_{R^{\sfc_1\cdots\sfc_4}} 
 = \begin{pmatrix}
 0 & \qquad 0 & \frac{1}{\sqrt{3!}}\,D_{\sfa \sfb_1\sfb_2\sfb_3} \\
 0 & \qquad 0 & 0 \\
 0 & \qquad 0 & 0 
 \end{pmatrix} ,
\\
 &\beta^{(2)}
 = \frac{1}{2}\,\beta^{\sfc_1\sfc_2}_{\gamma}\, \rho_{R^{\gamma}_{\sfc_1\sfc_2}} = \ \begin{pmatrix}
 0 & 0 & 0 \\
 -\beta^{\sfa\sfb}_{\alpha} & 0 & 0 \\
 0 & \frac{3}{\sqrt{3!}}\,\epsilon^{\beta\gamma}\,\delta^{[\sfa_1}_{\sfb}\,\beta^{\sfa_2\sfa_3]}_{\gamma} & 0 
 \end{pmatrix} ,
\\
 &\eta^{(4)}= -\frac{1}{4!}\,\eta^{\sfc_1\cdots \sfc_4}\,\rho_{R_{\sfc_1\cdots \sfc_4}} = \begin{pmatrix}
 0 & 0 & \qquad 0 \\
 0 & 0 & \qquad 0 \\
 -\frac{1}{\sqrt{3!}}\,\eta^{\sfa_1\sfa_2\sfa_3 \sfb} & 0 & \qquad 0 
 \end{pmatrix} .
\end{split}
\end{align}

\subsubsection{$n=5$: $\mathfrak{G} = E_{6(6)}$}

In this case, the generalized vielbein is given by
\begin{align}
 (E_M{}^A) \equiv \widehat{E}\, \Exp{\cD^{(4)}} \Exp{\cB^{(2)}} \Exp{v} 
\qquad \text{or}\qquad
 (E_M{}^A) \equiv \widehat{E}\, \Exp{\eta^{(4)}} \Exp{\beta^{(2)}} \Exp{v} 
\end{align}
with
\begin{align}
 &\widehat{E} = \begin{pmatrix}
 e_\sfm{}^\sfb & 0 & 0 & 0 \\
 0 & \delta_{\alpha}^{\beta}\,e^\sfm{}_\sfb & 0 & 0 \\
 0 & 0 & e^{\sfm_1\sfm_2\sfm_3}_{\,\,\sfb_1\sfb_2\sfb_3} & 0 \\
 0 & 0 & 0 & \delta_{\alpha}^{\beta}\,e^{\sfm_1\cdots \sfm_5}_{\,\,\sfb_1\cdots \sfb_5}
 \end{pmatrix} ,\qquad
 \Exp{v} = \begin{pmatrix}
 \delta_\sfa^\sfb & 0 & 0 & 0 \\
 0 & \cV_\alpha{}^{\beta}\,\delta^\sfa_\sfb & 0 & 0 \\
 0 & 0 & \delta^{\sfa_1\sfa_2\sfa_3}_{\sfb_1\sfb_2\sfb_3} & 0 \\
 0 & 0 & 0 & \cV_\alpha{}^{\beta} \,\delta^{\sfa_1\cdots \sfa_5}_{\sfb_1\cdots \sfb_5}
 \end{pmatrix} ,
\\
 &\cB^{(2)} = \frac{1}{2}\,B_{\sfc_1\sfc_2}^{\gamma}\,\rho_{R^{\sfc_1\sfc_2}_{\gamma}}
 = \begin{pmatrix}
 0 & \qquad B^{\beta}_{\sfa \sfb} & 0 & 0 \\
 0 & \qquad 0 & \frac{3}{\sqrt{3!}}\,\epsilon_{\alpha\gamma}\, \delta^\sfa_{[\sfb_1}\,B_{\sfb_2\sfb_3]}^{\gamma} & 0 \\
 0 & \qquad 0 & 0 & \sqrt{5}\,\delta^{\sfa_1\sfa_2\sfa_3}_{[\sfb_1\sfb_2\sfb_3}\,B^{\beta}_{\sfb_4\sfb_5]} \\
 0 & \qquad 0 & 0 & 0 
 \end{pmatrix} ,
\\
 &\cD^{(4)}= \frac{1}{4!}\,D_{\sfc_1\cdots\sfc_4}\,\rho_{R^{\sfc_1\cdots\sfc_4}}
 = \begin{pmatrix}
 0 & \qquad 0 & \ \frac{1}{\sqrt{3!}}\,D_{\sfa \sfb_1\sfb_2\sfb_3} & 0 \\
 0 & \qquad 0 & \ 0 & -\frac{5}{\sqrt{5!}}\,\delta^{\beta}_{\alpha}\,\delta^\sfa_{[\sfb_1}\,D_{\sfb_2\cdots \sfb_5]} \\
 0 & \qquad 0 & \ 0 & 0 \\
 0 & \qquad 0 & \ 0 & 0
 \end{pmatrix} ,
\\
 &\beta^{(2)}
 = \ \ \frac{1}{2}\,\beta^{\sfc_1\sfc_2}_{\gamma}\, \rho_{R^{\gamma}_{\sfc_1\sfc_2}} \ \ = \ \ \begin{pmatrix}
 0 & 0 & 0 & 0 \\
 -\beta^{\sfa\sfb}_{\alpha} & 0 & 0 & 0 \\
 0 & \frac{3}{\sqrt{3!}}\,\epsilon^{\beta\gamma}\,\delta^{[\sfa_1}_{\sfb}\,\beta^{\sfa_2\sfa_3]}_{\gamma} & 0 & 0 \\
 0 & 0 & \sqrt{5} \,\delta^{[\sfa_1\sfa_2\sfa_3}_{\sfb_1\sfb_2\sfb_3}\,\beta^{\sfa_4\sfa_5]}_{\alpha} & 0 
 \end{pmatrix} ,
\\
 &\eta^{(4)}= -\frac{1}{4!}\,\eta^{\sfc_1\cdots \sfc_4}\,\rho_{R_{\sfc_1\cdots \sfc_4}} = \begin{pmatrix}
 0 & 0 & 0 & \qquad 0 \\
 0 & 0 & 0 & \qquad 0 \\
 -\frac{1}{\sqrt{3!}}\,\eta^{\sfa_1\sfa_2\sfa_3 \sfb} & 0 & 0 & \qquad 0 \\
 0 & -\frac{5}{\sqrt{5!}}\,\delta_{\alpha}^{\beta}\,\delta^{[\sfa_1}_{\sfb}\,\eta^{\sfa_2\cdots \sfa_5]} & 0 & \qquad 0 
 \end{pmatrix} .
\end{align}

\subsubsection{$n=4$: $\mathfrak{G} = E_{7(7)}$}

In this case, the generalized vielbein is given by
\begin{align}
 (E_M{}^A) \equiv \widehat{E}\, \Exp{\cB^{(6)}} \Exp{\cD^{(4)}} \Exp{\cB^{(2)}} \Exp{v} 
\qquad \text{or}\qquad
 (E_M{}^A) \equiv \widehat{E}\, \Exp{\beta^{(6)}} \Exp{\eta^{(4)}} \Exp{\beta^{(2)}} \Exp{v} 
\end{align}
where
\begin{align}
 &\widehat{E} = {\footnotesize\begin{pmatrix}
 e_\sfm{}^\sfb & 0 & 0 & 0 & 0 \\
 0 & \delta_{\alpha}^{\beta}\,e^\sfm{}_\sfb & 0 & 0 & 0 \\
 0 & 0 & e^{\sfm_1\sfm_2\sfm_3}_{\,\,\sfb_1\sfb_2\sfb_3} & 0 & 0 \\
 0 & 0 & 0 & \delta_{\alpha}^{\beta}\,e^{\sfm_1\cdots \sfm_5}_{\,\,\sfb_1\cdots \sfb_5} & 0 \\
 0 & 0 & 0 & 0 & e^{-1}\,e^\sfm{}_\sfb
 \end{pmatrix}} ,
\quad
 \Exp{v} =
 {\footnotesize\begin{pmatrix}
 \delta_\sfa^\sfb & 0 & 0 & 0 & 0 \\
 0 & \cV_\alpha{}^{\beta}\,\delta^\sfa_\sfb & 0 & 0 & 0 \\
 0 & 0 & \delta^{\sfa_1\sfa_2\sfa_3}_{\sfb_1\sfb_2\sfb_3} & 0 & 0 \\
 0 & 0 & 0 & \cV_\alpha{}^{\beta} \,\delta^{\sfa_1\cdots \sfa_5}_{\sfb_1\cdots \sfb_5} & 0 \\
 0 & 0 & 0 & 0 & \delta^\sfa_\sfb 
 \end{pmatrix}} ,
\\
 &\cB^{(2)} = \frac{1}{2}\,B_{\sfc_1\sfc_2}^{\gamma}\,\rho_{R^{\sfc_1\sfc_2}_{\gamma}}
 ={\footnotesize\begin{pmatrix}
 0 & \qquad B^{\beta}_{\sfa \sfb} & 0 & 0 & 0 \\
 0 & \qquad 0 & \frac{3}{\sqrt{3!}}\,\epsilon_{\alpha\gamma}\, \delta^\sfa_{[\sfb_1}\,B_{\sfb_2\sfb_3]}^{\gamma} & 0 & 0 \\
 0 & \qquad 0 & 0 & \sqrt{5}\,\delta^{\sfa_1\sfa_2\sfa_3}_{[\sfb_1\sfb_2\sfb_3}\, B^{\beta}_{\sfb_4\sfb_5]} & 0
 \\
 0 & \qquad 0 & 0 & 0 & -\frac{1}{\sqrt{5!}}\,\epsilon_{\alpha\gamma}\,\epsilon^{\sfa_1\cdots \sfa_5 \sfc}\,B^{\gamma}_{\sfc \sfb} \\
 0 & \qquad 0 & 0 & 0 & 0
 \end{pmatrix}} ,
\\
 &\cD^{(4)} = \frac{1}{4!}\,D_{\sfc_1\cdots\sfc_4}\,\rho_{R^{\sfc_1\cdots\sfc_4}} =
 {\footnotesize\begin{pmatrix}
 0 & 0 & \frac{1}{\sqrt{3!}}\,D_{\sfa \sfb_1\sfb_2\sfb_3} & 0 & 0 \\
 0 & 0 & 0 & -\frac{5}{\sqrt{5!}}\,\delta^{\beta}_{\alpha}\,\delta^\sfa_{[\sfb_1}\,D_{\sfb_2\cdots \sfb_5]} & 0 \\
 0 & 0 & 0 & 0 & -\frac{1}{3!\sqrt{3!}}\, \epsilon^{\sfa_1\sfa_2\sfa_3 \sfc_1\sfc_2\sfc_3}\,D_{\sfc_1\sfc_2\sfc_3 \sfb} \\
 0 & 0 & 0 & 0 & 0 \\
 0 & 0 & 0 & 0 & 0
 \end{pmatrix}} , 
\\
 &\cB^{(6)}= \ \frac{1}{6!}\,B_{\sfc_1\cdots\sfc_6}^{\gamma}\,\rho_{R^{\sfc_1\cdots\sfc_6}_{\gamma}}=
 {\footnotesize\begin{pmatrix}
\ \ 0 & \quad \qquad 0 & \qquad \qquad 0 & \qquad -\frac{1}{\sqrt{5!}}\, B_{\sfa \sfb_1\cdots \sfb_5}^{\beta} & 0 \\
\ \ 0 & \quad \qquad 0 & \qquad \qquad 0 & \qquad 0 & \frac{1}{6!}\,\epsilon_{\alpha\gamma}\, \epsilon^{\sfc_1\cdots \sfc_6}\,B^{\gamma}_{\sfc_1\cdots \sfc_6} \,\delta^\sfa_\sfb \\
\ \ 0 & \quad \qquad 0 & \qquad \qquad 0 & \qquad 0 & 0 \\
\ \ 0 & \quad \qquad 0 & \qquad \qquad 0 & \qquad 0 & 0 \\
\ \ 0 & \quad \qquad 0 & \qquad \qquad 0 & \qquad 0 & 0
\end{pmatrix}}\,, 
\\
 &\beta^{(2)} \ 
 = \ \ \frac{1}{2}\,\beta^{\sfc_1\sfc_2}_{\gamma}\, \rho_{R^{\gamma}_{\sfc_1\sfc_2}} \ \ = \ \ 
 {\footnotesize\begin{pmatrix}
\ 0 & 0 & 0 & 0 & 0 \\
\ -\beta^{\sfa \sfb}_{\alpha} & 0 & 0 & 0 & 0 \\
\ 0 & \frac{3}{\sqrt{3!}}\,\epsilon^{\beta\gamma}\,\delta^{[\sfa_1}_{\sfb}\,\beta^{\sfa_2\sfa_3]}_{\gamma} & 0 & 0 & 0 \\
\ 0 & 0 & \sqrt{5} \,\delta^{[\sfa_1\sfa_2\sfa_3}_{\sfb_1\sfb_2\sfb_3}\,\beta^{\sfa_4\sfa_5]}_{\alpha} & 0 & 0 \\
\ 0 & 0 & 0 & -\frac{1}{\sqrt{5!}}\,\epsilon^{\beta\gamma}\,\epsilon_{\sfb_1\cdots \sfb_5 \sfc}\,\beta^{\sfc \sfa}_{\gamma} & 0
 \end{pmatrix}} ,
\\
 &\eta^{(4)}= -\frac{1}{4!}\,\eta^{\sfc_1\cdots \sfc_4}\,\rho_{R_{\sfc_1\cdots \sfc_4}} =
 {\footnotesize\begin{pmatrix}
 0 & 0 & 0 & 0 & 0 \\
 0 & 0 & 0 & 0 & 0 \\
 -\frac{1}{\sqrt{3!}}\,\eta^{\sfa_1\sfa_2\sfa_3 \sfb} & 0 & 0 & 0 & 0 \\
 0 & -\frac{5}{\sqrt{5!}}\,\delta_{\alpha}^{\beta}\,\delta^{[\sfa_1}_{\sfb}\,\eta^{\sfa_2\cdots \sfa_5]} & 0 & 0 & 0 \\
 0 & 0 & -\frac{1}{3!\sqrt{3!}}\,\epsilon_{\sfb_1\sfb_2\sfb_3 \sfc_1\sfc_2\sfc_3}\,\eta^{\sfc_1\sfc_2\sfc_3 \sfa} & 0 & 0
 \end{pmatrix}} , 
\\
 &\beta^{(6)}= \frac{1}{6!}\,\beta^{\sfc_1\cdots\sfc_6}_{\gamma}\, \rho_{R_{\sfc_1\cdots \sfc_6}^{\gamma}} =
 {\footnotesize\begin{pmatrix}
 0 & 0 & \quad 0 & \quad \qquad 0 & \quad \qquad 0 \\
 0 & 0 & \quad 0 & \quad \qquad 0 & \quad \qquad 0 \\
 0 & 0 & \quad 0 & \quad \qquad 0 & \quad \qquad 0 \\
 \frac{1}{\sqrt{5!}}\, \beta^{\sfa_1\cdots \sfa_5 \sfb}_{\alpha} & 0 & \quad 0 & \quad \qquad 0 & \quad \qquad 0 \\
 0 & \frac{1}{6!}\,\epsilon^{\beta\gamma}\,\epsilon_{\sfc_1\cdots \sfc_6}\,\beta_{\gamma}^{\sfc_1\cdots \sfc_6}\, \delta^\sfa_\sfb & \quad 0 & \quad \qquad 0 & \quad \qquad 0
\end{pmatrix}} \,.
\end{align}

\subsection{Ten-dimensional effective action}
\label{sec:type-IIB-action}

The type IIB supergravity action is also obtainable from solving the section condition, such that the ten-dimensional coordinates are given by $(x^{\hat{\mu}})\equiv (x^\mu,\,x^\sfm)$ ($\hat{\mu}=0,\dotsc,9$).
See Appendix \ref{app:action} for details. 
For instance, for the $E_{7(7)}$ EFT in the geometric parameterization, the action becomes
\begin{align}
 &\cL= \abs{\sfG}^{\frac{1}{2}}\,
 \biggl[R(\sfG) + \frac{1}{4}\,\sfG^{\hat{\mu}\hat{\nu}}\,\partial_{\hat{\mu}} m_{\alpha\beta}\,\partial_{\hat{\nu}} m^{\alpha\beta}
\nn\\
 &\qquad\qquad -\sfg^{\mu\nu}\,\Bigl(\frac{1}{2 \cdot 2!} \ m_{\alpha\beta}\,G^{\sfm_1\sfm_2,\,\sfn_1\sfn_2} \,\partial_\mu B^\alpha_{\sfm_1\sfm_2}\,\partial_\nu B^\beta_{\sfn_1\sfn_2} \nn \\
 & \qquad \qquad \qquad 
+\ \frac{1}{2 \cdot 4!} \ G^{\sfm_1\cdots \sfm_4,\,\sfn_1\cdots \sfn_4} \, \cG_{\mu,\,\sfm_1\cdots \sfm_4}\,\cG_{\nu,\,\sfn_1\cdots \sfn_4}
\nn\\
 &\qquad\qquad\qquad+\frac{1}{6!} \ m_{\alpha\beta}\,G^{\sfm_1\cdots \sfm_6,\,\sfn_1\cdots \sfn_6} \, \cG^\alpha_{\mu,\,\sfm_1\cdots \sfm_6}\,\cG^\beta_{\nu,\,\sfn_1\cdots \sfn_6}\Bigr) 
\nn\\
 &\qquad\qquad - \frac{1}{2 \cdot 3!} \ m_{\alpha\beta}\,G^{\sfm_1\sfm_2\sfm_3,\,\sfn_1\sfn_2\sfn_3} \, H^\alpha_{\sfm_1\sfm_2\sfm_3}\,H^{\beta}_{\sfn_1\sfn_2\sfn_3}
 \nn \\
&\qquad \qquad -\frac{1}{2 \cdot 5!} \ G^{\sfm_1\cdots \sfm_5,\,\sfn_1\cdots \sfn_5} \, G_{\sfm_1\cdots \sfm_5}\,G_{\sfn_1\cdots \sfn_5}\biggr] \,,
\end{align}
where the field strengths are 
\begin{align}
 &\cG_{\mu,\,\sfm_1\cdots \sfm_4} \equiv \partial_\mu D_{\sfm_1\cdots \sfm_4}-3\,\epsilon_{\gamma\delta}\,B^\gamma_{[\sfm_1\sfm_2\vert}\,\partial_\mu B^\delta_{\vert \sfm_3\sfm_4]}\,,
\\
 &\cG^\beta_{\mu,\,\sfm_1\cdots \sfm_6} \equiv \partial_\mu B^\beta_{\sfm_1\cdots \sfm_6} - 15\, B^\beta_{[\sfm_1\sfm_2\vert} \,\partial_\mu D_{\vert\sfm_3\cdots \sfm_6]} + 15\,\epsilon_{\gamma\delta}\,B^\beta_{[\sfm_1\sfm_2}\,B^\gamma_{\sfm_3\sfm_4\vert}\,\partial_\mu B^\delta_{\vert \sfm_5\sfm_6]} \,,
\label{eq:our-6-form}
\\
 &H^\alpha_{\sfm_1\sfm_2\sfm_3} \equiv 3\,\partial_{[\sfm_1} B^\alpha_{\sfm_2\sfm_3]}\,, \nn \\
& G_{\sfm_1\cdots \sfm_5} \equiv 5\, \partial_{[\sfm_1} C_{\sfm_2\cdots \sfm_5]}+30\,H^1_{[\sfm_1\sfm_2\sfm_3} C_{\sfm_4\sfm_5]} \,, 
\end{align}
and, for the ten-dimensional metric,
\begin{align}
 \bigl(\sfG_{\hat{\mu}\hat{\nu}}\bigr) \equiv \begin{pmatrix}
 \sfg_{\mu\nu} & 0 \cr 0 & G_{\sfm\sfn}
 \end{pmatrix}\,,
\end{align}
the associated Ricci scalar is given by
\begin{align}
 R(\sfG)&\equiv R(\sfg) +\sfg^{\mu\nu}\, \left[ \frac{1}{4}\,\partial_\mu G^{\sfm\sfn}\, \partial_\nu G_{\sfm\sfn}
 +\frac{1}{4}\, \partial_\mu\ln\abs{G}\,\partial_\nu\ln\abs{G} \right]
\nn\\
 &\ + R(G) + G^{\sfm\sfn}\, \left[ \frac{1}{4}\,\partial_\sfm \sfg^{\mu\nu}\,\partial_\sfn \sfg_{\mu\nu}
       + \frac{1}{4}\,G^{\sfm\sfn}\, \partial_\sfm \ln \abs{\sfg}\, \partial_\sfn \ln \abs{\sfg} \right] \,.
\end{align}
In the standard definitions of six-form potentials, the field strengths are given by
\begin{align}
 G^{(7)} &= \rmd C_{\text{(std.)}}^{(6)} + H^{(3)}\wedge C^{(4)}\,,
\\
 H^{(7)} &= \rmd B_{\text{(std.)}}^{(6)} + C^{(4)}\wedge \rmd C^{(2)} -\frac{1}{2}\,C^{(2)}\wedge C^{(2)}\wedge H^{(3)} + C^{(0)}\, G^{(7)} \,,
\end{align}
and the corresponding expression for $\cG^\alpha_{\mu,\,\sfm_1\cdots \sfm_6}$ should be
\begin{align}
 \cG^1_{\mu,\,\sfm_1\cdots \sfm_6} 
 &= \partial_\mu C^{\text{(std.)}}_{\sfm_1\cdots \sfm_6} + 15\, C_{[\sfm_1\cdots \sfm_4\vert}\, \partial_\mu B_{\vert\sfm_5\sfm_6]} \,,
\\
 \cG^2_{\mu,\,\sfm_1\cdots \sfm_6} 
 &= \partial_\mu B^{\text{(std.)}}_{\sfm_1\cdots \sfm_6} + 15\, C_{[\sfm_1\cdots \sfm_4\vert} \,\partial_\mu C_{\vert\sfm_5\sfm_6]} - 45\,C_{[\sfm_1\sfm_2}\,C_{\sfm_3\sfm_4\vert}\,\partial_\mu B_{\vert \sfm_5\sfm_6]} \,. 
\end{align}
Comparing these with the expression \eqref{eq:our-6-form}, we see that the conventional six-form potentials are related to the six-form potentials, $C_{\sfm_1\cdots \sfm_6}$ and $B_{\sfm_1\cdots \sfm_6}$, by
\begin{align}
 C^{\text{(std.)}}_{\sfm_1\cdots \sfm_6}
 &\equiv C_{\sfm_1\cdots \sfm_6}-15\,D_{[\sfm_1\cdots \sfm_4}\,B_{\sfm_5\sfm_6]} + 15\, B_{[\sfm_1\sfm_2}\,B_{\sfm_3\sfm_4}\, C_{\sfm_5\sfm_6]}\,,
\\
 B^{\text{(std.)}}_{\sfm_1\cdots \sfm_6}
 &\equiv B_{\sfm_1\cdots \sfm_6} - 15\,D_{[\sfm_1\cdots \sfm_4}\,C_{\sfm_5\sfm_6]} +30\,C_{[\sfm_1\sfm_2}\,C_{\sfm_3\sfm_4}\,B_{\sfm_5\sfm_6]}\,.
\end{align}
This completes the conventional, geometric parameterization. 

In the non-geometric parameterization, we obtain
\begin{align}
 &\cL = \abs{\widetilde{\sfG}}^{\frac{1}{2}}\,
 \biggl[R(\widetilde{\sfG}) + \frac{1}{4}\,\widetilde{\sfG}^{\hat{\mu}\hat{\nu}}\,\partial_{\hat{\mu}} \widetilde{m}_{\alpha\beta}\,\partial_{\hat{\nu}} \widetilde{m}^{\alpha\beta} 
\nn\\
 &\qquad\qquad - \frac{\widetilde{m}^{\alpha\beta}}{2}\,\bigl\langle Q_\alpha^{(1,\,2)},\,Q_\beta^{(1,\,2)}\bigr\rangle
- \frac{1}{2}\,\abs{P^{(1,\,4)}}^2
- \frac{\widetilde{m}^{\alpha\beta}}{2}\,\bigl\langle Q_\alpha^{(1,\,6)},\,Q_\beta^{(1,\,6)}\bigr\rangle \biggr]\,,
\end{align}
where we defined
\begin{align}
 Q_{\alpha,\,\hat{\mu}}{}^{\sfm \sfn} &\equiv \partial_{\hat{\mu}} \beta_\alpha^{\sfm \sfn}\,,\quad 
 P_{\hat{\mu}}{}^{\sfm_1\cdots \sfm_4}\equiv 
 \partial_{\hat{\mu}} \eta^{\sfm_1\cdots \sfm_4} + 3 \, \epsilon^{\gamma\delta}\, \beta_\gamma^{[\sfm_1\sfm_2}\,\partial_{\hat{\mu}} \beta_\delta^{\sfm_3\sfm_4]} \,,
\\
 Q_{\alpha,\,\hat{\mu}}{}^{\sfm_1\cdots \sfm_6}&\equiv \partial_{\hat{\mu}} \beta_\alpha^{\sfm_1\cdots \sfm_6} +15\,\beta_\alpha^{[\sfm_1\sfm_2}\,\partial_{\hat{\mu}} \eta^{\sfm_2\cdots \sfm_6]} + 15\,\epsilon^{\gamma\delta}\,\beta_\alpha^{[\sfm_1\sfm_2}\,\beta_\gamma^{\sfm_3\sfm_4}\,\partial_{\hat{\mu}} \beta_\delta^{\sfm_5\sfm_6]} \,,
\\
 \bigl\langle Q_\alpha^{(1,\,2)},\,Q_\beta^{(1,\,2)}\bigr\rangle
 &\equiv \frac{\widetilde{\sfG}^{\hat{\mu}\hat{\nu}}\,\widetilde{G}_{\sfm_1\sfm_2,\,\sfn_1\sfn_2}}{2!}\,Q_{\alpha,\,\hat{\mu}}{}^{\sfm_1\sfm_2}\, Q_{\beta,\,\hat{\nu}}{}^{\sfn_1\sfn_2} \,,
\\
 \abs{P^{(1,\,4)}}^2
 &\equiv \frac{\widetilde{\sfG}^{\hat{\mu}\hat{\nu}}\,\widetilde{G}_{\sfm_1\cdots \sfm_4,\,\sfn_1\cdots \sfn_4}}{4!}\, P_{\hat{\mu}}{}^{\sfm_1\cdots \sfm_4}\,P_{\hat{\nu}}{}^{\sfn_1\cdots \sfn_4}\,,
\\
 \bigl\langle Q_\alpha^{(1,\,6)},\,Q_\beta^{(1,\,6)}\bigr\rangle
 &\equiv \frac{\widetilde{\sfG}^{\hat{\mu}\hat{\nu}}\,\widetilde{G}_{\sfm_1\cdots \sfm_6,\,\sfn_1\cdots \sfn_6}}{6!} \, Q_{\alpha,\,\hat{\mu}}{}^{\sfm_1\cdots \sfm_6}\,Q_{\beta,\,\hat{\nu}}{}^{\sfn_1\cdots \sfn_6} \,.
\end{align}
The above action is manifestly $\SL(2)$-invariant. This action can also be expressed in the following form:
\begin{align}
 &\cL = \abs{\widetilde{\sfG}}^{\frac{1}{2}}\,
 \biggl[R(\widetilde{\sfG}) - \frac{1}{2}\,\Bigl(\abs{\rmd\phi}^2 
 +\Exp{\widetilde{\phi}} \abs{\cQ^{(1,\,2)}}^2
 +\Exp{-\widetilde{\phi}} \abs{\cQ^{(1,\,6)}}^2
\nn\\
 &\qquad\qquad\qquad\qquad\quad 
 +\Exp{-2\widetilde{\phi}} \abs{\cP^{(1,\,0)}}^2 
 +\Exp{-\widetilde{\phi}} \abs{\cP^{(1,\,2)}}^2
 + \abs{\cP^{(1,\,4)}}^2
 +\Exp{\widetilde{\phi}} \abs{\cP^{(1,\,6)}}^2 \Bigr)\biggr]\,,
\end{align}
where
\begin{align}
 \abs{\rmd\phi}^2\ \ &\equiv \widetilde{\sfG}^{\hat{\mu}\hat{\nu}}\,\partial_{\hat{\mu}}\phi\,\partial_{\hat{\nu}} \phi \,,
\\
 \abs{\cQ^{(1,\,p)}}^2
 &\equiv \frac{1}{p!}\, \widetilde{\sfG}^{\hat{\mu}\hat{\nu}}\,\widetilde{G}_{\sfm_1\cdots \sfm_p,\,\sfn_1\cdots \sfn_p}\, \cQ_{\hat{\mu}}{}^{\sfm_1\cdots \sfm_p}\,\cQ_{\hat{\nu}}{}^{\sfn_1\cdots \sfn_p}\,,
\\
 \abs{\cP^{(1,\,p)}}^2
 &\equiv \frac{1}{p!}\, \widetilde{\sfG}^{\hat{\mu}\hat{\nu}}\,\widetilde{G}_{\sfm_1\cdots \sfm_p,\,\sfn_1\cdots \sfn_p}\, \cP_{\hat{\mu}}{}^{\sfm_1\cdots \sfm_p}\,\cP_{\hat{\nu}}{}^{\sfn_1\cdots \sfn_p}\,,
\\
 \cQ_{\hat{\mu}}{}^{\sfm_1\sfm_2} 
 &\equiv Q_{\hat{\mu}}{}^{\sfm_1\sfm_2} \equiv \partial_{\hat{\mu}}\beta^{\sfm_1\sfm_2}\,,
\\
 \cQ_{\hat{\mu}}{}^{\sfm_1\cdots \sfm_6} 
 &\equiv Q_{\hat{\mu}}{}^{\sfm_1\cdots \sfm_6}-\gamma^{(0)}\,P_{\hat{\mu}}{}^{\sfm_1\cdots \sfm_6}\,, \quad 
 (Q_{\alpha,\,\hat{\mu}}{}^{\sfm_1\cdots \sfm_6})\equiv \begin{pmatrix}
 P_{\hat{\mu}}{}^{\sfm_1\cdots \sfm_6} \cr Q_{\hat{\mu}}{}^{\sfm_1\cdots \sfm_6}
 \end{pmatrix} \,,
\\
 \cP_{\hat{\mu}} 
 &\equiv P_{\hat{\mu}} \equiv \partial_{\hat{\mu}} \gamma^{(0)}\,,
\\
 \cP_{\hat{\mu}}{}^{\sfm_1\sfm_2} 
 &\equiv P_{\hat{\mu}}{}^{\sfm_1\sfm_2}-\gamma^{(0)}\,Q_{\hat{\mu}}{}^{\sfm_1\sfm_2}
 = \partial_{\hat{\mu}} \gamma^{\sfm_1\sfm_2}-\gamma^{(0)}\,\partial_{\hat{\mu}} \beta^{\sfm_1\sfm_2}\,,
\\
 \cP_{\hat{\mu}}{}^{\sfm_1\cdots \sfm_4} 
 &\equiv P_{\hat{\mu}}{}^{\sfm_1\cdots \sfm_4}
 =\partial_{\hat{\mu}} \eta^{\sfm_1\cdots \sfm_4} + 3 \, \epsilon^{\gamma\delta}\, \beta_\gamma^{[\sfm_1\sfm_2}\,\partial_{\hat{\mu}} \beta_\delta^{\sfm_3\sfm_4]}
\nn\\
 &=\partial_{\hat{\mu}} \gamma^{\sfm_1\cdots \sfm_4} - 6 \, \gamma^{[\sfm_1\sfm_2}\,\partial_{\hat{\mu}} \beta^{\sfm_3\sfm_4]}\,,
\\
 \cP_{\hat{\mu}}{}^{\sfm_1\cdots \sfm_6} 
 &\equiv P_{\hat{\mu}}{}^{\sfm_1\cdots \sfm_6}
 = \partial_{\hat{\mu}} \gamma^{\sfm_1\cdots \sfm_6} +15\, \beta^{[\sfm_1\sfm_2}\,\partial_{\hat{\mu}} \eta^{\sfm_3\cdots \sfm_6]} + 15\,\epsilon^{\gamma\delta}\,\beta^{[\sfm_1\sfm_2}\,\beta_\gamma^{\sfm_3\sfm_4}\,\partial_{\hat{\mu}} \beta_\delta^{\sfm_5\sfm_6]} 
\nn\\
 &= \partial_{\hat{\mu}} \tilde{\gamma}^{\sfm_1\cdots \sfm_6} - 15\, \gamma^{[\sfm_1\cdots \sfm_4}\, \partial_{\hat{\mu}}\beta^{\sfm_5\sfm_6]}\,,
\end{align}
and we defined $\tilde{\gamma}^{\sfm_1\cdots \sfm_6} \equiv \gamma^{\sfm_1\cdots \sfm_6} + 15\,\eta^{[\sfm_1\cdots \sfm_4}\,\beta^{\sfm_5\sfm_6]}+15\,\beta^{[\sfm_1\sfm_2}\,\beta^{\sfm_3\sfm_4}\,\gamma^{\sfm_5\sfm_6]}$\,. 
Finally, in the string frame, $(\sfG_{\textrm{str}})_{\hat{\mu}\hat{\nu}}\equiv \Exp{\frac{\widetilde{\phi}}{2}}\sfG_{\hat{\mu}\hat{\nu}}$, the above action becomes
\vskip0.5cm
\begin{tcolorbox}
\vskip-0.5cm
\begin{align}
 &\cL = \abs{\widetilde{\sfG}_{\textrm{str}}}^{\frac{1}{2}} \,
 \biggl[\Exp{-2\widetilde{\phi}} \Bigl(R(\widetilde{\sfG}_{\textrm{str}}) + 4\,\abs{\rmd \phi}_{\textrm{str}}^2 - \frac{1}{2}\, \abs{\cQ^{(1,\,2)}}_{\textrm{str}}^2\Bigr)
\nn\\
 &\qquad\qquad
 - \frac{1}{2} \Exp{-4\widetilde{\phi}} \Bigl( \abs{\cP^{(1,\,0)}}_{\textrm{str}}^2 
  + \abs{\cP^{(1,\,2)}}_{\textrm{str}}^2
  + \abs{\cP^{(1,\,4)}}_{\textrm{str}}^2
  + \abs{\cP^{(1,\,6)}}_{\textrm{str}}^2 \Bigr)
  \nn \\
  &\qquad \qquad - \frac{1}{2} \Exp{-6\widetilde{\phi}} \abs{\cQ^{(1,\,6)}}_{\textrm{str}}^2 
\biggr]\,. 
\label{eq:IIB-non-geometric}
\end{align}
\end{tcolorbox}
\noindent 
This action generalize the actions of $\beta$-supergravity and its extension obtained in \cite{Sakatani:2014hba}.

\section{Non-geometric Fluxes in Heterotic DFT}
\label{sec:HDFT}
In this section, we generalize the above constructions to the heterotic DFT, which incorporates the Yang-Mills theory with heterotic gauge group $G_{\rm YM} = \SO(32)$ or $E_8\times E_8$ to the $\ODD$ DFT in a T-duality covariant manner.  The $\ODDG$ gauged DFT provides an elegant framework for describing the heterotic DFT by combining the string NS-NS sector and gauge fields into a single $\ODDG$ multiplet \cite{Hohm:2011ex,Grana:2012rr}. A similar approach has been developed for studying the leading $\alpha'$-corrections in the heterotic DFT \cite{Bedoya:2014pma}. The main idea is to extend the heterotic gauge group $G_{\rm YM}$ by including the $G_{\rm LL} = \Spin(1,9)$ local Lorentz group: 
\begin{equation}
  G = G_{\rm YM} \times G_{\rm LL}\,. 
\end{equation}
The DFT spin-connection can be understood as the gauge field for $G_{\rm LL}$ acting on adjoint representations, so the heterotic Yang-Mills gauge fields and DFT spin-connection are treatable on an equal footing \cite{Andriot:2011iw,Garcia-Fernandez:2013gja,Baraglia:2013wua,Anderson:2014xha,delaOssa:2014cia,delaOssa:2014msa}. As our formalism works equally well for arbitrary gauge groups, we do not specify the gauge group $G$ explicitly. We will thus treat $G$ as an arbitrary Lie group until we need to work for the heterotic gauge group. 

Unlike the $\ODD$ or $E_{d(d)}$ cases, the $\ODDG$ is not a split real form, so its algebra contains non-compact Cartan generators. In section \ref{sec:general},  we tacitly assumed that the duality group $G$ to be a maximally non-compact group, thus we need to slightly modify the previous construction \cite{Lu:1998xt}. In this case, the  Iwasawa decomposition reads
\begin{equation}
  g = k\, a\, n\, \quad \in \quad \ODDG\,,
\label{IwasawaODDG}\end{equation}
where $k$ is an element of the maximal compact subgroup $\OoD\otimes\ODG$, $a$ is an element of the maximal non-compact Abelian subgroup, and $n$ is an element of the nilpotent subgroup generated by the positive (negative) root generators. Also, Cartan involution flips the sign of non-compact generators only. Note that the non-compact Cartan generators and positive (negative) root generators form a solvable Lie algebra, which is a subset of Borel subgroup. If we assume $G$ is a maximally non-compact group, the solvable Lie group is restored to Borel subgroup. Using (\ref{IwasawaODDG}) we will define non-geometric parameterization of generalized vielbein and non-geometric fluxes for the heterotic DFT.

\subsection{Parameterization of generalized vielbein}

The fundamental field variables of the heterotic DFT are furnished by $\ODDG$ generalized metric field $\cH$ and dilaton $d$ in a parameterization independent way. As the usual DFTs, the generalized metric is defined by a symmetric $\ODDG$ matrix satisfying
\begin{equation}
  \cH \cJ^{-1} \cH = \cJ\,,
\label{defODDG}\end{equation}
where $\cJ$ is the $\ODDG$ metric. In order to interpret the heterotic DFT as the heterotic supergravity, we need to impose a suitable parameterization of the generalized metric in terms of the supergravity fields. The simplest way is to solve (\ref{defODDG}), assuming that the upper-left conner is non-degenerate. However, such a parameterization is not unique due to the freedom of $\ODD$ transformation \footnote{The duality group for heterotic DFT with unbroken Yang-Mills gauge symmetry is given by just $\ODD$ rather than full $\ODDG$. This is because  there should be no mixing between NS-NS sector and Yang-Mills sector \cite{Hohm:2011ex,Grana:2012rr}. The extended duality group $\ODDG$ is a formal device to describe Yang-Mills sector within duality covariant framework.}.

The geometric parameterization, which yields the usual heterotic supergravity, is one possible choice among infinitely many viable parameterizations. The others are so-called non-geometric parameterizations in the sense that they cannot be represented in terms of the conventional supergravity fields. In this section, we shall focus on a particular non-geometric parameterization, which is associated to the $\mathbb{Z}_2$ part within the T-duality group, and refer to this as the non-geometric parameterization. 

One can introduce a local frame field $\cE_{\hM \hA}$ for heterotic DFT in terms of the gauged DFT \cite{Geissbuhler:2013uka,Berman:2013cli}. The local structure group is given by the maximal compact subgroup of $\ODDG$,
\begin{equation}
  K = \OoD \times \ODG \quad \subset \quad \ODDG\,.
\end{equation}
Geometrically, the physical degrees of freedom of heterotic DFT is represented by a local orthonormal frame field, so-called the generalized vielbein (or double-vielbein):
\begin{equation}
  \cE_{\hM \hA} =\{V_{\hM m}\,, \brV_{\hM \hbrm}\}.
\end{equation}
Here, $\hM$ is an $\ODDG$ vector index, $m$ is an $\OoD$ vector index, and $\hbrm$ is an $\ODG$ vector index. Under the local structure group, $V_{\widehat{M}}{}^{m}$ and $\bar{V}_{\widehat{M}}{}^{\hat{\bar{m}}}$ transforms
\begin{equation}
  V_{\hM}{}^{m} \to \Lambda^{m}{}_{n} V_{\hM}{}^{n}\,, \qquad \bar{V}_{\hM}{}^{\hat{\bar{m}}} \to \bar{\Lambda}^{\hat{\bar{m}}}{}_{\hat{\bar{n}}} \bar{V}_{\hM}{}^{\hat{\bar{n}}}
\label{local_Lorentz}\end{equation}
As we discussed in the last section, the double-vielbein is parameterized by the coset 
\begin{equation}
  \frac{\ODDG}{\OoD\times\ODG} \,.
\end{equation}

A necessary step in identifying the gauged DFT with heterotic supergravity is to fix the parameterization of generalized vielbein in terms of field variables of heterotic supergravity. To this end, it is necessary to decompose $\ODDG$ vector indices $\hM = \{M\,, \alpha\}$ and $\ODG$ frame indices $\hbrm = \{\brm\,, \bra\}$. We first decompose the $\ODDG$ metric and  $\ODG$ metric as
\begin{equation}
  \cJ_{\hM\hN} = \bpm \cJ_{MN} & 0 \\ 0& \frac{1}{\alpha'} \kappa_{\alpha\beta} \epm\,
\qquad \mbox{and} \qquad
\breta_{\hbrm\hbrn} = \bpm \breta_{\brm\brn} & 0 \\ 0 & \kappa_{\bra\brb}\epm\,.
\end{equation}
Here, $\cJ_{MN}$ is the $\ODD$ metric, while $\breta_{\brm\brn}$ is the $\OO(D-1,1)$ metric:
\begin{equation}
  \cJ_{MN} = \begin{pmatrix}
  0&\delta^\mu{}_{\nu}  \\
  \delta_{\mu}{}^{\nu}&0 
\end{pmatrix}
\,\qquad \mbox{and} \qquad
\eta_{mn} = -\bar{\eta}_{\bar{m}\bar{n}} = \text{diag}(-1,1,\cdots,1) \,,
\end{equation}
and $\kappa_{\bra\brb}$ is the Cartan-Killing form for the heterotic gauge group $G$ 
\begin{equation}
\kappa_{\bra\brb} = \tr \bigl(t^{\bra}\, t^{\brb}\bigr) \,,
\end{equation}
where $t^{\bra}$ denotes $\bra$-th generator in the adjoint representation
\begin{equation}
(t^{\bra})_{\brb\brc} = f_{\brb\brc}{}^{\bra}\,.
\end{equation}
Here, indices $\bra,\brb, \cdots = 1, \cdots, \dim G$, are adjoint gauge indices, and $\alpha,\beta,\cdots = 1, \cdots, \dim G$ are pull-back of $\bra,\brb,\cdots$ indices by introducing a matrix $(\phi^{\bar{a}})_{\alpha}$ that preserves the $\kappa_{\bar{a}\bar{b}}$.
Accordingly, we denote the pull-back of $\kappa_{\bar{a}\bar{b}}$ as $\kappa_{\alpha\beta}$: 
\begin{equation}
  \kappa_{\alpha\beta} = (\phi^{\bra})_{\alpha}\, (\phi^{\brb})_{\beta}\, \kappa_{\bra\brb}\,, \qquad (\phi^{\bra})_{\alpha} \in \OO({\rm dim}\,G)\,,
\end{equation}
and they are numerically equivalent. Furthermore, one can always fix $\phi^{\bar{a}}{}_{\alpha}$ as the identity matrix by using part of the local Lorentz transformation (\ref{local_Lorentz}), which is generated by $\bar{\Lambda}^{\bar{a}}{}_{\bar{b}}$ \cite{Bedoya:2014pma}. 

 It is important to note that $\kappa_{\bra\brb}$ is embedded into $\OG\subset \ODG$, which has negative-definite metric. 
Thus, in order to get the standard heterotic supergravity from the heterotic DFT through an explicit parameterization of the double-vielbein (or the generalized metric), we must impose a diagonal gauge-fixing of the two local Lorentz groups, which maps the barred quantities to unbarred quantities
\begin{equation}
  \bar{\eta} \to - \eta\,, \qquad  \kappa_{\bar{a}\bar{b}} \to - \kappa_{ab}\,.
\end{equation}
Hereafter, we will assume the diagonal gauge fixing condition and identify $\alpha,\beta,\gamma\cdots$ indices with $a,b,c,\cdots$ indices.

\subsubsection{Parameterization from coset representative}
We next construct the geometric and non-geometric parameterizations of double-vielbein through the Iwasawa decomposition for a non-split real form, as explained in the beginning of this section. 

The parameterization of generalized vielbein is constructed from the exponentiation of solvable Lie algebra as a generalization of  (\ref{eq:vielbein1})
\begin{equation}
  \cE = \exp[\mathfrak{g}^{s}]\,. 
\label{SolvLieAlg}\end{equation}
Here,  $\mathfrak{g}^{s}$ denotes the solvable Lie algebra which consists of the non-compact Cartan generators and the positive (negative)-root generators. The non-compact Cartan generator $H_m$ is given by the diagonal components of $\mathbf{gl}(D)$ generator $K^{m}{}_{n}$
\begin{equation}
 H_m := K^{m}{}_{m}\,.
\end{equation}
The matrix realization of $\mathbf{gl}(D)$ generator $K^{m}{}_{n}$ is given by
\begin{equation}
  (\rho_{K^{m}{}_n})_{\hP}{}^{\hQ} = \begin{pmatrix} \delta_{p}^{m}\,\delta_{n}^{q} & 0 & 0\\ 0 & 0 & 0 \\ 0 & 0 & -\delta^{p}_{n}\,\delta^{m}_{q} \end{pmatrix}\,, 
\label{positiveroot1}\end{equation}
and the matrix realization of $H_m$ is also given by
\begin{equation}
\begin{aligned}
  (\rho_{H_m})_{\hP}{}^{\hQ} := (\rho_{K_{m}{}^m})_{\hP}{}^{\hQ} = \begin{pmatrix} \delta_{p}^{m}\,\delta_{m}^{q} & 0 & 0\\ 0 & 0 & 0 \\ 0 & 0 & -\delta^{p}_{m}\, \delta^{m}_{q} \end{pmatrix}\,.
\end{aligned}
\end{equation}
The corresponding positive-root generators are realized as 
\begin{equation}
  (\rho_{R^{mn}})_{\hP}{}^{\hQ} = \begin{pmatrix} 0 & 0 & 2\,\delta_{pq}^{mn} \cr 0 & 0 & 0 \cr 0 & 0 & 0
 \end{pmatrix}\,,\qquad  
  (\rho_{R^m{}_{a}})_{\hP}{}^{\hQ} = \begin{pmatrix} 0 & \delta_{p}^{m}\,\delta_{a}^{d} & 0 \cr 0 & 0 & \kappa_{ca}\,\delta^{m}_{q} \cr 0 & 0 & 0
 \end{pmatrix}\,,
\label{positiveroot2}
\end{equation}
which satisfy the following Lie algebra
\begin{equation}
 \comm{\mathbf{H}}{K^{m}{}_{n}} = \mathbf{a}_{mn}\, K^{m}{}_{n} \,,\qquad 
 \comm{\mathbf{H}}{R^{mn}} = \mathbf{b}_{mn}\, R^{mn} \,, \qquad 
 \comm{\mathbf{H}}{R^{m}{}_{a}} = \mathbf{c}_m\, R^{m}{}_{a}\,.
\end{equation}
Here,  $\mathbf{a}_{mn}$, $\mathbf{b}_{mn}$ and $\mathbf{c}_{m}$ are positive roots for $D$-type (assume that $\dim G$ is even)
\begin{equation}
  \mathbf{a}_{mn} = \mathbf{e}_{m} - \mathbf{e}_{n}\,, \qquad \mathbf{b}_{mn} = \mathbf{e}_{m} + \mathbf{e}_{n} \,, \qquad \mathbf{c}_{m} = \mathbf{e}_{m}\,, \qquad (1\leq m < n \leq D)\,, 
\end{equation}
where $\mathbf{e}_{m} = (\underbrace{0,0,\cdots,0}_{m-1},1,0,0,\cdots ,0)$.
These positive root generators obey the commutation relations:
\begin{align}
\begin{split}
 [K^m{}_n,\,K^p{}_q]&= \delta_n^p\,K^m{}_q - \delta_q^m\,K^p{}_n\,, \qquad 
 [K^m{}_n,\,R^{pq}] =  \delta_n^p\, R^{mq} + \delta_n^q\, R^{pm}\,, \qquad
\\
 [K^m{}_n,\,R^p{}_{a}] &= \delta^p_n\, R^m{}_a \,,\qquad 
 \comm{R^{m}{}_{a}}{R^n{}_{b}} = \kappa_{ab}\, R^{mn}\,,
\qquad [R^{mn},\, R^p{}_{a} ] = 0 \,,
\end{split}
\end{align}

Using the above results, we construct the explicit geometric parameterization of the generalized vielbein. The coset representative $\cE_{\hM}{}^{\hA}$ of $\ODDG/\OO(1,D-1)\otimes\ODG$ is given by (\ref{SolvLieAlg}) with the noncompact positive root generators
\begin{equation}
 \cE_{\hM}{}^{\hA} = \Exp{h^m(x)\,\rho_{H_m}}\,\Exp{\sum_{m<n}h_m{}^n(x)\,\rho_{K^m{}_n}} \Exp{\frac{1}{2}\,B_{mn}(x)\,\rho_{R^{mn}}} \Exp{A_{m}{}^{a} \rho_{R^{m}{}_{a}}}\,.
\end{equation}
If we substitute the explicit matrix realizations of generators (\ref{positiveroot1}) and (\ref{positiveroot2}), we recover the usual geometric parameterization of generalized vielbein
\begin{equation}
 \cE_{\hM}{}^{\hB} = \big(E\,\Exp{\cB^{(2)}}\,\Exp{\cA}\big)_{\hM}{}^{\hB}
 =\begin{pmatrix} e_{\mu}{}^n & A_\mu{}^{a} & e_{\mu}{}^{p}\,B'_{pn} \cr 0 & \delta_{a}^{b} & \,(A^\rmT)_{a n} \cr 0 & 0 & (e^{-\rmT})^\mu{}_n
 \end{pmatrix}\,,
\end{equation}
where 
\begin{equation}
\begin{aligned}
 & E =\begin{pmatrix} e_{\mu}{}^{n} & 0 & 0 \cr 0 & \delta_{a}^{b} & 0 \cr 0 & 0 & (e^{-\rmT})^{\mu}{}_{n} \end{pmatrix}\,, \quad 
 \cB^{(2)} =\begin{pmatrix} 0 & 0 & B_{mn} \cr 0 & 0 & 0 \cr 0 & 0 & 0
 \end{pmatrix} \,,
\\
 &\cA = A_m{}^\alpha\,\rho_{R^m{}_\alpha} = \begin{pmatrix} 0 & A_m{}^{b} & 0 \cr 0 & 0 & \kappa_{ac}\,(A^\rmT)^{c}{}_{n} \cr 0 & 0 & 0
 \end{pmatrix}\,,
\\
 &\Exp{\cB^{(2)}} \Exp{\cA} =\begin{pmatrix} \delta_m^n & A_m{}^{b} & B'_{mn} \cr 0 & \delta_{a}{}^{b} & \kappa_{ac}\,(A^\rmT)^{c}{}_{n} \cr 0 & 0 & \delta^m_n
 \end{pmatrix} \, . 
\end{aligned}
\end{equation}
Here, 
\begin{equation}
 B'_{mn} := B_{mn} + \frac{1}{2}\,A_m{}^{a}\,\kappa_{ab}\,(A^\rmT)^{b}{}_n \, . 
 \end{equation}
Then, from the generalized vielbein, we get the geometric parameterization of generalized metric   using the defining relation $\cH = (\cE^{\rmT})_{\hM}{}^{\hA}\,\eta_{\hA\hB}\,\cE^{\hB}{}_{\hN}$,
\begin{equation}
\left(\cH_{\hM \hN} \right) = 
	\bpm
 		g + B' g^{-1} (B')^t + A \kappa A^t & A\kappa + B' g^{-1} A \kappa & B'g^{-1}
 		\\
 		\kappa A^t + \kappa A^t g^{-1} (B')^t & \kappa A^t g^{-1} A \kappa	+\tfrac{1}{\alpha'} \kappa & \kappa A^t g^{-1}
 		\\
		g^{-1} (B')^{t} & g^{-1} A \kappa & g^{-1}
		\epm \,. 
\label{geomGenMet}\end{equation}

Consider next the non-geometric parameterization. As discussed in the previous section, the associated non-geometric parameterization is constructed from the Cartan involution, which flips the sign of all the non-compact generators,
\begin{equation}
  K^{m}{}_{n} \to - K^{n}{}_{m}\,, \qquad R^{mn} \to - R_{mn}\,, \qquad R^{m}{}_{a} \to - R_{m}{}^{a}\,.
\end{equation}
The matrix realization of negative root generators are 
\begin{equation}
  (\rho_{R_{mn}})_{\hP}{}^{\hQ} = \begin{pmatrix} 0 & 0 & 0 \cr 0 & 0 & 0 \cr -2\,\delta^{pq}_{mn} & 0 & 0
 \end{pmatrix}\,,\qquad
  (\rho_{R_m{}^{a}})_{\hP}{}^{\hQ} = \begin{pmatrix} 0 & 0 & 0 \cr -\delta_c^a\,\delta_m^q & 0 & 0 \cr 0 & -\delta^p_m\,\kappa^{ad} & 0
 \end{pmatrix}\,,
\label{negativeroot}
\end{equation}
where $\kappa^{\alpha\beta}$ is the inverse matrix of $\kappa_{\alpha\beta}$.
It is straightforward to check that they satisfy
\begin{equation}
\begin{aligned}
 \comm{\mathbf{H}}{R_{mn}}  &= - \mathbf{b}_{mn}\, R_{mn}\,,&  
 \comm{\mathbf{H}}{R_{m}{}^{a}} &= - \mathbf{c}_{m}\, R_{m}{}^{a}\,,  \\
 \comm{K^m{}_{n}}{R_{pq}} &= -\delta_p^m\, R_{nq} - \delta_q^m\, R_{pn}\,,& \qquad
 [K^m{}_n,\,R_p{}^a]&= -\delta^m_p \, R_n{}^{a} \,, 
 \\
 \comm{R_{mn}}{R_{p}{}^{a}} &= 0\,, &\qquad  
 \comm{R_{m}{}^{a}}{R_{n}{}^{b}} &= -\kappa^{ab}\,R_{mn} \,,
\end{aligned}
\end{equation}

The non-geometric parameterization of the generalized vielbein is defined by the non-geometric coset representative $\cE_{\hM}{}^{\hA}$, which is expressed  in terms of negative root generators  
\begin{equation}
  \widetilde\cE_{\hM}{}^{\hA}(x) = \Exp{h^m(x)\,\rho_{H_m}} \Exp{\sum_{m<n}h_m{}^n(x)\,\rho_{K^m{}_n}} \Exp{\frac{1}{2}\,\beta^{mn}(x)\,\rho_{R_{mn}}} \Exp{-\tilde{A}^{m}{}_{a} \rho_{R_{m}{}^{a}}}\,.
\end{equation}
Using (\ref{negativeroot}), we find that
\begin{equation}
 \widetilde{\cE} 
 = E \Exp{\beta^{(2)}}\Exp{-\tilde{\cA}} 
 = \begin{pmatrix} \tilde{e}_{\mu}{}^{n} & 0 & 0 \\ (\tilde{A}^{\rmT})_{a}{}^{n} & \delta_{a}^{b} & 0 \\
 - (\tilde{e}^{-\rmT})^{\mu}{}_{k}\, \beta'^{km} & (\tilde{e}^{-\rmT})^{\mu}{}_{k}\, \tilde{A}^{k}{}_{c}\, \kappa^{cb} & (\tilde{e}^{-\rmT})^{\mu}{}_{m} \end{pmatrix} \, , 
\label{nongeomGV}\end{equation}
where
\begin{equation}
\begin{aligned}
 E &=\begin{pmatrix} \tilde{e}_{\mu}{}^n & 0 & 0 \cr 0 & \delta_{a}^{b} & 0 \cr 0 & 0 & (\tilde{e}^{-\rmT})^{\mu}{}_{n} \end{pmatrix}\,, \qquad \qquad 
 \mathbf{\beta}^{(2)} =\begin{pmatrix} 0 & 0 & 0 \cr 0 & 0 & 0 \cr -\beta^{mn} & 0 & 0
 \end{pmatrix} \,,
\\
 \tilde{\cA} &= \tilde{A}^{m}{}_{a}\,\rho_{R_{m}{}^{a}} = \begin{pmatrix} 0 & 0 & 0 \cr -(\tilde{A}^{\rmT})_{a}{}^{n} & 0 & 0 \cr 0 & -\tilde{A}^{m}{}_{c}\, \kappa^{cb} & 0
 \end{pmatrix}\,,
\\
 \Exp{\beta^{(2)}} \Exp{\cA} &=\begin{pmatrix} \delta_m^n & 0 & 0 \cr  (\tilde{A}^{\rmT})_{a}{}^{n} & \delta_{a}^{b} & 0 \cr - \beta'^{mn} & \tilde{A}^{m}{}_{c}\, \kappa^{cb} & \delta^m_n
 \end{pmatrix} \, .
\end{aligned}
\end{equation}
Here, 
\begin{equation}
 \beta'^{mn} := \beta^{mn} - \half \tilde{A}^{m}{}_{a} \, \kappa^{ab} \,(\tilde{A}^\rmT)_{b}{}^{n}
 \end{equation}
Likewise, the non-geometric parameterization of generalized metric is given by the defining condition $\cH_{\hM \hN} = (\cE^{\rmT})_{\hM}{}^{\hA}\,\eta_{\hA\hB}\,\cE^{\hB}{}_{\hN}$:
\vskip0.5cm
\begin{tcolorbox}
\vskip-0.1cm
\begin{equation}
\begin{aligned}
 (\cH_{\hM\hN}) &= 
	\begin{pmatrix}
  	\tilde{g} \ \ & \ \ \tilde{g} \tilde{A} & -\tilde{g}\,\beta'^\rmT
  	\\
	\tilde{A}^\rmT\, \tilde{g} \ \ & \ \ \tilde{A}^\rmT \, \tilde{g} \, \tilde{A} + \kappa & \tilde{A}^{\rmT} -\tilde{A}^{\rmT}\, \tilde{g}\, \beta'^\rmT
	\\
	-\beta' \, \tilde{g} \ \ & \ \  \tilde{A} - \beta' \,\tilde{g}\, \tilde{A} & ~~~\tilde{g}^{-1} + \beta' \, \tilde{g}\, \beta'^{\rmT} + \tilde{A}\, \kappa^{-1}\, \tilde{A}^\rmT~~~
\end{pmatrix}
\\
&= \bpm 1 & 0 & 0 \\ \tilde{A}^\rmT & 1 & 0 \\ -\beta' & \tilde{A}\,\kappa^{-1} & 1 \epm
   \bpm \tilde{g} \ & \ 0 \ & \ 0 \\ 0 \ & \  \kappa \ & \ 0 \\ 0 \ & \ 0  \ & \ \tilde{g}^{-1} \epm
   \bpm 1 & \tilde{A} & -\beta'^{\rmT} \\ 0 & 1 & \kappa^{-1}\,\tilde{A}^\rmT \\0 & 0 & 1\epm \,,
\end{aligned}
\label{nongeomGenMet}\end{equation}
\end{tcolorbox}
\noindent where $\tilde g=(\tilde g_{\mu\nu})$, $\beta'=(\beta'^{\mu\nu})$, $\tilde{A}=(\tilde{A}^\mu{}_{a})$, $\kappa^{-1}=(\kappa^{ab})$, and
\begin{equation}
 \eta_{\hA\hB} = \begin{pmatrix} \eta_{mn} & 0 & 0 \\ 0 & \kappa_{ab} & 0 \\ 0 & 0 & \eta^{mn} \end{pmatrix}\,.
\end{equation}
For the abelian reduction of the heterotic Yang-Mills group to the maximal Cartan subgroup, $U(1)^{16}$, we have $\kappa = I_{16\times16}$ and $f_{abc}=0$. 
Under this reduction, the non-geometric parameterization (\ref{nongeomGV}) coincides with the previous result in \cite{Blumenhagen:2014iua}.

Moreover, by comparing (\ref{geomGenMet}) with (\ref{nongeomGenMet}), one can confirm that the geometric parameterization and the non-geometric parameterization are related each other by field redefinitions. Denote the set of variables in geometric parameterization and the set of variables in non-geometric parameterization as
\begin{eqnarray}
\text{geometric:} \quad  \{g, B, A\} \qquad \mbox{and}
\qquad \text{non-geometric:} \quad \{\tilde{g},\beta,\tilde{A}\}\,, 
\end{eqnarray}
respectively. From the generalized metric, one then finds a relation between the geometric variables and the non-geometric variables as
\vskip0.5cm
\begin{tcolorbox}
\vskip-0.2cm
\begin{equation}
\begin{aligned}
    g &= \big(\tilde{g}^{-1} - \beta'^{\rm T}\big)^{-1} \tilde{g}^{-1} \big(\tilde{g}^{-1} - \beta'\big)^{-1}\,,
    \\
	B'&= - \big(\tilde{g}^{-1} - \beta'^{T}\big)^{-1} \beta'^{\rm T} \big(\tilde{g}^{-1} - \beta'\big)^{-1}\,,
	\\
	A &= - \big(\tilde{g}^{-1} - \beta'^{\rm T}\big)^{-1} \tilde{A}\,
\end{aligned}
\end{equation}
Likewise, 
\begin{equation}
\begin{aligned}
  \tilde{g} &=\big(g+B'\big)g^{-1} \big(g+B'^{\rm T}\big)\,,
  \\
  \beta'&= - \big(g+B'^{\rm T}\big)^{-1} B'^{\rm T} \big(g+B\big)^{-1}\,,
  \\
  \tilde{A} &= - \big(g+B'^{\rm T}\big)^{-1} A\,.
\end{aligned}\label{inverseMap}
\end{equation}
\end{tcolorbox}


\subsection{Non-geometric fluxes and action}
We now study the non-geometric fluxes for heterotic supergravity. In the gauged DFT viewpoint, various fluxes in gauged supergravity theories, viz. geometric fluxes, are regarded as components of the generalized spin connection of the gauged DFT in geometric parameterization. Similarly, non-geometric fluxes can be constructed by replacing the geometric parameterization of gauged DFT to non-geometric parameterization.\footnote{We present a systematic construction of heterotic non-geometric fluxes and action via semi-covariant approach in Appendix \ref{app:gauged-DFT}.}

The heterotic DFT action \cite{Hohm:2011ex} is given in terms of the generalized metric $\cH$ by
\begin{equation}
\begin{aligned}
  \cS_{\rm het}	= \int e^{-2d} \Big[& 4 \cH^{\hat{M}\hN} \partial_{\hM}\partial_{\hN}d - \partial_{\hM}\partial_{\hN}\cH^{\hM \hN} -4\cH^{\hat{M}\hN}\partial_{\hM}d \partial_{\hN} d + 4\partial_{\hM} \cH^{\hM \hN} \partial_{\hN} d 
  \\
  &+ \tfrac{1}{8} \cH^{\hM \hN }\partial_{\hM }\cH^{\hK \hL} \partial_{\hN}\cH_{\hK \hL} -\half \cH^{\hM \hN }\partial_{\hM}\cH^{\hK \hL } \partial_{\hK }\cH_{\hN \hL } \Big] \, . 
\end{aligned}
\end{equation}
Using the non-geometric parameterization of the generalized metric (\ref{nongeomGenMet}), we obtain the non-geometric action for heterotic supergravity in the string frame as 
\vskip0.5cm
\begin{tcolorbox}
\vskip-0.3cm
\begin{equation}
  S_{\rm het} = \int d^{D}x \sqrt{-g} e^{-2\phi} \Big( R(\tilde{\omega}) + 4 \partial^\mu \phi \partial_\mu \phi - \tfrac{1}{4} Q_{\mu}{}^{\nu\rho} Q^{\mu}{}_{\nu\rho}   -\tfrac{1}{4} \alpha' \tr\big(\tilde{F}_{\mu\nu}\tilde{F}^{\mu\nu}\big)\Big) + \cdots \,,
\label{nongeom_action}\end{equation}
\end{tcolorbox}
\noindent
where $R(\tilde{\omega})$ is the Ricci scalar with respect to the non-geometric frame field $\tilde{e}$ and its spin connection field $\tilde{\omega}$, and the ellipses denote terms of higher-order derivatives. In the non-geometric action (\ref{nongeom_action}), three kinds of fluxes are present: geometric flux $\tilde f_{mnp}$, non-geometric $Q$-flux and non-geometric gauge field strength $\widetilde{F}_{\mu\nu}$.

The geometric flux, $\tilde{f}^{m}{}_{np} = -2\tilde{e}_{[n}{}^{\mu} \tilde{e}_{p]}{}^{\nu} \partial_{\mu} \tilde{e}_{\nu}{}^{m}$, is given by the same form as geometric parameterization case. The dual spin connection ${\tilde{\omega}^m}_{np}$ is written in terms of the flux $\tilde{f}_{mnp}$
\begin{equation}
  \tilde{\omega}_{mnp} = \half (\tilde{f}_{mnp} +\tilde{f}_{nmp} -\tilde{f}_{pmn}) \,.
\end{equation}
The non-geometric $Q$-flux of the bi-vector field $\beta^{\mu\nu}$ is defined by
\begin{equation}
  Q_{\mu}{}^{\nu\rho} = \partial_{\mu} \beta^{\nu\rho} + \tfrac{1}{3} \alpha' \tilde{g}_{\mu\sigma}\Omega^{\sigma \nu\rho} \,,
\end{equation}
where $\Omega^{\mu\nu\rho}$ is the dual Chern-Simons three-form, defined by 
\begin{equation}
  \Omega^{\mu\nu\rho} = 3 \widetilde{A}^{[\mu}{}_{a}\partial^{\nu} \widetilde{A}^{\rho]a} - \widetilde{A}^{\mu a}\tilde{A}^{\nu b}\widetilde{A}^{\rho c}f_{abc} \,
\end{equation}
and $\partial^{\mu} = \tilde{g}^{\mu\nu}\partial_{\nu}$. 
Finally, the field strength of non-geometric gauge field $\tilde{A}^{\mu}$ are defined by
\begin{equation}
  \tilde{F}^{\mu\nu}{}_{a} = - 2\partial^{[\mu} \tilde{A}^{\nu]}{}_{a} + f_{abc} \tilde{A}^{\mu b} \tilde{A}^{\nu c}\,.
\end{equation}
Note that here we have assumed the simplifying ansatz first introduced in \cite{Andriot:2012wx}
\begin{equation}
  \beta^{\mu\nu} \partial_{\nu} = \tilde{A}^{\mu a} \partial_{\mu} = 0\,.
\end{equation}
The Bianchi identity for the $Q$-flux is given by 
\begin{eqnarray}
  \partial_{[\mu} Q_{\nu]}{}^{[\rho\sigma]} 
  &=& {\alpha' \over 12} {{\tilde F}_{\mu \nu}}^a {\tilde F}^{\rho \sigma \, a}
  + {\alpha' \over 6} {{\tilde F}_{[\mu}}^{\,\,\, [\rho \, a} {{\tilde F}_{\nu]}}^{\,\,\, \sigma ] \, a}
  + {\alpha' \over 3} {\tilde A}^a_{[\mu} \partial_{\nu]} {\tilde F}^{\rho \sigma \, a}
  + {\alpha' \over 3} {\tilde A}^{[\rho \, a} \partial_{[\mu} {{\tilde F}^{\sigma] \, a}}_{\, \, \, \,\,\, \nu ]}. 
\end{eqnarray}
If we set $\tilde{A}^a_\mu =0$, then this right-hand side vanishes and the result of  the conventional $\OO(d,d)$ DFT is reproduced.

\section{Exotic Branes and Non-geometric Fluxes}
\label{sec:applications}

Having constructed the effective actions for type IIA / M, type IIB, and heterotic supergravity theories, we now apply them to study exotic branes and  backgrounds with non-geometric fluxes. 

Exotic branes in the M-theory and type II string theory are first found and studied in \cite{Elitzur:1997zn,Blau:1997du,Hull:1997kb,Obers:1997kk,Obers:1998fb,Eyras:1999at} and their corresponding supergravity solutions are comprehensively constructed in \cite{LozanoTellechea:2000mc}. 
Properties of these solutions are revisited in \cite{deBoer:2010ud,deBoer:2012ma}, and it was noticed that the background fields are not single-valued and in particular that exotic brane backgrounds exhibit nontrivial monodromies under the $U$-duality transformations. As such, these backgrounds are referred to as $U$-folds. 

Exotic branes are defect branes of codimension-two. With $(x^1,\,x^2)$ denoting the coordinates in the two dimensional transversal space, the supergravity backgrounds, viz. U-folds, depend on the transverse space  $z\equiv x^1+\ii x^2\equiv r \Exp{\ii\theta}$ in terms of  the logarithmic function,
\begin{align}
 \rho(z)\equiv \rho_1+\ii\rho_2 = \ii \frac{\sigma}{2\pi}\,\log(r_\text{c}/z)= \frac{\sigma}{2\pi}\,\bigl[\theta+\ii \log(r_\text{c}/r)\bigr]  \,.
\end{align}
Here, $\sigma$ is a positive constant that depends on the brane and $r_\text{c}$ is an arbitrary positive constant.
\footnote{The arbitrary constant $r_\text{c}$ is the infrared regulator scale, setting a maximum radius of the background, at which the curvature diverges. This infrared singularity can be removed by introducing additional branes and interpreting the logarithmic function as a $r/r_\text{c}\to 0$ limit of a globally defined holomorphic function, as is well known for the backgrounds of seven-branes \cite{Greene:1989ya}. In that setup, the cutoff $r_\text{c}$ can be interpreted as the distance from the neighbouring brane \cite{Kikuchi:2012za}.} 
Note that, as one moves around the center counter-clockwise, the imaginary part, $\rho_2$, is single-valued while the real part, $\rho_1$, gets a shift, $\rho_1\to \rho_1+\sigma$. 
This shift causes the monodromy transformations in the defect-brane backgrounds. 

In \cite{deBoer:2010ud,deBoer:2012ma}, it was shown that the charges of defect branes are characterized by the monodromies. Therefore, the monodromy matrices are important physical observables. However, in  \cite{deBoer:2010ud,deBoer:2012ma}, the explicit form of monodromy matrices was shown only for the exotic $5^2_2$-brane in type II theories. 

In this paper, using the parameterization of the generalized metric in EFT, we obtain the explicit form of monodromy matrices for various exotic branes in the M-theory as well as type II theories compactified on a $d$-torus ($d\leq 7$). 
We confirm that the monodromy matrix of each brane is in one-to-one correspondence with the negative root generator of the exceptional group. This means that the monodromy is not in the geometric subgroup (i.e.~the gauge symmetry of the conventional supergravity) and so the background is non-geometric. 

Another definition for the charges of defect brane is given by the flux integral. As discussed in \cite{Hassler:2013wsa,Andriot:2014uda,Okada:2014wma,Sakatani:2014hba}, the charges of exotic branes are given by the flux integral of the non-geometric fluxes. As such, it is convenient to use the non-geometric parameterization of the generalized metric. Below, we show that the metric in an arbitrary exotic-brane background is single-valued in terms of the non-geometric fields, while it is multi-valued in terms of fields in the conventional parameterization. 
We also calculate the flux integrals of the non-geometric fluxes and identify the exotic branes as the magnetic sources of non-geometric fluxes. 

More recently, exotic-brane solutions in the heterotic supergravity have also been constructed in \cite{Sasaki:2016hpp}, where the monodromies of the backgrounds have been calculated but using the generalized metric of \cite{Hohm:2014sxa}. 
In this section, we study the same backgrounds in terms of the generalized metric obtained in section \ref{sec:HDFT}. 
We also show that, in the non-geometric parameterization, the metric becomes single valued and discuss non-geometric fluxes in these backgrounds. 

\subsection{Exotic branes in the heterotic DFT}

Consider first exotic branes in heterotic DFT. Heterotic supergravity admits three types of exotic-brane solutions \cite{Sasaki:2016hpp}, which inherit from symmetric, neutral and gauge NS5-brane solutions \cite{Rey:1989xj, Rey:1989xi, Rey:1991uu, Callan:1991ky}. Among these solutions, the symmetric solution makes use of the leading order $\alpha'$-corrections. Therefore, to analyze the symmetric exotic brane solution, we first need to retain the leading order $\alpha'$-corrections in the heterotic supergravity. The first order $\alpha'$-correction was constructed by combining the $\Spin(9,1)$ local Lorentz group with the $\SO(32)$ or $E_8\times E_8$ heterotic Yang-Mills gauge group \cite{Bergshoeff:1988nn,Bergshoeff:1989de}. The spin-connection is the gauge potential for the $\Spin(9,1)$ local Lorentz transformation, thus Yang-Mills gauge field $A_{\mu}$ and the spin-connection $\omega_\mu$ are treated on an equal footing. Therefore, at the level of $\alpha'$-corrections, the effective action is organized in terms of the modified spin-connection $\omega_{+\mu}{}^{ab}$ by adding the contribution of three-form field strength $H_{\mu ab}$, which is a pull-back of $H_{\mu\nu\rho}$ by the vielbein $e^{\mu}{}_{a}$:
\begin{equation}
  \omega_{\pm\mu}{}^{ab}(e, B, A) = \omega_{\mu}{}^{ab}(e) \pm \tfrac{1}{2} H_{\mu}{}^{ab}(e, B, A)\,. 
\label{omegap}\end{equation}
Here,  the $\alpha'$-corrected $H_{\mu\nu\rho}$ is defined by 
\begin{equation}
  H_{\mu\nu\rho}(e, B, A) = 3 \partial_{[\mu} B_{\nu\rho]} + \tfrac{1}{2} \alpha'\Omega^{A}{}_{\mu\nu\rho} -\tfrac{1}{2}\alpha' \Omega^{\omega_{+}}{}_{\mu\nu\rho}\,,
\end{equation}
and $\Omega^{A}$ and $\Omega^{\omega_{+}}$ are Chern-Simons three-forms of $A_{\mu}$ and $\omega_{+\mu}$, respectively. As the field strength of the deformed spin-connection is given by the deformed Riemann tensor, 
\begin{equation}
  R_{\pm\mu\nu}{}^{ab} = \partial_{\mu}\omega_{+\nu}{}^{ab} - \partial_{\nu}\omega_{\pm\mu}{}^{ab} + \omega_{\pm\mu}{}^{ac} \omega_{\pm\nu c}{}^{b} -\omega_{\pm\nu}{}^{ac} \omega_{\pm\mu c}{}^{b} \, , 
\end{equation}
the Riemann squared term in the leading $\alpha'$-correction is straightforwardly obtained from the kinetic term of Yang-Mills gauge field in (\ref{nongeom_action}).

Similarly, the leading-order $\alpha'$-corrections in heterotic DFT is obtainable from extending the Yang-Mills gauge group. As shown in section 3, the heterotic DFT gauge group $G$ is composed of $G=G_{\rm YM} \times G_{\rm LL}$, where $G_{\rm YM}$ is the heterotic Yang-Mills gauge group and $G_{\rm LL}$ is the $\OO(1,9)$ local Lorentz gauge group. The associated $\ODDG$ metric is also decomposed as
\begin{equation}
  \cJ_{\hat{M}\hat{N}} = \bpm \cJ_{MN} & 0 & 0\\ 0 & \frac{1}{\alpha'} {\kappa}_{ab} & 0 \\ 0 & 0 & -\frac{1}{\alpha'} \kappa_{\tilde{a}\tilde{b}} \epm\,, 
\end{equation}
where $\tilde{a},\tilde{b} \cdots$ are $\OO(\dim G_{\rm LL})$ vector indices. It is important to note the relative sign difference between the coefficients of $\kappa_{ab}$ and of $\kappa_{\tilde{a}\tilde{b}}$. Because of this difference, the traces of $G_{\rm YM}$ and the trace of $G_{\rm LL}$ always have the opposite sign. 
Furthermore, after the diagonal gauge fixing, the deformed spin connections, $\omega_{+}$ and $\omega_{-}$, are represented by the geometric parameterization of generalized spin connection $\Phi_{\bar{p}mn}$ and $\bar{\Phi}_{p\bar{m}\bar{n}}$, respectively. Once the leading-order $\alpha'$-correction is introduced into the geometric parameterization of the generalized metric, the symmetric part of $B'$ is replaced by
\begin{equation}
  B''_{\mu\nu}:= \mbox{sym}(B')_{\mu \nu} =  B_{\mu\nu} + \tfrac{1}{2}\alpha'\tr (A_{\mu}\, A_{\nu} ) - \tfrac{1}{2}\alpha'\tr (\omega_{+\mu}\,\omega_{+\nu})
\end{equation}
and the gauge field associated with the local Lorentz group is given by the generalized spin-connection
\begin{equation}
  \cA_{M [\brm\brn]} = \begin{pmatrix} 0\\e_{\mu}{}^{m}\,\brPhi_{m \brm\brn}\end{pmatrix}\,. 
\end{equation}
Likewise, the corresponding double-vielbein and generalized metric are also extended. Hereafter, we will construct several nontrivial solutions of the heterotic DFT that feature all these structures.

\subsubsection{Symmetric dual five-brane}
First, we construct the symmetric $5^2_2$-brane. Denote the coordinates of direction transverse to the brane as $(\rho,\theta,z,\psi)$.  The metric and Kalb-Ramond field are given by \cite{Sasaki:2016hpp}
\begin{equation}
 \rmd s^{2} = f( \rmd r^{2}+r^{2} \rmd \theta^{2}) + f K^{-1} (\rmd z^{2}+\rmd\psi^{2}) \, \qquad 
  \mbox{and} \qquad
  B = \sigma \theta K^{-1} \rmd z \wedge \rmd \psi \,,
\label{SymmetricSol522}\end{equation}
where $\sigma$ is a constant parameter, and $f$ and $K$ are defined as
\begin{equation}
  f = \sigma \log\frac{\mu}{r}\, \qquad \mbox{and} \qquad
  K= f^{2} + \sigma^{2}\theta^{2}\,. 
\end{equation}
The ansatz for Yang-Mills gauge field components are given by
\begin{equation}
\begin{aligned}
  A_{r} &= 0\,,
  \\
  A_{\theta} &= \frac{\sigma}{2} f^{-1} {\bf t}^{3}\,,
  \\
  A_{z} &= + \tfrac{1}{2} f^{-1} K^{-1} \Big[\big(f f' \sin \theta - f' \sigma \theta \cos \theta  \big) {\bf t}^{1} - \big(f f' \cos \theta + f'\sigma\theta \sin \theta \big) {\bf t}^{2}\Big]\,,
  \\
  A_{\psi} &= - \tfrac{1}{2} f^{-1}K^{-1} \Big[\big(f f'\cos \theta + f' \sigma \theta \sin \theta \big) {\bf t}^{1} + \big(f f' \sin \theta - f'\sigma \theta \cos \theta \big){\bf t}^{2}\Big] \,,
\end{aligned}\label{symmetric_A}
\end{equation}
where $f' = \partial_{r} f$, and ${\bf t}^{a} (a=1,2,3)$ are $SU(2)$ generators defined as
\begin{equation}
  {\bf t}^{1} = \begin{pmatrix} 0 & 0 & 0 & -1 \\ 0 & 0 & 1 & 0 \\ 0 & -1 & 0 & 0 \\ 1 & 0 & 0 & 0 \end{pmatrix} \,, \qquad {\bf t}^{2} = \begin{pmatrix} 0 & 0 & -1 & 0 \\ 0 & 0 & 0 & -1 \\ 1 & 0 & 0 & 0 \\ 0 & 1 & 0 & 0 \end{pmatrix} \,, \qquad {\bf t}^{3} = \begin{pmatrix} 0 & 1 & 0 & 0 \\ -1 & 0 & 0 & 0 \\ 0 & 0 & 0 & -1 \\ 0 & 0 & 1 & 0 \end{pmatrix} \,.
\end{equation}

For consistency, this solution have to satisfy the so-called symmetric embedding ansatz, which originates from the symmetric NS5-brane solution
\begin{equation}
  A_{\mu} = \omega_{+\mu}\,.
\label{EmbeddingAnsatz}\end{equation}
According to the definition of the deformed spin-connection (\ref{omegap}), the direct computation of $\omega_{+\mu}$ gives 
\begin{equation}
\begin{aligned}
  \omega_{+r} &= K^{-1} \sigma \theta f' \, {\bf  n}^{34}\,,
  \\
  \omega_{+\theta} &= (- {\bf n}^{12}  + {K}^{-1} \sigma f {\bf n}^{34}) + \frac{1}{2}f^{-1} \sigma {\bf t}^{3} \,,
  \\
 \omega_{+z} &= + K^{-3/2} [ \sigma \theta f'  {\bf t}^{1} + {1 \over 2} (f^2 - \sigma^2 \theta^2) (\log  f)'  {\bf t}^2] \,,
  \\
  \omega_{+\psi} &= - K^{-3/2} [ {1 \over 2} (f^2 - \sigma^2 \theta^2) (\log f)'  {\bf t}^{1} - \sigma \theta f'  {\bf t}^{2} ] \,,
\end{aligned}\label{symm_deformed_spco}
\end{equation}
where $\big({\bf n}^{ab}\big){}^{AB}$ denotes the $\mathbf{\rm SO}(4)$ generators.

As expressed, the above expressions appear to violate the embedding ansatz. However, one can show that the $A_{\mu}$ and $\omega_{+\mu}$ are related by gauge transformations. We now wish to find the explicit Yang-Mills gauge transformations and local Lorentz transformations which connects $A_{\mu}$ and $\omega_{+\mu}$. First, we take a gauge transformation for $A_{\mu}$,
\begin{equation}
  A'_{\mu} = - \partial_{\mu} L L^{-1} + L A_{\mu} L^{-1}\,,
\label{gaugeTrYM}\end{equation}
where the gauge parameter is chosen as
\begin{equation}
  L = \begin{pmatrix} \sin\theta & -\cos\theta & \qquad 0 & \qquad 0 \\ \cos\theta & \sin \theta & \qquad 0 & \qquad 0 \\ 0 & 0 & \qquad 1 & \qquad 0 \\ 0 & 0 & \qquad 0 & \qquad 1
  \end{pmatrix} \quad \in \quad SO(2)\,,
\end{equation}
then the $A'_{\mu}$ reads
\begin{equation}
\begin{aligned}
  A'_{r} &= 0 \,,
  \\
  A'_{\theta} &= - {\bf n}^{12} +\tfrac{1}{2} f^{-1} \sigma  {\bf t}^{3} \,,
  \\
  A'_{z} &= + {1 \over 2} K^{-1} f'  \big({\bf t}^{1}- f^{-1} \sigma\theta  {\bf t}^{2}\big)\,,
  \\
  A'_{\psi} &= - {1 \over 2} K^{-1} f' ( f^{-1} \sigma\theta {\bf t}^{1} + {\bf t}^{2} \big)\,.
\end{aligned}\label{A'}
\end{equation}

Next, we take a local Lorentz transformation for the $\omega_{+\mu}$,
\begin{equation}
  \omega'_{\mu} = - \partial_{\mu}\Lambda \Lambda^{-1} + \Lambda \omega_{+\mu} \Lambda^{-1}\,,
\end{equation}
where 
\begin{equation}
  \Lambda = \begin{pmatrix} 1 & \quad 0 & 0 & 0 \\ 0 & \quad 1 & 0 & 0 \\ 0 & \quad 0 & - K^{-1/2} \sigma\theta & K^{-1/2} f  \\ 0 & \quad 0 & - K^{-1/2} f  & - K^{-1/2} \sigma\theta  \end{pmatrix} \quad \in \quad SO(2)\,.
\end{equation}
After the gauge transformations,  $\omega'_{\mu}$ exactly matches with $A'_{\mu}$, and so the embedding ansatz is satisfied. The connections $A'_{\mu}$ and $\omega'_{\mu}$ are always combined according to the structure of leading-order $\alpha'$-corrections, they are canceled by the embedding ansatz (\ref{EmbeddingAnsatz}). Thus, the leading-order $\alpha'$-corrections do not contribute to the equations of motion, and the metric and $B_{\mu\nu}$ in (\ref{SymmetricSol522}) are reduced to the usual $5_{2}^{2}$-brane solution. However, it is important to note that the symmetric five-brane solution is not single-valued as the $\theta$ coordinate is encircled around the origin, $r = 0$. We will see later that this provides an example of $T$-fold.

Next, we consider the non-geometric solution given in terms of the fields $( \tilde{g}, \beta, \tilde{A} )$. Using the inverse map defined in (\ref{inverseMap}), we have a non-geometric solution whose metric $\tilde g$  and $\beta$ fields are
\begin{equation}
\begin{aligned}
  \tilde{g} = \begin{pmatrix} f & 0 & 0 & 0 \\ 0 & f\rho^{2} & 0 & 0\\ 0 & 0 & f^{-1} & 0 \\ 0 & 0 & 0& f^{-1}\end{pmatrix}\, \qquad \mbox{and } \qquad
  \beta = \sigma \theta \, {\partial_z} \wedge {\partial_\psi} \,,
\end{aligned}
\end{equation}
and $\tilde{A}^{\mu}$ components, which correspond to the gauge transformed gauge field (\ref{A'}), are
\begin{equation}
\begin{aligned}
  \tilde{A}^{r} &= 0\,,
  \\
  \tilde{A}^{\theta} &= \tfrac{1}{r^2} f^{-1} {\bf n}^{12} - {1 \over 2} {1 \over r^2} \sigma f^{-2} {\bf t}^{3} \,,
  \\
  \tilde{A}^{z} &= + {1 \over 2 r} \sigma f^{-1} {\bf t}^{1}\,,
  \\
  \tilde{A}^{\psi} &= -{1 \over 2r} \sigma f^{-1} {\bf t}^{2}\,.
\end{aligned}
\end{equation}
The metric and $\beta$ field are precisely the same as the non-geometric solution for the conventional $5^{2}_{2}$-brane. 

\subsubsection{Neutral and Gauge branes}
If we turn off the $\alpha'$-corrections in heterotic supergravity, the bosonic part is identical to the NS-NS-sector of the type II supergravity. Thus, the conventional $5^2_2$-brane is also the solution of heterotic supergravity. More generally, for the ansatz of vanishing heterotic gauge field, heterotic supergravity solutions without $\alpha'$-corrections is straightforwardly obtained from type II supergravity solutions. 

The gauge brane is also constructed without $\alpha'$-corrections, and the deformed spin-connection does not contribute.  Using one-form gauge transformation for simplicity, the Kalb-Ramond field can be set to be zero. The explicit solution is constructed in \cite{Sasaki:2016hpp}, and is given by 
\begin{equation}
  \rmd s^2 = h \big(\rmd r^{2}+r^{2} \rmd \theta^{2} \big) + e^{-4\phi_0} h \big(\rmd z^{2}+\rmd \psi^{2}\big)\,\qquad \mbox{and} \quad B = 0\,,
\end{equation}
where
\begin{equation}
  h(r) = e^{2\phi_{0}} - \frac{\alpha' \tilde{\sigma}^2}{2 r^2} \tilde{f}^{-1}\,,\qquad  \tilde{f} = \tilde{h}_0 - \frac{\tilde{\sigma}}{2}\log(r/\mu)
\label{GaugeBrane1}\end{equation}
and the gauge field components are 
\begin{equation}
\begin{aligned}
  A_{r} &= 0\,,
  \\
  A_{\theta} &= - \tfrac{\tilde{\sigma}}{4} \tilde{f}^{-1} {\bf t}^{3} \,,
  \\
  A_{z} &= \tfrac{1}{4 r} \tilde{\sigma} \tilde{f}^{-1} e^{-2\phi_{0}} \big(\cos\theta \, {\bf t}^{2} - \sin\theta \, {\bf t}^{1}\big)\,,
  \\
  A_{\psi} &= \tfrac{1}{4 r} \tilde{\sigma}\tilde{f}^{-1} e^{-2\phi_{0}}
  \big(\cos\theta \, {\bf t}^{1} + \sin\theta \, {\bf t}^{2}\big)\,. 
\end{aligned}
\end{equation}
Note that all $\theta$-dependences disappear once the same gauge transformation (\ref{gaugeTrYM}) is applied,
\begin{equation}
\begin{aligned}
  A'_{r} &= 0\,,
  \\
  A'_{\theta} &= - {\bf n}^{12} - {1 \over 4} \tilde{\sigma} \tilde{f}^{-1} \, {\bf t}^{3}\,,
  \\
  A'_{z} &= - {1 \over 4r} \tilde{\sigma} \tilde{f}^{-1} e^{-2\phi_{0}} \, {\bf t}^{1}\,,
  \\
  A'_{\psi} &= + {1 \over 4r} \tilde{\sigma} \tilde{f}^{-1} e^{-2\phi_{0}} \, {\bf t}^{2}\,. 
\end{aligned}\label{GaugeBrane2}
\end{equation}
As such, the gauge brane solution given by (\ref{GaugeBrane1}) and (\ref{GaugeBrane2}) does not depend on $\theta$. Therefore, this solution is single-valued when $\theta$ encircles the origin, and so it just describes a geometric background.

\subsubsection{Generalized metric and monodromy}
Consider next the generalized metric for the heterotic $5^2_2$-branes. While the geometric and non-geometric solutions are totally different, their associated generalized metrics are identical. We decompose the $\ODDG$ vector indices $\widehat{M}$ into $\{M,a,\tilde{a}\}$, where
\begin{equation}
\begin{aligned}
    M,N,\cdots &: ~ \ODD\, \text{vector indices}	\,,
    \\
    a,b,\cdots &: ~ \OO(\dim G_{\rm YM})\, \text{vector indices}\,,
    \\
    \tilde{a},\tilde{b},\cdots &: ~ \OO(\dim G_{\rm LL})\, \text{vector indices}\,. 
\end{aligned}
\end{equation}
The generalized metric is block-decomposed as\footnote{In this section, for simplicity, we use the different index ordering from (\ref{geomGenMet}) and (\ref{nongeomGenMet}).}
\begin{equation}
  \cH_{\hM\hN } = \begin{pmatrix} \cH_{MN} & \cH_{Mb} & \cH_{M\tilde{b}} 
  \\
  \cH_{a N} & \cH_{ab} & \cH_{a\tilde{b}}
  \\
  \cH_{\tilde{a}N} & \cH_{\tilde{a}b} & \cH_{\tilde{a}\tilde{b}}
  \end{pmatrix}\,.
\end{equation}
We shall now construct explicit form of the generalized metric for each type of five-brane solutions and, from them, deduce the corresponding monodromy matrix.

For symmetric solution, the corresponding generalized metric may be constructed from (\ref{SymmetricSol522}) and (\ref{A'}). Focusing only on $z$ and $\psi$ directions, we find the explicit form as
\begin{equation}
\begin{aligned}
\cH_{MN}^{\rm symm} &=   \begin{pmatrix}  f^{-1} & 0 & 0 & {\sigma\theta}{f}^{-1} 
  \\
  0 & f^{-1} & - {\sigma\theta}{f}^{-1} & 0 
  \\
  0 & - {\sigma\theta}{f}^{-1} & {K}f^{-1} & 0 
  \\
  {\sigma\theta}{f}^{-1} & 0 & 0 & {K}{f}^{-1} \end{pmatrix}\,,
  \\
  \cH_{M a}^{\rm symm} &= \begin{pmatrix} 
  +{1 \over 2} {f'}{f^{-2}} \, {\bf t}^{1} 
  \\
  - {1 \over 2} {f'}f^{-2} \, {\bf t}^{2}
  \\
  {1 \over 2} {f'}{f}^{-1} \big({\bf t}^{1} - {\sigma\theta}{f}^{-1} {\bf  t}^{2}\big)
  \\
 -{1 \over 2} {f'}{f}^{-1} \big( {\bf t}^{2} + {\sigma\theta}{f}^{-1} {\bf t}^{1}\big) \end{pmatrix}\,,
  \\
  \cH_{ab}^{\rm symm} &= \cH_{a\tilde{b}}^{\rm symm} = \cH_{\tilde{a}\tilde{b}}^{\rm symm} = -{1 \over 4 \rho^2} {\sigma^{2}}f^{-3 } \big( {\bf t}^{1}\otimes {\bf t}^{1} + {\bf t}^{2}\otimes {\bf t}^{2}\big)\,,
  \end{aligned}
\end{equation}
Again the $\cH_{\hM \hN}^{\rm symm}$ flux is not single-valued as the angular coordinate $\theta$ encircles the origin. Rather, it is transformed by an $\mathbf{O}(2,2)$ monodromy
\begin{equation}
  \cH^{\rm symm}(\theta + 2\pi) = (\Omega^{\rm symm})^T \cH^{\rm symm}(\theta) \Omega^{\rm symm}\,,
\end{equation}
where the monodromy matrix $\Omega^{\rm symm}$ is given by
\vskip0.5cm
\begin{tcolorbox}
\vskip-0.2cm
\begin{equation}
  \Omega^{{\rm symm}} = \begin{pmatrix} \mathbf{1}_{2} & 2\pi\sigma i\tau_{2} & 0
  \\
  0 & \mathbf{1}_{2} & 0 
  \\
  0 & 0 & \kappa
  \end{pmatrix} \in \mathbf{O}(2,2) \qquad \mbox{where} \qquad
  \tau_{2}=\begin{pmatrix} 0 & -i \\ i & 0\end{pmatrix}\,.
\end{equation}
\end{tcolorbox}
\noindent In fact, $\Omega^{\rm symm}$ is identical to the usual $5_2^2$-brane solution due to the embedding ansatz. This shows that symmetric $5^2_2$-brane background in heterotic supergravity is a T-fold. 
 
Consider next the generalized metric for neutral brane. As we discussed above, the gauge fields for neutral brane solution are trivial and the metric and Kalb-Ramond fields are identical to the usual $5^2_2$-brane solution. Thus, the corresponding generalized metric should be identical as well:
\begin{equation}
  \cH^{\rm neutral}_{MN} = \cH^{\rm symm}_{MN} \,, \qquad \cH^{\rm neutral}_{Ma}= \cH^{\rm neutral}_{ab}= \cH^{\rm neutral}_{\,a\tilde{b}} = \cH^{\rm neutral}_{\,\tilde{a}\tilde{b}} = 0\,, 
\end{equation}
and the monodromy matrix $\Omega^{\rm neutral}$ is given by
\vskip0.5cm
\begin{tcolorbox}
\vskip-0.2cm
\begin{equation}
  \Omega^{\rm neutral} = \Omega^{\rm symm}  \,.
\end{equation}
\end{tcolorbox}

Consider finally the gauge brane solution. This solution is $\theta$-independent,  as shown in (\ref{GaugeBrane1}) and (\ref{GaugeBrane2}), so the associated monodromy matrix is just the identity matrix:
\vskip0.5cm
\begin{tcolorbox}
\vskip-0.2cm
\begin{equation}
  \Omega^{\rm gauge} = \mathbf{1}\,.
\end{equation}
\end{tcolorbox}
\noindent
Therefore, the gauge brane solution is a geometric background.

\subsection{Exotic branes in the M-theory}

We now consider defect brane solutions in the eleven-dimensional M-theory \cite{LozanoTellechea:2000mc,Bergshoeff:2011se,deBoer:2012ma}. 
In this case, the solution depends not only on the holomorphic function $\rho(z)$ that behaves as $\ii \frac{\sigma}{2\pi}\,\log(r_\text{c}/z)$ near the center but also on another holomorphic function $f(z)$ that behaves as $f(z)\approx 1$ near the center. 
As observed in \cite{Eyras:1999at,LozanoTellechea:2000mc,Sakatani:2014hba}, for a given background of a conventional five-brane, we can easily find a background of its dual exotic branes through the following redefinitions:
\begin{align}
 \rho(z) \to -\rho^{-1}(z) \,,\qquad \rho_2\,\abs{f}^2\to \rho_2\,\abs{f}^2\,,\qquad \sigma\to \sigma^{-1}\,.
\label{eq:exotic-dual}
\end{align}
In the following, we shall study properties of exotic M5-brane backgrounds using our non-geometric parameterization in the M-theory section.

\subsubsection{$5^3$-brane}

Consider the M-theory compactified on a $d$-torus ($d\geq 3$) of radii $R_i$ along the $x^i$-directions. 
In this case, we have the defect M5-brane, which is the M5-brane (extended in $x^3,\dotsc ,x^7$-directions) smeared along $x^8,\,x^9,\,x^\sfM$-directions, and also the Kaluza-Klein vortex, which is the Kaluza-Klein monopole smeared along $x^8,\,x^9,\,x^\sfM$-directions. 
In addition, there exists an exotic $5^3$-brane (see Appendix \ref{app:exotic}). 
Below, we study each of them in detail. 

The background of the defect $\MM5(34567)$-brane is given by
\begin{align}
 \rmd s^2 = \rho_2^{-1/3}\,\bigl(\rho_2\,\abs{f}^2\,\rmd z\,\rmd \bar{z}+\rmd x^2_{03\cdots 7}\bigr) + \rho_2^{2/3} \, \rmd x^2_{89\sfM} \,, \quad A_{89\sfM} = \rho_1 \,,
\end{align}
where $\sigma= \sigma_{\MM5(34567)} \equiv l_{11}^3/R_8 R_9 R_\sfM$. 
Using the exotic duality transformation \eqref{eq:exotic-dual}, we obtain the background of $5^3(34567,89\sfM)$-brane as
\begin{align}
 \rmd s^2 = \Bigl(\frac{\rho_2}{\abs{\rho}^2}\Bigr)^{-\frac{1}{3}} \, \bigl(\rho_2\,\abs{f}^2\,\rmd z\,\rmd \bar{z}+\rmd x^2_{034567}\bigr) + \Bigl(\frac{\rho_2}{\abs{\rho}^2}\Bigr)^{\frac{2}{3}} \, \rmd x^2_{89\sfM} \,, 
\qquad
 A_{89\sfM} = -\frac{\rho_1}{\abs{\rho}^2} \,,
\label{eq:53-standard}
\end{align}
where $\sigma= \sigma_{5^3(34567,89\sfM)} \equiv R_8 R_9 R_\sfM/l_{11}^3$\,. 
Since $\rho_1$ is not single-valued, one can see that the metric is not single-valued. In fact, nontrivial monodromy arises only from this function. 

By comparing the two parameterization of a single generalized metric $\cM_{MN}$ and also using \eqref{eq:Omega-def2}, the $5^3$ brane background \eqref{eq:53-standard} can be rewritten in the non-geometric parameterization as
\begin{align}
 \rmd \widetilde{s}^2 = \rho_2^{1/3}\, \bigl(\rho_2\,\abs{f}^2\,\rmd z\,\rmd \bar{z}+\rmd x^2_{034567}\bigr) + \rho_2^{-2/3}\, \rmd x^2_{89\sfM} \,, 
\qquad
 \Omega^{89\sfM} = -\rho_1 \,,
\label{eq:53-non-geometric}
\end{align}
which coincides with the result obtained in \cite{Blair:2014zba}. 

As the multi-valuedness appears only from the function $\rho_1$, the metric in the non-geometric parameterization is single-valued. 
On the other hand, the tri-vector $\Omega^{89\sfM}$ has a monodromy, $\Omega^{89\sfM}\to \Omega^{89\sfM} - \sigma_{5^3(34567,89\sfM)}$, as one goes around the center. 
The monodromy matrix for the generalized metric,
\begin{align}
 \cM_{MN}\to \cM'_{MN}=(\Omega\,\cM\,\Omega^\rmT)_{MN}\qquad 
 (\theta\to \theta+2\pi)\,,
\end{align}
is given by 
\vskip0.5cm
\begin{tcolorbox}
\vskip-0.4cm
\begin{align}
 \Omega_{5^3(34567,89\sfM)} = \Exp{-\sigma_{5^3(34567,89\sfM)}\,\rho_{R_{89\sfM}}} \,. 
\label{monodromy-53}
\end{align}
\end{tcolorbox}
\noindent
For example, the monodromy matrix in $E_{7(7)}$ EFT becomes
\begin{align}
 &\Omega_{5^3(34567,89\sfM)} 
\nn\\
 &={\footnotesize\begin{pmatrix}
 \delta_a^b & 0 & 0 & \quad 0 \cr 
 3\sqrt{2}\,\sigma_{5^3(34567,89\sfM)}\,\delta^{a_1a_2 b}_{89\sfM} & \delta^{a_1a_2}_{b_1b_2} & 0 & \quad 0 \cr 
 0 & \sqrt{\frac{5!}{2}}\,\sigma_{5^3(34567,89\sfM)}\,\delta^{a_1a_2a_3a_4a_5}_{89\sfM b_1b_2} & \delta^{a_1\cdots a_5}_{b_1\cdots b_5} & \quad 0 \cr 
 0 & 0 & \frac{3}{\sqrt{5!}}\,\sigma_{5^3(34567,89\sfM)}\, \delta^a_{[8}\,\epsilon_{9\sfM] b_1\cdots b_5} & \quad \delta^a_b
 \end{pmatrix}}\,.
\end{align}
From the expression \eqref{monodromy-53}, one can see that the exotic $5^3$-brane is in one-to-one correspondence with the $E_{d(d)}$ generator $R_{abc}$, while the defect $\MM5$-brane is in one-to-one correspondence with the generator $R^{abc}$. 

We can extract the charge of $5^3(34567,89\ten)$-branes from the flux integral:
\vskip0.5cm
\begin{tcolorbox}
\vskip-0.4cm
\begin{align}
 Q_{5^3(34567,89\ten)} = - \sigma_{5^3(34567,89\ten)}^{-1} \oint_C \rmd \Omega^{89\ten}
 = - \sigma_{5^3(34567,89\ten)}^{-1} \oint_C \rmd x^{\hat{\mu}}\, S_{\hat{\mu}}{}^{89\ten} \,, 
\end{align}
\end{tcolorbox}
\noindent
where $C$ is a contour in the transverse two-dimensional space that enclose the exotic brane once counter-clockwise. 
In this sense, the exotic $5^3$-brane can be regarded as the magnetic source of the non-geometric $S^{(1,\,3)}$-flux. 

\subsubsection{$2^6$-brane}

For $d\geq 6$, there also arises another exotic $2^6$-brane. The background of the defect $\MM2(34)$-brane is given by
\begin{align}
 \underline{\text{ $\MM2$ :}}\quad 
 \rmd s^2 = \rho_2^{-2/3}\, \bigl(\rho_2\,\abs{f}^2\,\rmd z\,\rmd \bar{z}+\rmd x^2_{034}\bigr) + \rho_2^{1/3} \, \rmd x^2_{56789\ten} \,, 
\quad
 A_{5\cdots 9\ten} = \rho_1 \,,
\end{align}
where $\sigma =\sigma_{\MM2(34)} \equiv l_{11}^6/R_5 \cdots R_9 R_\ten$\,, From this, we can obtain the configuration of $2^6(34,56789\ten)$-brane using \eqref{eq:exotic-dual} as
\begin{align}
 \underline{\text{ $2^6$ :}}\quad 
 \rmd s^2 = \Bigl(\frac{\rho_2}{\abs{\rho}^2}\Bigr)^{-\frac{2}{3}}\, \bigl(\rho_2\,\abs{f}^2\,\rmd z\,\rmd \bar{z}+\rmd x^2_{034}\bigr) + \Bigl(\frac{\rho_2}{\abs{\rho}^2}\Bigr)^{\frac{1}{3}} \, \rmd x^2_{56789\ten} \,, \quad
 A_{5\cdots 9\ten} = -\frac{\rho_1}{\abs{\rho}^2} \,,
\label{eq:26-standard}
\end{align}
where $\sigma =\sigma_{2^6(34,56789\sfM)} \equiv R_5\cdots R_9 R_\sfM/l_{11}^6$\,. 

In the non-geometric parameterization, the configuration of $2^6(34,5\cdots 9\sfM)$-brane \eqref{eq:26-standard} can be rewritten as
\begin{align}
 \underline{\text{ $2^6$ :}}\quad 
 \rmd \widetilde{s}^2 = \rho_2^{2/3}\, \bigl(\rho_2\,\abs{f}^2\,\rmd z\,\rmd \bar{z}+\rmd x^2_{034}\bigr) + \rho_2^{-1/3}\, \rmd x^2_{5\cdots 9\sfM} \,, \qquad \ \  
 \Omega^{5\cdots 9\sfM} = - \rho_1 \,. \qquad
\label{eq:26-non-geom}
\end{align}
The multi-valuedness again appears only through $\Omega^{5\cdots 9\sfM}$ and the generalized metric undergoes the monodromy transformation 
\begin{align}
 \cM_{MN}\to \cM'_{MN}=(\Omega\,\cM\,\Omega^\rmT)_{MN}\, , 
 \end{align}
 where the monodromy matrix is given by 
 \vskip0.5cm
 \begin{tcolorbox}
 \vskip-0.4cm
 \begin{align} 
 \Omega_{2^6(34,56789\sfM)} = \Exp{\sigma_{2^6(34,56789\sfM)}\,\rho_{R_{5\cdots 9\sfM}}} \,. 
\end{align}
\end{tcolorbox}

Here again, we can count the charge of $2^6(34,56789\sfM)$-branes by the flux integral,
\vskip0.5cm
\begin{tcolorbox}
\vskip-0.5cm
\begin{align}
 Q_{2^6(34,56789\sfM)} = - \sigma_{2^6(34,56789\sfM)}^{-1} \oint_C \rmd \Omega^{5\cdots 9\sfM}
 = - \sigma_{2^6(34,56789\sfM)}^{-1} \oint_C \rmd x^{\hat{\mu}}\, S_{\hat{\mu}}{}^{5\cdots 9\sfM} \,, 
\end{align}
\end{tcolorbox}
\noindent
so the exotic $2^6$-brane may be regarded as the magnetic source of non-geometric $S^{(1,\,6)}$-flux. 

Note that, the conventional fields in the M2/M5-brane configuration and the non-geometric fields in the $2^6$/$5^3$-brane configuration are related each other in the following manner:
\vskip0.5cm
\begin{tcolorbox}
\vskip-0.3cm
\begin{align}
 & \ \ \ g_{\mu\nu}\rvert_{\text{conv.}} = g_{\mu\nu}\rvert_{\text{exotic}} \,,\qquad \quad \qquad
 \sfG_{ij}\rvert_{\text{conv.}} = \widetilde{\sfG}^{ij}\rvert_{\text{exotic}} \,,
\nn \\
 &A_{i_1i_2i_3}\rvert_{\text{conv.}} = -\Omega^{i_1i_2i_3}\rvert_{\text{exotic}}\,,\qquad 
 A_{i_1\cdots i_6}\rvert_{\text{conv.}} = -\Omega^{i_1\cdots i_6}\rvert_{\text{exotic}}\,,
 \nn
\end{align}
equivalently,
\begin{align}
 g_{\mu\nu}\rvert_{\text{conv.}} = g_{\mu\nu}\rvert_{\text{exotic}} \,,\qquad \qquad \qquad
 M_{MN}\rvert_{\text{conv.}} = \widetilde{M}^{MN}\rvert_{\text{exotic}}\,. 
\label{eq:correspondence}
\end{align}
\end{tcolorbox}
Here, note that $g_{\mu\nu}=\abs{G}^{\frac{1}{n-2}}\,\sfG_{\mu\nu}=\abs{\widetilde{G}}^{\frac{1}{n-2}}\,\widetilde{\sfG}_{\mu\nu}$.

\subsection{Exotic branes in the type IIB theory}

We finally consider defect-brane solutions in the type IIB supergravity \cite{LozanoTellechea:2000mc,Bergshoeff:2011se,deBoer:2012ma,Sakatani:2014hba}. 

\subsubsection{$5^2_2$-brane}

The type IIB $5^2_2$-brane is the exotic dual to the defect NS5-brane. 
From the defect $\NS5(34567)$-brane configuration,
\begin{align}
 \underline{\text{ $\NS5$ :}}\quad 
 \rmd s^2 = \rho_2\,\abs{f}^2\,\rmd z\,\rmd \bar{z} + \rmd x^2_{034567} + \rho_2\, \rmd x_{89}^2 \,, \qquad \Exp{2\phi} = \rho_2 \,, \qquad B_{89}= \rho_1 \,,
 \qquad
\end{align}
where $\sigma = \sigma_{\NS5} \equiv \ls^2/R_8 R_9$\,, using \eqref{eq:exotic-dual}, the $5^2_2$-brane configuration is obtained as
\begin{align}
 \underline{\text{ $5^2_2$ :}}\quad 
 \rmd s^2 = \rho_2\,\abs{f}^2\,\rmd z\,\rmd \bar{z} +\rmd x^2_{034567} + \frac{\rho_2}{\abs{\rho}^2}\,\, \rmd x_{89}^2 \,, \qquad
 \Exp{2\phi} = \frac{\rho_2}{\abs{\rho}^2}\,, \qquad 
 B_{89} = -\frac{\rho_1}{\abs{\rho}^2} \,, \qquad
\label{eq:522-conventional}
\end{align}
where $\sigma = \sigma_{5^2_2(34567,89)} \equiv R_8 R_9/\ls^2$\,. Applying the relation (\ref{eq:correspondence}) between the conventional parameterizations and the non-geometric parameterizations of the generalized metric in EFT, the $5^2_2$-brane configuration in the non-geometric parameterization is obtained as
\begin{align}
 \underline{\text{ $5^2_2$ :}}\quad
 \rmd \widetilde{s}^2 = \rho_2\,\abs{f}^2\,\rmd z\,\rmd \bar{z} + \rmd x^2_{034567} + \rho_2^{-1}\, \rmd x_{89}^2 \,, \qquad
 \Exp{2\widetilde{\phi}} = \rho_2^{-1} \,, \qquad
 \beta^{89}= -\rho_1 \,.  \ \ 
\end{align}
This coincides with the $5^2_2$-brane solution in the $\beta$-supergravity \cite{Hassler:2013wsa,Sakatani:2014hba} or the $\SL(5)$ EFT \cite{Blair:2014zba}. 
In the Einstein frame, this solution and the above NS5-brane solution in the conventional parameterization are also related by \eqref{eq:correspondence}. 
In fact, such relations persist to hold for all solutions to be considered below. 
Since there is no internal-coordinate dependence in the defect background, the potential part of the action does not contribute and the action has a symmetry under the map, $g_{\mu\nu}\to g_{\mu\nu}$ and $\cM_{MN}\to \widetilde\cM^{MN}$ in \eqref{eq:correspondence} \,. 
This is the reason why the identification \eqref{eq:correspondence} always connect the solutions in two different parameterizations. 

It is straightforward to check that the monodromy matrix for the generalized metric is given by
\vskip0.5cm
\begin{tcolorbox}
\vskip-0.3cm
\begin{align}
 \Omega_{5^2_2(34567,89)} = \Exp{-\sigma_{5^2_2(34567,89)}\,\rho_{R_1^{89}}} \,. 
\end{align}
\end{tcolorbox}
\noindent
We can explicitly check that the monodromy transformation is invariant under the T-duality. To see this, note that the $5^2_2(34567,89)$-brane is also present in the type IIA theory as the compactification of an (anti) $5^3(34567,89\sfM)$-brane. 
Therefore, in the type IIA (i.e.~the M-theory) section, the monodromy matrix is given by
\begin{align}
 \Omega_{5^2_2(34567,89)} = \Exp{\sigma_{5^3(34567,89\sfM)}\,\rho_{R_{89\sfM}}} \,. 
\end{align}
Note that the equality, $\sigma_{5^3(34567,89\sfM)}= R_8 R_9 R_\sfM/l_{11}^3 = R_8 R_9/\ls^2=\sigma_{5^2_2(34567,89)}$\,, is satisfied. 

We can count the charge of $5^2_2(34567,89)$-branes by the relevant flux integral,
\vskip0.5cm
\begin{tcolorbox}
\vskip-0.5cm
\begin{align}
 Q_{5^2_2(34567,89)} = - \sigma_{5^2_2(34567,89)}^{-1} \oint_C \rmd \beta^{89}
 = - \sigma_{5^2_2(34567,89)}^{-1} \oint_C \rmd x^{\hat{\mu}}\, Q_{\hat{\mu}}{}^{89} \,, 
\end{align}
\end{tcolorbox}
\noindent
so the exotic $5^2_2$-brane can be regarded as the magnetic source of the non-geometric $Q^{(1,\,2)}$-flux. This was previously noted in \cite{Hassler:2013wsa,Andriot:2014uda,Okada:2014wma}. 

\subsubsection{$p^{7-p}_3$-brane}
Consider next exotic D-branes. 

The background of defect $\DD p(3\cdots (p+2))$-brane is given by
\begin{align}
\begin{split}
 \underline{\text{ $\DD p$ :}}\qquad 
 &\rmd s^2
 = \rho_2^{-1/2}\, \bigl(\rho_2\,\abs{f}^2\,\rmd z\,\rmd \bar{z}+\rmd x^2_{03\cdots (p+2)}\bigr) + \rho_2^{1/2}\, \rmd x^2_{(p+3)\cdots 9} \,, \qquad  \qquad \qquad \qquad 
\\
 &\Exp{2\phi} = \rho_2^{\frac{3-p}{2}} \,,\quad 
 C_{(p+3)\cdots 9} = \rho_1 \,,
\end{split}
\end{align}
where $\sigma = \sigma_{\DD p(3\cdots (p+2))} \equiv \gs\, \ls^{7-p}/R_{p+3} \cdots R_9$\,. 
From this, we obtain the $p^{7-p}_3(3\cdots (p+2),(p+3)\cdots 9)$ configuration as
\begin{align}
\begin{split}
 \underline{\text{ $p^{7-p}_3$ :}}\qquad 
 &\rmd s^2 = \biggl(\frac{\rho_2}{\abs{\rho}^2}\biggr)^{-\frac{1}{2}}\, \bigl(\rho_2\,\abs{f}^2\,\rmd z\,\rmd \bar{z}+\rmd x^2_{03\cdots (p+2)}\bigr) + \biggl(\frac{\rho_2}{\abs{\rho}^2}\biggr)^{\frac{1}{2}} \, \rmd x^2_{(p+3)\cdots 9} \,, \qquad
\\
 &\Exp{2\phi} = \biggl(\frac{\rho_2}{\abs{\rho}^2}\biggr)^{\frac{3-p}{2}} \,, \qquad 
 C_{(p+3)\cdots 9} = -\frac{\rho_1}{\abs{\rho}^2} \,, 
\end{split}
\end{align}
where $\sigma = \sigma_{p^{7-p}_3(3\cdots (p+2),(p+3)\cdots 9)} \equiv R_{p+3} \cdots R_9/\gs\, \ls^{7-p}$\,. 

In the non-geometric parameterization, the $p^{7-p}_3$-brane configuration becomes
\begin{align}
\begin{split}
 \underline{\text{ $p^{7-p}_3$ :}}\qquad
 &\rmd \widetilde{s}^2 = \rho_2^{1/2}\,\bigl(\rho_2\,\abs{f}^2\,\rmd z\,\rmd \bar{z} + \rmd x^2_{03\cdots (p+2)}\bigr) + \rho_2^{-1/2}\, \rmd x^2_{(p+3)\cdots 9} \,, 
\hskip2.5cm \\
 &\Exp{2\widetilde{\phi}} = \rho_2^{\frac{p-3}{2}} \,, \qquad
 \gamma^{(p+3)\cdots 9} = -\rho_1 \,. 
\end{split}
\end{align}

From the above configurations, we obtain the monodromy matrices for $5^2_3$\,, $3^4_3$\,, and $1^6_3$-branes given by
\vskip0.5cm
\begin{tcolorbox}
\vskip-0.4cm
\begin{align}
 \Omega_{5^2_3(3\cdots 7,89)} 
 &=\Exp{-\sigma_{5^2_3(3\cdots 7,89)}\, \rho_{R^2_{89}}} \,, 
\\
 \Omega_{3^4_3(345,6\cdots 9)} 
 &=\Exp{\sigma_{3^4_3(3\cdots 5,6\cdots 9)}\, \rho_{R_{6789}}} \,, 
\\
 \Omega_{1^6_3(3,4\cdots 9)} 
 &=\Exp{- \sigma_{1^6_3(3,4\cdots 9)}\,\rho_{R_{4\cdots 9}^{1}}} \,.
\end{align}
\end{tcolorbox}
\noindent
Again, by circle compactification and T-duality transformation, the exotic $4^3_3$ and $2^5_3$-branes in the type IIA theory appear as the compactification of $5^3$, and $2^6$-brane, respectively, and their monodromy matrices are given by
\begin{align}
 \Omega_{4^3_3(3456,789)} 
 &=\Exp{-\sigma_{4^3_3(3456,789)}\, \rho_{R_{789}}} \,, 
\\
 \Omega_{2^5_3(34,5\cdots 9)} 
 &=\Exp{\sigma_{2^5_3(34,5\cdots 9)}\,\rho_{R_{5\cdots 9\sfM}}} \,.
\end{align}

The charge of $p^{7-p}_3(3\cdots (p+2),(p+3)\cdots 9)$-branes is counted by the flux integral,
\vskip0.5cm
\begin{tcolorbox}
\vskip-0.4cm
\begin{align}
 Q_{p^{7-p}_3(3\cdots (p+2),(p+3)\cdots 9)} &= - \sigma_{p^{7-p}_3(3\cdots (p+2),(p+3)\cdots 9)}^{-1} \oint_C \rmd \gamma^{(p+3)\cdots 9}
\nn\\
 &= - \sigma_{p^{7-p}_3(3\cdots (p+2),(p+3)\cdots 9)}^{-1} \oint_C \rmd x^{\hat{\mu}}\, P_{\hat{\mu}}{}^{(p+3)\cdots 9} \,, 
\end{align}
\end{tcolorbox}
\noindent
so the exotic $p^{7-p}_3$-brane can be regarded as the magnetic source of the non-geometric $P^{(1,\,7-p)}$-flux.

\subsubsection{$1^6_4$-brane}
Consider finally exotic fundamental string. 
The defect fundamental string configuration, $\FF1(3)$, is given by
\begin{align}
 \underline{\text{ $\FF1$ :}}\qquad 
 \rmd s^2 = \rho_2^{-1}\, \bigl(\rho_2\,\abs{f}^2\,\rmd z\,\rmd \bar{z} +\rmd x_{03}^2 \bigr)+ \rmd x^2_{4\cdots 9} \,, \qquad 
 \Exp{2\phi} = \rho_2^{-1} \,, \qquad 
 B_{4\cdots 9} = \rho_1 \,, \ \ 
\end{align}
where $\sigma = \sigma_{\FF1(3)} \equiv \gs^2\,\ls^6/R_4\cdots R_9$\,. Thus, the $1^6_4(3,4\cdots 9)$-brane configuration becomes
\begin{align}
 \underline{\text{ $1^6_4$ :}}\quad \rmd s^2= \frac{\abs{\rho}^2}{\rho_2} \, \bigl(\rho_2\,\abs{f}^2\,\rmd z\,\rmd \bar{z} +\rmd x_{03}^2 \bigr) + \rmd x^2_{4\cdots 9} \,, \qquad
 \Exp{2\phi} = \frac{\abs{\rho}^2}{\rho_2} \,, \qquad 
 B_{4\cdots 9}= -\frac{\rho_1}{\abs{\rho}^2} \,, 
\end{align}
where $\sigma = \sigma_{1^6_4(3,4\cdots 9)} \equiv R_4\cdots R_9/\gs^2\,\ls^6$\,. 

In the non-geometric parameterization, the above $1^6_4(3,4\cdots 9)$-brane configuration becomes
\begin{align}
 \underline{\text{ $1^6_4$ :}}\qquad 
 \rmd \widetilde{s}^2 = \rho_2\,\bigl(\rho_2\,\abs{f}^2\,\rmd z\,\rmd \bar{z} +\rmd x^2_{03}\bigr) + \rmd x_{4\cdots 9}^2 \,, \qquad
 \Exp{2\widetilde{\phi}} = \rho_2 \,, \qquad
 \beta^{4\cdots 9}= -\rho_1 \,. \ \ \ 
\end{align}

The monodromy matrix for the generalized metric is obtained as
\vskip0.5cm
\begin{tcolorbox}
\vskip-0.5cm
\begin{align}
 \Omega_{1^6_4(3,4\cdots 9)} = \Exp{- \sigma_{1^6_4(3,4\cdots 9)}\,\rho_{R^2_{4\cdots 9}}} \,. 
\end{align}
\end{tcolorbox}

We count the charge of $1^6_4(3,4\cdots 9)$-branes from the flux integral,
\vskip0.5cm
\begin{tcolorbox}
\vskip-0.5cm
\begin{align}
 Q_{1^6_4(3,4\cdots 9)} = - \sigma_{1^6_4(3,4\cdots 9)}^{-1} \oint_C \rmd \beta^{4\cdots 9}
 = - \sigma_{1^6_4(3,4\cdots 9)}^{-1} \oint_C \rmd x^{\hat{\mu}}\, Q_{\hat{\mu}}{}^{4\cdots 9} \,, 
\end{align}
\end{tcolorbox}
\noindent
and so the exotic $1^6_4$-brane can also be regarded as the magnetic source of non-geometric $Q^{(1,\,6)}$-flux. 

\section{Discussion}
\label{sec:discussion}

In this paper, we showed two parameterizations, conventional, geometric parameterization and non-conventional, non-geometric parameterization, of the generalized metric in extended field theories from an approach based on different decomposition of the Lie algebra for the duality transformation group. 
In this approach, the exotic duality between the two parameterization was identified with the generalized transpose of the generalized vielbein. 
We constructed the action of the extended field theories using the non-geometric parameterizations and, from them, obtained the effective actions for the non-geometric fluxes. 

In the type IIA and IIB theories, obtained from the EFT, the effective action involves the non-geometric $P^{(1,\,q)}$-fluxes and $Q^{(1,\,p)}$-fluxes with $p=2,6$, generalizing the action of the $\beta$-supergravity that includes only the $Q^{(1,\,2)}$-flux. We also constructed the effective action for heterotic theories by starting from the heterotic DFT in non-geometric parameterization. 

The non-geometric effective action we constructed in this work would open up many directions for future research. Here, we list some of them that we are currently investigating. 

\begin{itemize}
\item  One would like to investigate various non-geometric background directly from the non-geometric effective action. In particular, exotic brane backgrounds may be constructed directly from the non-geometric effective action. 
We would like to classify all 1/2-BPS backgrounds.
\item One also would like to understand the dynamics of defect branes in non-geometric background.  In particular, one wants to construct worldsheet conformal field theory approach to the non-geometric backgrounds. These would be the non-geometric counterpart of type II five-branes, whose near-horizon geometry is described by the exact conformal field theory of Kazama-Suzuki coset model times super-Liouville theory \cite{Rey:1989xi, Rey:1991uu, Callan:1991dj}. 
\item This effective action we constructed contains multitude of non-geometric fluxes. Therefore, it can describe the coexistence of different non-geometric fluxes. For example, we can describe the non-geometric backgrounds that correspond to a bound-state of various exotic branes. We would like to classify all 1/4-BPS non-BPS backgrounds. 
\item Apart from non-standard dilaton dependences, the type II non-geometric actions, \eqref{eq:IIA-non-geometric} or \eqref{eq:IIB-non-geometric}, has the structure similar to the conventional type II supergravity action. It is thus possible to find various classical solutions of that action that carry not only magnetic charges but also electric charges or dyonic charges. We note that a family of new classical solutions carrying an electric charge for a non-geometric potentials (such as $\beta^{mn}$) was already found in \cite{Sakatani:2014hba} from the action for a non-geometric flux. 
\item The non-geometric action in the $E_{d(d)}$ EFT with $d\leq 7$ does not contain all non-geometric fluxes associated with all classified exotic branes. In order to describe all non-geometric fluxes, we will need to consider the $E_{d(d)}$ EFT with higher $d$. The $E_{8(8)}$ case can be also considered in a similar way, using the results of \cite{Godazgar:2013rja}, but the cases with $d\geq 9$ will remain to be a challenging future program. 
\item The conserved charges in the conventional DFT are studied in \cite{Blair:2015eba,Park:2015bza,Naseer:2015fba}. There, the string winding charges are reproduced as the Noether charges associated with isometries along the dual directions. If we perform the same analysis in the $E_{8(8)}$ EFT, the charges of conventional branes and exotic branes will be reproduced in the same manner. Even in $E_{d(d)}$ EFT with $d\leq 7$, one may also reproduce the exotic brane charges as magnetic charges extending the analysis performed in DFT \cite{Bakhmatov:2016kfn}, where the non-geometric parameterization found in this paper will be useful. 
\end{itemize}
\section*{Acknowledgement}
We thank David Andriot, Machiko Hatsuda, Jeong-Hyuk Park, Shin Sasaki, and Shozo Uehara for useful discussions. 
We acknowledge organizers and participants of the CERN-CKC TH Institute `Duality Symmetries in String and M-Theories', and APCPT Focus Programs ``Liouville, Integrability and Branes (11), (12)'' and ``Duality and Novel Geometry in M-Theory'' for providing stimulating environment during this work. This work was supported in part by the National Research Foundation of Korea through the grants 2005-0093843, 2010-220-C00003 and 2012K2A1A9055280. 
\newpage

\appendix

\section{Notations}
\label{app:notations}

We use the following notations for anti-symmetrization:
\begin{align}
 &\delta^{a_1\cdots a_p}_{b_1\cdots b_p} = \delta^{a_1}_{[b_1} \cdots \delta^{a_p}_{b_p]} 
 = \frac{1}{p!}\,\bigl(\delta^{a_1}_{b_1} \cdots \delta^{a_p}_{b_p}\pm \text{permutations}\bigr)
\,, 
\\
 &e^{i_1i_2}_{\,\,a_1a_2} =(e^{-\rmT})^{i_1}{}_{[a_1}\,(e^{-\rmT})^{i_2}{}_{a_2]}\,, \quad 
 e\equiv \det e_i{}^a \,,
\\
 &G^{i_1\cdots i_n,\,j_1\cdots j_n} = \delta^{i_1\cdots i_n}_{k_1\cdots k_n}\,G^{k_1j_1}\cdots G^{k_nj_n} \,,\quad 
 \abs{G}\equiv \det G_{ij}\,,
\end{align}
where ${}^{-\rmT}$ represents a combination of the inverse and the matrix transpose. 
Similar expressions, such as $\delta^{i_1\cdots i_p}_{j_1\cdots j_p}$ or $\widetilde{G}^{i_1\cdots i_n,\,j_1\cdots j_n}$, are also used. 

Our coordinates are normalized such that the flat metric has the form,
\begin{align}
 \rmd s^2 = \delta_{ab}\,\rmd x^a\,\rmd x^b + \delta_{a_1a_2,\,b_1b_2}\,\rmd x^{a_1a_2}\,\rmd x^{b_1b_2} + \cdots \,. 
\end{align}
Then, if we redefine the coordinates in order to reduce the number of indices, we should introduce the following numerical factors:
\begin{align}
 y_{a_1\cdots a_p} \to z^{a_1\cdots a_{d-p}}\equiv \frac{1}{\sqrt{p!(d-p)!}}\, \epsilon^{a_1\cdots a_{d-p} b_1\cdots b_p}\,y_{b_1\cdots b_p} \,.
\end{align}

Various indices are summarized as follows. 
\begin{align}
 \text{M-theory: }\quad & (x^{\hat{\mu}}) = (x^\mu,\,x^i) \ \qquad (\hat{\mu}=0,\dotsc,9,\sfM,\quad i=n,\dotsc, 9,\sfM)\,,\\
 \text{type IIA: }\quad & (x^{\boldsymbol{\hat\mu}}) = (x^\mu,\,x^m) \qquad ({\boldsymbol{\hat\mu}}=0,\dotsc,9,\quad m=n,\dotsc, 9)\,,\\
 \text{type IIB: }\quad & (x^{\hat{\mu}}) = (x^\mu,\,x^{\sfm}) \ \qquad (\hat{\mu}=0,\dotsc,9,\quad \sfm=n,\dotsc, 9)\,,
\end{align}
where $\mu=0,\dotsc ,n-1$ and $n=11-d$\,. 

\subsection{$E_{d(d)}$ algebras: M-theory section}
\label{app:Edd-algebra}

The $E_{d(d)}$ algebra is given by \cite{Berman:2011jh}
\begin{align}
\begin{split}
 &\bigl[K^a{}_b,\, K^c{}_d\bigr] 
 = \delta^c_b\, K^a{}_d - \delta^{a}_{d}\,K^c{}_b\,,\quad \ \ \ 
 \bigl[K^a{}_b,\, R^{a_1a_2a_3} \bigr] 
 = 3\,\delta^{[a_1\vert}_{b}\,R^{a\vert a_2a_3]}\,, 
\\
 &\bigl[K^a{}_b,\, R_{a_1a_2a_3} \bigr] = - 3\,\delta^a_{[a_1\vert} \, R_{b\vert a_2a_3]}\,,\quad 
 \ \bigl[K^a{}_b,\, R^{a_1\cdots a_6} \bigr] 
 = 6\,\delta^{[a_1\vert}_b\, R^{a\vert a_2\cdots a_6]}\,,
\\
 &\bigl[K^a{}_b,\, R_{a_1\cdots a_6} \bigr] = - 6\,\delta^a_{[a_1\vert}\, R_{b\vert a_2\cdots a_6]}\,,\quad 
 \bigl[ R^{a_1a_2a_3},\,R^{a_4a_5a_6} \bigr] = R^{a_1\cdots a_6}\,, 
\\
 &\bigl[ R^{a_1a_2a_3},\,R_{b_1b_2b_3} \bigr] = 18\,\delta_{[b_1b_2}^{[a_1a_2}\, K^{a_3]}{}_{b_3]} - 2\,\delta_{b_1b_2b_3}^{a_1a_2a_3}\, D \,, 
\\
 &\bigl[ R^{a_1a_2a_3},\, R_{b_1\cdots b_6} \bigr] = 120 \,\delta_{[b_1b_2b_3}^{a_1a_2a_3}\,R_{b_4b_5b_6]} \,, 
\\
 & \bigl[ R_{a_1a_2a_3},\,R_{a_4a_5a_6} \bigr] = R_{a_1\cdots a_6}\,, \qquad \qquad 
\bigl[ R_{a_1a_2a_3},\,R^{b_1\cdots b_6} \bigr] = 120\,\delta^{[b_1b_2b_3}_{a_1a_2a_3}\,R^{b_4b_5b_6]} \,, 
\\
 &\bigl[R^{a_1\cdots a_6},\,R_{b_1\cdots b_6} \bigr] = - 4320\,\delta_{[b_1\cdots b_5}^{[a_1\cdots a_5}\, K^{a_6]}{}_{b_6]} + 480\,\delta_{b_1\cdots b_6}^{a_1\cdots a_6}\,D \,,
\end{split}
\end{align}
where $D\equiv \sum_a K^a{}_a$\,. 
Note that the normalizations of $R^{a_1\cdots a_6}$ and $R_{b_1\cdots b_6}$ are changed from those used in \cite{Berman:2011jh} by a factor 2 and $K^a{}_b$ corresponds to $\widetilde{K}^a{}_b\equiv K^a{}_b-\frac{1}{n-2}\,\delta^a_b\,\sum_\mu K^\mu{}_\mu$ of \cite{Berman:2011jh}. 

The commutators of the $E_{d(d)}$ generators with the central charges, $\bigl(P_a,\, Z^{a_1a_2} ,\, Z^{a_1\cdots a_5} ,\, W^a\equiv \frac{1}{7!}\,\epsilon_{a_1\cdots a_7}\,Z^{a_1\cdots a_7,\,a}\bigr)$, are given by \cite{Berman:2011jh}
\begin{align}
\begin{split}
 &\bigl[K^c{}_d,\, P_a\bigr] = - \delta^c_a\,P_d - \frac{\delta^c_d}{n-2}\, P_a\,, \quad \ \ \ \ \
 \bigl[K^c{}_d,\,Z^{a_1a_2}\bigr] = 2\,\delta^{[a_1\vert}_d\,Z^{c\vert a_2]} - \frac{\delta^c_d}{n-2} \,Z^{a_1a_2}\,,
\\
 &\bigl[K^c{}_d,\,Z^{a_1\cdots a_5}\bigr] = 5\,\delta^{[a_1\vert}_d\,Z^{c\vert a_2\cdots a_5]} - \frac{\delta^c_d}{n-2}\, Z^{a_1\cdots a_5}\,,
\\
 &\bigl[K^c{}_d,\, W^a\bigr] = \delta^a_d\, W^c + \delta^c_d\, W^a - \frac{\delta^c_d}{n-2}\, W^a \,,
\\
 &\bigl[R^{c_1c_2c_3},\,P_a\bigr] = 3\,\delta_a^{[c_1}\,Z^{c_2c_3]}\,, \qquad \qquad \quad
 \bigl[R^{c_1c_2c_3},\,Z^{a_1a_2}\bigr] = Z^{c_1\cdots c_3 a_1a_2}\,,
\\
 &\bigl[R^{c_1c_2c_3},\,Z^{a_1\cdots a_5}\bigr] = \epsilon^{a_1\cdots a_5[c_1c_2}\,W^{c_3]}\,,\quad \ \ 
 \bigl[R^{c_1\cdots c_6},\,P_a\bigr] = - 6\,\delta_a^{[c_1}\, Z^{c_2\cdots c_6]}\,,
\\
 &\bigl[R^{c_1\cdots c_6},\,Z^{a_1a_2}\bigr] = - 2\,\epsilon^{a_1a_2[c_1\cdots c_5} \,W^{c_6]}\,,\quad
 \bigl[R_{c_1c_2c_3},\,Z^{a_1a_2}\bigr] = 3!\,\delta^{a_1a_2}_{[c_1c_2}\,P_{c_3]}\,,
\\
 &\bigl[R_{c_1c_2c_3},\,Z^{a_1\cdots a_5}\bigr] = \frac{5!}{2}\,\delta^{[a_1a_2a_3}_{c_1c_2c_3}\,Z^{a_4a_5]}\,, \quad \
 \bigl[R_{c_1c_2c_3},\,W^a\bigr] 
 = \frac{9}{5!}\, \delta^a_{[c_1}\,\epsilon_{c_2c_3] b_1\cdots b_5}\,Z^{b_1\cdots b_5} \,,
\\
 &\bigl[R_{c_1\cdots c_6},\,Z^{a_1\cdots a_5}\bigr] = - 6\cdot 5!\,\delta^{a_1\cdots a_5 b}_{c_1\cdots c_5c_6}\,P_b \,, \quad 
  \bigl[R_{c_1\cdots c_6},\, W^a\bigr] = - 3 \, \epsilon_{c_1\cdots c_6 b}\,Z^{ba}\,.
\end{split}
\label{eq:Edd-momenta_M}
\end{align}

Using the normalization of the central charges for $E_{d(d)}$ with $d\leq 7$ (see \cite{Berman:2011jh}),
\begin{align}
 \bigl(Z_A\bigr)=\Bigl(P_a,\,\frac{Z^{a_1a_2}}{\sqrt{2}},\,\frac{Z^{a_1\cdots a_5}}{\sqrt{5!}},\,\frac{W^a}{3}\Bigr) \,,
\end{align}
the algebra \eqref{eq:Edd-momenta_M} together with \eqref{eq:l1rep-matrix} gives the following matrix representations for the $E_{d(d)}$ generators:
\begin{align}
 &(\rho_{K^c{}_d})_A{}^B =\begin{pmatrix}
  \delta_a^c\,\delta_d^b & 0 & 0 & 0 \cr 
  0 & -2\,\delta_d^{[a_1\vert}\,\delta^{c\vert a_2]}_{b_1b_2} & 0 & 0 \cr 
  0 & 0 & -5\,\delta_d^{[a_1\vert}\,\delta^{c\vert a_2\cdots a_5]}_{b_1\cdots b_5} & 0 \cr 
  0 & 0 & 0 & - 2\,\delta^{(a}_b\,\delta^{c)}_d 
 \end{pmatrix} + \frac{\delta^c_d}{n-2}\,\delta_A^B \,, 
\label{eq:rho-K-M}
\\
 &(\rho_{R^{c_1c_2c_3}})_A{}^B =\begin{pmatrix}
  0 & -\frac{3!}{\sqrt{2}}\,\delta^{c_1c_2c_3}_{a_1 b_1b_2} & 0 & 0 \cr 
  0 & 0 & -\frac{5!}{\sqrt{2\cdot 5!}}\,\delta^{c_1c_2c_3 a_1a_2}_{b_1b_2b_3b_4b_5} & 0 \cr 
  0 & 0 & 0 & -\frac{3}{\sqrt{5!}}\,\epsilon^{a_1\cdots a_5[c_1c_2}\,\delta^{c_3]}_b \cr 
  0 & 0 & 0 & 0 
 \end{pmatrix}\,,
\\
 &(\rho_{R^{c_1\cdots c_6}})_A{}^B =\begin{pmatrix}
  0 & \ \ \qquad 0 &\ \ \ \qquad \frac{6!}{\sqrt{5!}}\,\delta_{a b_1\cdots b_5}^{c_1\cdots c_6} & \qquad 0 \cr 
  0 & \ \ \qquad 0 &\ \ \ \qquad 0 & \qquad -\frac{2}{\sqrt{2}}\,\epsilon^{c_1\cdots c_6[a_1}\,\delta^{a_2]}_b \cr 
  0 & \ \ \qquad 0 &\ \ \ \qquad 0 & \qquad 0 \cr 
  0 & \ \ \qquad 0 &\ \ \ \qquad 0 & \qquad 0 
 \end{pmatrix}\,,
\\
 &(\rho_{R_{c_1c_2c_3}})_A{}^B =\begin{pmatrix}
  0 & 0 & 0 & 0 \cr 
  -\frac{3!}{\sqrt{2}}\,\delta^{a_1a_2 b}_{c_1c_2c_3} & 0 & 0 & 0 \cr 
  0 & -\frac{5!}{\sqrt{2\cdot 5!}}\,\delta^{a_1a_2a_3a_4a_5}_{c_1c_2c_3 b_1b_2} & 0 & 0 \cr 
  0 & 0 & -\frac{3}{\sqrt{5!}}\, \delta^a_{[c_1}\,\epsilon_{c_2c_3] b_1\cdots b_5} & 0
 \end{pmatrix}\,,
\\
 &(\rho_{R_{c_1\cdots c_6}})_A{}^B =\begin{pmatrix}
  0 & 0 & \qquad \qquad 0 & \qquad \qquad 0 \cr 
  0 & 0 & \qquad \qquad 0 & \qquad \qquad0 \cr 
  \frac{6!}{\sqrt{5!}}\,\delta^{a_1\cdots a_5 b}_{c_1\cdots c_5c_6} & 0 & \qquad \qquad 0 & \qquad \qquad 0 \cr 
  0 & -\frac{2}{\sqrt{2}}\,\delta_{[b_1}^a\epsilon_{b_2]c_1\cdots c_6} & \qquad \qquad 0 & \qquad \qquad 0 
 \end{pmatrix}\,. 
\label{eq:M-representation}
\end{align}
From the Cartan involution,
\begin{align}
 \theta(K^a{}_b)=-K^a{}_b\,,\quad 
 \theta(R^{c_1c_2c_3})= - R_{c_1c_2c_3} \,,\quad 
 \theta(R^{c_1\cdots c_6})= R_{c_1\cdots c_6} \,,
\label{eq:Cartan-involution-M}
\end{align}
we can see that the generalized transpose is indeed the same as the matrix transpose; $(\cdots)_A{}^B\to (\cdots)_B{}^A$. 

Concretely, for $E_{d(d)}$ with $d=4,5,6,7$, the central charges, equivalently, the generalized momenta are given by
\begin{align}
\begin{split}
 E_{7(7)}:&\quad \bigl({\cal Z}_A\bigr)=\Bigl(P_a,\,\frac{Z^{a_1a_2}}{\sqrt{2}},\,\frac{Z^{a_1\cdots a_5}}{\sqrt{5!}},\,\frac{W^a}{3}\Bigr) \,,
\\
 E_{6(6)}:&\quad \bigl({\cal Z}_A\bigr)=\Bigl(P_a,\,\frac{Z^{a_1a_2}}{\sqrt{2}},\,\frac{Z^{a_1\cdots a_5}}{\sqrt{5!}}\Bigr) \,,
\\
 \SO(5,5):&\quad \bigl({\cal Z}_A\bigr)=\Bigl(P_a,\,\frac{Z^{a_1a_2}}{\sqrt{2}},\,\frac{Z^{a_1\cdots a_5}}{\sqrt{5!}}\Bigr) \,,
\\
 \SL(5):&\quad \bigl({\cal Z}_A\bigr)=\Bigl(P_a,\,\frac{Z^{a_1a_2}}{\sqrt{2}}\Bigr) \,,
\end{split}
\end{align}
and the matrix representations for the $E_{d(d)}$ generators with $d=4,5,6$ are simply given by truncating those of the $E_{7(7)}$ generators. 

\subsection{$E_{d(d)}$ algebras: type IIB section}

In the type IIB section, the $E_{d(d)}$ generators are decomposed as \cite{Tumanov:2014pfa}
\begin{align}
 \bigl\{K^\sfa{}_\sfb ,\, 
 R_{\alpha\beta} ,\, 
 R^{\sfa_1\sfa_2}_{\alpha} ,\, 
 R_{\sfa_1\sfa_2}^{\alpha} ,\, 
 R^{\sfa_1\cdots \sfa_4} ,\, 
 R_{\sfa_1\cdots \sfa_4} ,\, 
 R^{\sfa_1\cdots \sfa_6}_{\alpha} ,\, 
 R_{\sfa_1\cdots \sfa_6}^{\alpha} \bigr\}\,, 
\end{align}
and the algebra is given by \cite{Tumanov:2014pfa}
\begin{align}
\begin{split}
 &\bigl[ K^\sfa{}_\sfb,\,K^\sfc{}_\sfd \bigr] 
 = \delta^\sfc_b\, K^\sfa{}_\sfd - \delta^\sfa_\sfd\, K^\sfc{}_\sfb\,, 
\\
 &\bigl[ K^\sfa{}_\sfb,\, R^{\sfa_1\sfa_2}_{\alpha} \bigr] 
 = 2\,\delta^{[\sfa_1\vert}_\sfb \, R^{\sfa\vert\sfa_2]}_{\alpha}\,, \hskip2.3cm 
  \bigl[ K^\sfa{}_\sfb,\, R_{\sfa_1\sfa_2}^{\alpha} \bigr] 
 = -2\,\delta_{[\sfa_1\vert}^\sfa\, R_{\sfb\vert\sfa_2]}^{\alpha}\,,
\\
 &\bigl[ K^\sfa{}_\sfb,\, R^{\sfa_1\cdots \sfa_4} \bigr] 
 = 4\,\delta^{[\sfa_1\vert}_\sfb \, R^{\sfa\vert\sfa_2\sfa_3\sfa_4]}\,, \hskip1.5cm
  \bigl[ K^\sfa{}_\sfb,\, R_{\sfa_1\cdots \sfa_4} \bigr] 
 = - 4\,\delta_{[\sfa_1\vert}^\sfa\, R_{\sfb\vert\sfa_2\sfa_3\sfa_4]}\,,
\\
 &\bigl[ K^\sfa{}_\sfb,\, R^{\sfa_1\cdots \sfa_6}_{\alpha} \bigr] 
 = 6\,\delta^{[\sfa_1\vert}_\sfb \, R^{\sfa\vert\sfa_2\cdots \sfa_6]}_{\alpha}\,, \hskip1.5cm
  \bigl[ K^\sfa{}_\sfb,\, R_{\sfa_1\cdots \sfa_6}^{\alpha} \bigr] 
 = - 6\,\delta_{[\sfa_1\vert}^\sfa \, R_{\sfb\vert\sfa_2\cdots \sfa_6]}^{\alpha}\,,
\\
 &\bigl[ R_{\alpha\beta},\,R_{\gamma\delta} \bigr] 
 = \delta^\sigma_{(\alpha}\,\epsilon_{\beta)\gamma} \,R_{\sigma\delta} + \delta^\sigma_{({\alpha}}\,\epsilon_{\beta)\delta} \,R_{\gamma\sigma} \,,
\\
 &\bigl[ R_{\alpha\beta},\,R^{\sfa_1\sfa_2}_{\gamma} \bigr] 
 = \delta^{\delta}_{({\alpha}}\,\epsilon_{\beta)\gamma}\, R^{\sfa_1\sfa_2}_{\delta}\,, \hskip2cm 
  \bigl[ R_{\alpha\beta},\,R_{\sfa_1\sfa_2}^{\gamma} \bigr] 
 = - \delta^{\gamma}_{({\alpha}}\,\epsilon_{\beta)\delta}\, R_{\sfa_1\sfa_2}^{\delta}\,,
\\
 &\bigl[ R_{\alpha\beta},\,R^{\sfa_1\cdots \sfa_6}_{\gamma} \bigr] 
 = \delta^{\delta}_{({\alpha}}\,\epsilon_{\beta)\gamma}\, R^{\sfa_1\cdots \sfa_6}_{\delta}\,, \hskip1.5cm
 \bigl[ R_{\alpha\beta},\,R_{\sfa_1\cdots \sfa_6}^{\gamma} \bigr] 
 = - \delta^{\gamma}_{({\alpha}}\, \epsilon_{\beta)\delta}\, R_{\sfa_1\cdots \sfa_6}^{\delta}\,,
\\
 &\bigl[ R_{\alpha}^{\sfa_1\sfa_2},\,R_{\beta}^{\sfb_1\sfb_2} \bigr] 
  = \epsilon_{\alpha\beta}\,R^{\sfa_1\sfa_2\sfb_1\sfb_2}\,,
\\
 &\bigl[ R_{\alpha}^{\sfa_1\sfa_2},\,R_{\sfb_1\sfb_2}^{\beta} \bigr] 
 = 4\,\delta_{\alpha}^{\beta}\,\delta^{[\sfa_1}_{[\sfb_1}\,K^{\sfa_2]}{}_{\sfb_2]} 
 - \frac{1}{2}\,\delta_{\alpha}^{\beta}\,\delta^{\sfa_1\sfa_2}_{\sfb_1\sfb_2}\,D 
 - 2\,\delta^{\sfa_1\sfa_2}_{\sfb_1\sfb_2}\,\epsilon^{\beta\gamma}\,R_{\alpha\gamma} \,,
\\
 &\bigl[ R_{\alpha}^{\sfa_1\sfa_2},\,R^{\sfb_1\cdots \sfb_4} \bigr] 
 = R^{\sfa_1\sfa_2\sfb_1\cdots \sfb_4}_{\alpha}\,,\hskip2cm
 \bigl[ R_{\alpha}^{\sfa_1\sfa_2},\,R_{\sfb_1\cdots \sfb_4} \bigr] 
  = 12\,\epsilon_{\alpha\beta}\,\delta^{\sfa_1\sfa_2}_{[\sfb_1\sfb_2}\,R_{\sfb_3\sfb_4]}^{\beta}\,,
\\
 &\bigl[ R_{\alpha}^{\sfa_1\sfa_2},\,R_{\sfb_1\cdots \sfb_6}^{\beta} \bigr] 
  = 30\,\delta_{\alpha}^{\beta}\,\delta^{\sfa_1\sfa_2}_{[\sfb_1\sfb_2}\,R_{\sfb_3\cdots \sfb_6]}\,,
\\
 &\bigl[ R^{\alpha}_{\sfa_1\sfa_2},\,R^{\beta}_{\sfb_1\sfb_2} \bigr] 
  = \epsilon^{\alpha\beta}\,R_{\sfa_1\sfa_2\sfb_1\sfb_2}\,, \hskip2cm 
  \bigl[ R^{\alpha}_{\sfa_1\sfa_2},\,R^{\sfb_1\cdots \sfb_4} \bigr] 
  = 12\,\epsilon^{\alpha\beta} \,\delta_{\sfa_1\sfa_2}^{[\sfb_1\sfb_2}\,R^{\sfb_3\sfb_4]}_{\beta}\,,
\\
 &\bigl[ R^{\alpha}_{\sfa_1\sfa_2},\,R_{\sfb_1\cdots \sfb_4} \bigr] 
 = R_{\sfa_1\sfa_2\sfb_1\cdots \sfb_4}^{\alpha}\,, \hskip2.2cm
  \bigl[ R^{\alpha}_{\sfa_1\sfa_2},\,R^{\sfb_1\cdots \sfb_6}_{\beta} \bigr] 
  = 30\,\delta^{\alpha}_{\beta}\,\delta_{\sfa_1\sfa_2}^{[\sfb_1\sfb_2}\,R^{\sfb_3\cdots \sfb_6]}\,,
\\
 &\bigl[ R^{\sfa_1\cdots \sfa_4},\,R_{\sfb_1\cdots \sfb_4} \bigr] 
  = 12\,\delta^{\sfa_1\cdots \sfa_4}_{\sfb_1\cdots \sfb_4}\,D - 96\,\delta^{[\sfa_1\sfa_2\sfa_3}_{[\sfb_1\sfb_2\sfb_3}\,K^{\sfa_4]}{}_{\sfb_4]}\,,
\\
 &\bigl[ R^{\sfa_1\cdots \sfa_4},\,R_{\sfb_1\cdots \sfb_6}^{\alpha} \bigr] 
 = 360\,\delta^{\sfa_1\cdots \sfa_4}_{[\sfb_1\cdots \sfb_4}\,R_{\sfb_5\sfb_6]}^{\alpha}\,,\hskip0.8cm
  \bigl[ R_{\sfa_1\cdots \sfa_4},\,R^{\sfb_1\cdots \sfb_6}_{\alpha} \bigr] 
 = 360\,\delta_{\sfa_1\cdots \sfa_4}^{[\sfb_1\cdots \sfb_4}\,R^{\sfb_5\sfb_6]}_{\alpha}\,,
\\
 &\bigl[ R_{\alpha}^{\sfa_1\cdots \sfa_6},\,R_{\sfb_1\cdots \sfb_6}^{\beta} \bigr] 
 = 1080\,\delta_{\alpha}^{\beta}\,\delta^{[\sfa_1\cdots \sfa_5}_{[\sfb_1\cdots \sfb_5}\,K^{\sfa_6]}{}_{\sfb_6]} 
 - 135 \,\delta_{\alpha}^{\beta}\,\delta^{\sfa_1\cdots \sfa_6}_{\sfb_1\cdots \sfb_6}\,D 
 - 180\,\delta^{\sfa_1\cdots \sfa_6}_{\sfb_1\cdots \sfb_6}\,\epsilon^{\beta\gamma}\,R_{\alpha\gamma} \,. 
\end{split}
\end{align}
Note that the definitions of $R^{\alpha}_{\sfa_1\sfa_2}$, $R_{\alpha}^{\sfa_1\sfa_2}$, $R_{\sfa_1\cdots \sfa_4}$, and $R^{\sfa_1\cdots \sfa_4}$ are changed from those used in \cite{Tumanov:2014pfa} by a minus sign and $K^a{}_b$ corresponds to $K^a{}_b-\frac{1}{n-2}\,\delta^a_b\,\sum_\mu K^\mu{}_\mu$ of \cite{Tumanov:2014pfa}. 

The commutators of the $E_{d(d)}$ generators with the central charges,
\begin{align}
 \bigl(P_\sfa,\, \cZ^\sfa_{\alpha},\, \cZ^{\sfa_1\sfa_2\sfa_3},\, \cZ^{\sfa_1\cdots \sfa_5}_{\alpha},\, \cZ^{\sfa_1\cdots \sfa_6,\,\sfa}\bigr) \,,
\end{align}
are given by \cite{Tumanov:2014pfa}
\begin{align}
\begin{split}
 &\bigl[ K^\sfc{}_\sfd,\,P_\sfa \bigr] 
 = - \delta^\sfc_a\,P_\sfd - \frac{\delta^\sfc_\sfd}{n-2} \,P_\sfa\,, \hskip1.5cm 
  \bigl[ K^\sfc{}_\sfd,\, \cZ^\sfa_{\alpha} \bigr] 
 = \delta^\sfa_\sfd\, \cZ^\sfc_{\alpha} - \frac{\delta^\sfc_\sfd}{n-2} \,\cZ^\sfa_{\alpha}\,,
\\
 &\bigl[ K^\sfc{}_\sfd,\,\cZ^{\sfa_1\sfa_2\sfa_3} \bigr] 
 = 3\,\delta^{[\sfa_1\vert}_\sfd\, \cZ^{\sfc\vert\sfa_2\sfa_3]} - \frac{\delta^\sfc_\sfd}{n-2}\, \cZ^{\sfa_1\sfa_2\sfa_3}\,,
\\
 &\bigl[ K^\sfc{}_\sfd,\, \cZ^{\sfa_1\cdots \sfa_5}_{\alpha} \bigr]
 = 5\,\delta^{[\sfa_1\vert}_\sfd\, \cZ^{\sfc\vert\sfa_2\cdots \sfa_5]}_{\alpha} - \frac{\delta^\sfc_\sfd}{n-2}\, \cZ^{\sfa_1\cdots \sfa_5}_{\alpha} \,,
\\
 &\bigl[ K^\sfc{}_\sfd,\,W^\sfa \bigr] 
 = \delta^\sfa_\sfd\,W^\sfc + \delta^\sfc_\sfd\,W^\sfa - \frac{\delta^\sfc_\sfd}{n-2}\, W^\sfa \,,
\\
 &\bigl[ R_{\gamma\delta},\,\cZ^\sfa_{\alpha} \bigr] 
 = \delta^{\beta}_{(\gamma} \,\epsilon_{\delta)\alpha} \, \cZ^\sfa_{\beta}\,, \hskip3cm
 \bigl[ R_{\gamma\delta} ,\, \cZ^{\sfa_1\cdots \sfa_5}_{\alpha} \bigr] 
 = \delta^{\beta}_{(\gamma} \,\epsilon_{\delta)\alpha} \,\cZ^{\sfa_1\cdots \sfa_5}_{\beta}\,,
\\
 &\bigl[ R_{\gamma}^{\sfc_1\sfc_2},\,P_\sfa \bigr] 
 = -\delta^{[\sfc_1}_\sfa\, \cZ^{\sfc_2]}_{\gamma}\,, \hskip2.9cm 
 \bigl[ R_{\gamma}^{\sfc_1\sfc_2},\, \cZ^\sfa_{\alpha} \bigr] 
 = \epsilon_{{\gamma}{\alpha}}\, \cZ^{\sfc_1\sfc_2 \sfa}\,, 
\\
 &\bigl[ R_{\gamma}^{\sfc_1\sfc_2},\,\cZ^{\sfa_1\sfa_2\sfa_3} \bigr] 
 = - \cZ^{\sfc_1\sfc_2 \sfa_1\sfa_2\sfa_3}_{\gamma} \,, \hskip1.7cm
  \bigl[ R_{\gamma}^{\sfc_1\sfc_2},\, \cZ^{\sfa_1\cdots \sfa_5}_{\alpha} \bigr] 
 = \epsilon_{\gamma\alpha} \,\epsilon^{\sfc_1\sfc_2[\sfa_1\cdots \sfa_4}\, W^{\sfa_5]}\,,
\\
 &\bigl[ R^{\sfc_1\cdots \sfc_4},\, P_\sfa \bigr] 
 = -2\,\delta^{[\sfc_1}_\sfa \, \cZ^{\sfc_2\sfc_3\sfc_4]}\,, \hskip1.8cm 
 \bigl[ R^{\sfc_1\cdots \sfc_4},\, \cZ^\sfa_{\alpha} \bigr] 
 = \cZ^{\sfc_1\cdots \sfc_4 \sfa}_{\alpha}\,,
\\
 &\bigl[ R^{\sfc_1\cdots \sfc_4},\, \cZ^{\sfa_1\sfa_2\sfa_3} \bigr] 
 = -\frac{3}{5}\, \epsilon^{\sfc_1\cdots \sfc_4[\sfa_1\sfa_2} \, W^{\sfa_3]}\,,
\\
 &\bigl[ R^{\sfc_1\cdots \sfc_6}_{\gamma},\, P_\sfa \bigr] 
 = 3 \,\delta^{[\sfc_1}_\sfa \, \cZ^{\sfc_2\cdots \sfc_6]}_{\gamma}\,, \hskip2.2cm
 \bigl[ R^{\sfc_1\cdots \sfc_6}_{\gamma},\, \cZ^\sfa_{\alpha} \bigr] 
 = \frac{1}{5}\,\epsilon_{\gamma\alpha}\,\epsilon^{\sfc_1\cdots \sfc_6}\, W^\sfa\,,
\\
 &\bigl[ R^{\gamma}_{\sfc_1\sfc_2},\, \cZ^\sfa_{\alpha} \bigr] 
  = 4\,\delta^{\gamma}_{\alpha}\, \delta_{[\sfc_1}^\sfa\, P_{\sfc_2]}\,, \hskip2.5cm 
 \bigl[ R^{\gamma}_{\sfc_1\sfc_2},\, \cZ^{\sfa_1\sfa_2\sfa_3} \bigr] 
 = 3!\,\epsilon^{\gamma\beta}\, \delta^{[\sfa_1\sfa_2}_{\sfc_1\sfc_2}\, \cZ^{\sfa_3]}_{\beta}\,,
\\
 &\bigl[ R^{\gamma}_{\sfc_1\sfc_2},\, \cZ^{\sfa_1\cdots \sfa_5}_{\alpha} \bigr] 
 = -20\,\delta^{\gamma}_{\alpha}\,\delta^{[\sfa_1\sfa_2}_{\sfc_1\sfc_2}\, \cZ^{\sfa_3\sfa_4\sfa_5]} \,, \hskip0.3cm 
  \bigl[ R^{\gamma}_{\sfc_1\sfc_2},\, W^\sfa \bigr] 
 = \frac{5}{24} \,\epsilon^{\gamma\delta}\,\epsilon_{\sfc_1\sfc_2 \sfd_1\cdots \sfd_4}\, \cZ^{\sfd_1\cdots \sfd_4 \sfa}_{\delta} \,,
\\
 &\bigl[ R_{\sfc_1\cdots \sfc_4},\, \cZ^{\sfa_1\sfa_2\sfa_3} \bigr] 
 = -48\,\delta^{\sfa_1\sfa_2\sfa_3}_{[\sfc_1\sfc_2\sfc_3}\,P_{\sfc_4]}\,, \hskip0.9cm 
 \bigl[ R_{\sfc_1\cdots \sfc_4},\, \cZ^{\sfa_1\cdots \sfa_5}_{\alpha} \bigr] 
 = -5!\,\delta^{[\sfa_1\cdots \sfa_4}_{\sfc_1\cdots \sfc_4}\,\cZ^{\sfa_5]}_{\alpha}\,, 
\\
 &\bigl[ R_{\sfc_1\cdots \sfc_4},\,W^\sfa \bigr] 
 = \frac{5}{2}\,\epsilon_{\sfc_1\cdots \sfc_4 \sfd_1\sfd_2}\, \cZ^{\sfd_1\sfd_2 \sfa} \,, 
\\
 &\bigl[ R_{\sfc_1\cdots \sfc_6}^{\gamma},\, Z^{\sfa_1\cdots \sfa_5}_{\alpha} \bigr] 
 = - 12\cdot 5!\,\delta^{\gamma}_{\alpha}\, \delta^{\sfa_1\cdots \sfa_5}_{[\sfc_1\cdots \sfc_5}\,P_{\sfc_6]}\,, \quad 
  \bigl[ R_{\sfc_1\cdots \sfc_6}^{\gamma},\, W^a \bigr] 
 = 5 \,\epsilon^{\gamma\delta} \,\epsilon_{\sfc_1\cdots \sfc_6}\, Z^\sfa_{\delta}\,.
\end{split}
\end{align}

Using the normalization of the generators,
\begin{align}
 (\cZ_A) \equiv \Bigl(P_\sfa,\,\frac{\cZ^\sfa_{\alpha}}{2},\,\frac{\cZ^{\sfa_1\sfa_2\sfa_3}}{2\sqrt{3!}},\,\frac{\cZ^{\sfa_1\cdots \sfa_5}_{\alpha}}{2\sqrt{5!}},\,\frac{W^\sfa}{10}\Bigr) \,,
\end{align}
we obtain the following matrix representations:
\begin{align}
 &(\rho_{K^\sfc{}_\sfd})_A{}^B ={\footnotesize\begin{pmatrix}
 \delta_\sfa^\sfc\,\delta_\sfd^\sfb & 0 & 0 & 0 & 0 \\
 0 & -\delta^\sfa_\sfd\,\delta^\sfc_\sfb\,\delta_{\alpha}^{\beta} & 0 & 0 & 0 \\
 0 & 0 & -3\,\delta^\sfc_{[\sfb_1\vert}\delta^{\sfa_1\sfa_2\sfa_3}_{\sfd\vert\sfb_2\sfb_3]} & 0 & 0 \\
 0 & 0 & 0 & -5\,\delta^\sfc_{[\sfb_1\vert}\delta^{\sfa_1\cdots \sfa_5}_{\sfd\vert\sfb_2\cdots \sfb_5]} \,\delta_{\alpha}^{\beta} & 0 \\
 0 & 0 & 0 & 0 & -2\,\delta^{(\sfa}_\sfd\,\delta^{\sfc)}_\sfb 
 \end{pmatrix}} + \frac{\delta^\sfc_\sfd}{n-2}\,\delta_A^B \,, 
\\
 &(\rho_{R_{\gamma\delta}})_A{}^B =
 {\footnotesize\begin{pmatrix}
 \ 0 & \ \ \ 0 & \qquad 0 & \quad \qquad 0 & \qquad \qquad 0 \ \\
 \ 0 & \ \ \ -\delta_{(\gamma}^{\beta}\,\epsilon_{\delta)\alpha} \,\delta^\sfa_\sfb & \qquad 0 & \quad \qquad 0 & \qquad \qquad 0 \ \\
 \ 0 & \ \ \ 0 & \qquad 0 & \quad \qquad 0 & \qquad \qquad 0 \ \\
 \ 0 & \ \ \ 0 & \qquad 0 & \quad \qquad -\delta^{\beta}_{(\gamma}\,\epsilon_{\delta)\alpha} \,\delta^{\sfa_1\cdots \sfa_5}_{\sfb_1\cdots \sfb_5} & \qquad \qquad 0 \ \\
 \ 0 & \ \ \ 0 & \qquad 0 & \quad \qquad 0 & \qquad \qquad 0 \
 \end{pmatrix}} ,
\\
 &(\rho_{R_{\gamma}^{\sfc_1\sfc_2}})_A{}^B =
 {\footnotesize\begin{pmatrix}
 0 & \ \ \ 2\,\delta_{\gamma}^{\beta} \delta^{\sfc_1\sfc_2}_{\sfa \sfb} & 0 & 0 & 0 \\
 0 & \ \ \ 0 & \frac{3!}{\sqrt{3!}}\,\epsilon_{\alpha\gamma} \,\delta^{\sfa \sfc_1\sfc_2}_{\sfb_1\sfb_2\sfb_3} & 0 & 0 \\
 0 & \ \ \ 0 & 0 & \frac{5!}{\sqrt{3!\,5!}}\,\delta^{\beta}_{\gamma}\,\delta^{\sfa_1\sfa_2\sfa_3 \sfc_1\sfc_2}_{\sfb_1\cdots \sfb_5} & 0 \\
 0 & \ \ \ 0 & 0 & 0 & -\frac{2}{\sqrt{5!}}\,\epsilon_{\alpha\gamma} \,\epsilon^{\sfa_1\cdots \sfa_5 [\sfc_1}\delta^{\sfc_2]}_\sfb \\
 0 & \ \ \ 0 & 0 & 0 & 0
 \end{pmatrix}} ,
\\
 &(\rho_{R^{\sfc_1\cdots \sfc_4}})_A{}^B =
 {\footnotesize\begin{pmatrix}
 0 & \ \ \ \ \ \ 0 & \quad \ \frac{4!}{\sqrt{3!}}\,\delta^{\sfc_1\cdots \sfc_4}_{\sfa \sfb_1\sfb_2\sfb_3} & \quad 0 & \ \ 0 \\
 0 & \ \ \ \ \ \ 0 & \quad \ 0 & \quad -\frac{5!}{\sqrt{5!}}\,\delta_{\alpha}^{\beta}\,\delta^{\sfa \sfc_1\cdots \sfc_4}_{\sfb_1\cdots \sfb_5} & \ \ 0 \\
 0 & \ \ \ \ \ \ 0 & \quad \ 0 & \quad 0 & \ \ -\frac{4}{\sqrt{3!}}\,\epsilon^{\sfa_1\sfa_2\sfa_3 [\sfc_1\sfc_2\sfc_3}\delta^{\sfc_4]}_\sfb \\
 0 & \ \ \ \ \ \ 0 & \quad \ 0 & \quad 0 & \ \ 0 \\
 0 & \ \ \ \ \ \ 0 & \quad \ 0 & \quad 0 & \ \ 0
 \end{pmatrix}} , 
\\
 &(\rho_{R^{\sfc_1\cdots \sfc_6}_{\gamma}})_A{}^B =
 {\footnotesize\begin{pmatrix}
 0 & \ \ \ \quad 0 & \qquad \qquad 0 &\qquad \ \ -\frac{6!}{\sqrt{5!}} \,\delta^{\beta}_{\gamma}\, \delta^{\sfc_1\cdots \sfc_6}_{\sfa \sfb_1\cdots \sfb_5} & \ \qquad 0 \ \ \ \\
 0 & \ \ \ \quad 0 & \qquad \qquad 0 & \qquad \ \ 0 & \qquad - \epsilon_{\gamma\alpha} \,\epsilon^{\sfc_1\cdots \sfc_6}\, \delta^\sfa_\sfb \ \ \ \\
 0 & \ \ \ \quad 0 & \qquad \qquad 0 & \qquad \ \ 0 &\ \qquad 0 \ \ \ \\
 0 & \ \ \ \quad 0 & \qquad \qquad 0 & \qquad \ \ 0 &\ \qquad 0 \ \ \ \\
 0 & \ \ \ \quad 0 & \qquad \qquad 0 & \qquad \ \ 0 &\ \qquad 0 \ \ \
\end{pmatrix}}\,, 
\\
 &(\rho_{R^{\gamma}_{\sfc_1\sfc_2}})_A{}^B =
 {\footnotesize\begin{pmatrix}
 0 & 0 & 0 & 0 & 0 \\
 -2\,\delta_{\alpha}^{\gamma} \delta^{\sfa \sfb}_{\sfc_1\sfc_2} & 0 & 0 & 0 & 0 \\
 0 & \frac{3!}{\sqrt{3!}}\,\epsilon^{\beta\gamma}\,\delta^{\sfa_1\sfa_2\sfa_3}_{\sfb \sfc_1\sfc_2} & 0 & 0 & 0 \\
 0 & 0 & \frac{5!}{\sqrt{3!\,5!}}\,\delta_{\alpha}^{\gamma}\,\delta^{\sfa_1\cdots \sfa_5}_{\sfb_1\sfb_2\sfb_3 \sfc_1\sfc_2} & 0 & 0 \\
 0 & 0 & 0 & -\frac{2}{\sqrt{5!}}\,\epsilon^{\beta\gamma}\,\epsilon_{\sfb_1\cdots \sfb_5 [\sfc_1}\delta_{\sfc_2]}^\sfa & 0
 \end{pmatrix}} ,
\\
 &(\rho_{R_{\sfc_1\cdots \sfc_4}})_A{}^B =
 {\footnotesize\begin{pmatrix}
 0 & 0 & 0 & \qquad 0 & \qquad \qquad 0 \\
 0 & 0 & 0 & \qquad 0 & \qquad \qquad 0 \\
 \frac{4!}{\sqrt{3!}}\,\delta^{\sfa_1\sfa_2\sfa_3 \sfb}_{\sfc_1\cdots \sfc_4} & 0 & 0 & \qquad 0 & \qquad \qquad 0 \\
 0 & \frac{5!}{\sqrt{5!}}\,\delta_{\alpha}^{\beta}\,\delta^{\sfa_1\cdots \sfa_5}_{\sfb \sfc_1\cdots \sfc_4} & 0 & \qquad 0 & \qquad \qquad 0 \\
 0 & 0 & \frac{4}{\sqrt{3!}}\,\epsilon_{\sfb_1\sfb_2\sfb_3 [\sfc_1\sfc_2\sfc_3}\delta_{\sfc_4]}^\sfa & \qquad 0 & \qquad \qquad 0
 \end{pmatrix}} , 
\\
 &(\rho_{R_{\sfc_1\cdots \sfc_6}^{\gamma}})_A{}^B =
 {\footnotesize\begin{pmatrix}
 0 & 0 & \qquad \quad 0 &\quad \qquad \qquad 0 & \quad \qquad \qquad 0 \\
 0 & 0 & \qquad \quad 0 & \quad \qquad \qquad 0 & \quad \qquad \qquad 0 \\
 0 & 0 & \qquad \quad 0 & \quad \qquad \qquad 0 & \quad \qquad \qquad 0 \\
 \frac{6!}{\sqrt{5!}} \,\delta^{\gamma}_{\alpha}\, \delta_{\sfc_1\cdots \sfc_6}^{\sfa_1\cdots \sfa_5 \sfb} & 0 & \quad \qquad 0 & \quad \qquad \qquad 0 & \qquad \qquad \quad 0 \\
 0 & - \epsilon^{\gamma\beta}\,\epsilon_{\sfc_1\cdots \sfc_6}\, \delta^\sfa_\sfb & \quad \qquad 0 & \quad \qquad \qquad 0 & \qquad \qquad \quad 0
\end{pmatrix}}\,. 
\end{align}
From the Cartan involution,
\begin{align}
\begin{split}
 &\theta(K^\sfa{}_\sfb)=-K^\sfa{}_\sfb\,,\qquad 
 \theta(R^{\sfa_1\cdots \sfa_6}_{\alpha})=-R_{\sfa_1\cdots \sfa_6}^{\alpha}\,, \qquad
 \theta(R_{\alpha}^{\sfa_1\sfa_2})=-R^{\alpha}_{\sfa_1\sfa_2}\,,
\\
 &\qquad \theta(R^{\sfa_1\cdots \sfa_4})=R_{\sfa_1\cdots \sfa_4} \, \qquad
 \theta(R_{\alpha\beta})= \epsilon_{\alpha\gamma} \,\epsilon_{\beta\delta}\,\delta^{\gamma\gamma'}\,\delta^{\delta\delta'}\,R_{\gamma'\delta'} \,,\quad 
\end{split}
\end{align}
we can see that the generalized transpose is indeed the same as the matrix transpose. 

\section{Calculation of the EFT action}
\label{app:action}
In this appendix, we summarize the construction of diverse EFTs. To construct the EFT action, we find it convenient to decompose the generalized metric as
\begin{align}
 \cM_{MN} \equiv (\vV\hat{\cM}\vV^{\rmT})_{MN} = \abs{G}^{\frac{1}{n-2}}\,M_{MN}\,\qquad \text{where} \qquad
M_{MN} \equiv (\vV\hat{M}\vV^{\rmT})_{MN}\,,
\end{align}
and define a connection
\begin{align}
 \omega_{\hat{\mu} M}{}^N \equiv (\vV^{-1})_M{}^K\,\partial_{\hat{\mu}} \vV_K{}^N \,,
\label{eq:omega-def}
\end{align}
where the index $\hat{\mu}$ runs over all conventional directions, $(\mu,\,i)$ in the M or Type IIA theory and $(\mu,\,\sfm)$ in the type IIB theory. 
We begin with showing the explicit form of $\widehat{M}_{MN}$, $\vV_M{}^N$, and $\omega_{{\hat{\mu}} M}{}^N$ for various EFTs and then construct the action in each case. 

\subsection{Redefinitions of coordinates}
\label{sec:redef}

For explicit computation of the effective action, it is more useful to redefine the coordinates and central charges to minimize the cluttering indices. 

\subsubsection*{M-theory section}

In the M-theory section of the $E_{d(d)}$ exceptional space with $n=6,5,4$, we redefine the coordinates as follows \cite{Berman:2011jh}:
\begin{align}
\begin{aligned}
 n=6:&\quad \bigl(y^M\bigr)= (x^i,\,y_{ij},\, z)&&(i,j=6,\dotsc,9,\sfM) \,,
\\
 n=5:&\quad \bigl(y^M\bigr)= (x^i,\, y_{ij},\, z^i)&&(i,j=5,\dotsc,9,\sfM) \,,
\\
 n=4:&\quad \bigl(y^M\bigr)= (x^i,\, y_{ij},\, z^{ij},\,z_i)\qquad&&(i,j=4,\dotsc,9,\sfM)\,.
\end{aligned}
\end{align}
Here, we defined
\begin{align}
 z^{i_1\cdots i_{d-5}}\equiv \frac{1}{\sqrt{(d-5)!\,5!}}\,\epsilon^{i_1\cdots i_{d-5} j_1\cdots j_5}\,y_{j_1\cdots j_5}\qquad (d=5,6,7)\,,
\end{align}
and correspondingly, we also redefine the central charge as
\begin{align}
 W_{a_1\cdots a_{d-5}}\equiv \frac{1}{5!}\,\epsilon_{a_1\cdots a_{d-5} b_1\cdots b_5}\, \cZ^{b_1\cdots b_5} \,. 
\end{align}
The generalized momenta after the redefinitions are given by
\begin{align}
 \bigl( \cZ_A\bigr)=\Bigl(P_a,\,\frac{ \cZ^{a_1a_2}}{\sqrt{2}},\,\frac{W^{a_1\cdots a_{d-5}}}{\sqrt{(d-5)!}}\Bigr) \,. 
\end{align}

The untwisted generalized metric is then given as follows:
\begin{align}
 \underline{\SL(5)}:\quad&(\widehat{M}_{MN}) = 
 \begin{pmatrix}
  G_{ij} & 0 \\
  0 & G^{i_1i_2,\, j_1j_2} 
 \end{pmatrix}\,,
\\
 \underline{\SO(5,5)}:\quad&(\widehat{M}_{MN}) = 
 \begin{pmatrix}
  G_{ij} & 0 & 0 \\
  0 & G^{i_1i_2,\, j_1j_2} & 0 \\
  0 & 0 & \abs{G}^{-1}
 \end{pmatrix}\,,
\\
  \underline{E_{6(6)}}:\quad&(\widehat{M}_{MN}) =
 \begin{pmatrix}
  G_{ij} & 0 & 0 \\
  0 & G^{i_1i_2,\, j_1j_2} & 0 \\
  0 & 0 & \abs{G}^{-1}\,G_{ij}
 \end{pmatrix}\,, 
\\
  \underline{E_{7(7)}}:\quad&(\widehat{M}_{MN}) =
 {\footnotesize\begin{pmatrix}
  G_{ij} & 0 & 0 & 0 \\
  \\
  0 & G^{i_1i_2,\, j_1j_2} & 0 & 0 \\
  \\
  0 & 0 & \abs{G}^{-1}\,G_{i_1i_2,\, j_1j_2} & 0 \\
  \\
  0 & 0 & 0 & \abs{G}^{-1}\,G^{ij}
 \end{pmatrix}} \,.
\end{align}
On the other hand, denoting the non-geometric fluxes as
\begin{align}
 S_{\hat{\mu}}{}^{i_1i_2i_3} \equiv \partial_{\hat{\mu}} \Omega^{i_1i_2i_3}\,\qquad \text{and} \qquad
 S_{\hat{\mu}}{}^{i_1\cdots i_6} \equiv \partial_{\hat{\mu}} \Omega^{i_1\cdots i_6}+10\,\Omega^{[i_1i_2i_3}\,\partial_{\hat{\mu}} \Omega^{i_4i_5i_6]}\,,
\end{align}
the twist matrix $\vV_M{}^N$ and $\omega_{{\hat{\mu}} M}{}^N$ given in \eqref{eq:omega-def} for the geometric or non-geometric parameterizations, respectively, are given as follows (we added tilde for $\vV$ and $\omega$ for the non-geometric parameterization):
\begin{align}
 &\underline{\SL(5)}: 
\nn\\
 \vV & = 
 \begin{pmatrix}
  \delta_i^j & - \frac{1}{\sqrt{2}}\, A_{i j_1j_2} \\
  0 & \delta^{i_1i_2}_{j_1j_2} 
 \end{pmatrix}\,,  \qquad \qquad \qquad 
 \widetilde{\vV} = \begin{pmatrix}
  \delta_i^j & 0  \\
  -\frac{1}{\sqrt{2}}\, \Omega^{i_1i_2 j} & \delta^{i_1i_2}_{j_1j_2} 
 \end{pmatrix} \,,
\\
 \bigl(\omega_{{\hat{\mu}} M}{}^N\bigr) & = 
 \begin{pmatrix}
  0 & - \frac{1}{\sqrt{2}}\, \partial_{\hat{\mu}} A_{i j_1j_2} \\
  0 & 0 
 \end{pmatrix}\,, \qquad  \qquad
  \bigl(\widetilde{\omega}_{{\hat{\mu}} M}{}^N\bigr) = 
 \begin{pmatrix}
  0 & 0 \\
 -\frac{1}{\sqrt{2}}\, S_{\hat{\mu}}{}^{i_1i_2 j} & 0 
 \end{pmatrix}\,, 
\\[5mm]
 &\underline{\SO(5,5)}: 
\nn\\
 \vV & =
 {\footnotesize\begin{pmatrix}
  \delta_i^j & - \frac{1}{\sqrt{2}}\, A_{i j_1j_2} & \frac{1}{4}\, A_{i k_1k_2}\, \widetilde{A}^{k_1k_2} \\
  0 & \delta^{i_1i_2}_{j_1j_2} & -\frac{1}{\sqrt{2}}\, \widetilde{A}^{i_1i_2} \\
  0 & 0 & 1
 \end{pmatrix}}\,, \quad
 \widetilde{\vV} ={\footnotesize\begin{pmatrix}
  \delta_i^j & 0 & 0 \\
  -\frac{1}{\sqrt{2}}\, \Omega^{i_1i_2 j} & \delta^{i_1i_2}_{j_1j_2} & 0 \\
  \frac{1}{4}\, \widetilde{\Omega}_{k_1k_2}\,\Omega^{k_1k_2 j} & -\frac{1}{\sqrt{2}}\, \widetilde{\Omega}_{j_1j_2} & 1
 \end{pmatrix}}\,, 
\\
 \bigl(\omega_{{\hat{\mu}} M}{}^N\bigr) 
  & = \begin{pmatrix}
  0 & \qquad - \frac{1}{\sqrt{2}}\, \partial_{\hat{\mu}} A_{i j_1j_2} & 0 \\
  0 & \qquad 0 & -\frac{1}{3!\,\sqrt{2}}\, \epsilon^{i_1i_2 k_1k_2k_3}\,\partial_{\hat{\mu}} A_{k_1k_2k_3} \\
  0 & \qquad 0 & 0
 \end{pmatrix} \,, 
\\ 
\bigl(\widetilde{\omega}_{{\hat{\mu}} M}{}^N\bigr) 
 & = \begin{pmatrix}
  0 & 0 & \quad \qquad 0 \\
  -\frac{1}{\sqrt{2}}\, S_{\hat{\mu}}{}^{i_1i_2 j} & 0 & \quad \qquad 0 \\
  0 & -\frac{1}{3!\,\sqrt{2}}\, \epsilon_{i_1i_2 k_1k_2k_3}\,S_{\hat{\mu}}{}^{k_1k_2k_3} & \quad \qquad 0
 \end{pmatrix} \,,
\\
 &\widetilde{A}^{i_1i_2}\equiv \frac{1}{3!}\,\epsilon^{i_1i_2j_1j_2j_3}\,A_{j_1j_2j_3}\,, \qquad \qquad
 \widetilde{\Omega}_{i_1i_2}\equiv \frac{1}{3!}\,\epsilon_{i_1i_2j_1j_2j_3}\,\Omega^{j_1j_2j_3}\,, 
\\[5mm]
 &\underline{E_{6(6)}} : 
\nn\\
 \vV & = {\footnotesize
 \begin{pmatrix}
  \delta_i^k & 0 & \delta_i^k\,A_6 \\
  0 & \delta^{i_1i_2}_{k_1k_2} & 0 \\
  0 & 0 & \delta_i^k
 \end{pmatrix} 
\begin{pmatrix}
  \delta_k^j & - \frac{1}{\sqrt{2}}\, A_{k j_1j_2} & \frac{1}{4}\, A_{k l_1l_2}\, \widetilde{A}^{l_1l_2 j} \\
  0 & \delta^{k_1k_2}_{j_1j_2} & -\frac{1}{\sqrt{2}}\, \widetilde{A}^{k_1k_2 j} \\
  0 & 0 & \delta_k^j
 \end{pmatrix}} \,,
\\
 \widetilde{\vV} & = {\footnotesize\begin{pmatrix}
  \delta_i^k & 0 & 0 \\
  0 & \delta^{i_1i_2}_{k_1k_2} & 0 \\
 \delta_i^k\,\Omega_6 & 0 & \delta_i^k
  \end{pmatrix} \begin{pmatrix}
  \delta_k^j & 0 & 0 \\
  - \frac{1}{\sqrt{2}}\, \Omega^{k_1k_2 j} & \delta^{k_1k_2}_{j_1j_2} & 0 \\
  \frac{1}{4}\, \widetilde{\Omega}_{k l_1l_2}\, \Omega^{l_1l_2 j} & -\frac{1}{\sqrt{2}}\, \widetilde{\Omega}_{k j_1j_2} & \delta_k^j
  \end{pmatrix} }\,,
\\
 \bigl(\omega_{{\hat{\mu}} M}{}^N\bigr) 
  &= {\footnotesize\begin{pmatrix}
  0 & \quad \qquad - \frac{1}{\sqrt{2}}\, \partial_{\hat{\mu}} A_{i j_1j_2} & \frac{1}{6!}\,\delta_i^j\, \epsilon^{i_1\cdots i_6}\,\cF_{{\hat{\mu}},\,i_1\cdots i_6} \\
  0 & \quad \qquad 0 & -\frac{1}{3!\sqrt{2}}\, \epsilon^{i_1i_2 j k_1k_2k_3}\,\partial_{\hat{\mu}} A_{k_1k_2k_3} \\
  0 & \quad \qquad 0 & 0
 \end{pmatrix}}\,, 
\\ 
 \bigl(\widetilde{\omega}_{{\hat{\mu}}M}{}^N\bigr) 
 & = {\footnotesize\begin{pmatrix}
  0 & 0 & \qquad 0 \\
  -\frac{1}{\sqrt{2}}\, S_{\hat{\mu}}{}^{i_1i_2 j} & 0 & \qquad 0 \\
  \frac{1}{6!}\,\delta^j_i\, \epsilon_{i_1\cdots i_6}\, S_{\hat{\mu}}{}^{i_1\cdots i_6} & -\frac{1}{3!\sqrt{2}}\, \epsilon_{i j_1j_2 k_1k_2k_3}\,S_{\hat{\mu}}{}^{k_1k_2k_3} & \qquad 0
 \end{pmatrix}} \,,
\\
\begin{split}
 \widetilde{A}^{i_1i_2 j} & \equiv \frac{1}{3!}\,\epsilon^{i_1i_2 j k_1k_2k_3}\, A_{k_1k_2k_3} \,, \qquad \qquad
 A_6\equiv \frac{1}{6!}\,\epsilon^{k_1\cdots k_6}\,A_{k_1\cdots k_6} \,,
\\
\widetilde{\Omega}_{i j_1j_2} & \equiv \frac{1}{3!}\,\epsilon_{i j_1j_2 k_1k_2k_3}\, \Omega^{k_1k_2k_3} \,,\qquad \qquad
 \Omega_6\equiv \frac{1}{6!}\,\epsilon_{k_1\cdots k_6}\,\Omega^{k_1\cdots k_6} \,,
\\
\cF_{{\hat{\mu}},\,i_1\cdots i_6} & \equiv \partial_{\hat{\mu}} A_{i_1\cdots i_6}-10\,A_{[i_1i_2i_3\vert}\,\partial_{\hat{\mu}} A_{\vert i_4i_5i_6]} \,,
\end{split}
\\[5mm]
 &\underline{E_{7(7)}} : 
\nn\\
 \vV & = {\footnotesize\begin{pmatrix}
 \delta_i^k & \qquad 0 & \qquad -\frac{1}{\sqrt{2}}\,\widetilde{A}_i{}^{k_1k_2} & \qquad 0 \\
 0 & \qquad \delta^{i_1i_2}_{k_1k_2} & \qquad 0 & \qquad \frac{1}{\sqrt{2}}\,(\widetilde{A}^{\rmT})^{i_1i_2}{}_k \\
 0 & \qquad 0 & \qquad \delta_{i_1i_2}^{k_1k_2} & \qquad 0 \\
 0 & \qquad 0 & \qquad 0 & \qquad \delta_k^i
 \end{pmatrix}} 
\nn\\
 &\qquad {\footnotesize \cdot \begin{pmatrix}
 \delta_k^j & \qquad - \frac{1}{\sqrt{2}}\, A_{k j_1j_2} & \frac{1}{4\sqrt{2}}\, A_{k l_1l_2}\,\widetilde{A}^{l_1l_2 j_1j_2} & - \frac{1}{24}\,A_{k l_1l_2}\,\widetilde{A}^{l_1l_2 l_1l_2} \,A_{l_1l_2 j} \\
 0 & \qquad \delta^{k_1k_2}_{j_1j_2} & -\frac{1}{2}\, \widetilde{A}^{k_1k_2 j_1j_2} & \frac{1}{4\sqrt{2}}\, \widetilde{A}^{k_1k_2 l_1l_2}\,A_{l_1l_2 j} \\
 0 & \qquad 0 & \delta_{k_1k_2}^{j_1j_2} & -\frac{1}{\sqrt{2}}\,A_{k_1k_2 j}\\
 0 & \qquad 0 & 0 & \delta_j^k
 \end{pmatrix}}\,,
\\
 \widetilde{\vV} &= {\footnotesize\begin{pmatrix}
 \delta_i^k & 0 & \qquad 0 & \qquad \qquad 0 \\
 0 & \delta^{i_1i_2}_{k_1k_2} & \qquad 0 & \qquad \qquad 0 \\
 -\frac{1}{\sqrt{2}}\,\widetilde{\Omega}_{i_1i_2}{}^k & 0 & \qquad \delta_{i_1i_2}^{k_1k_2} & \qquad \qquad 0\\
 0 & \frac{1}{\sqrt{2}}\,(\widetilde{\Omega}^{\rmT})^i{}_{k_1k_2} & \qquad 0 & \qquad \qquad \delta^i_k
 \end{pmatrix}} 
\nn\\
 &\qquad {\footnotesize \cdot \begin{pmatrix}
 \delta_k^j & 0 & 0 & \quad \qquad 0 \\
 -\frac{1}{\sqrt{2}}\, \Omega^{k_1k_2 j} & \delta^{k_1k_2}_{j_1j_2} & 0 & \quad \qquad 0 \\
 \frac{1}{4\sqrt{2}}\, \widetilde{\Omega}_{k_1k_2 l_1l_2}\,\Omega^{l_1l_2 j} & -\frac{1}{2}\, \widetilde{\Omega}_{k_1k_2 j_1j_2} & \delta_{k_1k_2}^{j_1j_2} & \quad \qquad 0\\
 -\frac{1}{24}\,\Omega^{k l_1l_2}\, \widetilde{\Omega}_{l_1l_2 l_1l_2}\,\Omega^{l_1l_2 j} & \frac{1}{4\sqrt{2}}\,\Omega^{k l_1l_2}\, \widetilde{\Omega}_{l_1l_2 j_1j_2} & -\frac{1}{\sqrt{2}}\,\Omega^{k j_1j_2} & \quad \qquad \delta^k_j
 \end{pmatrix}
 }\,,
\\
 \bigl(\omega_{{\hat{\mu}}M}{}^N\bigr) 
 &={\footnotesize\begin{pmatrix}
 0 & \qquad - \frac{\partial_{\hat{\mu}} A_{i j_1j_2}}{\sqrt{2}} & 
 -\frac{2\,\delta_{i}^{[j_1}\,\epsilon^{j_2]k_1\cdots k_6}\,\cF_{{\hat{\mu}},\,k_1\cdots k_6}}{6!\sqrt{2}} & 0 \\
 0 & \qquad 0 & -\frac{1}{2\cdot 3!}\,\epsilon^{i_1i_2 j_1j_2 k_1k_2k_3} \partial_{\hat{\mu}} A_{k_1k_2k_3} & \frac{2\,\delta_{j}^{[i_1}\,\epsilon^{i_2]k_1\cdots k_6}\,\cF_{{\hat{\mu}},\,k_1\cdots k_6}}{6!\sqrt{2}} \\
 0 & \qquad 0 & 0 & -\frac{1}{\sqrt{2}}\,\partial_{\hat{\mu}} A_{i_1i_2 j}\\
 0 & \qquad 0 & 0 & 0
 \end{pmatrix}} \,,
\\
 \bigl(\widetilde{\omega}_{{\hat{\mu}}M}{}^N\bigr) 
 & = {\footnotesize\begin{pmatrix}
 0 & 0 & 0 & \quad 0 \\
 -\frac{S_{\hat{\mu}}{}^{i_1i_2 j}}{\sqrt{2}} & 0 & 0 & \quad 0 \\
 -\frac{2\,\delta_{[i_1}\,\epsilon_{i_2]k_1\cdots k_6}^{j}\,S_{\hat{\mu}}{}^{k_1\cdots k_6}}{6!\sqrt{2}} & -\frac{1}{2\cdot 3!}\, \epsilon_{i_1i_2 j_1j_2 k_1k_2k_3}\,S_{\hat{\mu}}{}^{k_1k_2k_3} & 0 & \quad 0 \\
 0 & \frac{2\,\delta^{i}_{[j_1}\,\epsilon_{j_2]k_1\cdots k_6}\,S_{\hat{\mu}}{}^{k_1\cdots k_6}}{6!\sqrt{2}} & -\frac{1}{\sqrt{2}}\,S_{\hat{\mu}}{}^{i j_1j_2} & \quad 0
 \end{pmatrix}} \,,
\\
\begin{split}
\widetilde{A}^{i_1i_2 j_1j_2} &\equiv \frac{1}{3!}\,\epsilon^{i_1i_2 j_1j_2 k_1k_2k_3}\, A_{k_1k_2k_3} \,,\qquad \qquad
 \widetilde{\Omega}_{i_1i_2 j_1j_2}\equiv \frac{1}{3!}\,\epsilon_{i_1i_2 j_1j_2k_1k_2k_3}\, \Omega^{k_1k_2k_3} \,,
\\
\widetilde{A}_i{}^{j_1j_2} & \equiv \frac{2}{6!}\,\delta_i^{[j_1}\,\epsilon^{j_2]k_1\cdots k_6}\,A_{k_1\cdots k_6}\,,\qquad \qquad (\widetilde{A}^{\rmT})^{i_1i_2}{}_j\equiv \widetilde{A}_j{}^{i_1i_2} \,,
\\
 \widetilde{\Omega}_{i_1i_2}{}^j & \equiv \frac{2}{6!}\,\delta^j_{[i_1}\, \epsilon_{i_2]k_1\cdots k_6}\,\Omega^{k_1\cdots k_6}\,,\qquad \qquad (\widetilde{\Omega}^{\rmT})^i{}_{j_1j_2} \equiv \widetilde{\Omega}_{j_1j_2}{}^i \,,
\\
 \cF_{{\hat{\mu}},\,k_1\cdots k_6} & \equiv \partial_{\hat{\mu}} A_{k_1\cdots k_6} - 10 \, A_{[k_1k_2k_3\vert}\partial_{\hat{\mu}} A_{\vert k_4k_5k_6]}\,. 
\end{split}
\end{align}

\subsubsection*{Type IIB section}

In the type IIB section, we redefine the coordinates as follows:
\begin{align}
\begin{aligned}
 n=7:&\quad \bigl(y^M\bigr)= (x^\sfm,\, y_\sfm^\alpha,\, z)&& (\alpha=1,2,\ \sfm=7,8,9)\,,
\\
 n=6:&\quad \bigl(y^M\bigr)= (x^\sfm,\, y_\sfm^\alpha,\, z^\sfm)&& (\alpha=1,2,\ \sfm=6,\dotsc,9)\,,
\\
 n=5:&\quad \bigl(y^M\bigr)= (x^\sfm,\, y_\sfm^\alpha,\, z^{\sfm_1\sfm_2},\,z^\alpha)&& (\alpha=1,2,\ \sfm =5,\dotsc,9)\,,
\\
 n=4:&\quad \bigl(y^M\bigr)= (x^\sfm,\, y_\sfm^\alpha,\, y_{\sfm_1\sfm_2\sfm_3},\,z^{\alpha,\,\sfm},\,z_\sfm)\qquad &&(\alpha=1,2,\ \sfm =4,\dotsc,9)\,.
\end{aligned}
\end{align}
Here, we defined
\begin{align}
 z^{\sfm_1\cdots \sfm_{d-5}}&\equiv \frac{1}{\sqrt{(d-3)!\,3!}}\,\epsilon^{\sfm_1\cdots \sfm_{d-3} \sfn_1\sfn_2\sfn_3}\,y_{\sfn_1\sfn_2\sfn_3}\qquad (d=3,4,5)\,,
\\
 z^{\alpha,\,\sfm_1\cdots \sfm_{d-5}}&\equiv \frac{1}{\sqrt{(d-5)!\,5!}}\,\epsilon^{\sfm_1\cdots \sfm_{d-5} \sfn_1\cdots \sfn_5}\,y^\alpha_{\sfn_1\cdots \sfn_5}\qquad (d=5,6)\,,
\end{align}
and correspondingly, we also redefined the central charge as
\begin{align}
 W_{\sfa_1\cdots \sfa_{d-3}}&\equiv \frac{1}{3!}\,\epsilon_{\sfa_1\cdots \sfa_{d-3} \sfb_1\sfb_2\sfb_3}\, \cZ^{\sfb_1\sfb_2\sfb_3}\qquad (d=3,4,5)\,,
\\
 W_{\alpha,\,\sfa_1\cdots \sfa_{d-5}}&\equiv \frac{1}{5!}\,\epsilon_{\sfa_1\cdots \sfa_{d-5} \sfb_1\cdots\sfb_5}\, \cZ^{\sfb_1\cdots \sfb_5}\qquad (d=5,6)\,.
\end{align}
After these redefinitions, the generalized momenta are given by
\begin{align}
 \bigl(\cZ_A\bigr)&\equiv\Bigl(P_\sfa,\,\frac{\cZ^\sfa_{\alpha}}{2},\,\frac{W^{\sfa_1\cdots \sfa_{d-3}}}{2\sqrt{(d-3)!}}\Bigr) \qquad \qquad \qquad (d=4,5)\,, 
\\
 (\cZ_A) &\equiv \Bigl(P_\sfa,\,\frac{\cZ^\sfa_{\alpha}}{2},\,\frac{W^{\sfa_1\sfa_2}}{2\sqrt{2}},\,\frac{W_{\alpha}}{2}\Bigr) \qquad \qquad \qquad \ (d=6) \,,
\\
 (\cZ_A) &\equiv \Bigl(P_\sfa,\,\frac{\cZ^\sfa_{\alpha}}{2},\,\frac{\cZ^{\sfa_1\sfa_2\sfa_3}}{2\sqrt{3!}},\,\frac{W_{\alpha,\sfa}}{2},\,\frac{W^\sfa}{10}\Bigr) \qquad \quad (d=7)\,. 
\end{align}

The untwisted generalized metric is then given as follows:
\begin{align}
 \underline{\SL(5)}:\quad&(\widehat{M}_{MN}) = 
 \begin{pmatrix}
 G_{\sfm\sfn} & 0 & 0 \\
 0 & m_{\alpha\beta}\,G^{\sfm\sfn} & 0 \\
 0 & 0 & \abs{G}^{-1}
 \end{pmatrix} ,
\\
 \underline{\SO(5,5)}:\quad&(\widehat{M}_{MN}) = 
 \begin{pmatrix}
 G_{\sfm\sfn} & 0 & 0 \\
 0 & m_{\alpha\beta}\,G^{\sfm\sfn} & 0 \\
 0 & 0 & \abs{G}^{-1}\,G_{\sfm\sfn}
 \end{pmatrix} ,
\\
 \underline{E_{6(6)}}:\quad&(\widehat{M}_{MN}) =
 {\footnotesize\begin{pmatrix}
 G_{\sfm\sfn} & 0 & 0 & 0 \\
 0 & m_{\alpha\beta}\,G^{\sfm\sfn} & 0 & 0 \\
 0 & 0 & \abs{G}^{-1}\,G_{\sfm_1\sfm_2,\,\sfn_1\sfn_2} & 0 \\
 0 & 0 & 0 & m_{\alpha\beta}\,\abs{G}^{-1} 
 \end{pmatrix}} , 
\\
 \underline{E_{7(7)}}:\quad&(\widehat{M}_{MN}) =
 {\footnotesize\begin{pmatrix}
 G_{\sfm\sfn} & 0 & 0 & 0 & 0 \\
 0 & m_{\alpha\beta}\,G^{\sfm\sfn} & 0 & 0 & 0 \\
 0 & 0 & G^{\sfm_1\sfm_2\sfm_3,\,\sfn_1\sfn_2\sfn_3} & 0 & 0 \\
 0 & 0 & 0 & m_{\alpha\beta}\,\abs{G}^{-1}\,G_{\sfm\sfn} & 0 \\
 0 & 0 & 0 & 0 & \abs{G}^{-1}\,G^{\sfm\sfn} 
 \end{pmatrix}} . 
\end{align}
On the other hand, if we define the non-geometric fluxes as
\begin{align}
 \bigl(Q_{\alpha,\,{\hat{\mu}}}{}^{\sfm \sfn}\bigr) 
 &\equiv \begin{pmatrix} Q_{\hat{\mu}}{}^{\sfm \sfn} \cr P_{\hat{\mu}}{}^{\sfm \sfn} \end{pmatrix}
 \equiv \bigl(\partial_{\hat{\mu}} \beta_\alpha^{\sfm \sfn}\bigr) \,, 
\\
 P_{\hat{\mu}}{}^{\sfm_1\cdots \sfm_4}
 &\equiv 
 \partial_{\hat{\mu}} \eta^{\sfm_1\cdots \sfm_4} + 3 \, \epsilon^{\gamma\delta}\, \beta_\gamma^{[\sfm_1\sfm_2}\,\partial_{\hat{\mu}} \beta_\delta^{\sfm_3\sfm_4]} \,,
\\
 \bigl(Q_{\alpha,\,{\hat{\mu}}}{}^{\sfp_1\cdots \sfp_6} \bigr) 
 &\equiv \begin{pmatrix} P_{\hat{\mu}}{}^{\sfp_1\cdots \sfp_6} \cr Q_{\hat{\mu}}{}^{\sfp_1\cdots \sfp_6} \end{pmatrix}
 \equiv \bigl(\partial_{\hat{\mu}} \beta_\alpha^{\sfp_1\cdots \sfp_6} +15\,\beta_\alpha^{[\sfp_1\sfp_2}\,\partial_{\hat{\mu}} \eta^{\sfp_3\cdots \sfp_6]} + 15\,\epsilon^{\gamma\delta}\,\beta_\alpha^{[\sfp_1\sfp_2}\,\beta_\gamma^{\sfp_3\sfp_4}\,\partial_{\hat{\mu}} \beta_\delta^{\sfp_5\sfp_6]}\bigr) \,,
\end{align}
the twist matrix $\vV_M{}^N$ and $\omega_{{\hat{\mu}} M}{}^N$ are given as follows:
\begin{align}
 &\underline{\SL(5)}: 
\nn\\
 &\vV= \begin{pmatrix}
 \delta_\sfm^\sfn & \qquad B^\beta_{\sfm\sfn} & \frac{1}{2}\,B^\gamma_{\sfm\sfp}\,\widetilde{B}_\gamma^\sfp \\
 0 & \qquad \delta_\alpha^\beta\,\delta^\sfm_\sfn & \widetilde{B}_\alpha^\sfm \\
 0 & \qquad 0 & 1
 \end{pmatrix} , \qquad \qquad 
 \widetilde{B}_\alpha^\sfm\equiv \frac{1}{2}\,\epsilon_{\alpha\gamma}\,\epsilon^{\sfm\sfp_1\sfp_2}\,B^\gamma_{\sfp_1\sfp_2} \,,
\\
 &\widetilde{\vV} = \begin{pmatrix}
 \delta_\sfm^\sfn & 0 & \qquad 0 \\
 -\beta_\alpha^{\sfm\sfn} & \delta_\alpha^\beta\,\delta^\sfm_\sfn & \qquad 0 \\
 -\frac{1}{2}\,\beta_\alpha^{\sfm\sfn}\,\widetilde{\beta}^\beta_\sfn & \widetilde{\beta}^\beta_\sfn & \qquad 1
 \end{pmatrix} , \qquad \qquad 
 \widetilde{\beta}^\beta_n\equiv \frac{1}{2}\,\epsilon^{\beta\gamma}\,\epsilon_{\sfn\sfq_1\sfq_2}\,\beta_\gamma^{\sfq_1\sfq_2} \,,
\\
 &\bigl(\omega_{{\hat{\mu}}M}{}^N\bigr) 
 = \begin{pmatrix}
 0 & \qquad \quad \partial_{\hat{\mu}} B^\beta_{\sfm\sfn} & 0 \\
 0 & \qquad \quad 0 & \frac{1}{2}\,\epsilon_{\alpha\gamma}\,\epsilon^{\sfm \sfp_1\sfp_2}\partial_{\hat{\mu}} B^\gamma_{\sfp_1\sfp_2} \\
 0 & \qquad \quad 0 & 0 
 \end{pmatrix} \,, 
\\
 &\bigl(\widetilde{\omega}_{{\hat{\mu}}M}{}^N\bigr) 
 = 
 \begin{pmatrix}
 0 & 0 & \qquad 0 \\
 -Q_{\alpha,\,{\hat{\mu}}}{}^{\sfm \sfn} & 0 & \qquad 0 \\
 0 & \frac{1}{2}\,\epsilon^{\beta\gamma}\,\epsilon_{\sfn \sfq_1\sfq_2}\,Q_{\gamma,\,{\hat{\mu}}}{}^{\sfq_1\sfq_2} & \qquad 0 
 \end{pmatrix}\,, 
\\
 &\underline{\SO(5,5)}: 
\nn\\
 &\vV= \vV_4 \cdot \vV_2 \,, \qquad \text{where} \qquad
\\
 &\vV_2= \begin{pmatrix}
 \delta_\sfm^\sfn & \qquad B^\beta_{\sfm\sfn} & -\frac{1}{2}\,B^\gamma_{\sfm\sfp}\,\widetilde{B}_\gamma^{\sfp\sfn} \\
 0 & \qquad \delta_\alpha^\beta\,\delta^\sfm_\sfn & -\widetilde{B}_\alpha^{\sfm\sfn} \\
 0 & \qquad 0 & \delta_\sfm^\sfn
 \end{pmatrix} ,
\qquad \qquad 
 \vV_4=\begin{pmatrix}
 \delta_\sfm^\sfn & \quad 0 & D_4\,\delta_\sfm^\sfn \\
 0 & \quad \delta_\alpha^\beta\,\delta^\sfm_\sfn & 0 \\
 0 & \quad 0 & \delta_\sfm^\sfn
 \end{pmatrix} ,
\\
 &\widetilde{B}_\alpha^{\sfm \sfn} \equiv \frac{1}{2}\,\epsilon_{\alpha\gamma}\,\epsilon^{\sfm \sfn \sfp_1\sfp_2}\,B^\gamma_{\sfp_1\sfp_2} \,,\qquad \qquad 
 D_4\equiv \frac{1}{4!}\,\epsilon^{\sfm_1\cdots \sfm_4}\,D_{\sfm_1\cdots \sfm_4} \,,
\\
 &\widetilde{\vV} = \widetilde{\vV}_4\cdot \widetilde{\vV}_2\,, \qquad \text{where} 
\\
 &\widetilde{\vV}_2=\begin{pmatrix}
 \delta_\sfm^\sfn & \quad 0 & \qquad 0 \\
 -\beta_\alpha^{\sfm\sfn} & \quad \delta_\alpha^\beta\,\delta^\sfm_\sfn & \qquad 0 \\
 -\frac{1}{2}\,\widetilde{\beta}^\gamma_{\sfm\sfp}\,\beta_\gamma^{\sfp\sfn} & \quad \widetilde{\beta}^\beta_{\sfm\sfn} & \qquad \delta_\sfm^\sfn
 \end{pmatrix} , \qquad \quad 
 \widetilde{\vV}_4= \begin{pmatrix}
 \delta_\sfm^\sfn & \quad 0 & \quad 0 \\
 0 & \quad \delta_\alpha^\beta\,\delta^\sfm_\sfn & \quad 0 \\
 \eta_4\,\delta_\sfm^\sfn & \quad 0 & \quad \delta_\sfm^\sfn
 \end{pmatrix} ,
\\
 &\widetilde{\beta}^\beta_{\sfm \sfn} \equiv \frac{1}{2}\,\epsilon^{\beta\gamma}\,\epsilon_{\sfm \sfn \sfp_1\sfp_2}\,\beta_\gamma^{\sfp_1\sfp_2} \,,\qquad \qquad 
 \eta_4\equiv \frac{1}{4!}\,\epsilon_{\sfm_1\cdots \sfm_4}\,\eta^{\sfm_1\cdots \sfm_4} \,,
\\
 &\bigl(\omega_{{\hat{\mu}}M}{}^N\bigr) 
  = \ \begin{pmatrix}
 0 & \qquad \qquad \qquad \partial_{\hat{\mu}} B^\beta_{\sfm \sfn} & \qquad \frac{1}{4!}\,\delta_\sfm^\sfn\,\epsilon^{\sfp_1\cdots \sfp_4}\,\cG_{{\hat{\mu}},\,\sfp_1\cdots \sfp_4} \\
 0 & \qquad \qquad \qquad 0 & \qquad -\frac{1}{2}\,\epsilon_{\alpha\gamma}\,\epsilon^{\sfm \sfn \sfp_1\sfp_2}\,\partial_{\hat{\mu}} B^\gamma_{\sfp_1\sfp_2} \\
 0 & \qquad \qquad \qquad 0 & \qquad 0 
 \end{pmatrix} \,, 
\\
 &\bigl(\widetilde{\omega}_{{\hat{\mu}}M}{}^N\bigr) 
 = \begin{pmatrix}
 0 & 0 & \qquad \qquad 0 \\
 -Q_{\alpha,\,{\hat{\mu}}}{}^{\sfm \sfn} & 0 & \qquad \qquad 0 \\
 \frac{1}{4!}\,\delta_\sfm^\sfn\,\epsilon_{\sfp_1\cdots \sfp_4}\,P_{\hat{\mu}}{}^{\sfp_1\cdots \sfp_4} & \frac{1}{2}\,\epsilon^{\beta\gamma}\,\epsilon_{\sfm \sfn \sfp_1\sfp_2}\,Q_{\gamma,\,{\hat{\mu}}}{}^{\sfp_1\sfp_2} & \qquad \qquad 0 
 \end{pmatrix} \,,
\\
 &\cG_{{\hat{\mu}},\,\sfp_1\cdots \sfp_4}
 \equiv \partial_{\hat{\mu}} D_{\sfp_1\cdots \sfp_4} - 3\,\epsilon_{\gamma\delta}\,B^\gamma_{[\sfp_1\sfp_2\vert}\,\partial_{\hat{\mu}} B^\delta_{\vert \sfp_3\sfp_4]}\,,
\\
 &\underline{E_{6(6)}}: 
\nn\\
 &\vV= \vV_4 \cdot \vV_2 \,, \qquad \text{where} 
\\
 &\vV_2= {\footnotesize\begin{pmatrix}
 \delta_\sfm^\sfn & \qquad \qquad B^\beta_{\sfm\sfn} & \qquad \frac{1}{2\sqrt{2}}\,B^\gamma_{\sfm \sfp}\,\widetilde{B}_\gamma^{\sfp \sfn_1\sfn_2} & \frac{1}{12}\,B^\gamma_{\sfm \sfp_1}\,\widetilde{B}_\gamma^{\sfp_1 \sfp_2\sfp_3}\,B_{\sfp_2\sfp_3}^\beta 
\\
 0 & \qquad \qquad \delta_\alpha^\beta \,\delta^\sfm_\sfn & \frac{1}{\sqrt{2}}\,\widetilde{B}_\alpha^{\sfm \sfn_1\sfn_2} & \qquad \frac{1}{4}\,\widetilde{B}_\alpha^{\sfm \sfp_1\sfp_2}\, B_{\sfp_1\sfp_2}^\beta 
\\
 0 & \qquad \qquad 0 & \qquad \delta_{\sfm_1\sfm_2}^{\sfn_1\sfn_2} & \frac{1}{\sqrt{2}}\, B_{\sfm_1\sfm_2}^\beta 
\\
 0 & \qquad \qquad 0 & \qquad 0 & \delta_\alpha^\beta 
 \end{pmatrix}}
\\
 &\vV_4= \ {\footnotesize\begin{pmatrix}
 \delta_\sfm^\sfn & \qquad \qquad 0 & \qquad \qquad \frac{1}{\sqrt{2}}\,\widetilde{D}_\sfm^{\sfn_1\sfn_2} & \qquad \qquad \quad 0 \\
 0 & \qquad \qquad \delta_\alpha^\beta \,\delta^\sfm_\sfn & \qquad \qquad 0 & \qquad \qquad \quad -\delta_\alpha^\beta\,\widetilde{D}^\sfm \\
 0 & \qquad \qquad 0 & \qquad \qquad \delta_{\sfm_1\sfm_2}^{\sfn_1\sfn_2} & \qquad\qquad \quad 0 \\
 0 & \qquad \qquad 0 & \qquad\qquad 0 & \qquad \qquad \quad \delta_\alpha^\beta
 \end{pmatrix}} ,
\\
 &\widetilde{B}_\alpha^{\sfm\sfn_1\sfn_2} \equiv \frac{1}{2}\,\epsilon_{\alpha\gamma}\,\epsilon^{\sfm \sfn_1\sfn_2 \sfp_1\sfp_2}\,B^\gamma_{\sfp_1\sfp_2}\,, \quad 
 \widetilde{D}_\sfm^{\sfn_1\sfn_2} \equiv \frac{1}{3!}\,\epsilon^{\sfn_1\sfn_2 \sfp_1\sfp_2\sfp_3}\,D_{\sfm \sfp_1\sfp_2\sfp_3}\,,\quad
 \widetilde{D}^\sfm \equiv \frac{1}{4!}\,\epsilon^{\sfm \sfp_1\cdots \sfp_4}\,D_{\sfp_1\cdots \sfp_4}\,,
\\
 &\widetilde{\vV}= \widetilde{\vV}_4 \cdot \widetilde{\vV}_2 \,, \qquad \text{where}
\\
 &\widetilde{\vV}_2= {\footnotesize\begin{pmatrix}
 \delta_\sfm^\sfn & 0 & \qquad 0 & \qquad \qquad 0 \\
 -\beta_\alpha^{\sfm\sfn} & \delta_\alpha^\beta \,\delta^\sfm_\sfn & \qquad 0 & \qquad \qquad 0 \\
 -\frac{1}{2\sqrt{2}}\,\widetilde{\beta}^\beta_{\sfm_1\sfm_2 \sfp}\,\beta_\alpha^{\sfp\sfn} & \frac{1}{\sqrt{2}}\,\widetilde{\beta}^\beta_{\sfm_1\sfm_2 \sfn} & \qquad \delta_{\sfm_1\sfm_2}^{\sfn_1\sfn_2} & \qquad \qquad 0
\\
 -\frac{1}{12}\,\, \beta^{\sfp_1\sfp_2}_\alpha\,\widetilde{\beta}^\gamma_{\sfp_1\sfp_2\sfp_3}\,\beta_\gamma^{\sfp_3 \sfn} & \frac{1}{4}\,\beta^{\sfp_1\sfp_2}_\alpha\,\widetilde{\beta}^\beta_{\sfp_1\sfp_2 \sfn} & \qquad \frac{1}{\sqrt{2}}\, \beta^{\sfn_1\sfn_2}_\alpha &\qquad \qquad \delta_\alpha^\beta 
 \end{pmatrix}} ,
\\
 &\widetilde{\vV}_4={\footnotesize\begin{pmatrix}
\qquad \delta_\sfm^\sfn &\qquad \qquad 0 & \qquad \qquad 0  \ \ & \ \ \qquad \qquad 0 \\
\qquad 0 & \qquad \qquad \delta_\alpha^\beta \,\delta^\sfm_\sfn & \qquad \qquad 0 \ \ & \ \ \qquad \qquad 0 \\
\qquad \frac{1}{\sqrt{2}}\,\widetilde{\eta}_{\sfm_1\sfm_2}^\sfn & \qquad \qquad 0 & \qquad \qquad \delta_{\sfm_1\sfm_2}^{\sfn_1\sfn_2} \ \ & \ \ \qquad \qquad 0 \\
\qquad 0 & \qquad \qquad -\delta_\alpha^\beta\,\widetilde{\gamma}_\sfn & \qquad \qquad 0 \ \ & \ \ \qquad \qquad \delta_\alpha^\beta
 \end{pmatrix}} ,
\\
 &\widetilde{\beta}^\beta_{\sfm_1\sfm_2 \sfn} \equiv \frac{1}{2}\,\epsilon^{\beta\gamma}\,\epsilon_{\sfm_1\sfm_2 \sfn \sfp_1\sfp_2}\,\beta_\gamma^{\sfp_1\sfp_2}\,, \qquad 
 \widetilde{\eta}_{\sfm_1\sfm_2}^\sfn \equiv \frac{1}{3!}\,\epsilon_{\sfm_1\sfm_2 \sfp_1\sfp_2\sfp_3}\,\eta^{\sfn \sfp_1\sfp_2\sfp_3}\,,\qquad
 \widetilde{\gamma}_\sfn \equiv \frac{1}{4!}\,\epsilon_{\sfn \sfq_1\cdots \sfq_4}\,\gamma^{\sfq_1\cdots \sfq_4}\,,
\\
 &\bigl(\omega_{{\hat{\mu}}M}{}^N\bigr) 
 ={\footnotesize\begin{pmatrix}
 0 & \quad \qquad \qquad \partial_{\hat{\mu}} B^\beta_{\sfm\sfn} & \frac{1}{3!\sqrt{2}}\,\delta_\sfm^{[\sfp_1}\,\epsilon^{\sfp_2\sfp_3\sfp_4] \sfn_1\sfn_2}\,\cG_{{\hat{\mu}},\,\sfp_1\cdots \sfp_4} & 0 \\
 0 & \quad \qquad \qquad 0 & \frac{1}{2\sqrt{2}}\,\epsilon_{\alpha\gamma}\,\epsilon^{\sfm \sfn_1\sfn_2 \sfp_1\sfp_2}\,\partial_{\hat{\mu}} B^\gamma_{\sfp_1\sfp_2} & -\frac{1}{4!}\,\delta_\alpha^\beta\,\epsilon^{\sfm \sfp_1\cdots \sfp_4}\,\cG_{{\hat{\mu}},\,\sfp_1\cdots \sfp_4} \\
 0 & \quad \qquad \qquad 0 & 0 & \frac{1}{\sqrt{2}}\,\partial_{\hat{\mu}} B_{\sfm_1\sfm_2}^\beta \\
 0 & \quad \qquad \qquad 0 & 0 & 0 
 \end{pmatrix}} \,, 
\\
 &\bigl(\widetilde{\omega}_{{\hat{\mu}}M}{}^N\bigr) 
 = {\footnotesize\begin{pmatrix}
 0 & 0 & 0 & \quad 0 \\
 -Q_{{\hat{\mu}},\,\alpha}{}^{\sfm\sfn} & 0 & 0 & \quad 0 \\
 \frac{1}{3!\sqrt{2}}\,\delta^\sfn_{[\sfp_1}\,\epsilon_{\sfp_2\sfp_3\sfp_4] \sfm_1\sfm_2}\,P_{\hat{\mu}}{}^{\sfp_1\cdots \sfp_4} & \frac{1}{2\sqrt{2}}\,\epsilon^{\alpha\gamma}\,\epsilon_{\sfm_1\sfm_2 \sfn \sfp_1\sfp_2}\,Q_{{\hat{\mu}},\, \gamma}{}^{\sfp_1\sfp_2} & 0 & \quad 0 \\
 0 & -\frac{1}{4!}\,\delta_\alpha^\beta\,\epsilon_{\sfn \sfq_1\cdots \sfq_4}\,P_{\hat{\mu}}{}^{\sfq_1\cdots \sfq_4} & \frac{1}{\sqrt{2}}\,Q_{{\hat{\mu}},\,\alpha}{}^{\sfn_1\sfn_2} & \quad 0 
 \end{pmatrix}} \,,
\\
 &\cG_{{\hat{\mu}},\,\sfm_1\cdots \sfm_4}
  \equiv \partial_{\hat{\mu}} D_{\sfm_1\cdots \sfm_4}-3\,\epsilon_{\gamma\delta}\,B^\gamma_{[\sfm_1\sfm_2\vert}\,\partial_{\hat{\mu}} B^\delta_{\vert \sfm_3\sfm_4]}\,,
\\
 &\underline{E_{7(7)}}: 
\nn\\
 &\vV= \vV_6\cdot \vV_4\cdot \vV_2\,,
\\
 &\vV_2={\footnotesize\begin{pmatrix}
 \delta_\sfm^\sfn & B^\beta_{\sfm\sfn} & \frac{1}{2\sqrt{6}}\, B^\gamma_{\sfm \sfp}\,B^\sfp_{\gamma,\,\sfn_1\sfn_2\sfn_3} & -\frac{1}{36}\, B^\gamma_{\sfm \sfp}\,B^\sfp_{\gamma,\,\sfq_1\sfq_2\sfq_3}\,\widetilde{B}^{\beta,\, \sfq_1\sfq_2\sfq_3 \sfn} & -\frac{1}{144}\, \epsilon_{\delta\zeta}\,B^\gamma_{\sfm \sfp}\,B^\sfp_{\gamma,\,\sfq_1\sfq_2\sfq_3}\,\widetilde{B}^{\delta,\, \sfq_1\sfq_2\sfq_3 \sfr}\, B^\zeta_{\sfr \sfn} 
\\
 0 & \delta_\alpha^\beta\,\delta^\sfm_\sfn & \frac{1}{\sqrt{6}}\, B^\sfm_{\alpha,\,\sfn_1\sfn_2\sfn_3} & -\frac{1}{12}\, B^\sfm_{\alpha,\,\sfp_1\sfp_2\sfp_3}\,\widetilde{B}^{\beta,\, \sfp_1\sfp_2\sfp_3 \sfn} & -\frac{1}{36}\, \epsilon_{\gamma\delta}\,B^\sfm_{\alpha,\,\sfp_1\sfp_2\sfp_3}\,\widetilde{B}^{\gamma,\, \sfp_1\sfp_2\sfp_3 \sfq}\, B^\delta_{\sfq \sfn} 
\\
 0 & 0 & \delta^{\sfm_1\sfm_2\sfm_3}_{\sfn_1\sfn_2\sfn_3} & -\frac{1}{\sqrt{6}}\, \widetilde{B}^{\beta,\,\sfm_1\sfm_2\sfm_3 \sfn} & -\frac{1}{2\sqrt{6}}\, \epsilon_{\gamma\delta}\,\widetilde{B}^{\gamma,\, \sfm_1\sfm_2\sfm_3 \sfp}\, B^\delta_{\sfp \sfn} 
\\
 0 & 0 & 0 & \delta_\alpha^\beta\,\delta_\sfm^\sfn & \epsilon_{\alpha\gamma}\, B^\gamma_{\sfm\sfn} \\
 0 & 0 & 0 & 0 & \delta^\sfm_\sfn
 \end{pmatrix} } ,
\\
 &\vV_4= {\footnotesize\begin{pmatrix}
 \delta_\sfm^\sfn & \qquad \qquad 0 & \qquad \qquad \frac{1}{\sqrt{6}}\,D_{\sfm \sfn_1\sfn_2\sfn_3} & \qquad \qquad 0 & -\frac{1}{12}\,D_{\sfm \sfp_1\sfp_2\sfp_3}\,\widetilde{D}^{\sfp_1\sfp_2\sfp_3}_\sfn \\
 0 & \qquad \qquad \delta_\alpha^\beta\,\delta^\sfm_\sfn & \qquad \qquad 0 & \qquad \qquad \delta_\alpha^\beta\,\widetilde{D}^{\sfm\sfn} & 0 \\
 0 & \qquad \qquad 0 & \qquad \qquad \delta^{\sfm_1\sfm_2\sfm_3}_{\sfn_1\sfn_2\sfn_3} & \qquad \qquad 0 & -\frac{1}{\sqrt{6}}\,\widetilde{D}^{\sfm_1\sfm_2\sfm_3}_\sfn \\
 0 & \qquad \qquad 0 & \qquad \qquad 0 & \qquad \qquad \delta_\alpha^\beta\,\delta_\sfm^\sfn & 0 \\
 0 & \qquad \qquad 0 & \qquad \qquad 0 & \qquad \qquad 0 & \delta^\sfm_\sfn
 \end{pmatrix} } ,
\\
 &\vV_6= \ {\footnotesize\begin{pmatrix}
 \delta_\sfm^\sfn & \qquad \qquad 0 & \quad \qquad \qquad 0 & \qquad \qquad - B_6^\beta\,\delta^\sfn_\sfm & \qquad \qquad 0 \quad \\
 0 & \qquad \qquad \delta_\alpha^\beta\,\delta^\sfm_\sfn & \quad \qquad \qquad 0 & \qquad \qquad 0 & \qquad \qquad \epsilon_{\alpha\gamma}\,B_6^\gamma\, \delta^\sfm_\sfn \quad \\
 0 & \qquad \qquad 0 & \quad \qquad \qquad \delta^{\sfm_1\sfm_2\sfm_3}_{\sfn_1\sfn_2\sfn_3} & \qquad \qquad 0 & \qquad \qquad 0 \quad \\
 0 & \qquad \qquad 0 & \quad \qquad \qquad 0 & \qquad \qquad \delta_\alpha^\beta\,\delta_\sfm^\sfn & \qquad \qquad 0 \quad \\
 0 & \qquad \qquad 0 & \quad \qquad \qquad 0 & \qquad \qquad 0 & \qquad \qquad \delta^\sfm_\sfn \quad 
 \end{pmatrix} },
\\
 &B^\sfm_{\alpha,\,\sfn_1\sfn_2\sfn_3} \equiv 3\,\epsilon_{\alpha\gamma}\,\delta^\sfm_{[\sfn_1}\,B^\gamma_{\sfn_2\sfn_3]}\,, \quad
 \widetilde{B}^{\alpha,\,\sfm_1\sfm_2\sfm_3 \sfn}\equiv \frac{1}{2}\,\epsilon^{\sfm_1\sfm_2\sfm_3 \sfn \sfp_1\sfp_2}\,B^\alpha_{\sfp_1\sfp_2}\,,
\\
 &\widetilde{D}^{\sfm\sfn} \equiv \frac{1}{4!}\,\epsilon^{\sfm\sfn \sfp_1\cdots \sfp_4}\,D_{\sfp_1\cdots \sfp_4}\,,\quad
 \widetilde{D}^{\sfm_1\sfm_2\sfm_3}_\sfn \equiv \frac{1}{3!}\,\epsilon^{\sfm_1\sfm_2\sfm_3 \sfp_1\sfp_2\sfp_3}\,D_{\sfp_1\sfp_2\sfp_3 \sfn}\,,\quad 
 B^\alpha_6\equiv \frac{1}{6!}\,\epsilon^{\sfp_1\cdots \sfp_6}\,B^\alpha_{\sfp_1\cdots \sfp_6}\,,
\\
 &\widetilde{\vV}= \widetilde{\vV}_6\cdot \widetilde{\vV}_4\cdot \widetilde{\vV}_2\,, \qquad \text{where}
\\
 &\widetilde{\vV}_2= \! {\footnotesize\begin{pmatrix}
 \delta_\sfm^\sfn & 0 & 0 & 0 & 0 \\
 -\beta_\alpha^{\sfm\sfn} & \delta_\alpha^\beta\,\delta^\sfm_\sfn & 0 & 0 & 0 \\
 -\frac{1}{2\sqrt{6}}\,\widetilde{\beta}^{\gamma,\,\sfm_1\sfm_2\sfm_3}_\sfp\,\beta_\gamma^{\sfp\sfn} & \frac{1}{\sqrt{6}}\,\widetilde{\beta}^{\beta,\,\sfm_1\sfm_2\sfm_3}_\sfn & \delta^{\sfm_1\sfm_2\sfm_3}_{\sfn_1\sfn_2\sfn_3} & 0 & 0 \\
 -\frac{1}{36}\,\widetilde{\beta}_{\alpha,\,\sfm \sfp_1\sfp_2\sfp_3}\,\widetilde{\beta}^{\gamma,\,\sfp_1\sfp_2\sfp_3}_\sfq\,\beta_\gamma^{\sfq\sfn} & \frac{1}{12}\,\widetilde{\beta}_{\alpha,\,\sfm \sfp_1\sfp_2\sfp_3}\,\widetilde{\beta}^{\beta,\,\sfp_1\sfp_2\sfp_3}_\sfn & \frac{1}{\sqrt{6}}\,\widetilde{\beta}_{\alpha,\,\sfm \sfn_1\sfn_2\sfn_3} & \delta_\alpha^\beta\,\delta_\sfm^\sfn & 0 \\
 -\frac{1}{144}\,\epsilon^{\gamma\delta}\,\beta_\gamma^{\sfm\sfp}\,\widetilde{\beta}_{\delta,\,\sfp \sfq_1\sfq_2\sfq_3}\, \widetilde{\beta}^{\delta,\,\sfq_1\sfq_2\sfq_3}_\sfr\,\beta_\delta^{\sfr \sfn} & \frac{1}{36}\,\epsilon^{\gamma\delta}\,\beta_\gamma^{\sfm\sfp}\,\widetilde{\beta}_{\delta,\,\sfp \sfq_1\sfq_2\sfq_3}\, \widetilde{\beta}^{\beta,\,\sfq_1\sfq_2\sfq_3}_\sfn & \frac{1}{2\sqrt{6}}\,\epsilon^{\gamma\delta}\,\beta_\gamma^{\sfm\sfp}\,\widetilde{\beta}_{\delta,\,\sfp \sfn_1\sfn_2\sfn_3} & -\epsilon^{\beta\gamma}\,\beta_\gamma^{\sfm\sfn} & \delta^\sfm_\sfn
 \end{pmatrix} } ,
\\
 &\widetilde{\vV}_4= {\footnotesize\begin{pmatrix}
 \delta_\sfm^\sfn & \qquad \qquad 0 & \quad \qquad \qquad 0 & \quad \qquad \qquad 0 & \qquad \qquad \qquad 0 \\
 0 & \qquad \qquad \delta_\alpha^\beta\,\delta^\sfm_\sfn & \qquad \qquad \quad 0 & \qquad \qquad \quad 0 & \qquad \qquad \qquad 0 \\
 -\frac{1}{\sqrt{6}}\,\eta^{\sfm_1\sfm_2\sfm_3 \sfn} & \qquad \qquad 0 & \qquad \qquad \quad \delta^{\sfm_1\sfm_2\sfm_3}_{\sfn_1\sfn_2\sfn_3} & \qquad \qquad \quad 0 & \qquad \qquad \qquad 0 \\
 0 & \qquad \qquad -\delta_\alpha^\beta\,\widetilde{\eta}_{\sfm\sfn} & \qquad \qquad \quad 0 & \qquad \qquad \quad \delta_\alpha^\beta\,\delta_\sfm^\sfn & \qquad \qquad \qquad 0 \\
 \frac{1}{12}\,\widetilde{\eta}^\sfm_{\sfp_1\sfp_2\sfp_3}\,\eta^{\sfp_1\sfp_2\sfp_3 \sfn} & \quad \qquad \qquad 0 & \quad \qquad \qquad -\frac{1}{\sqrt{6}}\,\widetilde{\eta}^\sfm_{\sfn_1\sfn_2\sfn_3} & \quad \qquad \qquad 0 & \qquad \qquad \qquad \delta^\sfm_\sfn
 \end{pmatrix} } ,
\\
 &\widetilde{\vV}_6= {\footnotesize\begin{pmatrix}
 \delta_\sfm^\sfn & \qquad \qquad 0 & \qquad \qquad \quad 0 & \quad \qquad \qquad 0 & \qquad \qquad \qquad 0 \\
 0 & \qquad \qquad \delta_\alpha^\beta\,\delta^\sfm_\sfn & \quad \qquad \qquad 0 & \quad \qquad \qquad 0 & \qquad \qquad \qquad 0 \\
 0 & \qquad \qquad 0 & \quad \qquad \qquad \delta^{\sfm_1\sfm_2\sfm_3}_{\sfn_1\sfn_2\sfn_3} & \quad \qquad \qquad 0 & \qquad \qquad \qquad 0 \\
 -\beta^6_\alpha\,\delta_\sfm^\sfn & \qquad \qquad 0 & \quad \qquad \qquad 0 & \quad \qquad \qquad \delta_\alpha^\beta\,\delta_\sfm^\sfn & \qquad \qquad \qquad 0 \\
 0 & \qquad \qquad \epsilon^{\beta\gamma}\,\beta^6_\gamma\,\delta^\sfm_\sfn & \quad \qquad \qquad 0 & \quad \qquad \qquad 0 & \qquad \qquad \qquad \delta^\sfm_\sfn
 \end{pmatrix} },
\\
 &\beta_\sfn^{\beta,\,\sfm_1\sfm_2\sfm_3} \equiv 3\,\epsilon^{\beta\gamma}\,\delta_\sfn^{[\sfm_1}\,\beta_\gamma^{\sfm_2\sfm_3]}\,, \qquad \qquad 
 \widetilde{\beta}_{\alpha,\,\sfm_1\sfm_2\sfm_3 \sfn}\equiv \frac{1}{2}\,\epsilon_{\sfm_1\sfm_2\sfm_3 \sfn \sfp_1\sfp_2}\,\beta_\alpha^{\sfp_1\sfp_2}\,,
\\
 &\widetilde{\eta}_{\sfm\sfn} \equiv \frac{1}{4!}\,\epsilon_{\sfm\sfn \sfp_1\cdots \sfp_4}\,\eta^{\sfp_1\cdots \sfp_4}\,,\qquad
 \widetilde{\eta}_{\sfm_1\sfm_2\sfm_3}^\sfn \equiv \frac{1}{3!}\,\epsilon_{\sfm_1\sfm_2\sfm_3 \sfp_1\sfp_2\sfp_3}\,\eta^{\sfp_1\sfp_2\sfp_3 \sfn}\,,\qquad 
 \beta_\alpha^6\equiv \frac{1}{6!}\,\epsilon_{\sfp_1\cdots \sfp_6}\,\beta_\alpha^{\sfp_1\cdots \sfp_6}\,,
\\
 &\bigl(\omega_{{\hat{\mu}}M}{}^N\bigr) 
 ={\footnotesize\begin{pmatrix}
 0 & \partial_{\hat{\mu}} B^\beta_{\sfm\sfn} & \frac{1}{\sqrt{6}}\, \cG_{{\hat{\mu}},\,\sfm \sfn_1\sfn_2\sfn_3} & -\frac{1}{6!}\,\delta_\sfm^\sfn\,\epsilon^{\sfp_1\cdots \sfp_6}\,\cG^\beta_{{\hat{\mu}},\,\sfp_1\cdots \sfp_6} & 0 \\
 0 & 0 & \frac{3}{\sqrt{6}}\,\epsilon_{\alpha\gamma}\,\delta^\sfm_{[\sfn_1\vert}\partial_{\hat{\mu}} B^\gamma_{\vert \sfn_2\sfn_3]} & \frac{1}{4!}\,\delta_\alpha^\beta\,\epsilon^{\sfm\sfn \sfp_1\cdots \sfp_4}\,\cG_{{\hat{\mu}},\,\sfp_1\cdots \sfp_4} & \frac{1}{6!}\,\epsilon_{\alpha\gamma}\,\delta^\sfm_\sfn\,\epsilon^{\sfp_1\cdots \sfp_6}\,\cG^\gamma_{{\hat{\mu}},\sfp_1\cdots \sfp_6} \\
 0 & 0 & 0 & -\frac{1}{2\sqrt{6}}\,\epsilon^{\sfm_1\sfm_2\sfm_3 \sfn \sfp_1\sfp_2}\,\partial_{\hat{\mu}} B_{\sfp_1\sfp_2}^\beta & -\frac{1}{3!\sqrt{6}}\,\epsilon^{\sfm_1\sfm_2\sfm_3 \sfp_1\sfp_2\sfp_3}\,\cG_{{\hat{\mu}},\,\sfp_1\sfp_2\sfp_3\sfn} \\
 0 & 0 & 0 & 0 & \epsilon_{\alpha\gamma}\,\partial_{\hat{\mu}} B^\gamma_{\sfm\sfn} \\
 0 & 0 & 0 & 0 & 0 
 \end{pmatrix}} \,, 
\\
 &\bigl(\widetilde{\omega}_{{\hat{\mu}}M}{}^N\bigr) 
 ={\footnotesize\begin{pmatrix}
 0 & 0 & 0 & 0 & 0 \\
 -Q_{\alpha,\,{\hat{\mu}}}{}^{\sfm \sfn} & 0 & 0 & 0 & 0 \\
 -\frac{1}{\sqrt{6}}\,P_{\hat{\mu}}{}^{\sfm_1\sfm_2\sfm_3 \sfn} & \frac{3}{\sqrt{6}}\,\epsilon^{\beta\gamma}\,\delta^{[\sfm_1}_\sfn\,
 Q_{\gamma,\,{\hat{\mu}}}{}^{\sfm_2\sfm_3]} & 0 & 0 & 0 \\
 -\frac{1}{6!}\,\delta^\sfn_\sfm\,\epsilon_{\sfp_1\cdots \sfp_6}\,Q_{\alpha,\,{\hat{\mu}}}{}^{\sfp_1\cdots \sfp_6} & -\frac{1}{4!}\,\delta_\alpha^\beta\,\epsilon_{\sfm\sfn \sfp_1\cdots \sfp_4}\,P_{\hat{\mu}}{}^{\sfp_1\cdots \sfp_4} & \frac{1}{2\sqrt{6}}\,\epsilon_{\sfm \sfn_1\sfn_2\sfn_3 \sfp_1\sfp_2}\,Q_{\alpha,\, {\hat{\mu}}}{}^{\sfp_1\sfp_2} & 0 & 0 \\
 0 & \frac{1}{6!}\,\epsilon^{\beta\gamma}\,\delta^\sfm_\sfn\,\epsilon_{\sfp_1\cdots \sfp_6}\,Q_{\gamma,\,{\hat{\mu}}}{}^{\sfp_1\cdots \sfp_6} & -\frac{1}{3!\sqrt{6}}\,\epsilon_{\sfn_1\sfn_2\sfn_3 \sfp_1\sfp_2\sfp_3}\,P_{\hat{\mu}}{}^{\sfp_1\sfp_2\sfp_3 \sfm} & -\epsilon^{\beta\gamma}\,Q_{\gamma,\, {\hat{\mu}}}{}^{\sfm \sfn} & 0 
 \end{pmatrix}} \,, 
\\
 &\cG_{{\hat{\mu}},\,\sfm_1\cdots \sfm_4} \equiv \partial_{\hat{\mu}} D_{\sfm_1\cdots \sfm_4}-3\,\epsilon_{\gamma\delta}\,B^\gamma_{[\sfm_1\sfm_2\vert}\,\partial_{\hat{\mu}} B^\delta_{\vert \sfm_3\sfm_4]}\,,
\\
 &\cG^\beta_{{\hat{\mu}},\,\sfm_1\cdots \sfm_6} \equiv \partial_{\hat{\mu}} B^\beta_{\sfm_1\cdots \sfm_6} - 15\,B^\beta_{[\sfm_1\sfm_2}\,\partial_{\hat{\mu}} D_{\sfm_3\cdots \sfm_6]} + 15\,\epsilon_{\gamma\delta}\,B^\beta_{[\sfm_1\sfm_2}\,B^\gamma_{\sfm_3\sfm_4\vert}\,\partial_{\hat{\mu}} B^\delta_{\vert \sfm_5\sfm_6]} \,.
\end{align}

\subsection{External part}

For the external part, we focus on the following two-derivative terms:
\begin{align}
 \cL_{\text{EH}} = e\,R(g)\, \qquad \text{and}
\qquad
 \cL_{\text{scalar}}= \frac{e}{4\alpha_n}\,g^{\mu\nu}\,\partial_\mu \cM_{MN}\,\partial_\nu \cM^{MN} \,.
\end{align}

Recalling the relation, $g_{\mu\nu} = \abs{G}^{\frac{1}{n-2}}\,\sfg_{\mu\nu}$, the first term is given by
\begin{align}
 \cL_{\text{EH}}
 &= \abs{\sfg}^{\frac{1}{2}}\,\abs{G}^{1/2}\,R(\sfg) 
 + 2\,(n-1)\,\partial_\mu\bigl(e\,g^{\mu\nu}\,\partial_\nu \ln\abs{G}^{-\frac{1}{n-2}}\bigr)
\nn\\
 &+ \abs{\sfg}^{\frac{1}{2}}\,\abs{G}^{\frac{n/2}{n-2}}\,\frac{n-1}{4(n-2)}\,g^{\mu\nu}\, \partial_\mu\ln\abs{G} \, \partial_\nu\ln\abs{G} 
\nn\\
 &= \abs{\sfG}^{\frac{1}{2}}\,
 \Bigl[R(\sfg) + \frac{n-1}{4(n-2)}\,\sfg^{\mu\nu}\, \partial_\mu\ln\abs{G}\,\partial_\nu\ln\abs{G} \Bigr]\,,
\end{align}
where we defined $\abs{\sfG}^{\frac{1}{2}}= \abs{\sfg}^{\frac{1}{2}}\,\abs{G}^{\frac{1}{2}}$ and neglected the total derivative term at the second equality. 
For the scalar part, $\cL_{\text{scalar}}$, noting that the matrix $\vV$ has a block-wise upper/lower triangular form with constant diagonal elements, we obtain
\begin{align}
 \cL_{\text{scalar}}
 =\frac{e}{4\alpha_n}\,g^{\mu\nu}\,\partial_\mu \widehat{\cM}_{MN}\,\partial_\nu \widehat{\cM}^{MN}
 -\frac{e}{2\alpha_n}\,g^{\mu\nu}\,\widehat{\cM}^{MN}\,\widehat{\cM}_{PQ}\,\omega_{\mu M}{}^P\,\omega_{\mu N}{}^Q \,, 
\end{align}
where the first term simply becomes
\begin{align}
 &\frac{e}{4\alpha_n}\,g^{\mu\nu}\,\partial_\mu \widehat{\cM}_{MN}\,\partial_\nu \widehat{\cM}^{MN}
\nn\\
 &= \begin{cases}
 \abs{\sfG}^{\frac{1}{2}}\,
 \Bigl[ \frac{1}{4}\,\sfg^{\mu\nu}\,\partial_\mu G^{ij}\, \partial_\nu G_{ij}
        -\frac{1}{4(n-2)}\,\sfg^{\mu\nu}\, \partial_\mu\ln\abs{G}\,\partial_\nu\ln\abs{G} \Bigr] & \text{(M-theory)}\\
 \abs{\sfG}^{\frac{1}{2}}\,
 \Bigl[ \frac{1}{4}\,\sfg^{\mu\nu}\,\partial_\mu G^{\sfm\sfn}\, \partial_\nu G_{\sfm\sfn}
        -\frac{1}{4(n-2)}\,\sfg^{\mu\nu}\, \partial_\mu\ln\abs{G}\,\partial_\nu\ln\abs{G} \\
        \qquad +\frac{1}{4}\,\sfg^{\mu\nu}\,\partial_\mu m_{\alpha\beta}\,\partial_\nu m^{\alpha\beta} \Bigr] & \text{(type IIB)}
\end{cases}
\,.
\end{align}

We thus obtain
\begin{align}
 &\cL_{\text{EH}} + \cL_{\text{scalar}}
\nn\\
 &= \begin{cases}
 \abs{\sfG}^{\frac{1}{2}}\,
 \Bigl[R(\sfg) 
       +\frac{1}{4}\,\sfg^{\mu\nu}\,\partial_\mu G^{ij}\, \partial_\nu G_{ij}
       +\frac{1}{4}\,\sfg^{\mu\nu}\, \partial_\mu\ln\abs{G}\,\partial_\nu\ln\abs{G} \Bigr]
 + \cL_{\text{scalar}}^{(\text{mat})} & \text{(M-theory)}\\
 \abs{\sfG}^{\frac{1}{2}}\,
 \Bigl[R(\sfg) 
       +\frac{1}{4}\,\sfg^{\mu\nu}\,\partial_\mu G^{\sfm\sfn}\, \partial_\nu G_{\sfm\sfn}
       +\frac{1}{4}\,\sfg^{\mu\nu}\, \partial_\mu\ln\abs{G}\,\partial_\nu\ln\abs{G}\\
 \qquad\quad +\frac{1}{4}\,\sfg^{\mu\nu}\,\partial_\mu m_{\alpha\beta}\,\partial_\nu m^{\alpha\beta}
\Bigr]
 + \cL_{\text{scalar}}^{(\text{mat})} & \text{(type IIB)} 
\end{cases}\,,
\\
 &\cL_{\text{scalar}}^{(\text{mat})}
 \equiv -\frac{\abs{\sfG}^{\frac{1}{2}}}{2\alpha_n}\,\sfg^{\mu\nu}\,\widehat{M}^{MN}\,\widehat{M}_{PQ}\,\omega_{\mu M}{}^P\,\omega_{\nu N}{}^Q \,,
\end{align}
where we used, $e\,g^{\mu\nu}= \abs{\sfG}^{\frac{1}{2}}\,\sfg^{\mu\nu}$ and $\widehat{\cM}^{MN}\,\widehat{\cM}_{PQ} =\widehat{M}^{MN}\,\widehat{M}_{PQ}$\,. 

We can calculate the explicit form of $\cL_{\text{scalar}}^{(\text{mat})}$ as follows:
\begin{align}
 &\underline{\text{\textbf{M-theory section:}}}
\nn\\
 &\underline{\bullet\ \text{$\SL(5)$, $\SO(5,5)$ (geometric):}}
\nn\\
 &\cL_{\text{scalar}}^{(\text{mat})}
  = -\frac{\abs{\sfG}^{\frac{1}{2}}}{2\cdot 3!}\, \sfg^{\mu\nu}\,G^{i_1i_2i_3,\,j_1j_2j_3}\,\partial_\mu A_{i_1i_2i_3}\, \partial_\nu A_{j_1j_2j_3} \,,
\\
 &\underline{\bullet\ \text{$E_6$, $E_7$ (geometric):}}
\nn\\
 &\cL_{\text{scalar}}^{(\text{mat})}
  = - \abs{\sfG}^{\frac{1}{2}} \, \sfg^{\mu\nu}\,\Bigl(\frac{G^{i_1i_2i_3,\,j_1j_2j_3}}{2\cdot 3!}\,\partial_\mu A_{i_1i_2i_3}\, \partial_\nu A_{j_1j_2j_3} 
  +\frac{G^{i_1\cdots i_6,\,j_1\cdots j_6}}{2\cdot 6!}\,\cF_{\mu, i_1\cdots i_6}\,\cF_{\nu, j_1\cdots j_6}\Bigr) \,,
\\
 &\underline{\bullet\ \text{$\SL(5)$, $\SO(5,5)$ (non-geometric):}}
\nn\\
 &\cL_{\text{scalar}}^{(\text{mat})}
 = -\frac{\abs{\widetilde{\sfG}}^{\frac{1}{2}}}{2\cdot 3!}\, \widetilde{\sfg}^{\mu\nu}\,\widetilde{G}_{i_1i_2i_3,\,j_1j_2j_3}\,S_\mu{}^{i_1i_2i_3}\, S_\nu{}^{j_1j_2j_3} \,,
\\
 &\underline{\bullet\ \text{$E_6$, $E_7$ (non-geometric):}}
\nn\\
 &\cL_{\text{scalar}}^{(\text{mat})}
 = - \abs{\widetilde{\sfG}}^{\frac{1}{2}}\, \widetilde{\sfg}^{\mu\nu}\,
 \Bigl(\frac{\widetilde{G}_{i_1i_2i_3,\,j_1j_2j_3}}{2\cdot 3!}\,S_\mu{}^{i_1i_2i_3}\,S_\nu{}^{j_1j_2j_3} 
 +\frac{\widetilde{G}_{i_1\cdots i_6,\,j_1\cdots j_6}}{2\cdot 6!}\,S_\mu{}^{i_1\cdots i_6}\,S_\nu{}^{j_1\cdots j_6}\Bigr) \,,
\\
 &\underline{\text{\textbf{type IIB section:}}}
\nn\\
 &\underline{\bullet\ \text{$\SL(5)$ (geometric):}}
\nn\\
 &\cL_{\text{scalar}}^{(\text{mat})}
  = -\frac{\abs{\sfG}^{\frac{1}{2}}}{2\cdot 2!}\, \sfg^{\mu\nu}\,m_{\alpha\beta}\,G^{\sfm_1\sfm_2,\,\sfn_1\sfn_2}\,\partial_\mu B^\alpha_{\sfm_1\sfm_2}\, \partial_\nu B^\beta_{\sfn_1\sfn_2} \,,
\\
 &\underline{\bullet\ \text{$\SO(5,5)$, $E_6$ (geometric):}}
\nn\\
 &\cL_{\text{scalar}}^{(\text{mat})} 
 = -\frac{\abs{\sfG}^{\frac{1}{2}}}{2}\,\sfg^{\mu\nu}\,\Bigl(\frac{m_{\alpha\beta}\,G^{\sfm_1\sfm_2,\,\sfn_1\sfn_2}}{2!}\,\partial_\mu B^\alpha_{\sfm_1\sfm_2}\,\partial_\nu B^\beta_{\sfn_1\sfn_2} 
 +\frac{G^{\sfm_1\cdots \sfm_4,\,\sfn_1\cdots \sfn_4}}{4!} \, \cG_{\mu,\sfm_1\cdots \sfm_4}\,\cG_{\nu,\sfn_1\cdots \sfn_4}\Bigr) \,,
\\
 &\underline{\bullet\ \text{$E_7$ (geometric):}}
\nn\\
 &\cL_{\text{scalar}}^{(\text{mat})} 
 = -\frac{\abs{\sfG}^{\frac{1}{2}}}{2}\,\sfg^{\mu\nu}\,\Bigl(\frac{m_{\alpha\beta}\,G^{\sfm_1\sfm_2,\,\sfn_1\sfn_2}}{2!}\,\partial_\mu B^\alpha_{\sfm_1\sfm_2}\,\partial_\nu B^\beta_{\sfn_1\sfn_2} 
 +\frac{G^{\sfm_1\cdots \sfm_4,\,\sfn_1\cdots \sfn_4}}{4!} \, \cG_{\mu,\sfm_1\cdots \sfm_4}\,\cG_{\nu,\sfn_1\cdots \sfn_4} 
\nn\\
 &\qquad\qquad\qquad\qquad\quad +\frac{m_{\alpha\beta}\,G^{\sfm_1\cdots \sfm_6,\,\sfn_1\cdots \sfn_6}}{6!} \, \cG^\alpha_{\mu,\sfm_1\cdots \sfm_6}\,\cG^\beta_{\nu,\sfn_1\cdots \sfn_6}\Bigr) \,,
\\
 &\underline{\bullet\ \text{$\SL(5)$ (non-geometric):}}
\nn\\
 &\cL_{\text{scalar}}^{(\text{mat})}
  = -\frac{\abs{\widetilde{\sfG}}^{\frac{1}{2}}}{2\cdot 2!}\, \widetilde{\sfg}^{\mu\nu}\,\widetilde{m}^{\alpha\beta}\,\widetilde{G}_{\sfm_1\sfm_2,\,\sfn_1\sfn_2}\, Q_{\alpha,\mu}{}^{\sfm_1\sfm_2}\, Q_{\beta,\nu}{}^{\sfn_1\sfn_2} \,,
\\
 &\underline{\bullet\ \text{$\SO(5,5)$, $E_6$ (non-geometric):}}
\nn\\
 &\cL_{\text{scalar}}^{(\text{mat})} 
 = -\frac{\abs{\widetilde{\sfG}}^{\frac{1}{2}}}{2}\, \widetilde{\sfg}^{\mu\nu}\,
 \Bigl(\frac{\widetilde{m}^{\alpha\beta}\,\widetilde{G}_{\sfm_1\sfm_2,\,\sfn_1\sfn_2}}{2!}\,Q_{\alpha,\mu}{}^{\sfm_1\sfm_2}\, Q_{\beta,\nu}{}^{\sfn_1\sfn_2} 
 +\frac{\widetilde{G}_{\sfm_1\cdots \sfm_4,\,\sfn_1\cdots \sfn_4}}{4!}\, P_\mu{}^{\sfm_1\cdots \sfm_4}\,P_\nu{}^{\sfn_1\cdots \sfn_4}\Bigr) \,,
\\
 &\underline{\bullet\ \text{$E_7$ (non-geometric):}}
\nn\\
 &\cL_{\text{scalar}}^{(\text{mat})} 
 = -\frac{\abs{\widetilde{\sfG}}^{\frac{1}{2}}}{2}\, \widetilde{\sfg}^{\mu\nu}\,
 \Bigl(\frac{\widetilde{m}^{\alpha\beta}\,\widetilde{G}_{\sfm_1\sfm_2,\,\sfn_1\sfn_2}}{2!}\,Q_{\alpha,\mu}{}^{\sfm_1\sfm_2}\, Q_{\beta,\nu}{}^{\sfn_1\sfn_2} 
 +\frac{\widetilde{G}_{\sfm_1\cdots \sfm_4,\,\sfn_1\cdots \sfn_4}}{4!}\, P_\mu{}^{\sfm_1\cdots \sfm_4}\,P_\nu{}^{\sfn_1\cdots \sfn_4} 
\nn\\
 &\qquad\qquad\qquad\qquad\quad +\frac{\widetilde{m}^{\alpha\beta}\,\widetilde{G}_{\sfm_1\cdots \sfm_6,\,\sfn_1\cdots \sfn_6}}{6!} \, Q_{\alpha,\mu}{}^{\sfm_1\cdots \sfm_6}\,Q_{\beta,\nu}{}^{\sfn_1\cdots \sfn_6}\Bigr) \,.
\end{align}

\subsection{Internal (potential) part}

The internal part, or the potential part, consists of three terms
\begin{align}
 \cL_{\text{pot}} &\equiv \cL_{\text{pot}}^{(1)} + \cL_{\text{pot}}^{(2)} + \cL_{\text{pot}}^{(3)} \,,
\\
 \cL_{\text{pot}}^{(1)}
 &\equiv e \frac{1}{4\alpha_n}\,\cM^{MN}\,\partial_M\cM^{KL}\,\partial_N \cM_{KL} \,,
\\
 \cL_{\text{pot}}^{(2)}
 &\equiv -e \frac{1}{2}\,\cM^{MN}\,\partial_N \cM^{KL}\,\partial_L \cM_{MK} \,,
\\
 \cL_{\text{pot}}^{(3)}
 &\equiv e \,\Bigl(\partial_M \ln e\,\partial_N\cM^{MN} + \cM^{MN}\,\partial_M e\,\partial_N e + \frac{1}{4}\,\cM^{MN}\,\partial_M g^{\mu\nu}\,\partial_N g_{\mu\nu}\Bigr) \,.
\end{align}
Here, we choose the canonical section, $(\partial_M)=(\partial_\sfi\,,0,\dotsc,0)$, where the index $\sfi$ represents $i$ in the M-theory section or $\sfm$ in the type IIB section. 
In this case, the first and the third terms can be obtained as follows:
\begin{align}
 \cL_{\text{pot}}^{(1)}
 &= e\frac{1}{4\alpha_n}\,\cM^{\sfi \sfj}\,\partial_\sfi \widehat{M}^{KL}\,\partial_\sfj \widehat{M}_{KL}
  -e \frac{1}{2\alpha_n}\,\cM^{\sfi \sfj}\,\widehat{M}^{MN}\,\widehat{M}_{PQ}\,\omega_{\sfi M}{}^P\,\omega_{\sfj N}{}^Q 
\nn\\
 &=\abs{\sfG}^{\frac{1}{2}}\, 
 \Bigl[ \frac{1}{4}\,G^{\sfi \sfj}\,\partial_\sfi G^{\sfk \sfl}\, \partial_\sfj G_{\sfk \sfl}
       -\frac{1}{4(n-2)}\,G^{\sfi \sfj}\, \partial_\sfi\ln\abs{G} \, \partial_\sfj\ln\abs{G} 
  -\frac{1}{2\alpha_n}\,G^{\sfi \sfj}\,\widehat{M}^{MN}\,\widehat{M}_{PQ}\,\omega_{\sfi M}{}^P\,\omega_{\sfj N}{}^Q \Bigr] \,,
\\
 \cL_{\text{pot}}^{(3)}
 &= \abs{\sfG}^{\frac{1}{2}}\,\biggl( \frac{n\,(n-3)}{4(n-2)^2}\,G^{\sfi \sfj}\, \partial_{\sfi}\ln \abs{G} \, \partial_{\sfj}\ln \abs{G} - \frac{n}{2(n-2)}\,G^{\sfi \sfj}\,G^{\sfk \sfl}\, \partial_{\sfi}\ln \abs{G}\,\partial_{\sfl}G_{\sfj \sfk} 
\nn\\
 &\quad + \frac{1}{4}\,G^{\sfi \sfj}\, \partial_\sfi \ln \abs{\sfg} \, \partial_\sfj \ln \abs{\sfg}
  + \frac{1}{2}\,\partial_\sfi \ln \abs{\sfg}\, \partial_\sfj G^{\sfi \sfj} 
  + \frac{1}{2}\,G^{\sfi \sfj}\,\partial_\sfi \ln\abs{G}\,\partial_\sfj \ln \abs{\sfg}
  + \frac{1}{4}\,G^{\sfi \sfj}\,\partial_\sfi \sfg^{\mu\nu}\,\partial_\sfj \sfg_{\mu\nu}\biggr) \,.
\end{align}

On the other hand, as we show later, the second term $\cL_{\text{pot}}^{(2)}$ can be written as
\begin{align}
 \cL_{\text{pot}}^{(2)}
 &= -e\frac{1}{2}\,\widehat{\cM}^{MN}\,\partial_N \widehat{\cM}^{KL}\,\partial_L \widehat{\cM}_{MK} +\Delta\cL_{\text{pot}}^{(2)} \,,
\label{eq:def-Delta-Lpot2}
\\
 &= \begin{cases}
 \abs{\sfG}^{\frac{1}{2}}\, 
    \Bigl[ -\frac{1}{2}\,G^{ij} \,\partial_{k}G_{il}\,\partial_{j}G^{kl} 
         + \frac{1}{2(n-2)^2}\,G^{ij}\, \partial_{i}\ln\abs{G} \, \partial_{j}\ln\abs{G} 
\\
 \qquad\quad + \frac{1}{n-2}\,G^{ij}\,G^{kl}\, \partial_{i}\ln\abs{G} \,\partial_{k}G_{jl} \Bigr]
  +\Delta\cL_{\text{pot}}^{(2)} & \text{(M-theory)}
\\
 \abs{\sfG}^{\frac{1}{2}}\, 
    \Bigl[ -\frac{1}{2}\,G^{\sfm\sfn} \,\partial_\sfp G_{\sfm\sfq}\,\partial_\sfn G^{\sfp\sfq} 
         + \frac{1}{2(n-2)^2}\,G^{\sfm\sfn}\, \partial_\sfm\ln\abs{G} \, \partial_\sfn\ln\abs{G} 
\\
 \qquad\quad + \frac{1}{n-2}\,G^{\sfm\sfn}\,G^{\sfp\sfq}\, \partial_\sfm\ln\abs{G} \,\partial_\sfp G_{\sfn\sfq} 
 + \frac{1}{4}\,G^{\sfm\sfn}\,\partial_\sfm m_{\alpha\beta}\,\partial_\sfn m^{\alpha\beta} \Bigr]
  +\Delta\cL_{\text{pot}}^{(2)} & \text{(type IIB)}
\end{cases} \,,
\end{align}
where $\Delta\cL_{\text{pot}}^{(2)}$ does not include derivatives of metric. 

We thus obtain the potential as 
\begin{align}
 \cL_{\text{pot}} 
 &= \begin{cases}
 \abs{\sfG}^{\frac{1}{2}}\, \Bigl[ R(G) +\frac{1}{4}\,G^{ij}\,\partial_i \sfg^{\mu\nu}\,\partial_j \sfg_{\mu\nu}
       + \frac{1}{4}\,G^{ij}\, \partial_i \ln \abs{\sfg}\, \partial_j \ln\abs{\sfg}
\\
 \qquad\ -\frac{1}{2\alpha_n}\,G^{ij}\,\widehat{M}^{MN}\,\widehat{M}_{PQ}\,\omega_{i M}{}^P\,\omega_{j N}{}^Q\Bigr] 
 +\Delta\cL_{\text{pot}}^{(2)} & \text{(M-theory)}
\\
 \abs{\sfG}^{\frac{1}{2}}\, \Bigl[ R(G) +\frac{1}{4}\,G^{\sfm\sfn}\,\partial_\sfm \sfg^{\mu\nu}\,\partial_\sfn \sfg_{\mu\nu}
       + \frac{1}{4}\,G^{\sfm\sfn}\, \partial_\sfm \ln \abs{\sfg}\, \partial_\sfn \ln\abs{\sfg}
\\
 \qquad\ + G^{\sfm\sfn}\,\bigl(\frac{1}{4}\,\partial_\sfm m_{\alpha\beta}\,\partial_\sfn m^{\alpha\beta}
 -\frac{1}{2\alpha_n}\,\widehat{M}^{MN}\,\widehat{M}_{PQ}\,\omega_{\sfm M}{}^P\,\omega_{\sfn N}{}^Q\bigr)\Bigr] 
 +\Delta\cL_{\text{pot}}^{(2)} & \text{(type IIB)}
\end{cases} \,,
\end{align}
where we used the formula
\begin{align}
 R(G)&= 
 \frac{1}{4}\,G^{\sfi \sfj}\,\partial_{\sfi}G_{\sfk \sfl}\,\partial_{\sfj}G^{\sfk \sfl}
 -\frac{1}{2}\,G^{\sfi \sfj}\,\partial_{\sfk}G_{\sfi \sfl} \partial_{\sfj}G^{\sfk \sfl} 
 +\frac{1}{4}\, G^{\sfi \sfj}\, \partial_{\sfi}\ln\abs{G} \, \partial_{\sfj}\ln\abs{G} 
\nn\\
 &\quad -\frac{1}{2}\,G^{\sfi \sfj}\,G^{\sfk \sfl}\, \partial_{\sfi}\ln\abs{G} \,\partial_{\sfk}G_{\sfj \sfl} 
 -\frac{1}{\abs{\sfG}^{\frac{1}{2}}}\,\partial_\sfi\bigl[\abs{\sfG}^{\frac{1}{2}}\,G^{\sfi \sfj}\,G^{\sfk \sfl}\,\bigl(\partial_\sfj G_{\sfk \sfl}-\partial_\sfk G_{\sfl \sfj}\bigr)\bigr]
\nn\\
&= 
 \frac{1}{4}\,G^{\sfi \sfj}\,\partial_{\sfi}G_{\sfk \sfl}\,\partial_{\sfj}G^{\sfk \sfl}
 -\frac{1}{2}\,G^{\sfi \sfj}\,\partial_{\sfk}G_{\sfi \sfl}\,\partial_{\sfj}G^{\sfk \sfl} 
 +\frac{1}{4}\, G^{\sfi \sfj}\, \partial_{\sfi}\ln\abs{G} \, \partial_{\sfj}\ln\abs{G} 
\nn\\
 &\quad 
 -\frac{1}{2}\,G^{\sfi \sfj}\,G^{\sfk \sfl}\, \partial_{\sfi}\ln\abs{G} \,\partial_{\sfk}G_{\sfj \sfl} 
 + \frac{1}{2}\,G^{\sfi \sfj}\, \partial_\sfi\ln\abs{\sfg}\, \partial_\sfj\ln\abs{G} 
 + \frac{1}{2}\,\partial_\sfi\ln \abs{\sfg}\, \partial_\sfj G^{\sfi \sfj} 
\nn\\
 &\quad -\frac{1}{\abs{\sfG}^{\frac{1}{2}}}\,\partial_\sfi\bigl[\abs{\sfG}^{\frac{1}{2}}\,G^{\sfi \sfj}\,G^{\sfk \sfl}\,\bigl(\partial_\sfj G_{\sfk \sfl}-\partial_\sfk G_{\sfl \sfj}\bigr)\bigr] \,,
\end{align}
and dropped the boundary term.

\paragraph{\underline{Calculation of $\cL_{\text{pot}}^{(2)}$}\\}

Here, we show equation \eqref{eq:def-Delta-Lpot2} and determine the explicit form of $\Delta\cL_{\text{pot}}^{(2)}$. 

First, let us calculate $\cL_{\text{pot}}^{(2)}$ in the case of the conventional parameterization. 
In this case, noticing $\vV_M{}^\sfi=\delta_M^\sfi=(\vV^{-1})_M{}^\sfi$ and $(\vV^{\rmT})^\sfi{}_M=\delta^\sfi_M=(\vV^{-\rmT})^\sfi{}_M$, we obtain
\begin{align}
 \cL_{\text{pot}}^{(2)} 
 &= -\frac{e}{2}\,\bigl(\vV^{-\rmT} \widehat{\cM}^{-1} \vV^{-1}\bigr)^{M\sfi}\,\partial_\sfi \bigl(\vV^{-\rmT} \widehat{\cM}^{-1} \vV^{-1}\bigr)^{K\sfj}\,\partial_\sfj \bigl(\vV \widehat{\cM} \vV^{\rmT}\bigr)_{MK}
\nn\\
 &= -\frac{e}{2}\, \widehat{\cM}^{M\sfi}\,\partial_\sfi \widehat{\cM}^{K\sfj}\,\partial_\sfj \widehat{\cM}_{MK} 
    +\frac{e}{2}\,\widehat{\cM}^{\sfi\sfl}\,\widehat{\cM}^{\sfj\sfk}\,\widehat{\cM}_{PQ}\, \omega_{\sfi\sfj}{}^P\,\omega_{\sfk\sfl}{}^Q\,.
\end{align}
Thus, comparing this with \eqref{eq:def-Delta-Lpot2}, we obtain
\begin{align}
 \Delta\cL_{\text{pot}}^{(2)} = \frac{e}{2}\,\widehat{\cM}^{\sfi\sfl}\,\widehat{\cM}^{\sfj\sfk}\,\widehat{\cM}_{PQ}\, \omega_{\sfi\sfj}{}^P\,\omega_{\sfk\sfl}{}^Q
 = \frac{\abs{\sfG}^{\frac{1}{2}}}{2}\,G^{\sfi\sfl}\,G^{\sfj\sfk}\,\widehat{M}_{PQ}\, \omega_{\sfi\sfj}{}^P\,\omega_{\sfk\sfl}{}^Q\,.
\end{align}

We next calculate $\cL_{\text{pot}}^{(2)}$ in the non-geometric parameterization.
In this case, we use the simplifying assumption \cite{Andriot:2011uh} that requires any derivatives contracted with the dual potentials vanishes (e.g.~$\beta^{mn}\,\partial_m =0$). 
In our notation, it can be expressed as
\begin{align}
 \bigl(\cdots \mathbb{V}_{M}{}^\sfi\bigr) \,\partial_\sfi = \bigl(\cdots \delta_{M}{}^\sfi\bigr) \,\partial_\sfi \,,\quad 
 \partial_\sfi\bigl(\cdots \mathbb{V}_{M}{}^\sfi\bigr) =\partial_\sfi\bigl(\cdots \delta_{M}{}^\sfi\bigr) \quad
 \bigl(\mathbb{V}=\vV\text{ or }\vV^{-1}\bigr)\,,
\end{align}
where the ellipsis represent arbitrary tensors or derivatives.
Using the simplifying assumption, we obtain
\begin{align}
 \cL_{\text{pot}}^{(2)}
 &= -\frac{e}{2}\,\cM^{M\sfi}\,\partial_\sfi \cM^{K\sfj}\,\partial_\sfj \cM_{MK} 
\nn\\
 &= -\frac{e}{2}\,\bigl(\vV^{-\rmT} \widehat{\cM}^{-1} \vV^{-1}\bigr)^{M\sfi}\,\partial_\sfi \bigl(\vV^{-\rmT} \widehat{\cM}^{-1} \vV^{-1}\bigr)^{K\sfl}\,\partial_\sfl \cM_{MK}
\nn\\
 &= -\frac{e}{2}\, (\vV^{-\rmT})^{M}{}_\sfk\,\widehat{\cM}^{\sfk \sfi}\,\partial_\sfi \bigl[(\vV^{-\rmT})^K{}_\sfl\,\widehat{\cM}^{\sfl \sfj}\bigr]\,\partial_\sfj \cM_{MK}
  = -\frac{e}{2}\, \widehat{\cM}^{\sfk\sfi}\,\partial_\sfi \widehat{\cM}^{\sfl\sfj} \,\partial_\sfj \widehat{\cM}_{\sfk \sfl} 
\nn\\
 &=-\frac{e}{2}\,\widehat{\cM}^{MN}\,\partial_N \widehat{\cM}^{KL}\,\partial_L \widehat{\cM}_{MK} \,.
\label{eq:L-pot-2-non-geom}
\end{align}
where, in the third equality, we used the simplifying assumption and $\widehat{\cM}^{P\sfi}=\delta^P_\sfj\,\widehat{\cM}^{\sfj \sfi}$, and in the fourth equality, we used $(\vV^{-\rmT})^{M}{}_\sfj=\delta^M_\sfj$ and $\cM_{\sfk\sfl}=\widehat{\cM}_{\sfk\sfl}$ which are generally satisfied in the non-geometric parameterization. 
Comparing \eqref{eq:L-pot-2-non-geom} with \eqref{eq:def-Delta-Lpot2}, we obtain $\Delta\cL_{\text{pot}}^{(2)} =0$ in the non-geometric parameterization. 

\paragraph{\underline{Summary of the potential $\cL_{\text{pot}}$}\\}

To summarize, we obtained
\begin{align}
 \cL_{\text{pot}} 
 &= \begin{cases}
 \abs{\sfG}^{\frac{1}{2}}\, \Bigl[ R(G) +\frac{1}{4}\,G^{ij}\,\partial_i \sfg^{\mu\nu}\,\partial_j \sfg_{\mu\nu}
       + \frac{1}{4}\,G^{ij}\, \partial_i \ln \abs{\sfg}\, \partial_j \ln \abs{\sfg}\Bigr] 
 +\cL_{\text{pot}}^{(\text{mat})} & \text{(M-theory)}
\\
 \abs{\sfG}^{\frac{1}{2}}\, \Bigl[ R(G) +\frac{1}{4}\,G^{\sfm\sfn}\,\partial_\sfm \sfg^{\mu\nu}\,\partial_\sfn \sfg_{\mu\nu}
       + \frac{1}{4}\,G^{\sfm\sfn}\, \partial_\sfm \ln \abs{\sfg}\, \partial_\sfn \ln \abs{\sfg}
\\
\qquad\quad +\frac{1}{4}\,G^{\sfm\sfn}\,\partial_\sfm m_{\alpha\beta}\,\partial_\sfn m^{\alpha\beta} \Bigr] 
 +\cL_{\text{pot}}^{(\text{mat})} & \text{(type IIB)}
\end{cases}
\,. 
\end{align}
Here, $\cL_{\text{pot}}^{(\text{mat})}$ is given as follows:
\begin{align}
 &\underline{\bullet\ \text{geometric parameterization:}}
\nn\\
 &\cL_{\text{pot}}^{(\text{mat})}
  = -\frac{\abs{\sfG}^{\frac{1}{2}}}{2}\,
  \Bigl(\frac{1}{\alpha_n}\,G^{\sfi\sfj}\,\widehat{M}^{MN} \,\omega_{\sfi M}{}^P\,\omega_{\sfj N}{}^Q
       - G^{\sfi\sfl}\,G^{\sfj\sfk}\,\omega_{\sfi\sfj}{}^P\,\omega_{\sfk\sfl}{}^Q\Bigr)\,\widehat{M}_{PQ} \,,
\\
 &\underline{\bullet\ \text{non-geometric parameterization:}}
\nn\\
 &\cL_{\text{pot}}^{(\text{mat})}
  = -\frac{\abs{\widetilde{\sfG}}^{\frac{1}{2}}}{2\alpha_n}\,\widetilde{\sfG}^{\sfi\sfj}\,\widehat{M}^{MN}\,\widehat{M}_{PQ} \,\omega_{\sfi M}{}^P\,\omega_{\sfj N}{}^Q \,.
\end{align}
More explicit form of $\cL_{\text{pot}}^{(\text{mat})}$ in each case is given as follows:
\begin{align}
 &\underline{\text{\textbf{M-theory section:}}}
\nn\\
 &\underline{\bullet\ \text{$\SL(5)$, $\SO(5,5)$, $E_6$ (geometric):}}
\nn\\
 &\cL_{\text{pot}}^{(\text{mat})}
 = -\frac{\abs{\sfG}^{\frac{1}{2}}}{2\cdot 4!}\, G^{i_1\cdots i_4,\,j_1\cdots j_4}\,F_{i_1\cdots i_4}\,F_{j_1\cdots j_4}\,,
\\
 &\underline{\bullet\ \text{$E_7$ (geometric):}}
\nn\\
 &\cL_{\text{pot}}^{(\text{mat})}
 = - \abs{\sfG}^{\frac{1}{2}} \,\Bigl(\frac{1}{2\cdot 4!}\,G^{i_1\cdots i_4,\,j_1\cdots j_4}\, F_{i_1\cdots i_4}\,F_{j_1\cdots j_4}
    +\frac{1}{2\cdot 7!}\, G^{i_1\cdots i_7,\,j_1\cdots j_7}\,F_{i_1\cdots i_7}\,F_{j_1\cdots j_7}\Bigr)\,, 
\\
 &\underline{\bullet\ \text{$\SL(5)$, $\SO(5,5)$ (non-geometric):}}
\nn\\
 &\cL_{\text{pot}}^{(\text{mat})} 
 = -\frac{\abs{\widetilde{\sfG}}^{\frac{1}{2}}}{2\cdot 3!}\, \widetilde{G}^{ij}\,\widetilde{G}_{i_1i_2i_3,\,j_1j_2j_3}\,S_i{}^{i_1i_2i_3}\, S_j{}^{j_1j_2j_3}\,,
\\
 &\underline{\bullet\ \text{$E_6$, $E_7$ (non-geometric):}}
\nn\\
 &\cL_{\text{pot}}^{(\text{mat})} 
 = - \abs{\widetilde{\sfG}}^{\frac{1}{2}}\, \widetilde{G}^{ij}\,
 \Bigl(\frac{\widetilde{G}_{i_1i_2i_3,\,j_1j_2j_3}}{2\cdot 3!}\,S_i{}^{i_1i_2i_3}\,S_j{}^{j_1j_2j_3}
 + \frac{\widetilde{G}_{i_1\cdots i_6,\,j_1\cdots j_6}}{2\cdot 6!}\,S_i{}^{i_1\cdots i_6}\,S_j{}^{j_1\cdots j_6} \Bigr)\,,
\\
 &\underline{\text{\textbf{type IIB section:}}}
\nn\\
 &\underline{\bullet\ \text{$\SL(5)$, $\SO(5,5)$ (geometric):}}
\nn\\
 &\cL_{\text{pot}}^{(\text{mat})}
  = -\frac{\abs{\sfG}^{\frac{1}{2}}}{2\cdot 3!}\, m_{\alpha\beta}\,G^{\sfm_1\sfm_2\sfm_3,\,\sfn_1\sfn_2\sfn_3}\, H^\alpha_{\sfm_1\sfm_2\sfm_3}\,H^{\beta}_{\sfn_1\sfn_2\sfn_3} \,,
\\
 &\underline{\bullet\ \text{$E_{6(6)}$, $E_{7(7)}$ (geometric):}}
\nn\\
 &\cL_{\text{pot}}^{(\text{mat})}
 = -\frac{\abs{\sfG}^{\frac{1}{2}}}{2}\,\Bigl(\frac{m_{\alpha\beta}\,G^{\sfm_1\sfm_2\sfm_3,\,\sfn_1\sfn_2\sfn_3}}{3!}\, H^\alpha_{\sfm_1\sfm_2\sfm_3}\,H^{\beta}_{\sfn_1\sfn_2\sfn_3}
    +\frac{G^{\sfm_1\cdots \sfm_5,\,\sfn_1\cdots \sfn_5}}{5!} \, G_{\sfm_1\cdots \sfm_5}\,G_{\sfn_1\cdots \sfn_5}\Bigr) \,,
\\
 &\underline{\bullet\ \text{$\SL(5)$ (non-geometric):}}
\nn\\
 &\cL_{\text{pot}}^{(\text{mat})} 
 = -\frac{\abs{\widetilde{\sfG}}^{\frac{1}{2}}}{2\cdot 2!}\, \widetilde{m}^{\alpha\beta}\,\widetilde{G}^{\sfm\sfn}\,\widetilde{G}_{\sfm_1\sfm_2,\,\sfn_1\sfn_2}\, Q_{\alpha,\,\sfm}{}^{\sfm_1\sfm_2}\, Q_{\beta,\,\sfn}{}^{\sfn_1\sfn_2} \,,
\\
 &\underline{\bullet\ \text{$\SO(5,5)$, $E_{6(6)}$ (non-geometric):}}
\nn\\
 &\cL_{\text{pot}}^{(\text{mat})} 
  = -\frac{\abs{\widetilde{\sfG}}^{\frac{1}{2}}}{2}\, \widetilde{G}^{\sfm\sfn}\,
 \Bigl(\frac{\widetilde{m}^{\alpha\beta}\,\widetilde{G}_{\sfm_1\sfm_2,\,\sfn_1\sfn_2}}{2!}\,Q_{\alpha,\,\sfm}{}^{\sfm_1\sfm_2}\, Q_{\beta,\,\sfn}{}^{\sfn_1\sfn_2} 
 -\frac{\widetilde{G}_{\sfm_1\cdots \sfm_4,\,\sfn_1\cdots \sfn_4}}{4!}\, P_\sfm{}^{\sfm_1\cdots \sfm_4}\,P_\sfn{}^{\sfn_1\cdots \sfn_4}\Bigr) \,,
\\
 &\underline{\bullet\ \text{$E_{7(7)}$ (non-geometric):}}
\nn\\
 &\cL_{\text{pot}}^{(\text{mat})} 
  = -\frac{\abs{\widetilde{\sfG}}^{\frac{1}{2}}}{2}\, \widetilde{G}^{\sfm\sfn}\,
 \Bigl(\frac{\widetilde{m}^{\alpha\beta}\,\widetilde{G}_{\sfm_1\sfm_2,\,\sfn_1\sfn_2}}{2!}\,Q_{\alpha,\,\sfm}{}^{\sfm_1\sfm_2}\, Q_{\beta,\,\sfn}{}^{\sfn_1\sfn_2} 
 -\frac{\widetilde{G}_{\sfm_1\cdots \sfm_4,\,\sfn_1\cdots \sfn_4}}{4!}\, P_\sfm{}^{\sfm_1\cdots \sfm_4}\,P_\sfn{}^{\sfn_1\cdots \sfn_4}
\nn\\
 &\qquad\qquad\qquad\qquad\qquad +\frac{\widetilde{m}^{\alpha\beta}\,\widetilde{G}_{\sfm_1\cdots \sfm_6,\,\sfn_1\cdots \sfn_6}}{6!} \, Q_{\alpha,\,\sfm}{}^{\sfm_1\cdots \sfm_6}\,Q_{\beta,\,\sfn}{}^{\sfn_1\cdots \sfn_6}\Bigr) \,.
\end{align}
where we defined
\begin{align}
 F_{i_1\cdots i_4}&\equiv 4\,\partial_{[i_1} A_{i_2i_3i_4]}\,,\qquad \qquad 
 F_{i_1\cdots i_7}\equiv 7\,\partial_{[i_1} A_{i_2\cdots i_7]} + \tfrac{35}{2} \, A_{[i_1i_2i_3}\,F_{i_4i_5i_6i_7]} \,,
\\
 H^\alpha_{\sfm_1\sfm_2\sfm_3}&\equiv 3\,\partial_{[\sfm_1} B^\alpha_{\sfm_2\sfm_3]}\,,
\\
 G_{\sfm_1\cdots \sfm_5}&\equiv 5\, \partial_{[\sfm_1} D_{\sfm_2\cdots \sfm_5]}-15\,\epsilon_{\gamma\delta}\,B^\gamma_{[\sfm_1\sfm_2}\,\partial_{\sfm_3} B^\delta_{\sfm_4\sfm_5]} 
\nn\\
 &= 5\, \partial_{[\sfm_1} C_{\sfm_2\cdots \sfm_5]}+30\,H^1_{[\sfm_1\sfm_2\sfm_3}\,C_{\sfm_4\sfm_5]} \,.
\end{align}

\subsection{Summary}

In this appendix section, we evaluated several external terms in the EFT action,
\begin{align}
 \cL_{\text{EH}}+\cL_{\text{scalar}} 
 = e\,R(g) + \frac{e}{4\alpha_n}\,g^{\mu\nu}\,\partial_\mu \cM_{MN}\,\partial_\nu \cM^{MN} \,,
\end{align}
and the potential part, $\cL_{\text{pot}}$. 
Combining these, we obtain
\begin{align}
 \cL=&\abs{\sfG}^{\frac{1}{2}}\,
 \Bigl[R(\sfg) 
       +\frac{1}{4}\,\sfg^{\mu\nu}\,\partial_\mu G^{\sfi\sfj}\, \partial_\nu G_{\sfi\sfj}
       +\frac{1}{4}\,\sfg^{\mu\nu}\, \partial_\mu\ln\abs{G}\,\partial_\nu\ln\abs{G} 
\nn\\
 &\qquad\ + R(G) +\frac{1}{4}\,G^{\sfi\sfj}\,\partial_\sfi \sfg^{\mu\nu}\,\partial_\sfj \sfg_{\mu\nu} + \frac{1}{4}\,G^{\sfi\sfj}\, \partial_\sfi \ln \abs{\sfg}\, \partial_\sfj \ln \abs{\sfg}\Bigr] 
 + \cL_{\text{scalar}}^{(\text{mat})} +\cL_{\text{pot}}^{(\text{mat})}
\nn\\
 &\equiv \abs{\sfG}^{\frac{1}{2}}\, R(\sfG) + \cL_{\text{scalar}}^{(\text{mat})} +\cL_{\text{pot}}^{(\text{mat})} \,. 
\end{align}
For example, for the $E_{7(7)}$ EFT in the geometric parameterization, this becomes
\begin{align}
 \cL&= \abs{\sfG}^{\frac{1}{2}} \,\biggl[ R(\sfG) 
 - \frac{G^{i_1i_2i_3,\,j_1j_2j_3}}{2\cdot 3!}\,\sfg^{\mu\nu}\,\partial_\mu A_{i_1i_2i_3}\, \partial_\nu A_{j_1j_2j_3} 
  -\frac{G^{i_1\cdots i_6,\,j_1\cdots j_6}}{2\cdot 6!}\,\sfg^{\mu\nu}\,\cF_{\mu, i_1\cdots i_6}\,\cF_{\nu, j_1\cdots j_6}\biggr]
\nn\\
 &\quad - \abs{\sfG}^{\frac{1}{2}} \,\Bigl(\frac{1}{2\cdot 4!}\,G^{i_1\cdots i_4,\,j_1\cdots j_4}\, F_{i_1\cdots i_4}\,F_{j_1\cdots j_4}
    +\frac{1}{2\cdot 7!}\, G^{i_1\cdots i_7,\,j_1\cdots j_7}\,F_{i_1\cdots i_7}\,F_{j_1\cdots j_7}\Bigr)\,.
\end{align}

\section{Double-vielbein formalism for gauged DFT} 
\label{app:gauged-DFT}

\subsection{Parameterization from defining properties of double-vielbein}
The previous result from the Iwasawa decomposition provides the upper or lower triangular parameterization of the generalized vielbein. However, the triangulation breaks the full local structure group into the diagonal subgroup. If we decompose $\ODG$ as $\OO(D-1,1) \times \OG$, then we choose the diagonal gauge-fixing by identifying the two local Lorentz groups,
\begin{equation}
  \OO(D-1,1) \times \OO(1,D-1) \qquad \rightarrow \qquad \OO(D-1,1)_{D}\,.
\end{equation}

Here, we shall construct the geometric parameterization and the non-geometric parameterization directly from the defining conditions of double-vielbein. This approach does not require any gauge-fixing condition and ensures manifest $\OO(1,D-1)\times\ODG$ covariance.
Analogous to the ordinary $\ODD$ case, double-vielbein for $\ODDG$ gauged DFT satisfies the following defining properties \cite{Jeon:2011cn},
\begin{equation}
\begin{aligned}
  &V_{\hM p} V^{\hM}{}_{q}=\eta_{pq}\,,\qquad \qquad 
\brV_{\hM \hbrp} \brV^{\hM }{}_{\brq}=\hat\breta_{\hbrp\hat\brq}\,,
\\
&V_{\hM p} \brV^{\hM}{}_{\hat\brq}=0\,,\qquad V_{\hM p}V_{\hN }{}^{p}+\brV_{\hM \hbrp} \brV_{\hN}{}^{\hbrp} = \hat\cJ_{\hM \hN }\,, 
\end{aligned}\label{defV}
\end{equation}
where $\eta_{mn}$ and $\hat\breta_{\hbrp\hbrq}$ are $\OO(1,D-1)$ and $\OO(D-1,1+ \dim G)$ metric, respectively. The double-vielbein is then decomposed as
\begin{equation}
  V_{\hM}{}^{m} = \bpm V_{M}{}^{m} \\ V_{\alpha}{}^{m} \epm\, \qquad \mbox{and} \qquad 
\brV_{\hM}{}^{\hat{\brm}} = \bpm \brV_{M}{}^{\brm} & \brV_{M}{}^{\bra} \\ \brV_{\alpha}{}^{\brm} & \brV_{\alpha}{}^{\bra} \epm\,.
\label{vielbeindecomp}\end{equation}
Note that the usual geometric parameterization is obtained by assuming that the upper-half blocks of $V_M{}^{m}$ and $\brV_{M}{}^{\brm}$ are non-degenerate and by identifying them as a pair of conventional vielbeins \cite{Jeon:2011cn}.  However, the non-degeneracy assumption can be relaxed in a consistent manner. 

Suppose that the upper-half blocks of $V^{m}$ and $\brV^{\brm}$ are given by
\begin{equation}
  V^{\mu m} = (e^{-1})^{\mu m} + \beta'^{\mu\nu} e_{\nu}{}^{m} \, \qquad \mbox{and}
  \qquad 
  \brV^{\mu \brm} = (\bre^{-1})^{\mu m} + \beta'^{\mu\nu} \bre_{\nu}{}^{m}\,,
\label{upperblocks}\end{equation}
where $e_{\mu}{}^{m}$ and $\bre_{\mu}{}^{\brm}$ are two copies of the $D$-dimensional vielbein corresponding to the same metric $g_{\mu\nu}$
\begin{equation}
  e_{\mu}^{m} e_{\nu}{}^{n} \eta_{mn} = - \bre_{\mu}{}^{\brm} \bre_{\nu}{}^{\brn} \breta_{\brm\brn} = g_{\mu\nu}\,,
\end{equation}
and $\beta'$ is an arbitrary tensor. Then, $V^{\mu m}$ and $\brV^{\mu \brm}$ are not guaranteed to be non-degenerate. Substituting the previous decomposition ansatz (\ref{vielbeindecomp}) and (\ref{upperblocks}) into the defining properties (\ref{defV}), we find the most general parameterization that satisfy all the algebraic constraints (\ref{defV}) for $V_{\hat{M}}{}^{m}$
\begin{equation}
\begin{aligned}
V_{M}{}^m &= \tfrac{1}{\sqrt{2}} \begin{pmatrix} e_\mu{}^{m} + B'_{\mu\nu} \big((e^{-1})^{\nu m} - \beta'^{\nu\rho}\, e_{\rho}{}^{m}\big)\\ (e^{-1})^{\mu m} - \beta'^{\mu\nu} e_{\nu}{}^{m} \end{pmatrix}\,, 
\\
V_{\alpha}{}^{m} &= \tfrac{1}{\sqrt{2}} \begin{pmatrix}
	\kappa_{\alpha\beta}(A^{\rmT})^{\beta}{}_{\mu}  \big((e^{-1})^{\mu m} - \beta'^{\mu\nu} e_{\nu}{}^{m}\big) - \kappa_{\alpha\beta} (\tilde{A}^{\rmT})^{\beta\mu} e_{\mu}{}^{m}
\end{pmatrix}\,, 
\end{aligned}\label{parameterizationV}
\end{equation}
and for $\brV_{\hM}{}^{\hbrm}$ 
\begin{equation}
\begin{aligned}
\brV_{M}{}^{\brm} &= \tfrac{1}{\sqrt{2}} \begin{pmatrix} \bre_\mu{}^{\brm} + B'_{\mu\nu} \big((\bre^{-1})^{\nu \brm} - \beta'^{\nu\rho}\, \bre_{\rho}{}^{\brm}\big)\\ (\bar{e}^{-1})^{\mu \brm} - \beta'^{\mu\nu} \bre_{\nu}{}^{\brm}   \end{pmatrix}\,, 
\\
\brV_{\alpha}{}^{\brm} &= \tfrac{1}{\sqrt{2}} \begin{pmatrix}
    (A^{\rmT})_{\alpha\mu} \big((\bre^{-1})^{\mu \brm} - \beta'^{\mu\nu} \bre_{\nu}{}^{\brm}\big) - (\tilde{A}^{\rmT})_{\alpha}{}^{\mu}  \bre_{\mu}{}^{\brm}
\end{pmatrix}\,, 
\\
\brV_{M}{}^{\bra} &= \begin{pmatrix}
	-A_{\mu}{}^{\bra}+ B'_{\mu\nu} \tilde{A}^{\nu \bra}\\ \tilde{A}^{\mu\alpha} (\phi^{\rmT})_{\alpha}{}^{\bar{a}}
\end{pmatrix}\,, \qquad \brV_{\alpha}{}^{\bra} = \phi^{\bra}{}_{\alpha} + \phi^{\bar{a}}{}_{\beta} (\tilde{A}^{\rmT})^{\beta\mu} A_{\mu \alpha} \, . 
\end{aligned}\label{parameterizationbrV}
\end{equation}
Here, $B'_{\mu\nu}$ and $\beta'^{\mu\nu}$ are defined as
\begin{equation}
\begin{aligned}
  	B'_{\mu\nu} = B_{\mu\nu} + \tfrac{1}{2} \alpha' A_{\mu}{}^{\alpha} (A^{\rm T})_{\alpha\nu}\,,
	\\
	\beta'^{\mu\nu} = \beta^{\mu\nu} - \tfrac{1}{2} \alpha' \tilde{A}^{\mu}{}^{\alpha} 
	(\tilde{A}^{\rmT})_{\alpha}{}^{\nu}\,,	
\end{aligned}
\end{equation}
in which $B_{\mu\nu}$ and $\beta^{\mu\nu}$ are antisymmetric tensors. 

However, if we assume that each blocks of $V_{M}{}^{m}$ and $\bar{V}_{M}{}^{\bar{m}}$ are non-degenerate, this solution is over-parameterized. The physical degrees of freedom are determined by the coset
\begin{equation}
  \frac{\ODDG}{\OoD\times\ODG} \,,
\end{equation}
and the associated number of degrees of freedom is given by
\begin{equation}
  \half(2 D + G) (2 D + G - 1) - \half D (D - 1) - \half (D + G) (D + G - 1) = D^2+ DG \, , 
\end{equation}
where $G$ denotes $\dim G$. The $D^2$ components arise from the $\big\{g_{\mu\nu}, B_{\mu\nu}\big\}$ or $\big\{\tilde g_{\mu\nu}, \beta^{\mu\nu}\big\}$, and $DG$ components arise from the $A_{\mu}{}^{\bra}$ or $\tilde{A}^{\mu\bar{a}}$. Thus, only $\{g, B, A\}$ or $\{ \tilde g, \beta, \tilde{A}\}$ are sufficient to make up the parameterization.

The geometric parameterization, which is for the conventional heterotic supergravity \cite{Lee:2015kba}, is obtained by turning off $\beta^{\mu\nu}$ and $\tilde{A}^{\mu \bra}$,
\begin{equation}
	V_{M}{}^m = \tfrac{1}{\sqrt{2}} \begin{pmatrix} e_\mu{}^{m} + B'_{\mu\nu} (e^{-1})^{\nu m} \\ (e^{-1})^{\mu m}  \end{pmatrix}\,, \qquad
	V_{\alpha}{}^{m} = \tfrac{1}{\sqrt{2}}(A^{\rmT})_{\alpha\mu} (e^{-1})^{\mu m} \,, 
\label{geomparameterizationV}
\end{equation}
and 
\begin{equation}
\begin{aligned}
	\brV_{M}{}^{\brm} = \tfrac{1}{\sqrt{2}} \begin{pmatrix} \bre_\mu{}^{\brm} + B'_{\mu\nu} (\bre^{-1})^{\nu \brm} \\ (\bre^{-1})^{\mu \brm} \end{pmatrix}\,, \qquad
	&\brV_{\alpha}{}^{\brm} = \tfrac{1}{\sqrt{2}}(A^{\rmT})_{\alpha\mu} (\bre^{-1})^{\mu \brm} \,, 
	\\
	\brV_{M}{}^{\bra} =-\sqrt{\alpha'} \bpm A_{\mu}{}^{\alpha} (\phi^{\rmT})_{\alpha}{}  ^{\bar{a}} \\ 0\epm\,,  \qquad &\brV_{\alpha}{}^{\bra} = \tfrac{1}{\sqrt{\alpha'}} (\phi^{\bra})_{\alpha} \,.
\end{aligned}\label{geomparameterizationbrV}
\end{equation}
Under the non-degeneracy assumption,  one can show through a field redefinition that the geometric parameterization is essentially the same as the most general solution (\ref{parameterizationV}) and (\ref{parameterizationbrV}). On the other hand, if we assume that some of components of $V^{\mu m}$ or $\bar{V}^{\mu \bar{m}}$ are vanishing, we can define an another class of non-geometric background, which cannot be related by field redefinition from geometric parameterization \cite{Lee:2013hma}. 

Using the relation the projection operators and double-vielbein:
\begin{equation}
	P_{\hM\hN} =  V_{\hM}{}^{m} \eta_{mn} (V^{\rm T})^{n}{}_{\hat{N}}\, \qquad \mbox{and} \qquad 
	\brP_{\hM\hN} = \brV_{\hM}{}^{\brm} \bar{\eta}_{\bar{m}\bar{n}} (\brV^{\rm T})^{\bar{n}}{}_{\hN} + \brV_{\hM}{}^{\bra} \kappa_{\bar{a}\bar{b}}(\brV^{\rm T})^{\brb}{}_{\hN}\,,
\end{equation}
we construct a geometric parameterization for the projection operators as
\begin{equation}
P = 
	\half \bpm
		g + \alpha' A\, \kappa \,A^t + B' g^{-1} (B')^t & A\kappa + B' g^{-1} A\kappa & \mathbf{1} + B'g^{-1}
		\\
		\kappa A^t + \kappa\, A^t g^{-1} (B')^t & \kappa \, A^t g^{-1} A\,\kappa & \kappa \,A^t g^{-1}
		\\
		\mathbf{1} + g^{-1} (B')^{t} & g^{-1} A \,\kappa & g^{-1}
	\epm\,,
\end{equation}
and
\begin{equation}
\brP = 
	\half \bpm
		- g - \alpha' A\, \kappa\, A^t - B' g^{-1} (B')^t & - A\kappa - B' g^{-1} A\kappa & \mathbf{1} - B'g^{-1}
		\\
		-\kappa A^t - \kappa A^t g^{-1} (B')^t &  -\kappa A^t g^{-1} A \kappa- \tfrac{2}{\alpha'} \kappa & -\kappa A^t g^{-1}
		\\
		\mathbf{1} - g^{-1} (B')^{t} & - g^{-1} A \kappa & - g^{-1}
	\epm\,.
\end{equation}
Here, we used $\cK_{\alpha\beta} = - (t^{\bar{a}})^{\rm T}_{\alpha} \kappa_{\bar{a}\bar{b}} t^{\bar{b}}{}_{\beta}$.
In this parameterization, it follows that the projection operators satisfy the complete relation, $\cJ = P + \brP$ and that the generalized metric defined by $\cH = P- \brP$ takes the form:
\begin{equation}
\cH = 
	\bpm
 		g + B' g^{-1} (B')^t + A \kappa A^t & A\kappa + B' g^{-1} A \kappa & B'g^{-1} 
 		\\
 		\kappa A^t + \kappa A^t g^{-1} (B')^t & \kappa A^t g^{-1} A \kappa	+\tfrac{1}{\alpha'} \kappa & \kappa A^t g^{-1} 
 		\\
		g^{-1} (B')^{t} & g^{-1} A \kappa & g^{-1} 
	\epm\,. 
\label{geomGenMet1}\end{equation}

Consider next the non-geometric parameterization. As for the geometric parameterization, it is simply given by turning off $B_{\mu\nu}$ and $A_{\mu}{}^{\bra}$ while keeping $\beta$ and $\tilde{A}$ in (\ref{parameterizationV}) and (\ref{parameterizationbrV}):
\begin{equation}
\begin{aligned}
	V_{M}{}^m &= \tfrac{1}{\sqrt{2}} \begin{pmatrix} e_\mu{}^{m}\\ (e^{-1})^{\mu m} - \beta'^{\mu\nu} e_{\nu}{}^{m}  \end{pmatrix}\,, 
	\\
	V_{\alpha}{}^{m} &= - \tfrac{1}{\sqrt{2}} \kappa_{\alpha\beta} (\tilde{A}^{\rmT})^{\beta\mu} e_{\mu}{}^{m} \,, 
\end{aligned}\label{NongeomParametrizationV}
\end{equation}
and 
\begin{equation}
\begin{aligned}
	\brV_{M}{}^{\brm} &= \tfrac{1}{\sqrt{2}} \begin{pmatrix} \bre_\mu{}^{\brm} \\ (\bre^{-1})^{\mu \brm} - \beta'^{\mu\nu} \bre_{\nu}{}^{\brm}  \end{pmatrix}\,, 
	\qquad 
	&\brV_{\alpha}{}^{\brm} &= - \tfrac{1}{\sqrt{2}}\kappa_{\alpha\beta} (\tilde{A}^{\rmT})^{\beta\mu} \bre_{\mu}{}^{\brm}\,, 
	\\
& \qquad \brV_{M}{}^{\bra} = \begin{pmatrix}
	0 \\ \sqrt{\alpha'} \tilde{A}^{\mu \alpha} (\phi^{\rmT})_{\alpha}{}^{\bar{a}}
	\end{pmatrix}\,, 
	\qquad 
	&\brV_{\alpha}{}^{\bra} &= \tfrac{1}{\sqrt{\alpha'}} \phi^{\bra}{}_{\alpha} \,.
\end{aligned}\label{NongeomParametrizationbrV}
\end{equation} 
The corresponding projection operators are constructed as
\begin{equation}
  P =  \tfrac{1}{2}\begin{pmatrix}
  	\tilde g & - \tilde g \tilde{A} \kappa & \mathbf{1} - \tilde g \beta'^{\rm T} 
	\\
  	-\kappa\tilde{A}^{\rm T} \tilde g & \kappa\tilde{A}^{\rm T} \tilde g \tilde{A}\kappa & - \kappa\tilde{A}^{\rm T} + \kappa\tilde{A}^{\rm T} \tilde g \beta'^{\rm T}
	\\
	\mathbf{1} - \beta' \tilde g & - \tilde{A} \kappa + \beta' \tilde g \tilde{A} \kappa & \tilde g^{-1} + \beta' \tilde g \beta'^{\rm T} + \alpha' \tilde{A} \kappa \tilde{A}^{\rm T}
\end{pmatrix}\,,
\end{equation}
and
\begin{equation}
  \brP =  \tfrac{1}{2}\begin{pmatrix}
  	- \tilde g & \tilde g \tilde{A}\kappa & \mathbf{1} + \tilde g \beta'^{\rm T}
	\\
	\kappa \tilde{A}^{\rm T} \tilde g & - \kappa\tilde{A}^{\rm T} \tilde g \tilde{A}\kappa - \tfrac{2}{\alpha'} \kappa & \kappa\tilde{A}^{\rm T} -\kappa \tilde{A}^{\rm T} \tilde g \beta'^{\rm T}
	\\
	\mathbf{1} + \beta' \tilde g & \tilde{A} \kappa - \beta' \tilde g \tilde{A}\kappa & -\tilde g^{-1} - \beta' \tilde g \beta'^{\rm T} - \alpha' \tilde{A} \kappa \tilde{A}^{\rm T}
\end{pmatrix}\,.
\end{equation}
Once again, in this parameterization, it follows that the complete relation $\cJ = P + \bar{P}$ is satisfied and that the the generalized metric $\cH = P - \bar{P}$ is expressed by 
\begin{equation}
  \cH =  \begin{pmatrix}
  	\tilde g & - \tilde g \tilde{A}\kappa & - \tilde g \beta'^{\rm T}
	\\
	- \kappa\tilde{A}^{\rm T} \tilde g & \kappa\tilde{A}^{\rm T} \tilde g \tilde{A}\kappa +\tfrac{1}{\alpha'} \kappa & - \kappa\tilde{A}^{\rm T} +\kappa\tilde{A}^{\rm T} \tilde g \beta'^{\rm T}
  	\\
	- \beta' \tilde g & -\tilde{A} \kappa + \beta' \tilde g \tilde{A} \kappa & \tilde g^{-1} + \beta' \tilde g \beta'^{\rm T} + \alpha' \tilde{A} \kappa \tilde{A}^{\rm T}
\end{pmatrix}\,.
\label{nongeomparaH}\end{equation}
One notes that this result is consistent with the parameterization in terms of the Iwasawa decomposition given in (\ref{nongeomGenMet}).

We should remark that, ultimately, the double-vielbein formalism is imperative. 
For the bosonic case, the geometric parameterization and the non-geometric parameterization of double-vielbein, (\ref{geomparameterizationV}) and (\ref{NongeomParametrizationbrV}), respectively, are equivalent to the previous result constructed by coset representative, as they should. Even though these two approaches are consistent for the bosonic case, for introducing supersymmetry, the double-vielbein formalism is the most adequate approach \cite{Jeon:2011sq,Jeon:2012hp, Lee:2015kba}.

\subsection{Connection and Curvature}
The gauge symmetry for gauged DFT is given by a twisted generalized Lie derivative which is defined by
\begin{equation}
\begin{aligned}
  (\hcL_{X} V)^{\hM}{}_{\hN} &= (\hcLo_{X} V)^{\hM}{}_{\hN} - f^{\hM}{}_{\hP\hQ}X^{\hP} V^{\hQ}{}_{\hN}  - f_{\hN\hP}{}^{\hQ} X^{\hP}V^{\hM}{}_{\hQ}\,,
\\
\hcL_{X} d &= \hcLo_{X} d \,.
\end{aligned}\label{symm}
\end{equation}
The $\hcLo_{X}$ is the ordinary generalized Lie derivative defined in the un-gauged DFT by
\begin{equation}
\begin{aligned}
  (\hcLo_{X} V )^{\hM}{}_{\hN} &= X^{\hP} \partial_{\hP} V^{\hM}{}_{\hN} + (\partial^{\hM} X_{\hP} - \partial_{\hP} X^{\hM}) V^{\hP}{}_{\hN} + (\partial_{\hN} X^{\hP} - \partial^{\hP} X_{\hN}) V^{\hM}{}_{\hP}\,,
\\
\hcLo_{X} d &= X^{\hM} \partial_{\hM} d - \half \partial_{\hM} X^{\hM}\,,
\end{aligned}\label{}
\end{equation}
where $f_{\hM\hN\hP}$ are the structure constants for Yang-Mills gauge group.  The gauge parameter $X^{\hM}$ consists of ordinary generalized Lie derivative part and a Yang-Mills gauge symmetry part in an $\ODDG$ covariant way.

As for the covariant differential operator of the gauge transformations (\ref{symm}), we present a covariant derivative which can be applied to any arbitrary $\ODDG$, ${\rm \Spin}({D-1},1)$ and ${\rm \Spin}(1,{D-1}+\rm{dim}~G)$ representations as follows 
\begin{equation}
  \hat\cD_{\hM} := \partial_{\hM}  + \Gamma_{\hM} + \Phi_{\hM} + \brPhi_{\hM}\, . 
\label{MasterDerivative}\end{equation}
where $\Phi_{\hM mn}$ and $\brPhi_{\hM\hbrm\hbrn}$ are spin-connections and $\Gamma_{\hM\hN\hP}$ is semi-covariant connection which are constructed in gauged DFT \cite{Berman:2013cli}
\begin{equation}
  \Gamma_{\hP\hM\hN} = 
	\Gamma^{\scriptscriptstyle{0}}{}_{\hP\hM\hN} 
	+ \left(\delta_{P}{}^{\hQ} P_{\hM}{}^{\hR} P_{\hN}{}^{\hS} 
	+ \delta_{\hP}{}^{\hQ} \brP_{\hM}{}^{\hR} \brP_{\hN}{}^{\hS} \right) f_{\hQ\hR\hS} 
	-\tfrac{2}{3} \left(\cP + \bar{\cP}\right)_{\hP\hM\hN}{}^{\hQ\hR\hS} f_{\hQ\hR\hS}\,.
\label{conn}\end{equation}
where $\Gamma^{\scriptscriptstyle{0}}{}_{PMN}$ is the connection for ordinary DFT \cite{Jeon:2011cn},
\begin{equation}
\begin{aligned}
  \Gamma^{\scriptscriptstyle{0}}{}_{\hP\hM\hN}  = & 
	2(P\partial_{\hP} P \brP )_{[\hM\hN]} 
	+ 2 (\brP_{[\hM}{}^{\hQ} \brP_{\hN]}{}^{\hR} 
	- P_{[\hM}{}^{\hQ} P_{\hN]}{}^{\hR} ) \partial_{\hQ} P_{\hR \hP} 
\\& 
	- \tfrac{4}{D-1} \big(\brP_{P[\hM} \brP_{\hN]}{}^{\hQ} + P_{\hP[\hM} P_{\hN]}{}^{\hQ}) 
		\big(\partial_{\hQ}d + (P\partial^{\hR} P \brP\big)_{[\hR\hQ]}\big)\,,
\label{oldconn}\end{aligned}\end{equation}
and $\cP_{\hP\hM\hN}{}^{\hQ\hR\hS}$ and $\bar{\cP}_{\hP\hM\hN}{}^{\hQ\hR\hS}$ are rank-six projection operators 
\begin{equation}
\begin{aligned}
    \cP_{\hP\hM\hN}{}^{\hS\hQ\hR}:=& P_{\hP}{}^{\hS}P_{[\hM}{}^{[\hQ}P_{\hN]}{}^{\hR]}+\tfrac{2}{D-1}P_{\hP[\hM}P_{\hN]}{}^{[\hQ}P^{\hR]\hS}\,,
\\
\bcP_{\hP\hM\hN}{}^{\hS\hQ\hR}:=&\brP_{\hP}{}^{\hS}\brP_{[\hM}{}^{[\hQ}\brP_{\hN]}{}^{\hR]}+\frac{2}{D-1}\brP_{\hP[\hM}\brP_{\hN]}{}^{[\hQ}\brP^{\hR]\hS}\,,	
\end{aligned}\label{}
\end{equation}
which are symmetric and traceless,
\begin{equation}
\begin{aligned}
  {\cP_{\hP\hM\hN\hQ\hR\hS}=\cP_{\hQ\hR\hS\hP\hM\hN}=\cP_{\hP[\hM\hN]\hQ[\hR\hS]}\,,}
~~&~~
	{\bcP_{\hP\hM\hN\hQ\hR\hS}=\bcP_{\hQ\hR\hS\hP\hM\hN}=\bcP_{\hP[\hM\hN]\hQ[\hR\hS]}\,,} 
\\
{\cP^{\hP}{}_{\hP\hM\hQ\hR\hS}=0\,,~~~~\,P^{\hP\hM}\cP_{\hP\hM\hN\hQ\hR\hS}=0\,,}
~~&~~
	{\bcP^{\hP}{}_{\hP\hM\hQ\hR\hS}=0\,,~~~~\,\brP^{\hP\hM}\bcP_{\hP\hM\hN\hQ\hR\hS}=0\,.}
\label{symP6}\end{aligned}\end{equation}
Here the superscript `0' indicates a quantity defined in the un-gauged DFT.

The spin-connections are defined by using the semi-covariant derivative
\begin{equation}
\begin{aligned}
  \Phi_{\hM mn} &= V^{\hN}{}_{m} \partial_{\hM} V_{\hN n} + \Gamma_{\hM \hN\hP} V^{\hN}{}_{m} V^{\hP}{}_{n} \,, 
  \\
  \brPhi_{\hM \hbrm\hbrn} &= \brV^{\hN}{}_{\hbrm} \partial_{\hM} \brV_{\hN \hbrn} + \Gamma_{\hM \hN\hP} \brV^{\hN}{}_{\hbrm} V^{\hP}{}_{\hbrn}\,.
\label{defspinconn}\end{aligned}
\end{equation}
Although these are not gauge covariant, we can project out to the tensor part
\begin{equation}
\begin{aligned}
  &\Phi_{\brp mn}\,,\qquad \Phi_{\bra mn}\,,\qquad \Phi_{[pmn]}\,,\qquad \Phi^{p}{}_{pm}\,,
\\
&\brPhi_{p \brm\brn}\,,\qquad \brPhi_{p \brm\bra}\,,\qquad \brPhi_{p \bra\brb}\,,\qquad \brPhi_{[\brp\brm\brn]}\,,\qquad \brPhi_{[\brp\brm\bra]}\,,\qquad \brPhi_{[\brp\bra\brb]}\,,
\\ 
&\brPhi_{[\bra\brb\brc]}\,,\qquad \brPhi^{\hbrp}{}_{\hbrp \brm}\,,\qquad \brPhi^{\hbrp}{}_{\hbrp \bra}\,.	
\end{aligned}\label{covspinconnections}
\end{equation}
These will be the building block that the formalism uses. Various covariant quantities can be generated by using these spin-connections and their derivatives \cite{Berman:2013cli}.

The heterotic DFT action is given by the generalized curvature tensor from semi-covariant curvature tensor $S_{\hM\hN\hP\hQ}$ as
\begin{equation}
  S_{\hM\hN\hP\hQ} = \half \big(R_{\hM\hN\hP\hQ} + R_{\hP\hQ\hM\hN} - \Gamma^{\hR}{}_{\hM\hN} \Gamma_{\hR\hP\hQ} \big)\,,
\label{curvature}\end{equation}
where $R_{\hM\hN\hP\hQ}$ is defined from the standard commutator of the covariant derivatives
\begin{equation}
  R_{\hM\hN\hP\hQ} = \partial_{\hM}\Gamma_{\hN\hP\hQ} - \partial_{\hN}\Gamma_{\hM\hP\hQ} + \Gamma_{\hM\hP}{}^{\hR} \Gamma_{\hN\hR\hQ} -  \Gamma_{\hN\hP}{}^{\hR} \Gamma_{\hM\hR\hQ} + f_{\hR\hM\hN} \Gamma^{\hR}{}_{\hP\hQ}\,.
\label{}\end{equation}
Then, the generalized curvature scalar is defined by contraction of $S_{\hM\hN\hP\hQ}$ with the projection operators
\begin{equation}
\begin{aligned}
  S := &2 P^{\hM\hN} P^{\hP\hQ} S_{\hM\hP\hN\hQ} 
\\
	=& 
	2 \big( 2 \partial^{m} \Phi^{n}{}_{mn} 
	- \Phi^{m}{}_{m}{}^{p} \Phi^{n}{}_{np} 
	-\tfrac{3}{2} \Phi^{[mnp]} \Phi_{mnp} 
	-\half \Phi^{\brp mn} \Phi_{\brp mn} 
	-\half \Phi^{\bra mn} \Phi_{\bra mn} 
\\&
	 - f_{p mn } \Phi^{pmn} 
	 - f_{\brp mn } \Phi^{\brp mn}
	 - f_{\bra mn } \Phi^{\bra mn}\big) \,.
\end{aligned}\label{generalizedcurvaturescalar}
\end{equation}

\subsection{Nongeometric fluxes and action}
There are several approaches for constructing differential geometry of the gauged DFT \cite{Geissbuhler:2013uka, Berman:2013cli}. Here, we follow the so called semi-covariant formalism \cite{Berman:2013cli} which is well-suited for supersymmetry \footnote{See appendix \ref{app:gauged-DFT} for the concise review of double-vielbein formalism for gauged DFTs.}

To define non-geometric fluxes, we adopt the non-geometric parameterizations of double-vielbein obtained in (\ref{NongeomParametrizationV}) and (\ref{NongeomParametrizationbrV}),  and substitute them to the definition of generalized spin connection (\ref{defspinconn}). Not all of the components of generalized spin connection are involved for defining heterotic DFT action. The relevant components of generalized spin-connection should be invariant under the generalized diffeomorphism for gauged DFT or twisted generalized Lie derivative. They define the non-geometric fluxes 
\begin{equation}
\begin{aligned}
	\Phi^{m}{}_{mn} =& + \tfrac{1}{\sqrt{2}} \big( (e^{-1})^{\mu m}\omega_{\mu mn} -2 \partial_{n} \phi\big)\,,
	\\
	\Phi_{\brm np} =& + \tfrac{1}{\sqrt{2}}\big(\omega_{\brm np} - \half Q_{\rho}{}^{\mu\nu} \bar{e}^{\rho}{}_{\brm} e_{\mu n} e_{\nu p} + Q_{\rho}{}^{\mu\nu} e^{\rho}{}_{n} e_{\mu p} \bre_{\nu \brm} \big)\,,
	\\
	\Phi_{\bra mn} =& -\half \tilde{F}^{\mu\nu}{}_{\bra} e_{\mu m} e_{\nu n}\,,
	\\
	\Phi_{[mnp]} =& + \tfrac{1}{\sqrt{2}}\big(\omega_{[mnp]} - \half Q_{\rho}{}^{\mu\nu} e^{\rho}{}_{[m} e_{|\mu| n} e_{|\nu| p]} \big)\,.
\end{aligned}\label{nongeom_sc}
\end{equation}
In the above expression, the components of generalized spin connection comprise three kinds of fluxes that were introduced in (\ref{nongeom_action}).

Consider now the non-geometric action of heterotic DFT in terms of the non-geometric fluxes. The action is given by the generalized curvature scalar $S$, which is defined in (\ref{generalizedcurvaturescalar}) in terms of the generalized spin-connections:
\begin{equation}
  \cS_{\rm het} = \int e^{-2d} \,2S\,,
\end{equation}
where 
\begin{equation}
\begin{aligned}
  S = 	~& 2 \partial^{m} \Phi^{n}{}_{mn} 
	- \Phi^{m}{}_{m}{}^{p} \Phi^{n}{}_{np} 
	-\tfrac{3}{2} \Phi^{[mnp]} \Phi_{mnp} 
	-\half \Phi^{\brp mn} \Phi_{\brp mn} 
	-\half \Phi^{\bra mn} \Phi_{\bra mn} \,.
\end{aligned}\label{paraspin}
\end{equation}
By substituting (\ref{nongeom_sc}) into this action, one can show that (\ref{paraspin}) is equivalent to the previous non-geometric heterotic action (\ref{nongeom_action}).


\section{Exotic branes}
\label{app:exotic}

A defect brane refers to a codimension-two configuration in type II string theory. 
Denote them by
\begin{align}
 b^{(d,\,c)}_{\mathbf{n}}(n_1\cdots n_b,\,m_1\cdots,m_c,\,\ell_1\cdots \ell_d)\,,
\end{align}
for the configuration wrapped or smeared over the 7-torus \cite{Obers:1998fb,deBoer:2010ud,deBoer:2012ma} and thus has the mass:
\begin{align}
 M_{b^{(d,\,c)}_{\mathbf{n}}}
 = \frac{1}{\gs^{\mathbf{n}}\,\ls}\,\Bigl(\frac{R_{n_1}\cdots R_{n_b}}{\ls^b}\Bigr)\,\Bigl(\frac{R_{m_1}\cdots R_{m_c}}{\ls^c}\Bigr)^2\, \Bigl(\frac{R_{\ell_1}\cdots R_{\ell_d}}{\ls^d}\Bigr)^3\,.
\end{align}
Here, $R_{i}$ is the compactification radius in the $x^i$-direction and $\gs$ is the string coupling constant and $b^c_{\mathbf{n}}\equiv b^{(d=0,\,c)}_{\mathbf{n}}$ and $b_{\mathbf{n}}\equiv b^{(d=0,\,c=0)}_{\mathbf{n}}$. 

In this paper, we consider compactification on shrinking tori. As an example, consider a $5^2_2(34567,89)$-brane \eqref{eq:522-conventional} in the $E_{7(7)}$ EFT. In this case, $x^\sfm$ ($\sfm=4,\dotsc,9$) are compactified on a six-torus and $x^3$ direction is a noncompact direction. 
In this case, the ``$5^2_2(34567,89)$-brane'' is a one-dimensional extended object with the tension,
\begin{align}
 T = \frac{1}{2\pi\gs^2\,\ls^2}\,\Bigl(\frac{R_{4}\cdots R_{9}}{\ls^5}\Bigr)\,\Bigl(\frac{R_{8} R_{9}}{\ls^2}\Bigr)^2 \,. 
\end{align}
We will still call it a point-like $5^2_2(34567,89)$-brane as its mass becomes that of the usual $5^2_2(34567,89)$-brane after further compactifying the $x^3$-direction. 

\begin{table}[h]
\begin{alignat}{2}
 &\underline{\text{type IIA theory}}& &\underline{\text{M-theory}}
\nn\\
 &\left.
\begin{array}{l}
 0_1=\DD0 \\
 \PP(n_1)
\end{array}\right\}\ (M_{\PP(n)}=R_n^{-1})& \quad\leftrightarrow\qquad & 
 \PP \ \left\{
\begin{array}{l}
 \PP(\sfM) \\
 \PP(n_1)
\end{array}\right.\ (M_{\PP(n)}=R_n^{-1}) 
\nn\\
 &\left.
\begin{array}{l}
 1_0(n) =\FF1(n) \\
 2_1(n_1n_2)=\DD2(n_1n_2)
\end{array}\right\}& \quad\leftrightarrow\qquad & 
 \MM2=2_3 \ \left\{
\begin{array}{l}
 \MM2(n\sfM) \\
 \MM2(n_1n_2)
\end{array}\right. 
\nn\\
 &\left.
\begin{array}{l}
 4_1(n_1\cdots n_4)=\DD4\\
 5_2(n_1\cdots n_5)=\NS5 
\end{array}\right\}& \quad\leftrightarrow\qquad & 
 \MM5=5_6 \ \left\{
\begin{array}{l}
 \MM5(n_1\cdots n_4\sfM) \\
 \MM5(n_1\cdots n_5)
\end{array}\right. 
\nn\\
 &\left.
\begin{array}{l}
 6_1 (n_1\cdots n_6)=\DD6\\
 5^1_2 (n_1\cdots n_5,\,n_6)=\KKM \\
 6^1_3(n_1\cdots n_5,\,n_7)
\end{array}\right\}& \quad\leftrightarrow\qquad & 
 \KKM=6^1_9 \ \left\{
\begin{array}{l}
 \KKM(n_1\cdots n_6,\,\sfM) \\
 \KKM(n_1\cdots n_5\sfM,\,n_6) \\
 \KKM(n_1\cdots n_6,\,n_7) 
\end{array}\right. 
\nn\\
 &\left.
\begin{array}{l}
 5^2_2(n_1\cdots n_4) \\
 4^3_3(n_1\cdots n_4,\,m_1m_2m_3) 
\end{array}\right\}& \quad\leftrightarrow\qquad & 
 5^3_{12} \ \left\{
\begin{array}{l}
 5^3(n_1\cdots n_5,\,m_1m_2\sfM) \\
 5^3(n_1\cdots n_4\sfM,\,m_1m_2m_3)
\end{array}\right. 
\nn\\
 &\left.
\begin{array}{l}
 2^5_3(n_1n_2,\,m_1\cdots m_5) \\
 1^6_4(n_1,\,m_1\cdots m_6)
\end{array}\right\}& \quad\leftrightarrow\qquad & 
 2^6_{15}\ \left\{
\begin{array}{l}
 2^6(n_1n_2,\,m_1\cdots m_5\sfM) \\
 2^6(n_1\sfM,\,m_1\cdots m_6)
\end{array}\right. 
\nn\\
 &\left.
\begin{array}{l}
 0^7_3(,\,3\cdots 9) \\
 0^{(1,\,6)}_4(,\,n_1\cdots n_6,\,m_1)
\end{array}\right\}& \quad\leftrightarrow\qquad & 
 0^{(1,\,7)}_{18} \ \left\{
\begin{array}{l}
 0^{(1,\,7)}(,\,3\cdots 9,\,\sfM) \\
 0^{(1,\,7)}(,\,n_1\cdots n_6\sfM,\,m_1)
\end{array}\right. 
\nn
\end{alignat}
\caption{\sl Defect branes in the type IIA theory/$T^7$ and the M-theory/$T^8$.
\label{table:defects-M-IIA}}
\end{table}
A list of defect branes in the type IIA theory compactified on a seven-torus is collected in Table \ref{table:defects-M-IIA}. 
As shown in the table, each defect brane of the type IIA theory can be regarded as a reduction of a defect brane of the M-theory compactified on an eight-torus. 
By using the relation
\begin{align}
 \ls = R_{\sfM}^{-1/2}\,l_{11}^{3/2}\,\qquad \text{and} \qquad 
 \gs = R_{\sfM}^{3/2}\,l_{11}^{-3/2} \,, 
\end{align}
where $R_{\sfM}$ is the radius in the M-theory direction and $l_{11}$ is the Planck length in eleven dimensions, a $b^{(d,\,c)}_{\mathbf{n}}$-brane in the type IIA theory can be identified with a defect $b^{(d,\,c)}_{\tilde{\mathbf{n}}}$-brane in the M-theory with the mass,
\begin{align}
 M_{b^{(d,\,c)}_{\tilde{\mathbf{n}}}}&= \frac{1}{R_{\sfM}}\,\Bigl(\frac{R_{\sfM}}{l_{11}}\Bigr)^{\tilde{\mathbf{n}}}\,\Bigl(\frac{R_{n_1}\cdots R_{n_b}}{R_{\sfM}^b}\Bigr)\,\Bigl(\frac{R_{m_1}\cdots R_{m_c}}{R_{\sfM}^c}\Bigr)^2\, \Bigl(\frac{R_{\ell_1}\cdots R_{\ell_d}}{R_{\sfM}^d}\Bigr)^3
\nn\\
 &= \frac{\bigl(R_{n_1}\cdots R_{n_b}\bigr)\,\bigl(R_{m_1}\cdots R_{m_c}\bigr)^2\,\bigl(R_{\ell_1}\cdots R_{\ell_d}\bigr)^3}{l_{11}^{\tilde{\mathbf{n}}}} \,,
\\
 \tilde{\mathbf{n}}&\equiv 3\,\Bigl(\frac{b+2c+3d-\mathbf{n}+1}{2}\Bigr) \,.
\end{align}
Here, the indices $n_i,\,m_i,\,\ell_i$ run over $3,\dotsc,9,\sfM$, where $\sfM$ represents the M-theory direction. We also used the non-trivial identity, $\tilde{\mathbf{n}}=b+2c+3d+1$, satisfied by all M-theory branes. 
Note that the subscript $\tilde{\mathbf{n}}$ of $b^{(d,\,c)}_{\tilde{\mathbf{n}}}$ is usually suppressed.

\end{document}